%% file: hugs_proc.tex
\documentstyle[mprocl,epsfig,pstricks,amsmath,rotating,mytools,latexsym,fancybox,pst-node,pst-text,pst-3d]{article}

\input{psfig}
\def\bea{\begin{eqnarray}}
\def\eea{\end{eqnarray}}

\begin{document}
\newcommand{\nonett}[9]
{
\setlength{\unitlength}{0.8mm}
\begin{picture}(150.00,90.00)
\put(10.00,45.00){\vector(1,0){70.00}}
\put(45.00,10.00){\vector(0,1){70.00}}
\put(110.00,45.00){\vector(1,0){30.00}}
\put(125.00,30.00){\vector(0,1){30.00}}
\put(82.50,42.50){\makebox(5.00,5.00){$I_3$}}
\put(142.50,42.50){\makebox(5.00,5.00){$I_3$}}
\put(127.50,60.00){\makebox(5.00,5.00){\quad\ S\quad\ Singlet }}
\put(35.50,78.00){\makebox(5.00,5.00){\quad\ S }}
\put(45.50,78.00){\makebox(5.00,5.00){\quad\ \quad\ Octet }}
\put(45.00,45.00){\circle*{2.00}}
\put(70.00,45.00){\circle*{2.00}}
\put(20.00,45.00){\circle*{2.00}}
\put(32.50,70.00){\circle*{2.00}}
\put(57.50,70.00){\circle*{2.00}}
\put(32.50,20.00){\circle*{2.00}}
\put(57.50,20.00){\circle*{2.00}}
\put(125.00,45.00){\circle*{2.00}}
\put(45.00,45.00){\circle{0.00}}
\put(45.00,45.00){\circle{5.00}}
\put(32.50,70.00){\line(1,0){25.00}}
\put(57.50,70.00){\line(1,-2){12.50}}
\put(70.00,45.00){\line(-1,-2){12.50}}
\put(57.50,20.00){\line(-1,0){25.00}}
\put(32.50,20.00){\line(-1,2){12.50}}
\put(20.00,45.00){\line(1,2){12.50}}
\put(60.00,70.00){\makebox(15.00,5.00)[l]{#2}}
\put(15.00,70.00){\makebox(15.00,5.00)[r]{#1}}
\put(15.00,15.00){\makebox(15.00,5.00)[r]{#3}}
\put(6.75,37.50){\makebox(13.25,5.00)[r]{#5}}
\put(60.00,15.00){\makebox(15.00,5.00)[l]{#4}}
\put(70.00,37.50){\makebox(13.75,5.00)[l]{#6}}
\put(47.50,37.50){\makebox(12.50,5.00)[l]{#7}}
\put(47.50,47.50){\makebox(12.50,5.00)[l]{#8}}
\put(127.50,47.50){\makebox(12.50,5.00)[l]{#9}}
\end{picture}
}
\def\CI{\cos{\theta_T}}
\def\SI{\sin{\theta_T}}
\def\CV{\cos{\theta_V}}
\def\SV{\sin{\theta_V}}
\def\CF{\cos{\theta_{PS}}}
\def\SF{\sin{\theta_{PS}}}
\def\CFT{\cos^2{\theta_{PS}}}
\def\SFT{\sin^2{\theta_{PS}}}
\def\Eskip{ }
\def\sfrac#1#2{\mbox{$\frac{#1}{\sqrt{#2}}$}}%
\def\ssfrac#1#2{\mbox{$\frac{\sqrt{#1}}{\sqrt{#2}}$}}%
\def\cf{\cos\theta_\eta}%
\def\sf{\sin\theta_\eta}%
\def\ci{\cos\theta_f}%
\def\si{\sin\theta_f}%
\def\ck{\cos\theta_K}%
\def\sk{\sin\theta_K}%
\def\xybar#1#2{\mbox{$\mbox{#1}\bar{\mbox{#2}}$}}%
\def\xbary#1#2{\mbox{$\bar{\mbox{#1}}\mbox{#2}$}}%
\def\xxbar#1{\xybar{#1}{#1}}%
\def\xbarx#1{\xbary{#1}{#1}}%
\def\qqbar{\xxbar{q}}%
\def\nnbar{\xxbar{n}}%
\def\uubar{\xxbar{u}}%
\def\ddbar{\xxbar{d}}%
\def\ssbar{\xxbar{s}}%
\def\ccbar{\xxbar{c}}%
\def\bbbar{\xxbar{b}}%
\def\ttbar{\xxbar{t}}%
\def\ppbar{\xxbar{p}}%
\def\pbarp{\xbarx{p}}%
\def\pbard{\xbary{p}{d}}%
\def\pbarn{\xbary{p}{n}}%
\def\pbarN{\xbary{p}{N}}%
\def\nbarn{\xbarx{n}}%
\def\nbarp{\xbary{p}}%
\def\nbard{\xbary{d}}%
\def\nbarN{\xbary{N}}%
\newcommand{\ra}        {\mbox{$\rightarrow$}}
\newcommand{\bc}        {\begin{center}}
\newcommand{\ec}        {\end{center}}
\newcommand{\bi}        {\begin{enumerate}}
\newcommand{\ei}        {\end{enumerate}}
\newcommand{\be}        {\begin{equation}}
\newcommand{\ee}        {\end{equation}}
\newcommand{\gam}       {\mbox{$\gamma$}}
\newcommand{\p}         {\mbox{$\pi$}}
\newcommand{\piz}       {\mbox{$\pi^0$}}
\newcommand{\pip}       {\mbox{$\pi^+$}}
\newcommand{\kp}        {\mbox{$\rm K^+$}}
\newcommand{\km}        {\mbox{$\rm K^-$}}
\newcommand{\kn}        {\mbox{$\rm K^0$}}
\newcommand{\knb}       {\mbox{$\rm\bar K^0$}}
\newcommand{\pim}       {\mbox{$\pi^-$}}
\newcommand{\pipm}      {\mbox{$\pi^{\pm}$}}
\newcommand{\pimp}      {\mbox{$\pi^{\mp}$}}
\newcommand{\etg}       {\mbox{$\eta$}}
\newcommand{\etp}       {\mbox{$\eta^{\prime}$}}
\newcommand{\omg}       {\mbox{$\omega$}}
\newcommand{\rh}        {\mbox{$\rho$}}
\newcommand{\rhp}        {\mbox{$\rho^+$}}
\newcommand{\rhm}        {\mbox{$\rho^-$}}
\newcommand{\rhz}        {\mbox{$\rho^0$}}
\newcommand{\rhpm}       {\mbox{$\rho^{\pm}$}}
\newcommand{\rhmp}       {\mbox{$\rho^{\mp}$}}
\newcommand{\pbp}       {\mbox{$\bar{p}p$}}
\newcommand{\ssb}       {\mbox{$\bar{s}s$}}
\newcommand{\NNb}       {\mbox{$\overline{N}N$}}
\newcommand{\kkb}       {\mbox{$\rm\overline{K}K$}}
\newcommand{\pbar}       {\mbox{$\bar{p}$}}

%
%
%
%
\title{\large\bf GLUEBALLS, HYBRIDS, PENTAQUARKS:\\
\vskip 2mm
\large Introduction to Hadron Spectroscopy \\
and Review of Selected Topics}

\author{Eberhard Klempt}

\address{Helmholtz--Institut f\"ur Strahlen- und Kernphysik\\
der Universit\"at Bonn\\
Nu\ss allee 14 -16, 53115 Bonn, Germany\\
E-mail: klempt@iskp.uni-bonn.de}

\maketitle

\abstracts{
Hadron spectroscopy has received revitalised
interest due to the discovery of states with unexpected properties.
The BABAR collaboration found a  $\rm D_{sJ}(2317)$ (likely scalar) meson,
accompanied by a second state, the $\rm D_{sJ}(2463)$ with preferred
spin $J=1$, discovered at CLEO.  
Both are found at an unexpectedly low mass and are narrow. 
A further narrow resonance, discovered 
at BELLE in its decay into
$\pi\pi$J/$\psi$, might be a $\rm DD^*$ molecule or a
$c\bar c$ meson in which the
color field, concentrated in a flux tube, is excited. A $q\bar q$
state with excited gluon field is called hybrid, such 
excitations are expected from QCD. The mass of the state 
at the $\rm DD^*$ threshold underlines
the importance of meson--meson interactions or four--quark dynamics
at the opening of new thresholds. BES reports a signal in
radiative J/$\psi$ decays into a proton and an antiproton which
has the properties as predicted for $N\bar N$ quasi-nuclear
bound states.
And last not least the
$\Theta^+(1540)$, seen in several experiments, shows that the
'naive' quark model needs to be extended. 
There is also considerable progress in understanding the dynamics of
quarks in more conventional situations even though the view presented 
here is not
uncontested. In meson and in baryon spectroscopy, evidence is emerging
that one--gluon--exchange does not provide the appropiate means to
understand low--energy QCD; instan\-ton--induced interactions yield
much more insight. In particular the rich spectrum of baryon resonances
is very well suited to test dynamical quark models using constituent
quarks, a confinement potential plus some residual interactions.  
The baryon spectrum favors definitely 
instanton--induced interactions over long--range one--gluon exchange.
A still controversial issue is the question if
glueballs and hybrids exist. There is the possibility that
two $q\bar q$ scalar states and a glueball form three observed
resonances by mixing. However, there is also rather conclusive evidence
against this interpretation. Mesons with exotic quantum numbers
have been reported but there is no reason why they should be hybrids:
a four--quark interpretation is enforced for the $\pi_1(1400)$ and 
not ruled out in the other cases. 
\phantom{This report is based on a lecture series which had the intent
 in the other cases. }
This report is based on a lecture series which had the intent to
indroduce young scientists into hadron spectroscopy. The attempt
is made to transmit basic ideas standing behind some models, without
any formulae. Of course, these models require (and deserve) a much
deeper study. However, it may be useful to explain in a
simple language some of the ideas behind the formalisms.  Often, 
a personal view is presented which
is not shared by many experts working in the field. In the last
section, the attempt is made to combine the findings into a picture of
hadronic interactions and to show some of the consequences the picture
entails and to suggest further experimental and theoretical work.
\phantom{This report is based on a lecture series which had the intent
and so in the other cases. }
{\bf\large This file contains the abstract only. 
The full text with 27 Tables and 81 figures
including a tar file is available at 
www.uni-bonn.de/ek/   hugs\_proc.tar \qquad }
}
\vfill
{
\bc
18th Annual Hampton University Graduate Studies\\
Jefferson Lab, Newport News, Virginia\\
         June 2-20, 2003\\
\ec
}    
\cleardoublepage
\input{hugs_acknowledgement.tex}

\cleardoublepage
\tableofcontents
\cleardoublepage
\input{Chapter1_proc.tex}

\cleardoublepage
\input{Chapter2_proc.tex}

\cleardoublepage
\input{Chapter3_proc.tex}

\cleardoublepage
\input{Chapter4_proc.tex}

\cleardoublepage
\input{Chapter5_proc.tex}

\cleardoublepage
\input{Chapter6_proc.tex}

\cleardoublepage

\input{hugs_bib.tex}
\end{document}

%% file: hugs_acknowledgement.tex
\section*{Acknowledgments}
First of all I would like to thank the organizers for this
nice workshop. I enjoyed the friendly atmosphere and the 
scientific environment provided by 
the Jefferson Laboratory, and I enjoyed giving 
lectures to an interested audiences of young scientists.
\par
It is a pleasure for me to acknowledge the numerous discussions
with friends and colleagues from various laboratories and
universities. In naming some of them I will undoubtedly forget
some very important conversations; nevertheless I would like
to mention fruitful discussions with R. Alkofer, A.V. Anisovich, 
V.V. Anisovich, C. Amsler, Chr. Batty, F. Bradamante, D. Bugg, 
S. Capstick, S.U. Chung, F.E. Close, D. Diakonov, W. Dunwoodie, 
St. Dytman, A. Dzierba,
W. D\"unnweber, P. Eugenio, A. F\"assler, M. F\"assler, 
H. Fritzsch, U. Gastaldi,
K. Goeke, D. Herzog, N. Isgur, K. K\"onigsmann, 
F. Klein, S. Krewald, H. Koch, G.D. Lafferty, R. Landua, D.B. Lichtenberg,
M. Locher, R.S. Longacre, V.E. Markushin, 
A. Martin, U.-G. Mei\ss ner,
C.A. Meyer, V. Metag, B. Metsch, 
L. Montanet, W. Ochs, M. Ostrick, Ph. Page, M. Pennington, 
K. Peters, H. Petry, M.V. Polyakov, 
J.M. Richard, A. Sarantsev, B. Schoch, E.S. Swanson,
W. Schwille, A.P. Szczepaniak, 
J. Speth, L. Tiator, P. Tru\"ol, 
U. Thoma, Chr. Weinheimer, W. Weise, U. Wiedner, 
H. Willutzki, A. Zaitsev, and C. Zupancic. 
\par
The work described in this report is partly based on experiments
I had to pleasure to participate in. Over the time I have had 
the privilege to work with a large number of PhD students. The results
were certainly not achieved without their unflagged enthusiasm
for physics. I would like to mention 
O. Bartholomy,
J. Brose,
V. Crede,
K.D. Duch,
A. Ehmanns,
I. Fabry,
M. Fuchs,
G. G\"artner,
J. Junkersfeld,
J. Haas, 
R. Hackmann,
M. Heel,
Chr. Heimann,
M. Herz,
G. Hilkert, 
I. Horn,
B. Kalteyer,
F. Kayser,
R. Landua,
J. Link,
J. Lotz,
M. Matveev, 
K. Neubecker,
H. v.Pee,
K. Peters,
B. Pick,
W. Povkov, 
J. Reifenr\"other,
G. Reifenr\"other,
J. Reinnarth, 
St. Resag, 
E. Sch\"afer,
C. Schmidt,
R. Schulze,
R. Schneider,
O. Schreiber,
S. Spanier,
Chr. Stra\ss burger,
J.S. Suh,
T. Sczcepanek, 
U. Thoma,
F. Walter,
K. Wittmack, 
H. Wolf,
R.W. Wodrich,
M. Ziegler.
\par
Very special thanks go to my colleague Hartmut Kalinowsky with whom
I have worked with jointly for more than 30 years. Anyone 
familiar with one of the experiments I was working on knows
his contributions are invaluable to our common effort. To him my
very personal thanks.

%% file: Chapter1_proc.tex
\section{\label{section1}Getting started}
\subsection{\label{section1.1}Historical remarks}
\subsubsection{Nuclear interactions}
In the beginning of the 1930'ties, three particles
were known from which all matter is
built: protons and neutrons form the nuclei and their charges are
neutralized by very light electrons. The binding forces between
electrons and nuclei were reasonably well understood as
electromagnetic interaction but nobody knew why protons and
neutrons stick together forming nuclei. Protons and neutrons have
similar masses; Heisenberg suggested they be consider as one
particle called nucleon. He proposed a new quantum
number, isospin $I$, with $I=1/2,I_3=1/2$ for protons and
$I=1/2,I_3=-1/2$ for neutrons. Pauli had suggested that a massless
weakly interacting neutrino ($\nu$) should exist, but it
was considered to be undetectable.
\par
In 1935 Hideki Yukawa published an article~\cite{Yukawa:1935xg}
in which he proposed a 
field theory of nuclear forces to explain their short range and
predicted the existence of a meson, called $\pi$-meson or pion.
While the Coulomb potential is given by 
\be 
V_{QED} = \frac{e}{4\pi\epsilon_0 r} 
\ee 
and originates from the exchange of photons
with zero mass, Yukawa proposed that strong interactions may be
described by the exchange of a particle having a mass of about 100\,MeV
leading to a potential with a range 1/$m_{\pi}$: 
\be 
V_{strong} =
\frac{g}{4\pi r}\cdot e^{-m_{\pi}r}. 
\ee 
The pion was discovered
by C. Powell in 1947~\cite{Lattes:mx}, and two years later 
Yukawa received the
Nobel prize. We now know that there are 3 pions,
$\pi^+$, $\pi^0$, and $\pi^-$. This is an isospin triplet with
$I=1$ and the third component being $I_3=1,0$, or $-1$.
\par
\subsubsection{Resonances in strong interactions}
\begin{figure}[h!]
\vspace*{-15mm}
\bc
\epsfig{file=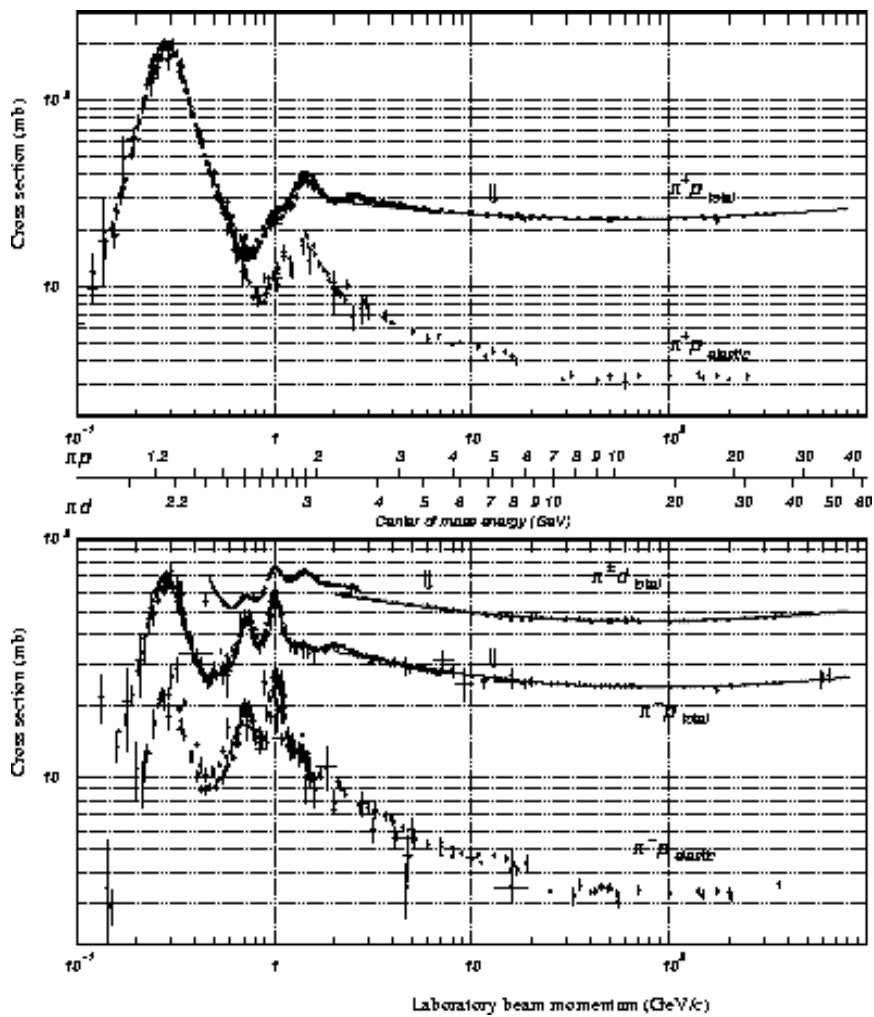,width=0.9\textwidth}
\ec
\vspace*{-2mm}
\caption{Cross sections for $\pi N$ scattering 
(from~\protect\cite{Hagiwara:fs})}.
\label{piN}
\end{figure}

In 1952 E. Fermi and collaborators measured
the cross section for $\pi^+ p\to\pi^+ p$
and found it steeply rising~\cite{Anderson:nw}. Modern data 
(extracted from~\cite{Hagiwara:fs}) on
$\pi N\to\pi N$ scattering are shown in figure~\ref{piN}.
Cross sections are given for elastic and
charge exchange scattering, with maximum cross sections
\bc
\begin{tabular}{cc}
$ \sigma_{tot,\pi^+ p} = 210$\,mb & $ \sigma_{tot,\pi^- p} = 70$\,mb\\
$ \sigma_{el,\pi^+ p} = 210$\,mb  & $ \sigma_{el,\pi^- p} = 23$\,mb.
\end{tabular}
\ec
The largest cross section occurs at an invariant mass of 1230\,MeV.
It is a resonance, called $\Delta(1232)$. It can be observed in four
different charge states $\Delta^{++}(1232)$,
$\Delta^{+}(1232)$, $\Delta^{0}(1232)$, $\Delta^{-}(1232)$. Like in
the case of the nucleon these states are put into an isospin
multiplet with $I=3/2$, and $I_3=3/2 , I=1/2, I=-1/2$ and $I=-3/2$,
respectively.
\par
The $\Delta(1232)$ has a  width of 150\,MeV and thus a
lifetime $\tau = \hbar /\Gamma\sim 0.45\ 10^{-23}$\,s. This
is really a short time;
within $\tau$ a particle travels about 1\,fm at the speed
of light.
\vspace*{-3mm}
\par
\subsubsection{Clebsch Gordan coefficients}
The peak cross sections for $\Delta(1232)$ production in $\pi^+$p,
$\pi^-$p elastic scattering and for $\pi^-$p charge exchange
differ substantially. Moreover, there are further peaks, some of which
show up only in $\pi^-$p scattering and not in $\pi^+$p. Here we
can see the utility of the isospin concept. Pions have isospin 1,
nucleons isospin 1/2. In strong interactions isospin is conserved:
we can form two scattering amplitudes, one with isospin 3/2
(leading to the $\Delta(1232)$ and to $\Delta^*$ resonances) and one
with isospin 1/2 which contains resonance excitations of the
nucleon. $\pi^+$p is always isospin 3/2 (since $I_3=3/2$);
$\pi^-$p can either have $I=1/2$ or $I=3/2$. The ratio is given by
Clebsch--Gordan coefficients, the same as used to add angular
momenta.
\par
To understand the height of the various cross sections, these
are compared to squared Clebsch--Gordan coefficients. It is a
very useful exercise to look up these coefficients in a table and
to compare the numbers with those listed explicitly below. The
cross sections for $\Delta(1232)$ production scale as predicted
from (squared) Clebsch--Gordan coefficients\,!
\par
{\small
\bc
\hspace*{-3mm}
\begin{tabular}{lccccr}
$\sigma_{tot,\pi^+ p}$ &=& $\sigma_{\pi^+ p\to\pi^+ p}$ &=&
210\,mb & $\propto 1\times 1$ \\
\multicolumn{6}{l}{$\rm CG_{(I=1/2,I_3=1/2)+(I=1,I_3=1)\to (I=3/2,I_3=3/2)}\times
CG_{(I=3/2,I_3=3/2)\to (I=1/2,I_3=1/2)+(I=1,I_3=1)}$} \\
$\sigma_{el,\pi^+ p}$ &=& $\sigma_{\pi^+ p\to\pi^+ p}$ &=&
210\,mb & $\propto 1\times 1$ \\
\multicolumn{6}{l}{$\rm CG_{(I=1/2,I_3=1/2)+(I=1,I_3=1)\to (I=3/2,I_3=3/2)}\times
CG_{(I=3/2,I_3=3/2)\to (I=1/2,I_3=1/2)+(I=1,I_3=1)}$}\\
$\sigma_{tot,\pi^- p}$ &=& $\sigma_{\pi^- p\to\pi^- p} +
\sigma_{\pi^- p\to\pi^0 n}$ &=& 70\,mb &$\propto 1/3\times 2/3 + 1/3\times 1/3$\\
\multicolumn{6}{l}{$\rm CG_{(I=1/2,I_3=1/2)+(I=1,I_3=-1)\to (I=3/2,I_3=-1/2)} +
CG_{(I=1/2,I_3=-1/2) + (I=1,I_3=0)+(I=3/2,I_3=-1/2)}$}\\
$\sigma_{el,\pi^- p}$ &=& $\sigma_{\pi^- p\to\pi^- p}$  &=&  23\,mb &
$\propto 1/3 \times 1/3$  \\
\multicolumn{6}{l}{$\rm CG_{(I=1/2,I_3=1/2)+(I=1,I_3=-1)\to (I=3/2,I_3=-1/2)}\times
CG_{(I=3/2,I_3=-1/2)\to (I=1/2,I_3=1/2)+(I=1,I_3=-1)}$}\\
\end{tabular}
\ec
}

\subsubsection{The particle zoo}
In the same year 1949, Rochester and Butler~\cite{Rochester:mi}
found reactions of the
type ${\rm\pi^- p \ra\ \Lambda K_s^0}$ where both particles had
long lifetimes, on the order of $10^{-10}\,s$. This seems to be a
short time but when we consider the $\rm K_s^0$ as a composite particle
like positronium (the $\rm e^+e^-$ analogue of hydrogen atoms)
consisting of a quark and an antiquark rotating with the velocity
of light and a radius of 0.5\,fm, then the  $\rm K_s^0$ lives for
$\sim 10^{13}$ revolutions. A fantastically long life time\,! The
earth has only encircled the sun for $5\cdot 10^{9}$ revolutions.
The surprise was that these particles are produced via strong
interactions and in pairs (in this case a $\rm K_s^0$ and a
$\Lambda$). This phenomenon was called associated production by
Pais in 1952~\cite{Pais:mi}. 
Both particles decay by weak interaction. To explain
this strange behavior, production by strong interactions and decay
via weak interactions, a new additive quantum number 
was introduced called
strangeness $S$. Strangeness is produced as $S$ and anti-$S$ (or
$\bar S$) pairs and conserved in strong interactions. The
$\Lambda$, e.g., carries strangeness $S=-1$ and does not decay via strong
interactions. Instead, as all strange particles, it decays 
via weak interactions,
$\rm\Lambda\to N\pi$, with a long life time.

\par
The first idea was to consider the proton, neutron, and the $\Lambda$
as building blocks in nature but more and more strongly
interacting particles were discovered and the notion 'the particle
zoo' was created. In particular there are 3 pions having a mass of
135\,MeV, the $\pi^+ , \pi^-$ and the $ \pi^0$; the $\eta (547)$ and
$ \eta^{\prime} (958)$; and four Kaons  K$^+$, K$^-$, K$^0_s$,
K$^0_l$ having a mass close to 500\,MeV. All these particles have
spin 0. The three $\rho^{+-0} (770)$, the $\omega (782)$, the
$\Phi(1020)$ and four  $\rm K^*(892)$ have spin 1 and there are
particles with spin 2, i.e. the $\rm  a_2(1320)$ and  $\rm  f_2(1270)$,
among others. These particles have, like photons, integer spins;
they are bosons obeying Bose symmetry: the wave function of,
let us say $n_{\pi^0}$ neutral pions, must be
symmetric with respect to the exchange of any pair of $\pi^0$'s. All
these particles can be created or destroyed as the number of bosons
is not a conserved quantum number.
\par
Protons, neutrons and $\Lambda$'s are different: they have spin 1/2
and obey Fermi statistics. A nuclear wave function must be
antisymmetric when two protons are exchanged. These particles are
called baryons. The number of baryons is a conserved
quantity. More baryons were discovered, $\Sigma^{+}$,
$\Sigma^{0}$, $\Sigma^{-}$
with three charge states, pair $\Xi^{0}$, $\Xi^{-}$, the
$\Delta (1232)$ and many more.
\par
Of course not all of these
particles can be 'fundamental'. In 1964, Gell-Mann suggested
the quark model~\cite{Gell-Mann:nj}. 
He postulated the existence of three quarks
called up $(u)$, down $(d)$, and strange $(s)$. All baryons can then 
be classified as bound systems of three quarks, and mesons as bound states
of one quark and one antiquark. With the quarks $u, d, s$ we expect
families (with identical spin and parities) of 9 mesons. This is
indeed the case. Baryons consist of 3 quarks, $qqq$.
We might expect families of 27 baryons but this is wrong;
the Pauli principle reduces the number of states.
\subsubsection{Color}
In the quark model, the $\Delta^{++}(1232)$ consists of three $u$ quarks
with parallel spin, all in an $S$--wave. Quarks are Fermions and
the Pauli principle requires the wave function to be antisymmetric
with respect to the exchange of two identical quarks. Hence the three quarks
cannot be identical. A possible solution is the introduction of
a further quark property, the so--called color. 
There are three colors, red, green, and blue. A baryon then can be
written as the determinant of three 
lines $q_1$, $q_2$, $q_3$, and three rows, flavor, spin, color.
The determinant has the desired antisymmetry. When color was
introduced~\cite{Greenberg:pe} it was to ensure antisymmetry of 
baryon wave functions. It was only with the advent of quantum
chromodynamics that color became a source of gluon fields and 
resumed a decisive dynamical role. Colored quarks and
gluons interact via
exchange of gluons in the same way as charges interact via exchange of
photons. 
\par
\subsubsection{Units}
We use $\hbar = c = 1$ in these lecture notes.
The fine structure constant is defined as
\be
\alpha = \frac{e^2}{4\pi\epsilon_0\hbar c} = 1/137.036.
\ee
The factor $4\pi\epsilon_0\hbar c$ depends on the units chosen;
but one does not need to remember units. If you
take a formula from a textbook, look up the Coulomb
potential and replace it with $eV_{Coulomb} = \alpha /r$.
If there is a factor $4\pi\epsilon_0$ or $\epsilon_0$
in a formula, replace it by 1 whenever it occurs. (But remember
wether  $4\pi\epsilon_0$ or $\epsilon_0$ is equal to 1\,!)
If there is electron charge $e$ the interaction will be $e^2$,
which is $\alpha$.
\par
A second important number to remember is
\be
\hbar c = 197.327 (\sim 200) \ {\mathrm MeV\,fm}.
\ee
If your final number does not havethe units you want, multiply
with $\hbar c$ and with $c$ until you get the
right units. It sounds like a miracle, but this
technique works.

\subsection{\label{section1.2}Mesons and their quantum numbers}
Quarks have spin $S=1/2$ and baryon number $B=1/3$, antiquarks
$S=1/2$ and $B=-1/3$. Quarks and antiquarks
couple to $B=0$ and spin $S =1$ or $S = 0$. Conventional
mesons can be described as $q\bar q$ systems
and thus have the following properties.
\par
The parity due to angular momentum is $P = (-1)^{L}$. Quarks
have intrinsic parity  which we define to be $P = 1$; antiquarks 
have opposite parity (this follows from the Dirac
equation). The parity of a $q\bar q$ meson is hence given by
\be
P = (-1)^{L+1}.
\ee
Neutral mesons with no strangeness are eigenstates of the charge
conjugation operator, sometimes called $C$--parity,
\be
C = (-1)^{L+S}
\ee
which is only defined for
neutral mesons. A proton and neutron form a isospin doublet
with $I=1/2$, $I_3=\pm 1/2$ for (p,n). The three pions have
isospin $I=1$. From these quantum numbers we can define the
$G$-parity
\be
G = (-1)^{L+1+I}
\ee
which is approximately conserved in strong
interactions. However, chiral symmetry is not an exact symmetry,
and $G$--parity can be violated like in $\eta\to 3\pi$ or
in $\omega\to 2\pi$ decays.
\par
We can use these quantum numbers to characterize a
meson by its $J^{PC}$ values. These are measured quantities. We
may also borrow the spectroscopic notation
$^{2s+1}L_{J}$ from atomic physics. 
Here, $s$ is the total spin of the two quarks, $L$
their relative orbital angular momentum and $J$ the total angular
momentum.
\par
The mesons with lowest mass have $L=0$ and the two quark spins
have opposite directions: $\vec s_1+\vec s_2=0$. This leads to
quantum numbers $J^{PC}=0^{-+}$. These mesons form the  nonet
of pseudoscalar mesons
\vspace*{-5mm}
\begin{center}
\nonett{${\rm K^0 \ {  (d\bar s)} }$}{${\rm K^+ \ {  (u\bar s)}}$}
{${\rm K^- \ {  (s\bar u)}}$}{$\overline{{\rm K^0 \ {  (s\bar d)}}}$}
{$\pi^- \ {  (d\bar u)} $}{$\pi^+ \ {  (-u\bar d)}$}{$\pi^0$}
{$\eta_8$}{$\eta_1$}
\end{center}
\vspace*{-5mm}
\noindent
with $\pi^0 = \frac{1}{\sqrt 2}\left|u\bar u - d\bar d\right>$, 
$\eta_8 = \frac{1}{\sqrt 6}\left|u\bar u + d\bar d - 2s\bar s\right>$, 
and $\eta_1 =  \frac{1}{\sqrt 3}\left|u\bar u + d\bar d + s\bar s\right>$.
\subsubsection{Mixing angles}
The two states $\eta_8$ and $\eta_1$ have 
identical quantum numbers, hence they can mix; the mixing angle
is denoted by $\Theta_{ps}$:
\bc
\renewcommand{\arraystretch}{1.2}
\vskip 2mm
\begin{tabular}{ccccc}
$\eta$         & = & $\cos\Theta_{ps}\,\eta_8$& - 
& $\sin\Theta_{ps}\,\eta_1$ \\
$\eta^{\prime}$& = 
& $\sin\Theta_{ps}\,\eta_8$& + 
& $\cos\Theta_{ps}\,\eta_1$
\end{tabular}
\vskip 2mm
\renewcommand{\arraystretch}{1.0}
\ec
Apart from these well established $\bar qq$ mesons, 
other kinds of mesons could also exist: glueballs, mesons
with no (constituent) $\bar qq$ content, hybrids in
which the binding fields between $\bar q$ and $q$
are excited, or multiquark states like  $\bar q\bar qqq$
or meson--meson molecular--type states.
As we will see, these are predicted by theory.
We may then cautiously extend the mixing scheme to
include a possible glueball content

\renewcommand{\arraystretch}{1.2}
\bc
\vskip 2mm
\begin{tabular}{ccccccc}
$\eta$ &= & ${X_{\eta}}\cdot
\frac{1}{\sqrt 2}\left|\,u\bar u + d\bar d\,\right>$ &+
& ${Y_{\eta}}\cdot\left|\,s\bar s\,\right>$ &+
& ${Z_{\eta}}\cdot\left|\,{\rm glue}\,\right>$ \\
$\eta^{\prime}$ &= &${X_{\eta^{\prime}}}\cdot
\frac{1}{\sqrt 2}\left|\,u\bar u + d\bar d\,\right>$ &+
& ${Y_{\eta^{\prime}}}\cdot\left|\,s\bar s\,\right>$
&+ & ${Z_{\eta^{\prime}}}\cdot\left|\,{\rm glue}\,\right>$ \\
&&{light quark} &&{strange quark}
&&{inert}
\end{tabular}
\vskip 2mm
\ec
\renewcommand{\arraystretch}{1.0}

\def\nnb{n\bar n}
\def\ssb{s\bar s}

Experimentally it turns out that ${Z_{\eta}=Z_{\eta^{\prime}}\sim
0}$. The pseudoscalar glueball, if it exists, does not mix
strongly with the ground--state pseudoscalar mesons~\cite{Benayoun:2003we}.
\subsubsection{Mixing angles, examples}
We now write down the wave functions for a few special mixing angles.
We define $\nnb = 1/\sqrt 2(u\bar u +d\bar d)$. 

\begin{tabular}{lll}
$\Theta_{PS} =  0^{\circ} $&$  
\ket{\eta } = \sqrt\frac{1}{3}\,\ket{\nnb - \sqrt{2}\ssb} $&$
\ket{\eta^{\prime}} = \sqrt\frac{2}{3}\,\ket{\nnb + \frac{1}{\sqrt 2}\ssb} $\\
$\Theta_{PS} = -11.1^{\circ} $&$ 
\ket{\eta } = \frac{1}{\sqrt 2}\,\ket{\nnb  - \ssb}$&$ 
\ket{\eta^{\prime}} = \frac{1}{\sqrt 2}\,\ket{\nnb  + \ssb}  $\\
$\Theta_{PS} = -19.3^{\circ} $&$ 
\ket{\eta } = \sqrt\frac{2}{3}\,\ket{\nnb - \frac{1}{\sqrt 2}\ssb} $&$ 
\ket{\eta^{\prime}} = \sqrt\frac{1}{3}\,\ket{\nnb + \sqrt{2}\ssb} $\\
$\Theta_{PS} = \Theta_{ideal} = 35.3^{\circ} $&$  
\ket{\eta }       = \ket{s\bar s}$&$
\ket{\eta^{\prime}} = \ket{n\bar n} $\\
\end{tabular}

For $\Theta_{PS} = 0$ we retain the octet and singlet wave functions.
$\Theta_{PS} = -11.1^{\circ}$ is used often in older literature;
with this mixing angle, $\eta$ and $\eta^{\prime}$ have the
same strangeness content. For
$\Theta_{PS} = -19.3^{\circ}$, the wave function is similar to the
octet/singlet wave functions, except for the sign. The $\bar ss$
component in the $\eta^{\prime}$ is now twice as strong as in the $\eta$.
Finally $\Theta_{PS} = 35.3^{\circ}$ gives a decoupling of the
$\bar ss$ from $(u\bar u + d\bar d)$. This is the {\it ideal} mixing
angle. For most meson nonets the mixing angle is approximately ideal. 
Exceptions are the nonet of pseudoscalar and scalar mesons.
\par
For $L=0$ and $\vec s_1+\vec s_2=\vec S$ and $ S=1$ we get the
nonet of vector mesons with $J^{PC} = 1^{--}$. Additionally
there can be orbital angular momentum between the quark and antiquark,
and $\vec L$ and $\vec S$ can combine to $J=2$ thus forming
the nonet of tensor mesons with $J^{PC} = 2^{++}$.
\par
The vector and tensor mixing angles, $\Theta_{V}$ and
$\Theta_{T}$, are both close to $35.3^{\circ}$. Hence we have
\vspace*{-2mm}
\bc
\begin{tabular}{cccccc}
$ \omega$&=&$  \frac{1}{\sqrt 2}\left(u\bar u + d\bar d\right)$&
$ \Phi  $&=&$  s\bar s$\\
$ f_2(1270)$&=&$  \frac{1}{\sqrt 2}\left(u\bar u + d\bar d\right)$&
$ f_2(1525)$&=&$  s\bar s$\\
\end{tabular}
\ec

Quarks and antiquarks can have any orbital angular momentum.
Combined with the spin, there is a large variety of mesons which
can be formed. Not all of them are known experimentally. Their
masses provide constraints for the forces which tie together
quarks and antiquarks.
\subsubsection{The Gell-Mann-Okubo mass formula}
We now assume that mesons have a common mass $M_0$ plus
the mass due to its two flavor-dependent quark masses 
$M_{q_1}$ and  $M_{q_2}$.
Then the masses can be written as

\vspace*{-5mm}
\begin{minipage}[t]{0.48\textwidth}
\begin{eqnarray*}
M_{\pi} & =& M_0 + 2M_q \\
M_{K}   & =& M_0 + M_q + M_s\\
M_{\eta}& =& M_8\cos^2\Theta + M_1\sin^2\Theta\\
M_{\eta^{\prime}} &=& M_8\sin^2\Theta + M_1\cos^2\Theta
\end{eqnarray*}
\end{minipage}
\begin{minipage}[t]{0.48\textwidth}
\begin{eqnarray*}
M_1 = M_0 + 4/3 M_q + 2/3 M_s\\
M_8 = M_0 + 2/3 M_q + 4/3 M_s
\end{eqnarray*}
\end{minipage}
From these equations we derive the linear mass formula:
\be
\tan^2\Theta = \frac{3M_{\eta} + M_{\pi} - 4M_{K}}
{4M_{K} - 3M_{\eta^{\prime}} - M_{\pi}}
\ee
The Gordon equation is quadratic in mass, hence we may
also try the quadratic mass formula~\cite{Okubo:1961jc}:
\be
\tan^2\Theta = \frac{3M^{2}_{\eta} + M_{\pi}^2 - 4M_{K}^2}
{4M_{K}^2 - 3M^{2}_{\eta^{\prime}} - M_{\pi}^2}
\ee
A better justification for the  quadratic mass formula
is the fact that in first order of chiral symmetry breaking,
squared meson masses are linearly related to quark masses.
\\

\renewcommand{\arraystretch}{1.3}
\bc
\begin{tabular}{ccc}
\hline\hline
          Nonet members & $\Theta_{\rm linear}$ & $\Theta_{\rm quad} $\\
$\pi , K , \eta^{\prime} , \eta $ & $ -23^{\circ}$ & $ -10^{\circ}$ \\
$\rho , K^* , \Phi\ , \omega $    & $ 36^{\circ}$ & $ 39^{\circ}$ \\
$a_2(1320) , K^{*}_{2}(1430) , f_2(1525) , f_2(1270)$ &
$ 26^{\circ}$ & $ 29^{\circ}$ \\
\hline\hline
\end{tabular}
\ec
\renewcommand{\arraystretch}{1.0}

\subsubsection{Naming scheme}
In table~\ref{scheme} a summary of light mesons
for intrinsic orbital angular momenta up to 4 is given.
\subsubsection{Regge trajectories}
\begin{figure}[h!]
\bc
\epsfig{file=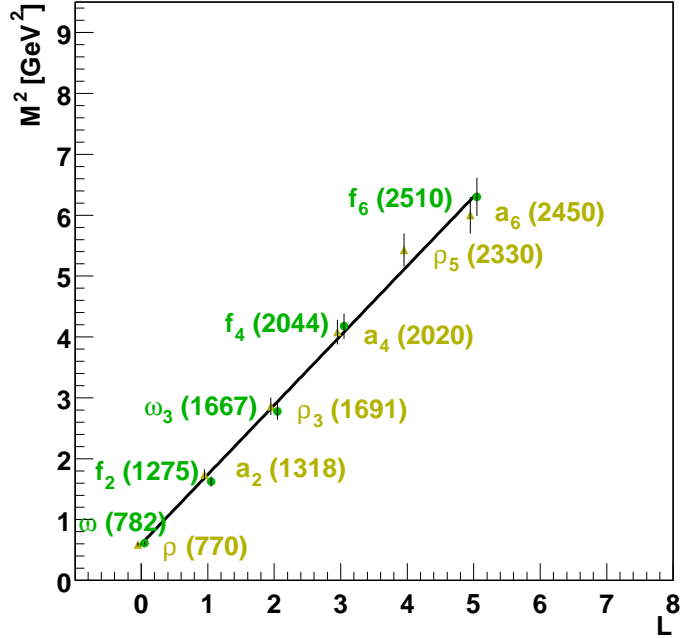,width=0.7\textwidth}
\ec
\vspace*{-5mm}
\caption{
Regge trajectory for mesons with $J=L+1$.
}
\label{mesons}
\end{figure}

The squared meson masses are linearly dependent on the total angular
momentum J, meson resonances lie on 
{\it Regge trajectories}~\cite{Regge:mz}. 
Figure~\ref{mesons} shows such a plot for
light $\bar nn$ mesons with $J=L+1$. The mesons belong to nonets with
an approximately ideal mixing angle; the masses of the
$\bar uu\pm\bar dd$ mesons are degenerate. Their mean value is
used. The dotted line represents a fit to the
meson masses taken from the PDG~\cite{Hagiwara:fs}; the error in the fit is
given by the PDG errors and a second systematic error of 30 MeV is added
quadratically. The slope is determined to  1.142 GeV$^2$.
\par
Table~\ref{scheme} contains only the ground states, but as in atomic
physics there are also {\it radial excitations},
states with wave functions having nodes. The $\rho$ meson for example
is the $1^3S_1\rho(770)$ ground state. It has a
radial excitation denoted as $2^3S_1\rho(1440)$. The 
state $1^3D_1\rho(1700)$ has the same measurable quantum numbers 
$J^{PC}=1^{--}$. Generally we cannot determine the internal structure
of a meson with 
\begin{table}[h!]
\caption{Naming scheme for light mesons}
\vskip 2mm
\begin{center}
\renewcommand{\arraystretch}{1.1}
\begin{tabular}{lllccccc}
\hline\hline
    &   &   $ J^{PC}$ &$ ^{2s+1}L_{J}$& I=1 & I=0  $ (n\bar{n})$
    & I=0 $ (s\bar{s})$  & I=1/2    \\
 L=0    & S=0   &$ 0^{-+}$ &$ ^{1}S_{0}$
& $ \pi$ & $ \eta$&$ \eta'$  & $ K$ \\
    & S=1   &$ 1^{--}$ &$ ^{3}S_{1}$
& $ \rho$ & $ \omega$&$ \Phi$& $ K^*$ \\
 L=1    & S=0   &   $ 1^{+-}$ &$ ^{1}P_{1}$
& $ b_1$ & $ h_1$   & $ h_1'$           & $ K_1$ \\
    & S=1   &   $ 0^{++}$ &$ ^{3}P_{0}$
& $ a_0$ & $ f_0$ & $ f_0'$         & $ K_0^*$ \\
    &   &   $ 1^{++}$ &$ ^{3}P_{1}$
& $ a_1$ & $ f_1$ & $ f_1'$ & $ K_1$ \\
    &   &   $ 2^{++}$ &$ ^{3}P_{2}$
& $ a_2$ & $ f_2$ & $ f_2'$ & $ K_2^*$ \\
 L=2    & S=0   & $ 2^{-+}$ &$ ^{1}D_{2}$
& $ \pi_2$ & $ \eta_2$ & $ \eta_2'$ & $ K_2$ \\
    & S=1   & $ 1^{--}$ &$ ^{3}D_{1}$
& $ \rho$  & $ \omega$ & $ \Phi$ & $ K_1^*$\\
    &   & $ 2^{--}$ &$ ^{3}D_{2}$
& $ \rho_2$ & $ \omega_2$ & $ \Phi_2$ & $ K_2$  \\
    &   & $ 3^{--}$ &$ ^{3}D_{3}$
& $ \rho_3$ & $ \omega_3$ & $ \Phi_3$ & $ K_3^*$\\
 L=3    & S=0   &   $ 3^{+-}$ &$ ^{1}F_{3}$
& $ b_3$ & $ h$   & $ h'_3$           & $ K_3$ \\
    & S=1   &   $ 2^{++}$ &$ ^{3}F_{2}$
& $ a_2$ & $ f_2$ & $ f_2'$         & $ K_2^*$ \\
    &   &   $ 3^{++}$ &$ ^{3}F_{3}$
& $ a_3$ & $ f_3$ & $ f_3'$ & $ K_3$ \\
    &   &   $ 4^{++}$ &$ ^{3}F_{4}$
& $ a_4$ & $ f_4$ & $ f_4'$ & $ K_4^*$ \\
 L=4    & S=0   & $ 4^{-+}$ &$ ^{1}G_{2}$
& $ \pi_4$ & $ \eta_4$ & $ \eta_4'$ & $ K_4$ \\
    & S=1   & $ 3^{--}$ &$ ^{3}G_{1}$
& $ \rho_3$  & $ \omega_3$ & $ \Phi_3$ & $ K_3^*$\\
    &   & $ 4^{--}$ &$ ^{3}G_{2}$
& $ \rho_4$ & $ \omega_4$ & $ \Phi_4$ & $ K_4$  \\
    &   & $ 5^{--}$ &$ ^{3}G_{3}$
& $ \rho_5$ & $ \omega_5$ & $ \Phi_5$ & $ K_5^*$\\
\hline\hline
\renewcommand{\arraystretch}{1.0}
\end{tabular}
\end{center}
\label{scheme}
\vspace*{-5mm}
\end{table}
\clearpage
\noindent
given quantum numbers $J^{PC}$ and a given mass.
A spectroscopic assignment requires models but a given state can be
assigned to a spectroscopic state on the basis of its decays or 
due to its mass. Of course, 
mixing is possible between different internal configurations.
\par
\begin{figure}[h]
\hspace{-.2cm}
\centerline{\epsfig{file=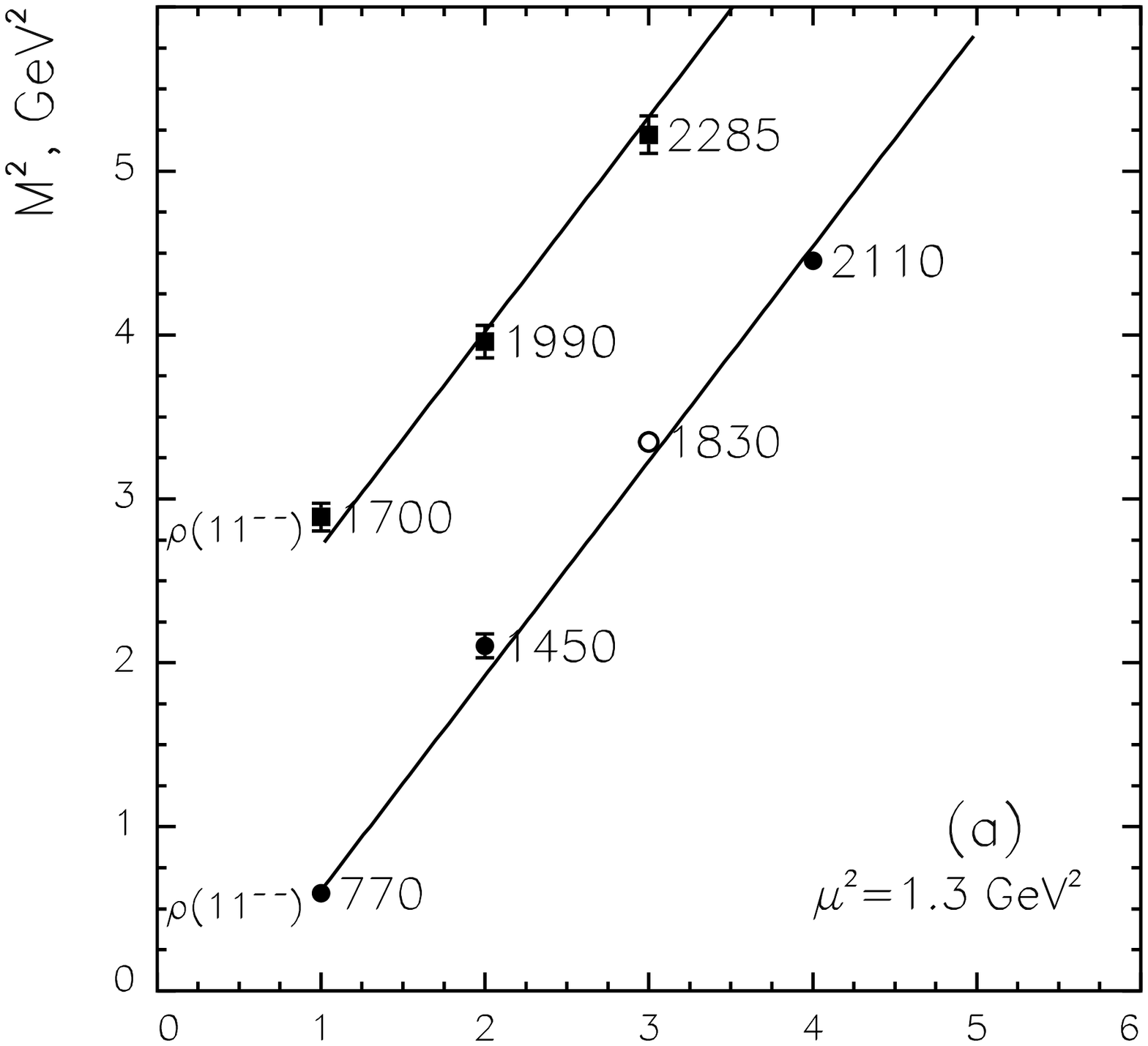,width=0.4\textwidth}\hspace{-1.0cm}
            \epsfig{file=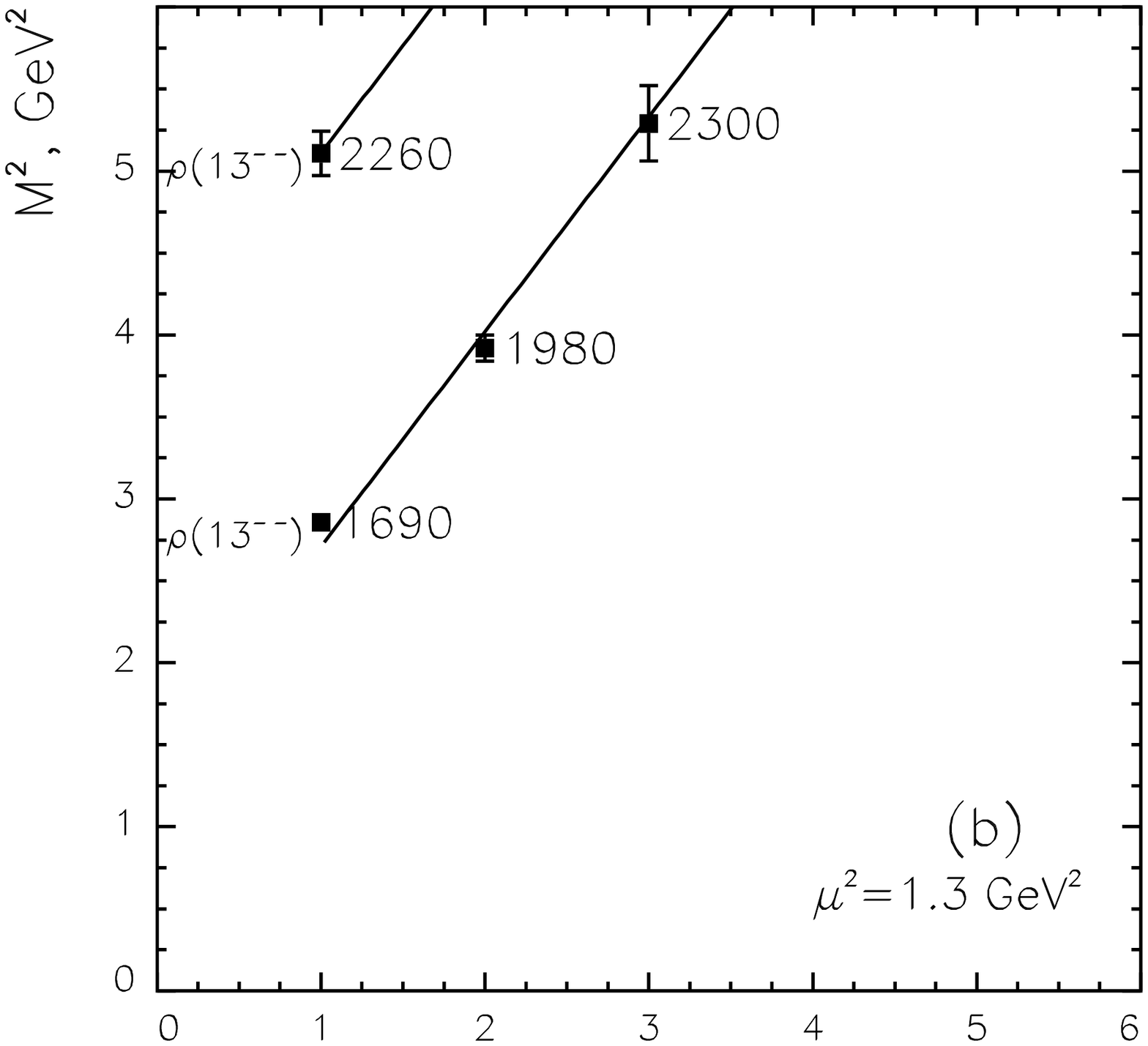,width=0.4\textwidth}\hspace{-1.0cm}
            \epsfig{file=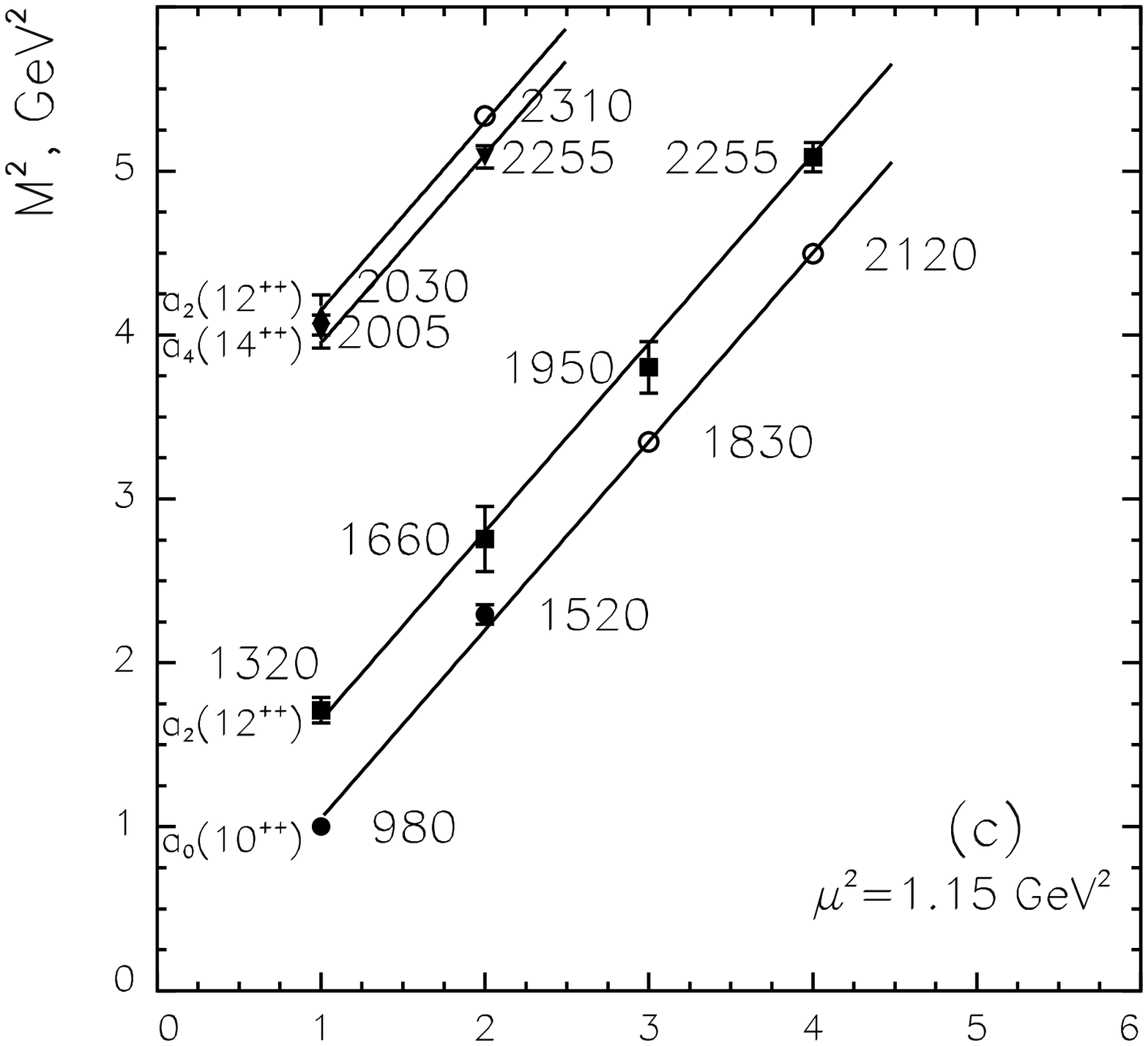,width=0.4\textwidth}}

\vspace*{-7mm}

$\hspace{9.4cm}^3F_4\;q\bar q(L=3)$

$\hspace{9.4cm}^3F_2\;q\bar q(L=3)$

$\hspace{12mmm}^3D_1\;q\bar q(L=2)\hspace{17mm}^3G_3\;q\bar q(L=4)\hspace{18mm}^3P_2\;q\bar q(L=1)$

$\hspace{12mmm}^3S_1\;q\bar q(L=0)\hspace{18mm}^3D_3\;q\bar q(L=2)\hspace{18mm}^3P_0\;q\bar q(L=1)$
\caption{\label{va}
Mesons of identical $J^{PC}$ fall onto linear trajectories with
a slope $(1.15-1.30)$ GeV$^2$ which is similar to the Regge slope
found in figure~\protect\ref{mesons}. Note that many resonances
shown here are not listed by the PDG. Hence some interpretation
of the data is required. (The data are from~\protect\cite{Anisovich:2000kx}).
}
\end{figure}

In figure~\ref{va} squared meson masses are plotted all having the
same quantum numbers $J^{PC}$. The $\rho(770)$ as $1^3S_1$ is followed by
the $2^3S_1$ and $3^3S_1$. A new sequence is started for the
$1^3D_1$, $2^3D_1$, and $3^DS_1$. Likewise, $L=2$ and $L=4$ can couple to
form two $J^{PC}=3^{--}$ series. Figure~\ref{va} is
taken from~\cite{Anisovich:2000kx}.

\vspace*{-4mm}
\vspace*{-4mm}
\subsection{\label{section1.3}Charmonium and bottonium}
\subsubsection{Discovery of the J/$ \psi$}
A narrow resonance was discovered in 1974 in two reactions. 
At Brookhaven National Laboratory (BNL) in Long Island, New York, 
the process proton + Be $ \to e^+e^-$ + anything
was studied~\cite{Aubert:js}; at the Stanford University, the new
resonance was observed in the SPEAR storage ring in 
$ e^+e^-$ annihilation to $ \mu^+\mu^- , e^+e^-$ and
into hadrons~\cite{Augustin:1974xw} 
This discovery initiated the ``November revolution
of particle physics''.
\par
Electrons and positrons are rarely produced in hadronic reactions.
Dalitz pairs (from $\pi^0\to\gamma  e^+e^-$) have very low
invariant masses; the probability to produce an  $e^+e^-$ 
pair having a large invariant mass is very small. 
In the BNL experiment $e^+e^-$ pairs with large invariant masses 
were observed. The two particles were identified by Cerenkov radiation 
and time--of--flight, their momenta were measured in two
spectrometers. In the $e^+e^-$
invariant mass distribution, $M^2=(\sum E_i)^2-(\sum \vec p_i)^2$,
a new resonance showed up which was named $J$. 
This type of experiment is called a {\it production experiment}.
In production experiments the width of a narrow peak is given by
the resolution of the detector. 
\par
SPEAR was a  $e^+e^-$ storage ring in which  $e^+$ and $e^-$ pairs
annihilated into virtual photons. These virtual photons couple to
mesons (having the same quantum numbers as photons), to $J^{PC}=1^{--}$
vector mesons. The momenta of $e^+$ and $e^-$ are the same in
magnitude but opposite in sign. Vector mesons are formed when the
total energy W, often called $\sqrt s$, coincides with their mass.
W is varied while scanning the resonance. The width of a narrow peak
is given by the accuracy with which the beams can be tuned.
This type of experiment is called a {\it formation experiment}.

Figure~\ref{jpsi} shows 
(see http://bes.ihep.ac.cn/besI\&II/physics/JPSI/index.html)
a modern scan 
of the beam energy across a resonance 
which was named $\psi$. It is the same resonance as the $J$ particle,
hence its name J/$\psi$. The width of the 
resonance is less than the spread of the beam energies, which is 
less than 1 MeV. Due to the production mode via an intermediate
virtual photon, 
spin, parity and charge conjugation
are $ J^{PC} = 1^{--}$ (as the $ \rho$-meson). The
J/$\psi$ is a vector meson. 

\subsubsection{Width of the J/$ \psi$}
\begin{figure}[h!]
\begin{minipage}[c]{0.48\textwidth}
\epsfig{file=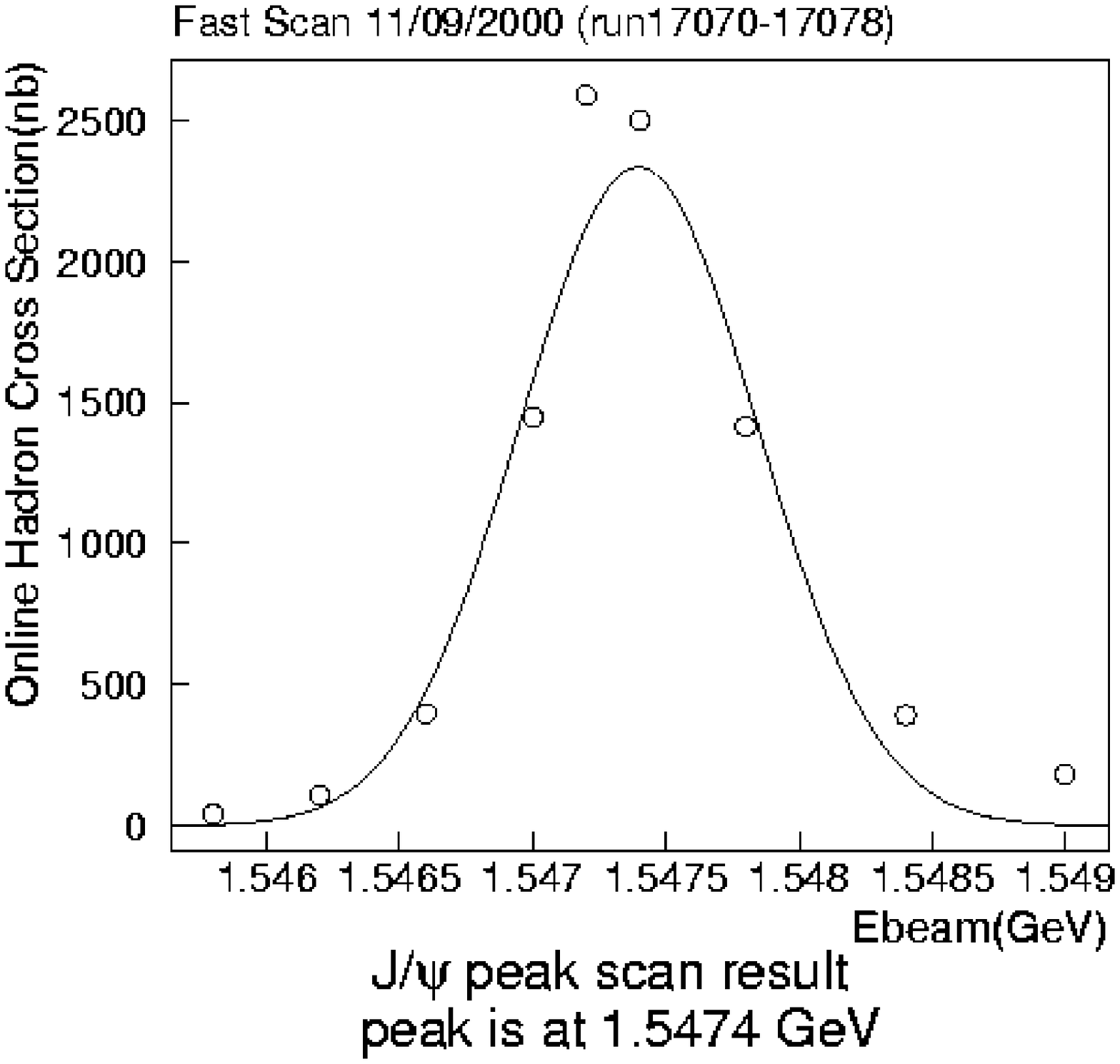,width=0.95\textwidth,clip=on}
\caption{Scan of the beam energies in $e^+e^-$ scattering
through the J/$ \psi$ resonance region;
(see http://bes.ihep.ac.cn/besI\&II/physics/
JPSI/index.html).}
\label{jpsi}
\end{minipage}
\begin{minipage}[cr]{0.5\textwidth}
\vspace*{-7mm}

center--of--mass system (cms), $E$ the
cms energy, and $ \Gamma$ the total width. The first term of the
right--hand side of (\ref{jeq}) is the usual Breit--Wigner 
function describing a
resonant behavior. The second part sums over the spin components
in the final state and averages over the spin components in the
initial state. $s_1=s_2=1/2$ are electron and positron spin; $J=1$
is the J/$ \psi$ total angular momentum.
\par
In case of specific reactions, like $ e^+e^-\to\psi\to e^+e^-$, we
have to replace the total width in the numerator by
\be
\Gamma^2 \to \Gamma_{initial}\Gamma_{final} = \Gamma^2_{e^+e^-},
\ee
where $ \Gamma_{e^+e^-}$ is the partial width for the decay into
$e^+e^-$. Then:
\end{minipage}
\end{figure}

The natural width of the J/$ \psi$ is too narrow to
be determined from figure~\ref{jpsi} but it is
related to the total cross section. The cross section can be
written in the form 

\be
\sigma (E) =
(4\pi)(\lambda^2/4\pi^2)
\frac{\Gamma^2/4}
{\left[\left(E-E_R\right) + \Gamma^2/4\right)]}\frac{2J+1}{(2s_1+1)(2s_2+1)}
\label{jeq}
\ee
with $ \lambda/2\pi = 1/p = 2/E$ being the de Broglie wave length
of $ e^+$ and $ e^-$ in the 
\par
\noindent
\be
\sigma (E_{e^+e^-\to\psi\to e^+e^-}) =
(3\pi)(\lambda^2/4\pi^2)
\frac{\Gamma^2_{e^+e^-}/4}{\left(E-E_R\right) + \Gamma^2/4}
\ee
The number of $e^+e^-$ pairs is proportional to
\be
\int_0^{\infty} \sigma(E)dE = \frac{3\pi^2}{2}\frac{\lambda^2}{4\pi^2}
\frac{\Gamma^2_{e^+e^-}}{\Gamma^2}\Gamma .
\ee
After substituting $\tan\theta = 2(E-E_R)/\Gamma$
the integration can be carried out and results in
\be
\int_0^{\infty} 
\sigma(E_{e^+e^-\to e^+e^-,\mu^+\mu^-,\mathrm{hadrons}})dE =
\frac{6\pi^2}{E_R^2\Gamma}
\Gamma_{e^+e^-}
\Gamma_{(e^+e^-,\mu^+\mu^-,\mathrm{hadrons})}.
\ee
The total width is given by the sum of the partial
decay widths:
\be
\Gamma = \Gamma_{e^+e^-} + \Gamma_{\mu^+\mu^-} + \Gamma_{hadrons}
\ee
Imposing $ \Gamma_{e^+e^-} = \Gamma_{\mu^+\mu^-}$ yields
3 equations and thus 3 unknown widths.
\par
The J/$\psi$ has a mass $ (3096.87\pm 0.04)$\,MeV and
a width $ 87\pm 5$\,keV. We may compare this to the
$\rho$ mass, 770\,MeV, and its width 150\,MeV.
Obviously the J/$ \psi$ is extremely narrow. This can be
understood by assuming that the J/$ \psi$ is a bound
state of a new kind of quarks called charmed quarks $c$, and that 
\bc
J/$\psi = c\bar c$.
\ec
\par
The Okubo--Zweig--Ishida (OZI) rule then explains why the
J/$ \psi$ is so narrow.

\subsubsection{The OZI rule and flavor tagging}
\begin{figure}[h!]
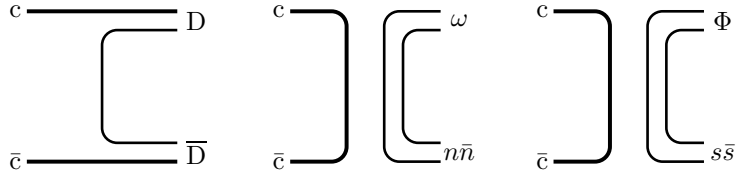

\vspace*{-6mm}
\begin{center}
\psset{xunit=.5cm,yunit=.5cm,linearc=.2}
\pspicture(-1,-.1)(19,5)
\psline[linewidth=1.5pt](0,4)(4,4)
\psline[linewidth=1.5pt](0,0)(4,0)
\psline[linewidth=1pt](4,0.5)(2,0.5)(2,3.5)(4,3.5)
\rput[cl](4.5,3.75){$\mathrm{D}$}
\rput[cl](4.5,.25){$\overline{\mathrm{D}}$}
\rput[cr](-.3,4){$\mathrm{c}$}
\rput[cr](-0.3,0){$\bar{\mathrm{c}}$}
\rput(7,0){
\psline[linewidth=1.5pt](0,4)(1.5,4)(1.5,0)(0,0)
\psline[linewidth=1pt](4,4)(2.5,4)(2.5,0)(4,0)
\psline[linewidth=1pt](4,0.5)(3,0.5)(3,3.5)(4,3.5)
\rput[cl](4.5,3.75){$\omega$}
\rput[cl](4.5,.25){$n\bar n$}
\rput[cr](-.3,4){$\mathrm{c}$}
\rput[cr](-0.3,0){$\bar{\mathrm{c}}$}
}
\rput(14,0){
\psline[linewidth=1.5pt](0,4)(1.5,4)(1.5,0)(0,0)
\psline[linewidth=1pt](4,4)(2.5,4)(2.5,0)(4,0)
\psline[linewidth=1pt](4,0.5)(3,0.5)(3,3.5)(4,3.5)
\rput[cl](4.5,3.75){$\Phi$}
\rput[cl](4.5,.25){$s\bar s$}
\rput[cr](-.3,4){$\mathrm{c}$}
\rput[cr](-0.3,0){$\bar{\mathrm{c}}$}
}
\endpspicture
\end{center}
\caption{Decays of charmonium states into $\rm\bar DD$ are allowed
only above the  $\rm\bar DD$ threshold, the J$/\psi$ (and the
$\eta_c$ and $\chi$ states) can decay only into light quarks. A
$\omega$ or $\Phi(1020)$ signal determines the $\bar uu +\bar dd$ and
$\bar ss$ component, respectively, of the recoiling meson. The
thick lines represent charmed quarks. } \label{ddecay}
\end{figure}

A  low--lying $c\bar c$ bound state cannot decay into two $D$ mesons
having {\it open} charm (see figure~\ref{ddecay}).
The J/$\psi$ must annihilate
completely and new particles have to be created out of the vacuum.
Such processes are suppressed; the four-momentum of the 
$c\bar c$ bound state must convert into gluons carrying
a large four-momentum, and we will see  in section~\ref{section3} 
that the coupling of quarks to gluons with large four-momenta is small. 
Hence the J/$\psi$ is narrow. This OZI rule
can be exploited to tag the flavor of mesons produced in
J/$\psi$ decays in cases where one of the two mesons has a known
flavor content. If it is a $\bar uu +\bar dd$ meson, 
like the $\omega$, the recoiling meson
couples with its $\bar uu +\bar dd$ component. If a
$\Phi(1020)$ meson is produced, the recoiling meson
couples with its $\bar ss$ component. Thus the flavor structure
of mesons can be determined. This was done for the
$\eta$ and $\eta^{\prime}$ mesons and led to the
pseudoscalar mixing angle as discussed 
above~\cite{Coffman:1988ve,Jousset:1988ni}.
\par
\begin{figure}[h!]
\centerline{\epsfig{file=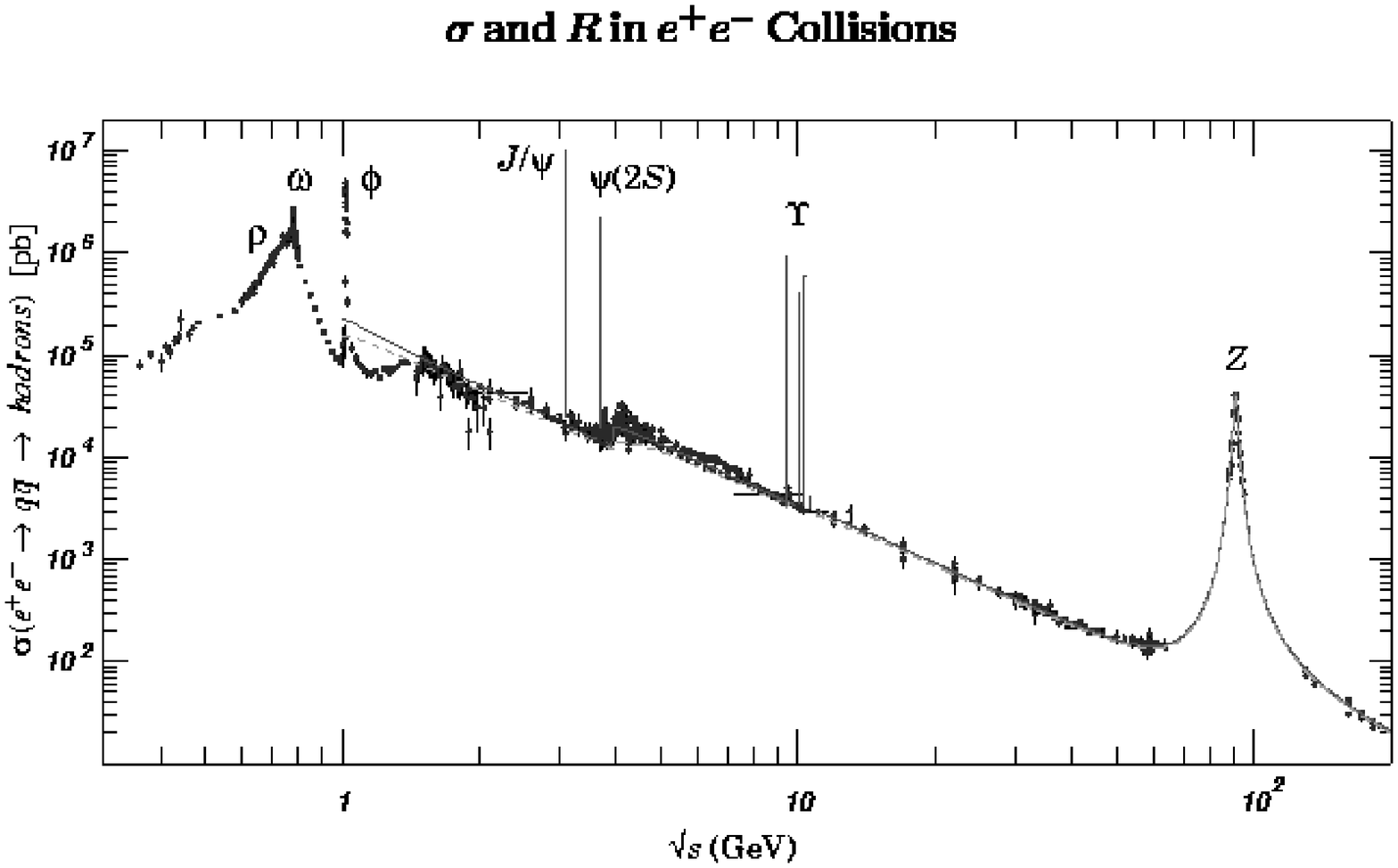,width=0.9\textwidth,clip=on}}
\vspace*{-3mm}
\caption{$e^+e^-$ hadronic cross section (from~\protect\cite{Hagiwara:fs}).}
\label{crosse+e-}
\vspace*{-3mm}
\end{figure}

\begin{figure}[h!]
\centerline{\epsfig{file=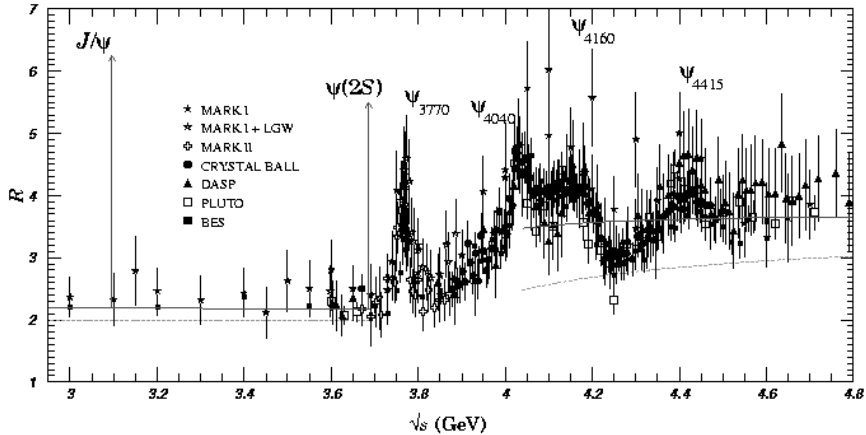,width=0.9\textwidth,clip=on}}
\caption{$e^+e^-\to$ hadrons versus $\mu^+\mu^-$ in the $c\bar c$
region (from~\protect\cite{Hagiwara:fs}).}
\label{cross-ratio}
\end{figure}

The cross section for $e^+e^-$ annihilation into hadrons 
shown in figure~\ref{crosse+e-} gives access to a variety of physics questions.
At low energies the cross section is dominated by $\rho , \omega$
and $\Phi$ mesons. Then it falls off according to
$(4\pi\alpha^2/3\,s)\cdot3\cdot\sum Q^2_i$ where $Q_i$ are the quark charges
and the sum extends over all quarks which can be created at the
given energy. The factor $3$ accounts for the three different
colors. There are two narrow $c\bar c$ states, the
J/$\psi$ and the $\psi(2S)$, and three narrow $b\bar b$
states, the $\Upsilon, \Upsilon(2S)$, and $\Upsilon(3S)$. 
At 90\,GeV the $Z^0$
resonance, the neutral weak interaction boson, is observed.
The ratio $R$ of the cross section for
$e^+e^-\to$ hadrons to that for $\mu^+\mu^-$ is given by
$R = 3\times \sum Q_i^2$ and increases above quark-antiquark
thresholds. In figure \ref{cross-ratio} we see that $R$ increases
from the value 2.2 below the J/$\psi$ to 3.7 above. For quarks with
charges 2/3 we expect an increase of $3\times(2/3)^2=4/9\simeq 0.44$.
To account for the observed increase we have to include
$\tau^+\tau^-$ production, setting in
at about the open $\bar cc$ threshold. The $\tau$ charge is $1$, so
$R$ should increase by 1.44, apart from corrections due to gluon
radiation. 
A closer look reveals additional peaks in the cross sections.
These resonaces decay into mesons with {  open charm}
like $D^+D^-$ or $D^0\overline{D^0}$.
Their mass is $D^{\pm} = 1870$\,MeV and $D^{0} = 1865$\,MeV,
respectively. They carry open charm, their quark content is
$c\bar d$ and $d\bar c$, and $c\bar u$ and $u\bar c$,
respectively, and 
their quantum numbers are $J^{PC}=0^{-+}$. The states map the
Kaon states onto the charm sector.

\subsubsection{J/$\psi$ decays to $e^+e^-$}
In J/$\psi$ decays to $e^+e^-$ the intermediate state is a single
virtual photon. This resembles QED in positronium atoms and
indeed, one can adopt the transition rate from positronium to $\bar
cc$ decays. The van Royen Weisskopf equation~\cite{VanRoyen:nq}
reads
\be
\Gamma (J/\psi\to e^+e^-) = \frac{16\pi\alpha^2 Q^2}{M_V^2}|\psi(0)|^2
\ee
Here $Q^2$ is the squared sum of contributing quark charges (see
table~\ref{charges}).

\begin{table}[h!]
\caption{Photo--coupling of vector mesons.}
\bc
\begin{tabular}{lccc}
\hline\hline
Meson     & wave function                                     & $Q^2$ &\\
$\rho^0$: & $\frac{1}{\sqrt 2}\left(u\bar u - d\bar d\right)$ &
$\left[\frac{1}{\sqrt 2}\left(2/3 - (-1/3)\right)\right]^2$ & 1/2 \\
$\omega$: & $\frac{1}{\sqrt 2}\left(u\bar u + d\bar d\right)$ &
$\left[\frac{1}{\sqrt 2}\left(2/3 + (-1/3)\right)\right]^2$ & 1/18 \\
$\Phi$:   & $\left(s\bar s\right)$  & $\left(1/3\right)^2$  & 1/9 \\
J/$\psi$: & $\left(c\bar c\right)$  & $\left(2/3\right)^2$  & 4/9 \\
\hline\hline
\end{tabular}
\ec
\label{charges}
\end{table}
This is an important result: photons couple to $\rho$
(in amplitude) 3 times stronger than to $\omega$. The hypothesis
that photons couple to hadrons dominantly via intermediate
vector mesons is known as {\it vector meson dominance}. 
As a side remark; the N\,N\,$\omega$ coupling is (again in
amplitude) $\sim 3.5$ times stronger than the N\,N\,$\rho$ coupling.
\subsubsection{Charmonium states in radiative decays}
\begin{figure}[h!]
\bc
\epsfig{file=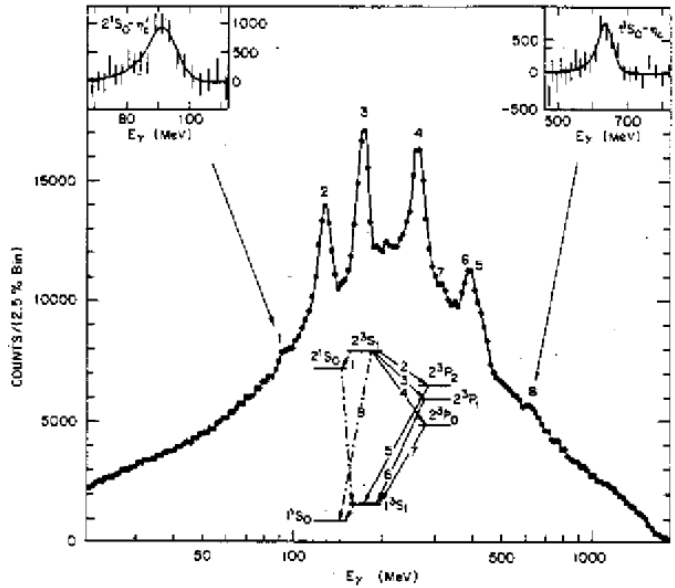,width=0.70\textwidth,clip=on}
\ec
\caption{Radiative transitions between charmonium levels
 (from~\protect\cite{Gaiser:1985ix}).}
\label{cb-rad}
\end{figure}

In $e^+e^-$ annihilation, only mesons with quantum numbers
$J^{PC} = 1^{--}$ are formed. But we also expect charmonium
states to exist with other quantum numbers, in particular states
with positive $C$--parity. The transitions can be searched for 
in radiative transitions from the $\psi (2S)$ state. 
Figure~\ref{cb-rad} shows
the inclusive photon spectrum from the $\psi(2S)$ 
states~\cite{Gaiser:1985ix}. A series
of narrow states is seen identifying the masses of intermediate
states. The level scheme assigns the lines to specific transitions
as expected from charmonium models. The width of the lines
is given by the experimental resolution of the detector; the
charmonium states are {\it produced}. The lowest mass state, the
$1^1S_0$ state, is called $\eta_c$. It is the $\bar cc$ analogue of
\begin{figure}[b!]
\bc
\epsfig{file=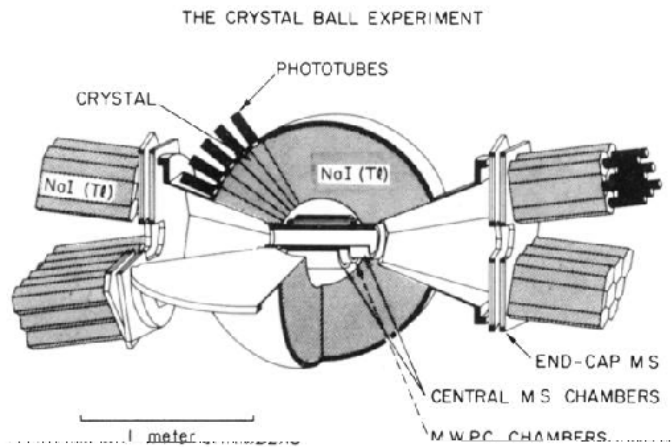,width=0.70\textwidth,clip=on}
\ec
\caption{The Crystal Ball detector~\protect\cite{Gaiser:1985ix}.}
\label{cball}
\end{figure}
the $\eta$. Its radial excitation is called $2^1S_0$ or
$\eta_c^{\prime}$. The small peak assigned to it turned out to be
fake in later experiments. The $^3P_J$ states are the lowest--mass
$P$-states and are now called $\chi$-- or $\chi_{cJ}(1P)$ states.
The photons from these
transitions were detected in the Crystal Ball detector, 
a segmented scintillation
counter shown in figure~\ref{cball}  
which at that time was installed at the SPEAR storage ring.

\subsubsection{Charmonium states in $\bar pp$ annihilation}
Due to the quantum numbers of the virtual photon, 
$\chi$ states can only be {\it produced},  not {\it formed},
in $e^+e^-$ annihilation; formation in this process is restricted to
vector mesons. The $\bar pp$ annihilation process
is much richer due to the finite size and compositeness of
the collision partners, hence $\chi$ states can also be observed in a
formation experiment. The instrumental width with which a resonance
can be seen is limited only by the momentum resolution
of the beam and not by the detector resolution. The detector is needed
only to identify the number of $\chi$ states. 
However, $\bar pp$ annihilation  is dominated
by multi--meson final states; the total cross section is on the order
of $\sigma_{hadronic} = 100mb$. When $\bar cc$ states are to be
{\it formed}, the 3 quarks and antiquarks have to annihilate
and then a $\bar cc$ pair has to be created. This is an unlikely
process, the cross section is correspondingly small, $\sigma_{c\bar c}
= 1\mu b$. Hence, the background below the signal is huge.
These problems can be overcome by insisting that the $\bar cc$
state should decay into the J/$\psi$ or $\eta_c$ which then decays into
two electrons or photons, respectively, having a high invariant mass.
In this way, the background can be greatly suppressed. The
small branching ratio for J/$\psi$ or $\eta_c$ decays into the
desired final state reduces the rate even further, high count rates
are mandatory.
\par
\begin{figure}[h!]
\bc
\epsfig{file=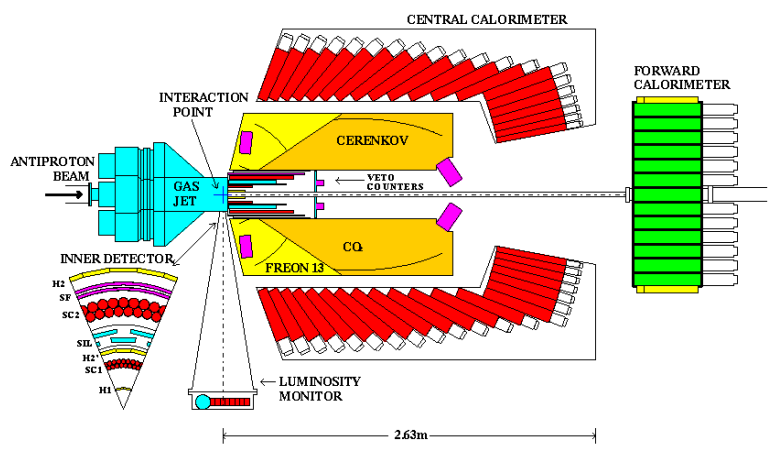,width=0.8\textwidth,clip=on}
\ec
\caption{\label{fnal}
Experiment E760/E835 at FNAL~\protect\cite{Ambrogiani:2001jw}.}
\end{figure}

An experiment of this type~\cite{Ambrogiani:2001jw}
 was carried out at Fermilab (see
figure~\ref{fnal}).
Antiprotons were produced by bombarding a target with high--energy
protons. The antiprotons are cooled in phase space in a storage
ring and are then accelerated to study $\bar pp$ collisions
at extremely high energies. A fraction of the antiprotons
were used for medium--energy physics: $8\cdot 10^{11}$ antiprotons
circulated in the Fermilab accumulator ring with a
frequency $f_{rev}=0.63$MHz. At each revolution antiprotons are
passed through a hydrogen gas jet target, with $\rho_{jet}= 3\cdot
10^{14}$H$_2/cm^3$, 
which results in a  luminosity $ L=N_{\bar p}f_{rev}\rho_{jet}$
of $2\cdot 10^{31}$/cm$^2$s. The energy of the antiproton
beam, and thus the invariant
mass of the $\bar pp$ system, can be tuned very precisely according to
$\sqrt{s} = m_p\cdot\sqrt{2(1+m_pE_{\bar p})}$.
The luminosity is an important concept; the observed rate $R$
of events is related to the luminosity by $R=\sigma\cdot L$.
The hadronic background is produced with
$R=2\cdot 10^{6}$/s.

\begin{figure}[h!]
\begin{minipage}[t]{0.46\textwidth}
\hspace*{5mm}\epsfig{file=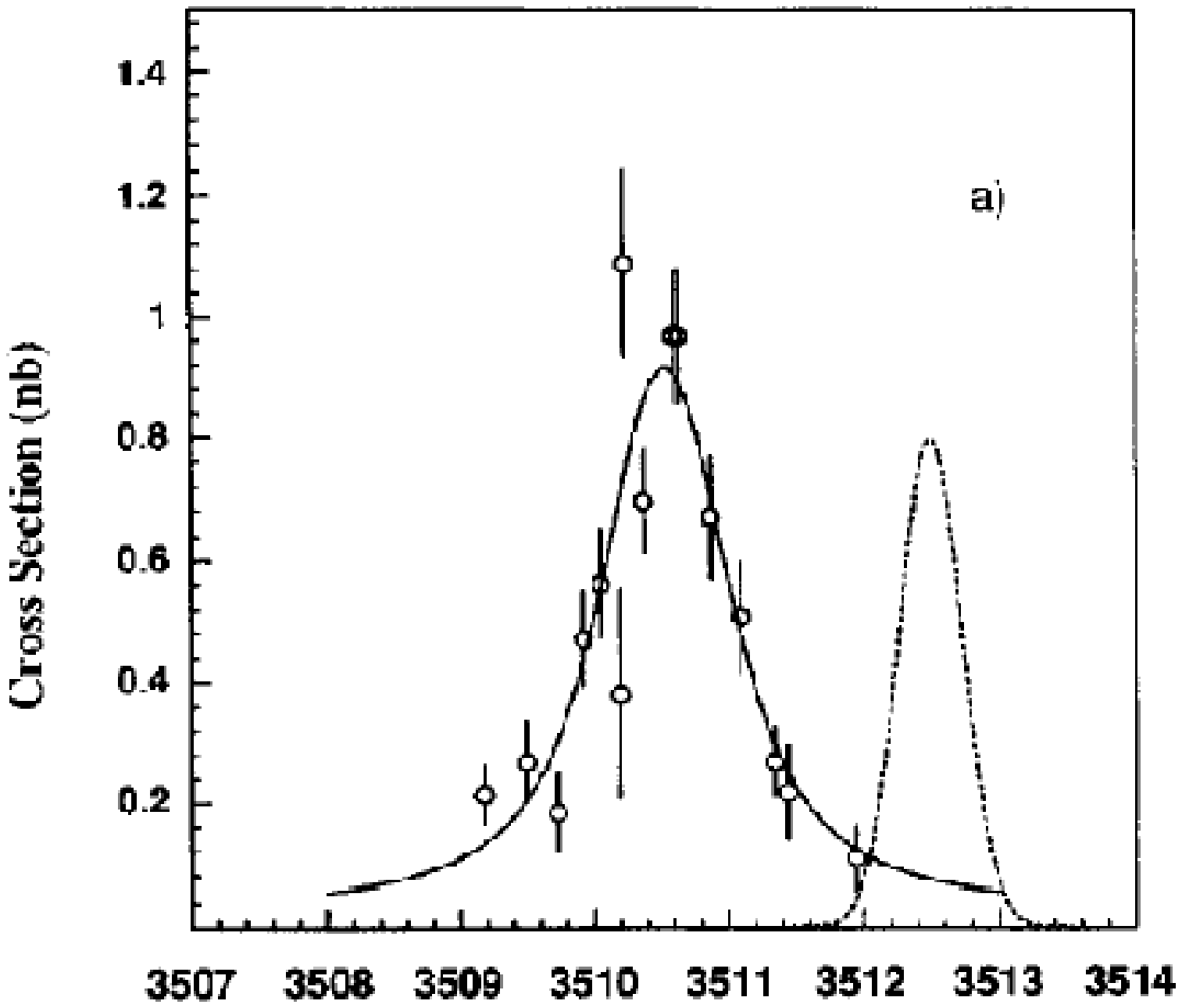,width=\textwidth,clip=}
\end{minipage}
\begin{minipage}[t]{0.45\textwidth}
\hspace*{5mm}\epsfig{file=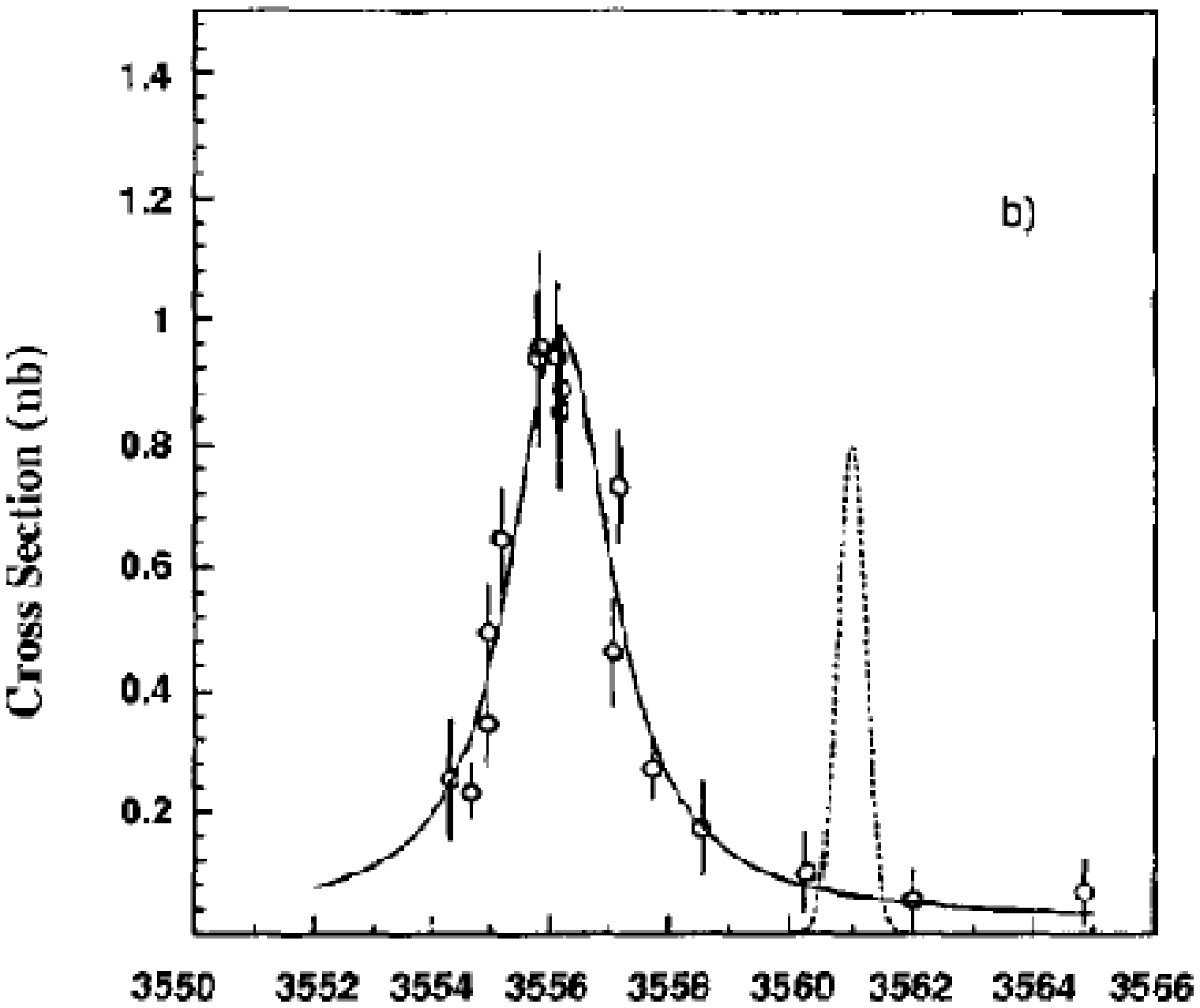,width=\textwidth,clip=}
\end{minipage}
\caption{The number of J/$\psi$ as a function of the
$\bar pp$ mass in the
$\chi_1$ (a) and $\chi_2$ (b) mass regions~\protect\cite{Ambrogiani:2001jw}.}
\label{cb-scan}
\end{figure}

Figure~\ref{cb-scan} shows scans of the $\chi_{c1}(1P)$
and $\chi_{c1}(2P)$ regions. The experimental resolution,
given by the precision of the beam momentum, is shown as
dashed line. The observed distributions are broader: the
natural widths of the states due to their finite
life time can now be observed.

\subsubsection{New players at Bejing and Cornell}
Charmonium physics came out of the focus of the community. 
However a new $e^+e^-$ collider ring was constructed at
Bejing, and is
\begin{figure}[h!]
\begin{tabular}{cc}
\hspace*{2mm}\epsfig{file=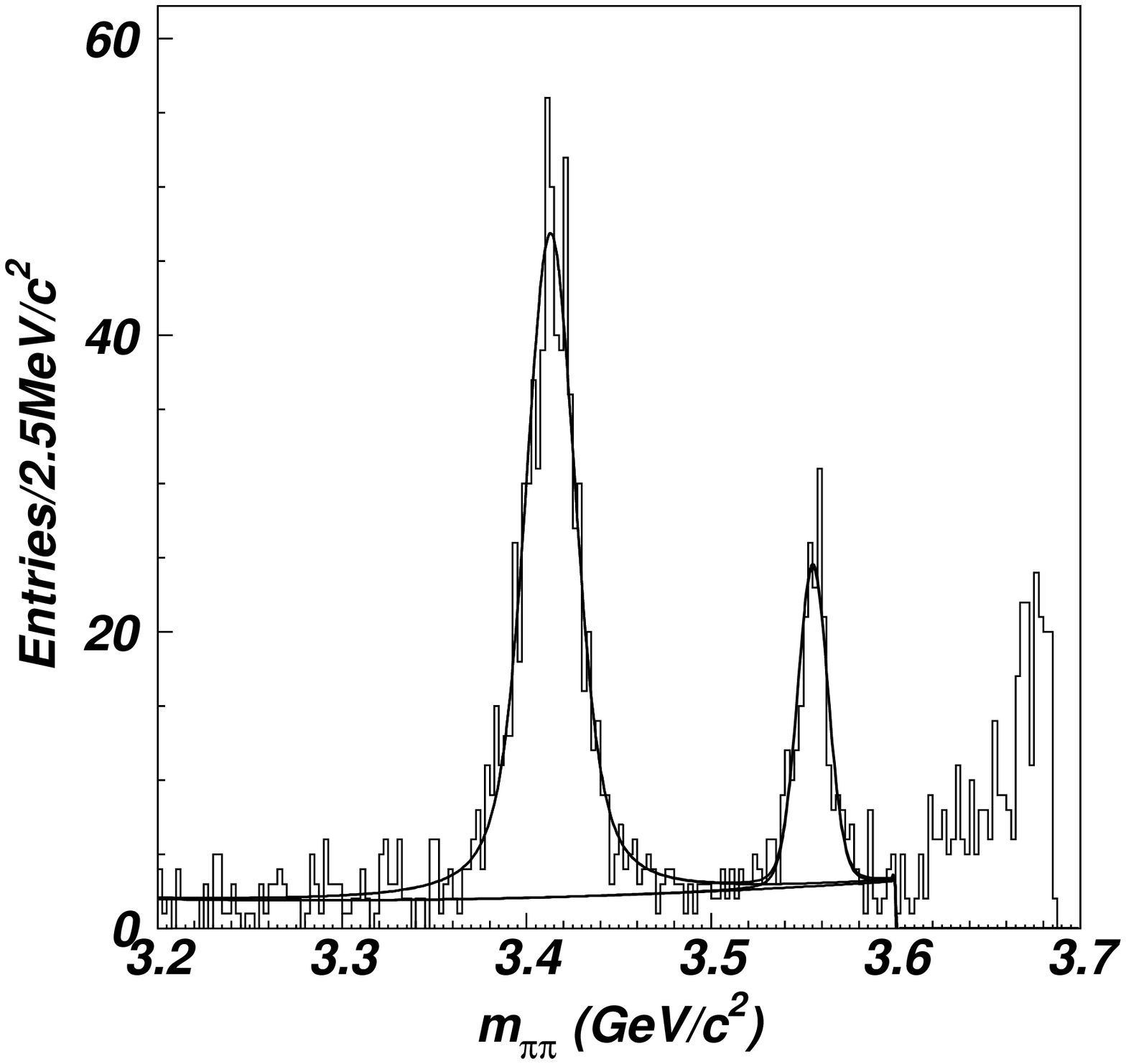,width=0.45\textwidth,clip=}&
\hspace*{-4mm}
\epsfig{file=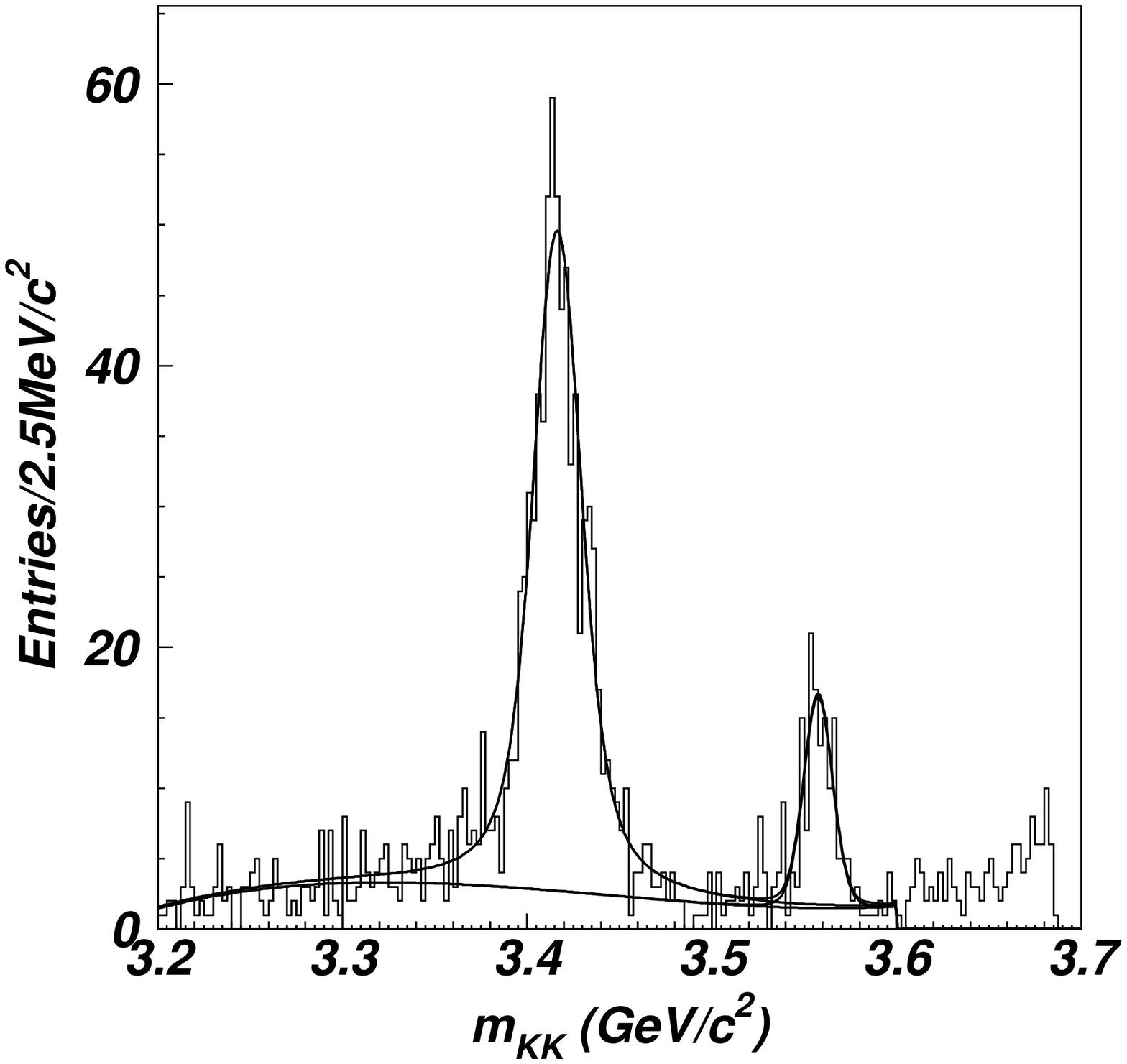,width=0.45\textwidth,clip=}
\end{tabular}
\caption{Radiative transitions of the $\psi (2S)$
charmonium state~\protect\cite{Bai:2001ha}.}
\label{bebs}
\end{figure}
producing results. The BES detector measures charged and
neutral particles, therefore reactions like $\psi(2S)\rightarrow
\gamma \pi^+ \pi^-$ and $\gamma K^+ K^-$ 
can be studied~\cite{Bai:2001ha} (see
figure~\ref{bebs}). From this data, spin, parities and decay
branching ratios of the $\chi$--states can be determined. At
present, the  $e^+e^-$ collider ring at Cornell is reduced in
energy but upgraded in luminosity and will take data very soon in
the charmonium region with extremely high precision.
\par
\begin{figure}[h!]
\bc
\epsfig{file=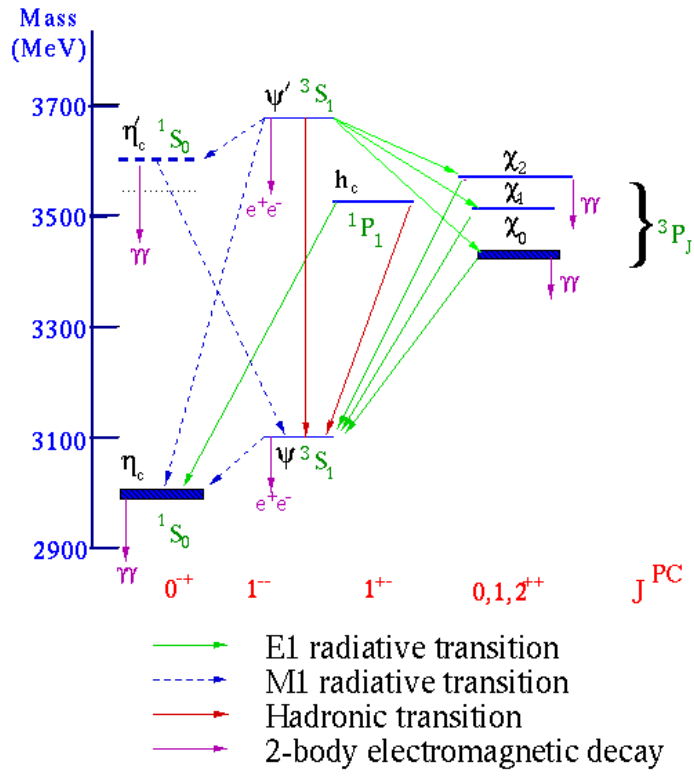,width=0.75\textwidth,clip=}
\ec
\vspace*{-5mm}
\caption{The charmonium level scheme and
radiative and hadronic transitions~\protect\cite{Hagiwara:fs}.}
\label{ccbarlevel}
\end{figure}

Figure~\ref{ccbarlevel} summarizes the charmonium
levels and the transitions between them. The level scheme clearly
resembles positronium. 
\vspace*{-5mm}
\subsection{\label{section1.4}D and B mesons}
The charmed quark enriches the spectrum of mesons and baryons
by new classes of hadrons with open and hidden charm. Similarly,
even more hadrons can be formed with the bottom quark $b$ coming
into the play. Mesons with one charmed quark are called $D$-mesons,
mesons with one charmed and one strange quark are known as $D_s$,
and mesons with one bottom quark are $B$-mesons. These have spin $S=0$ and
angular momentum $l=0$, and they are pseudoscalar mesons. Vector mesons
with $L=0, S=1$ are called $D^*$
or $B^*$. The $D^*_2$ has $S=1$ and $L=1$ coupling to $J=2$. Mesons
with one bottom and one charmed quark are called $B_c$. Mesons with hidden
charm are the J/$\psi$, $\psi (2S)$, .., the $\chi_{cJ}(1P)$ states
(with $S=1$) or $h_{c}(1P)$ (with $S=0$).
\par
\begin{figure}[h!]
\begin{tabular}{cc}
\epsfig{file=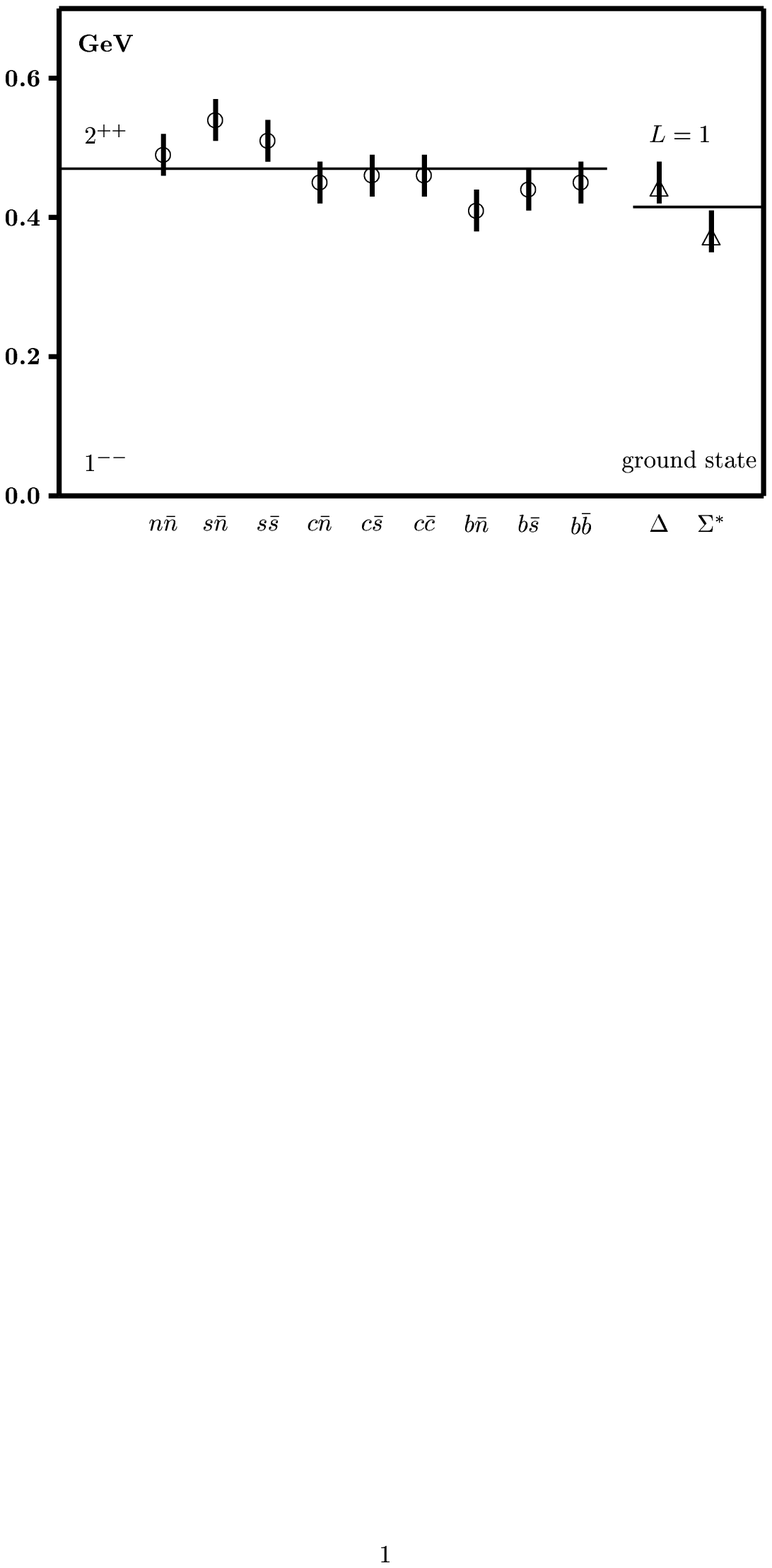,width=0.49\textwidth,clip=}&
\epsfig{file=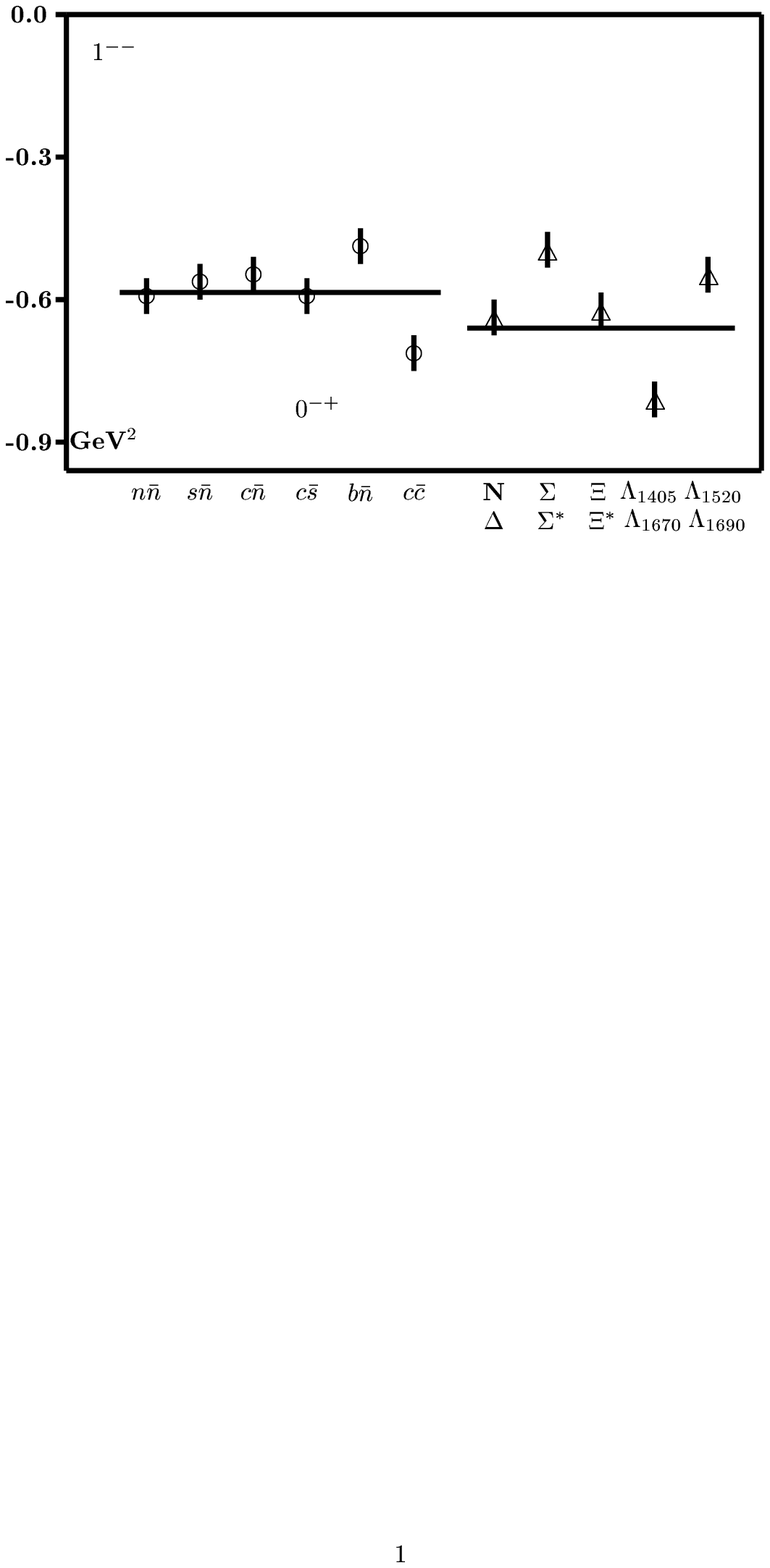,width=0.49\textwidth,clip=}
\end{tabular}
\caption{\label{forces}
Flavor independence of the strong forces:
Left panel: Mass differences for $L=1$ excitations 
of mesons and baryons (for systems   
with aligned spin). Right panel:                   
Mass square differences between PS and
V mesons for L=0 octet-decuplet
and $L=1$ singlet-octet baryons.
}
\vspace*{-2mm}
\end{figure}

Heavy baryons have been been discovered as well, like
the $\Lambda_c$ or the $\Lambda_b$ with one charmed (bottom) quark,
$\Sigma_c$ or $\Sigma_b$, and so on. These mesons and baryons
have very different masses, but the forces between quarks
of different flavor are the same\,! This can be seen in figure~\ref{forces}.
The left panel shows the mass gap for $L=1$ excitations with 
quark spins aligned, 
to the $L=0$ meson or baryon ground states
(again with aligned quark spins).
On the right panel, the 'magnetic' mass splitting between
states with aligned and not aligned spins is plotted. In this case, the
differences in mass square are plotted. Note that $M_1^2-M_2^2 =
(M_1-M_2)\cdot (M_1+M_2)$. If this is a constant, $M_1-M_2$ scales
with $1/M$.
\subsection{\label{section1.5}The new states}
Exciting discoveries were made last year. Several
strikingly narrow resonances were observed, at unexpected
masses or with exotic quantum numbers. These were
mostly mesons and will be discussed here. Among the new
states are two baryon resonance,
called $\Theta^+(1540)$ and $\Xi^{--}(1862)$. Their
discussion is deferred to section~\ref{section5}.
\par
The detectors involved in the discovery of the new meson resonances
all provide very good momentum resolution (charged
particles are tracked through a magnetic field), photon detection,
and particle identification. Even though the correct discrimination
of kaons against the pion background or identification of single
photons not originating from $\pi^0$ decay deserves focused
attention, we will assume here that the final state particles are
unambiguously identified.
\par

\subsubsection{The BABAR resonance}
\def\Dz{D^0}
\def\DsTT{D_{sJ}^*(2317)^+}
\def\DsTO{D_s^{*}(2112)^+}
The primary aim of the BABAR (and BELLE) experiments is
the study of CP violation in the $b\bar b$ system. The colliding
$e^+e^-$ beams produce however many final states; the study reported here
was done by searching inclusively (i.e. independent of other particles
\begin{figure}[h!]
\begin{minipage}[c]{0.56\textwidth}
\includegraphics[width=\textwidth]{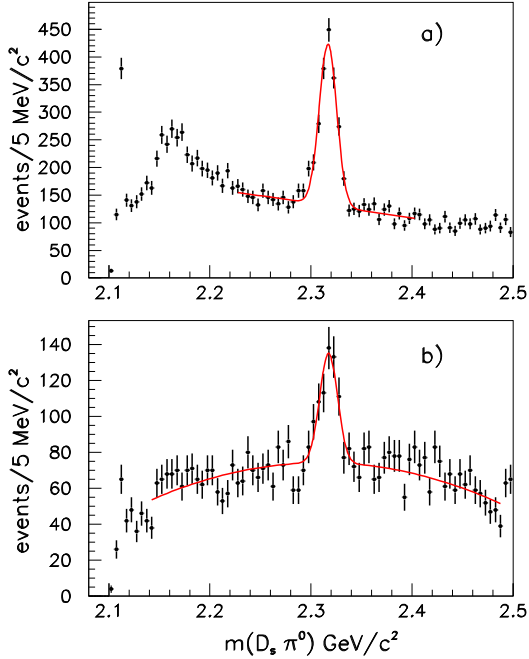}
\end{minipage}
\begin{minipage}[c]{0.4\textwidth}
\caption{\label{fig:dspiz} 
The $D_s^+\piz$ mass distribution
for (a) the decay $D_s^+  \to \rm K^+ K^- \pi^+$
and (b) the decay $D_s^+  \to \rm K^+ K^- \pi^+ \pi^0$.
The fits to the mass distributions as described in the
text are indicated by the curves~\protect\cite{Aubert:2003fg}.
}
\end{minipage}
\end{figure}
also produced in the same event) for events with two charged kaons, one
charged pion, and at 2 or 4 photons which can be combined to one
or two $\pi^0$. Calculating the $\rm K^+ K^- \pi^+$ 
(or $\rm K^+ K^- \pi^+ \pi^0$) invariant mass reveals 
contributions from the $D_s^+$. 
Then, the $D_s^+\pi^0$ invariant mass spectra are calculated and
shown in figure~\ref{fig:dspiz}. The two spectra refer to two different
decay modes of the  $D_s^+$.
The fit yields a mass $(2316.8\pm 0.4)$\,MeV 
and $(2317.6\pm 1.3)$\,MeV respectively and a width
estimated to be less than 10\,MeV~\cite{Aubert:2003fg}.
\par
The angular momentum of the $D_{sJ}(2317)^+$ is not known 
but due to its low mass $J=0$ seems to be most likely.
\subsubsection{The CLEO resonance}
The CLEO collaboration confirmed the $D_s^+(2317)$ and
observed a further $D_{sJ}$ resonance~\cite{Besson:2003cp}
called $D_{sJ}(2463)^+$. The mass difference spectrum 
$\Delta M(D_s^*\pi^0) = M(D_s^*\pi^0) - M(D_s^*)$ 
\begin{figure}[h!]
\begin{minipage}[c]{0.48\textwidth}
  \includegraphics*[width=\textwidth]{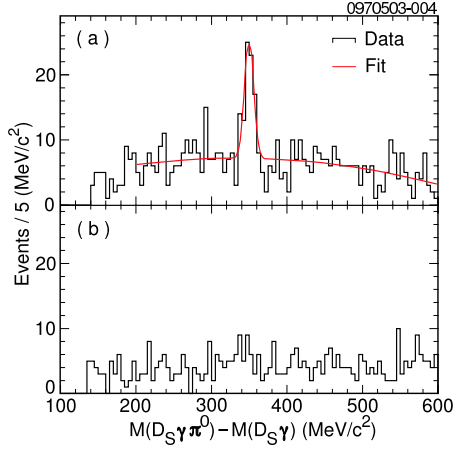}
\end{minipage}
\begin{minipage}[c]{0.4\textwidth}
\hskip 5mm  \caption{(a) The mass difference spectrum 
           $\Delta M(D_s^*\pi^0) = M(D_s\gamma\pi^0) - M(D_s\gamma)$
           for combinations where the $D_s\gamma$ system is consistent
           with $D_s^*$ decay 
           (b) The corresponding spectrum 
           where $D_s\gamma$ combinations are selected from the 
           $D_s^*$ side band regions which are defined as 
           $20.8 < |\Delta M(D_s \gamma) - 143.9\,\mbox{MeV/c}^2| < 33.8\,
           \mbox{MeV/c}^2$ (from~\protect\cite{Besson:2003cp}).
           }
  \label{fig:deltamstar}
\end{minipage}
\end{figure}
is shown in figure~\ref{fig:deltamstar} where the
$D_s^*$ is defined by its $D_s\gamma$ decay mode.
The signal is kinematically linked to the $D_{sJ}(2317)^+$; the two
resonances contribute mutually to a peaked background. A correlation
study demonstrates the existence of both resonances. The authors argue
that, likely, $J=1$. 

\subsubsection{The BELLE resonance}
\newcommand{\Mbc}{M_{\rm bc}}
\newcommand{\jp}{J/\psi}
\newcommand{\pipi}{\pi^{+}\pi^{-}}
\newcommand{\DE}{\Delta E}
The BELLE resonance is observed~\cite{Choi:2003ue} 
in the exclusive decay process $B^{\pm}\to\rm K^{\pm}X^0, 
X^0\to\pipi\jp$.  
The data was collected with the $e^+e^-$ beams set to 
the $\Upsilon(4S)$ resonance which decays into two B mesons. 
\par
The beam energy is more precisely known than the momenta of the decay
particles. Therefore, 
B mesons decaying to $\rm K^+\pipi\jp$
are reconstructed using the beam-energy constrained mass 
$\Mbc$ and  the energy difference $\DE$
\begin{eqnarray}
\Mbc\equiv\sqrt{(E_{\rm beam}^{\rm CM})^2-(p_B^{\rm CM})^2}
\qquad &\qquad
\DE\equiv E_B^{\rm CM} - E_{\rm beam}^{\rm CM},
\end{eqnarray}   
where $E_{\rm beam}^{\rm CM}$ is the beam 
energy in the CM system, 
and $E_B^{\rm CM}$ and $p_B^{\rm CM}$ are the CM energy and
momentum of the $B$ candidate. 
\par
\begin{figure}[htb]
\includegraphics[angle=270,width=\textwidth]{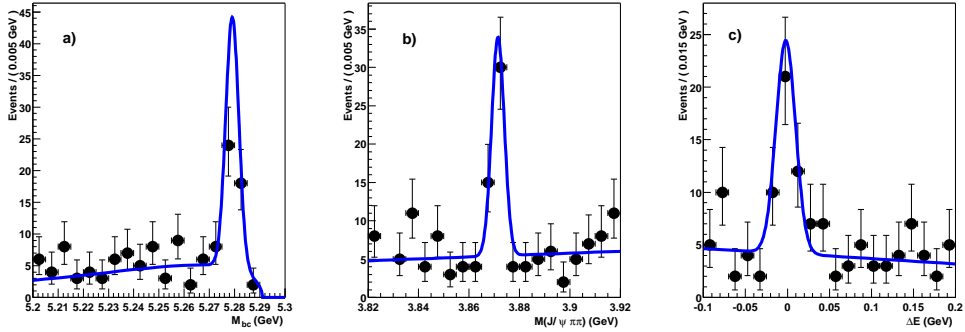}
\caption{\label{fig:X_pipijp_fit}
Signal-band projections of {\bf (a)} $\Mbc$,
{\bf (b)} $M_{\pipi\jp}$ and {\bf (c)} $\DE$ for the
$X(3872)\to\pipi\jp$ signal region with the results of the
un-binned fit superimposed~\protect\cite{Choi:2003ue}. 
}
\vspace*{-5mm}
\end{figure}

Figures~\ref{fig:X_pipijp_fit}(a), 
(b) and (c) show
the $\Mbc$, $M_{\pipi\jp}$, and $\DE$ distributions, respectively.  
The superimposed curves indicate the results of a fit giving 
 a mass of $(3872.0\pm 0.6 {\rm (stat)} \pm 0.5{\rm 
(syst)})$~MeV. 
\vskip 3mm
The result was confirmed at Fermilab by the CDFII collaboration
in proton antiproton collisions at $\sqrt{s}=1.96 TeV$~\cite{Acosta:2003zx}.

\subsubsection{The BES resonance}
\begin{figure}[htb]
\begin{minipage}[c]{0.48\textwidth}
\caption{\label{fig:2pg_thresh_fit}
(a) The near threshold $M_{\ppbar}-2m_p$ distribution for
the $\gamma\ppbar$ event sample.  The dashed
curve is a background function.  The dotted curve
indicates how the acceptance varies with $\ppbar$ invariant
mass; the dashed curve shows the fitted background function.
(b) The  $M_{\ppbar}-2m_p$ distribution with events weighted
by $q_0/q$~\protect\cite{Bai:2003sw}. }
\end{minipage}
\vspace*{-3mm}
\begin{minipage}[c]{0.48\textwidth}
\includegraphics*[width=\textwidth]{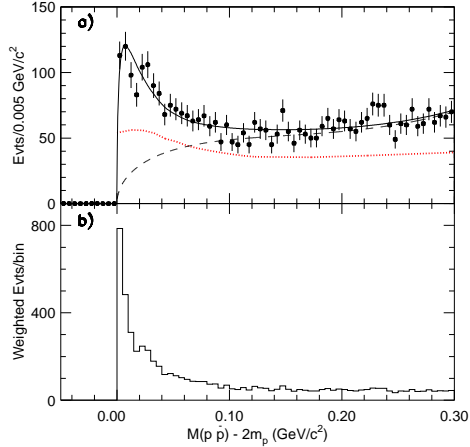}
\end{minipage}
\end{figure}

\newcommand{\jpsi}{J/\psi}
At BES a narrow enhancement is observed~\cite{Bai:2003sw} in radiative
$\jpsi\to\gamma\bar pp$ decays. Its mass is very
close to $2m_p$. No similar structure is seen in $\jpsi\to \pi^0\bar
pp$ decays. Figure~\ref{fig:2pg_thresh_fit} shows the $p\bar p$ mass
distribution without (a) and with (b) phase space correction.
The strong contribution at threshold suggests that $p$ and
$\bar p$ should be in an S-wave. 
If fit with a Breit Wigner function the peak mass is below $2m_p$, at
$M = (1859 ^{~+3}_{-10}~{\rm (stat)} ^{~+5}_{-25}~{\rm (sys))}~{\rm 
MeV}/c^2$, and the total width is $\Gamma < 30 $~MeV/$c^2$ at the 90\%
confidence level. Since charge conjugation  must be positive,
the most likely quantum numbers are $J^{PC}=0^{-+}$. The decay
angular distribution is not incompatible with this conjecture.

\vspace*{-3mm}

\vspace*{-10mm}
\subsubsection{Discussion}
\begin{figure}[h!]
\begin{minipage}[c]{0.70\textwidth}
\epsfig{file=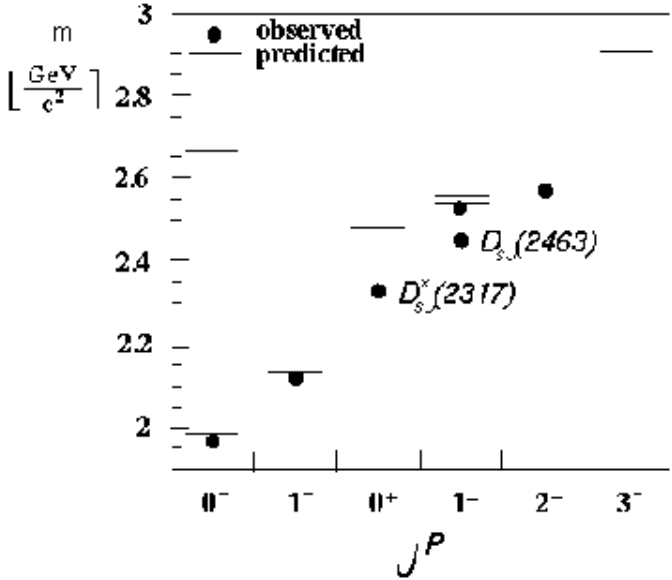,width=\textwidth,clip=on}
\end{minipage}
\begin{minipage}[c]{0.28\textwidth}
\caption{\label{fig:ds}
The mass spectrum of  $D_{sJ}^+$ resonances and their 
quantum numbers (mostly to be confirmed). The two 
states $^3P_1$ and $^1P_1$ differ only by their C-parity which is
undefined for states with open charm and/or strangeness. Hence the
two states mix to form the two mass eigenstates. The predictions
of Godfrey and Isgur ~\protect\cite{Godfrey:xj} are represented 
by lines and the data by $\bullet$. 
}
\end{minipage}
\vspace*{-5mm}
\end{figure}

The two new resonances, $D_{s0}(2317)^+$ and $D_{s1}(2463)^+$, 
belong to the family of $D_{sJ}^+$ resonances where some members are
already known~\cite{Hagiwara:fs}: 
the pseudoscalar ground state  $D_{s}^+$ at 1970\,MeV,
the vector state $D_{s1}^*(2112)^+$ and two states with orbital
angular momentum one, the  $D_{s1}(2536)^+$ and  $D_{s2}(2573)^+$.
Figure~\ref{fig:ds} shows the mass spectrum of  $D_{sJ}^+$ resonances.
We expect 2 states with $L=0$ and $S=0$ or $S=1$, 
and four states with $L=1$ and $S=0$ or $S=1$ where
$S=1$ and $L=1$ couple to $J=0, 1, 2$. 
\clearpage
\noindent
In addition higher mass
resonances, radial and higher orbital excitations can
be seen to be predicted. The new states do not fit well to the
expected masses and there is an intense discussion why the
masses of the new mesons are so low. 
For mesons with one heavy and one light
quark one may assume that the light quark spin and orbital angular
momenta couple to $j_{\ell}=l_{\ell}+s_{\ell}$, which then couples
to the spin of the heavy quark. The heavy quark 
is supposed to be so heavy that
its motion can be neglected. Models based on this assumption are
called {\it heavy quark effective theories} (HQET).
Within this frame the masses can be reproduced reasonably 
well~\cite{Dai:2003yg} but other
approaches are certainly not excluded.
\par
The BELLE resonance has a mass of $3872.0$~MeV and is thus far
above the $\rm D\bar D$ threshold. Its decay mode shows that it
is a state with hidden charm, that it contains a $c\bar c$ pair. 
The $\psi(3770)$ has a full
width of 24\,MeV; the BELLE resonance with its higher mass should 
be wider. 
It is not, it is narrow\,! This is unexpected. A hint for a solution
may lie in the fact that its mass 
is very close to the $M_{D} + M_{D^*}$ mass threshold. There are
speculations that the resonance might be a $c\bar cg$ state where
the  $c$ quark and the $\bar c$ antiquark couple to a color octet
which then is color--neutralized by the gluon 
field~\cite{Close:2003sg}. Such objects are called hybrids. 
\par
The BES resonance is a meson with strong coupling to
proton plus antiproton.  It is narrow,too. It might decay into
multi-meson final states with ample phase space but it does not
(otherwise it would be a broad resonance). Hence it is interpreted as
a $p\bar p$ bound state~\cite{Datta:2003iy}. 
While $p\bar p$ bound states close to the
$p\bar p$ threshold having very high intrinsic orbital angular
momenta  might survive annihilation, a narrow state with
pseudoscalar quantum numbers seems very unlikely to 
exist~\cite{Klempt:ap}. Th. Walther pointed out that the mass
distribution might be faked by bremsstrahlung~\cite{Walcher}. 

\vspace*{-2mm}
\subsection{\label{section1.6}Baryons}
\par
Symmetries play a decisive role in the classification of baryon
resonances. The baryon wave function
can be decomposed into a color wave function, which is
antisymmetric with respect to the exchange of two quarks, the
spatial and the spin-flavor wave function. The second ket in the
wave function
\begin{eqnarray}
\rm  |qqq> = |colour>_A \cdot &\rm |space;\ spin, flavour>_S
 \\
&\hspace*{-10mm}\rm O(6) \qquad SU(6) \nonumber
\end{eqnarray}
must be symmetric.
The SU(6) part can be decomposed into SU(3)$\otimes$SU(2).
\subsubsection{The spatial wave function}
The motion of three quarks at positions $r_i$ can be described using
Jacobean coordinates:
\vspace*{-5mm}
\begin{eqnarray}
\qquad r_1 - r_2  \label{rho}\\
\qquad r_1 + r_2 - 2r_3\label{lambda}   \\
\qquad r_1 + r_2 + r_3  \label{cms}
\end{eqnarray}
Equation~(\ref{cms}) describes the baryon center--of--mass motion
and is not relevant for the internal dynamics of the 3--quark
system. There remain two separable motions, called $\rho$ and
$\lambda$, where the first one is antisymmetric and the second symmetric
with respect to the exchange of quarks 1 and 2.

\subsubsection{SU(3) and SU(6)}
From now on, we restrict ourselves to light flavors i.e. to $up, down$
and $strange$ quarks. The flavor wave function is then given by SU(3)
and allows a decomposition
\begin{equation}
\rm 3 \otimes\ 3 \otimes\ 3 = 10_S \oplus\ 8_M \oplus\ 8_M \oplus\ 1_A,
\label{su3}
\end{equation}
into a decuplet symmetric w.r.t. the exchange of any two quarks and 
an antisymmetric singlet and two octets of mixed symmetry. The
two octets have different SU(3) structures and only one of them
fulfills the symmetry requirements in the total wave functions.
Remember that the SU(3) multiplets contain six particle families:
\bc
\renewcommand{\arraystretch}{1.4}
\begin{tabular}{ccccccc}
\hline
\hline
SU(3) &  N  & $\Delta$ & $\Lambda$ & $\Sigma$ & $\Xi$ & $\Omega$ \\
 1    & no  &  no      &   yes     &   no     &   no  &    no    \\
 8    & yes &  no      &   yes     &   yes    &   yes &    no    \\
10    & no  &  yes     &   no      &   yes    &  yes  &    yes   \\
\hline
\hline
\end{tabular}
\renewcommand{\arraystretch}{1.0}
\ec
\par
The spin-flavor wave function
can be classified according to SU(6).
\begin{equation}
\rm 6 \otimes\ 6 \otimes\ 6 = 56_S \oplus\ 70_M \oplus\ 70_M \oplus\ 20_A
\label{su6}
\end{equation}
In the ground state the spatial wave function is symmetric,
and the spin-flavor wave function has to be symmetric too.
Then, spin and flavor can both be symmetric; this is the case
for the decuplet. Spin and flavor wave functions can individually
have mixed symmetry, with symmetry in the combined spin-flavor wave
function. This coupling represents the baryon octet. The 56-plet thus
decomposes into a decuplet with spin 3/2 (four spin projections) plus
an octet with
spin 1/2 (two spin projections) according to
\begin{equation}
56 = {^4}10\ \oplus\  ^{2}8.
\end{equation}
\par
\begin{figure}[h!]
\begin{center}
\setlength{\unitlength}{0.7mm}
\begin{picture}(150.00,90.00)
\put(-10.00,45.00){\vector(1,0){70.00}}
\put(20.00,10.00){\vector(0,1){70.00}}
\put(62.50,42.50){\makebox(5.00,5.00){$\bf  I_3$}}
\put(18.50,85.00){\makebox(5.00,5.00){\bf Octet }}
\put(20.50,78.00){\makebox(5.00,5.00){\bf  S}}
\put(20.00,45.00){\circle*{2.00}}
\put(45.00,45.00){\circle*{2.00}}
\put(-5.00,45.00){\circle*{2.00}}
\put(7.50,70.00){\circle*{2.00}}
\put(32.50,70.00){\circle*{2.00}}
\put(7.50,20.00){\circle*{2.00}}
\put(32.50,20.00){\circle*{2.00}}
\put(20.00,45.00){\circle{0.00}}
\put(20.00,45.00){\circle{5.00}}
\put(7.50,70.00){\line(1,0){25.00}}
\put(32.50,70.00){\line(1,-2){12.50}}
\put(45.00,45.00){\line(-1,-2){12.50}}
\put(32.50,20.00){\line(-1,0){25.00}}
\put(7.50,20.00){\line(-1,2){12.50}}
\put(-5.00,45.00){\line(1,2){12.50}}
\put(-10.00,70.00){\makebox(15.00,5.00)[r]{n}}
\put(35.00,70.00){\makebox(15.00,5.00)[l]{p}}
\put(14.00,47.50){\makebox(12.50,5.00)[l]{$\bf \Lambda$}}
\put(22.00,38.50){\makebox(12.50,5.00)[l]{$\bf \Sigma^0$}}
\put(-16.25,39.00){\makebox(13.25,5.00)[r]{$\bf \Sigma^-$}}
\put(44.00,39.00){\makebox(13.75,5.00)[l]{$\bf \Sigma^+$}}
\put(-7.00,15.00){\makebox(15.00,5.00)[r]{$\bf \Xi^-$}}
\put(35.00,15.00){\makebox(15.00,5.00)[l]{$\bf \Xi^0$}}
\put(120.50,85.00){\makebox(5.00,5.00){\bf Decuplet }}
\put(165.50,42.50){\makebox(5.00,5.00){$\bf I_3$}}
\put(123.50,78.00){\makebox(5.00,5.00){\bf S}}
\put(93.00,45.00){\vector(1,0){70.00}}
\put(123.00,-10.00){\vector(0,1){90.00}}
\put(123.00,45.00){\circle*{2.00}}
\put(148.00,45.00){\circle*{2.00}}
\put(98.00,45.00){\circle*{2.00}}
\put(85.50,70.00){\circle*{2.00}}
\put(110.50,70.00){\circle*{2.00}}
\put(135.50,70.00){\circle*{2.00}}
\put(160.50,70.00){\circle*{2.00}}
\put(110.50,20.00){\circle*{2.00}}
\put(135.50,20.00){\circle*{2.00}}
\put(123.00,-5.00){\circle{4.80}}
\put(123.00,-5.00){\circle{4.40}}
\put(123.00,-5.00){\circle{4.00}}
\put(123.00,-5.00){\circle{3.60}}
\put(123.00,-5.00){\circle{3.20}}
\put(123.00,-5.00){\circle{2.80}}
\put(123.00,-5.00){\circle{2.40}}
\put(123.00,-5.00){\circle*{2.00}}
\put(85.50,70.00){\line(1,0){75.00}}
\put(123.00,-5.00){\line(-1,2){37.50}}
\put(123.00,-5.00){\line(1,2){37.50}}
\put(73.00,72.00){\makebox(15.00,5.00)[r]{$\bf \Delta^-$}}
\put(100.00,72.00){\makebox(15.00,5.00)[r]{$\bf \Delta^0$}}
\put(133.00,72.00){\makebox(15.00,5.00)[l]{$\bf \Delta^+$}}
\put(160.00,72.00){\makebox(15.00,5.00)[l]{$\bf \Delta^{++}$}}
\put(84.75,39.00){\makebox(13.25,5.00)[r]{$\bf \Sigma^{*-}$}}
\put(125.50,39.00){\makebox(12.50,5.00)[l]{$\bf \Sigma^{*0}$}}
\put(148.00,39.00){\makebox(13.75,5.00)[l]{$\bf \Sigma^{*+}$}}
\put(100.50,17.50){\makebox(12.50,5.00)[l]{$\bf \Xi^{*-}$}}
\put(138.50,17.50){\makebox(12.50,5.00)[l]{$\bf \Xi^{*0}$}}
\put(128.50,-6.00){\makebox(12.50,5.00)[l]{$\bf \Omega^-$}}
\end{picture}
\end{center}
\caption{Octet and decuplet baryons}
\label{octet+decu}
\end{figure}
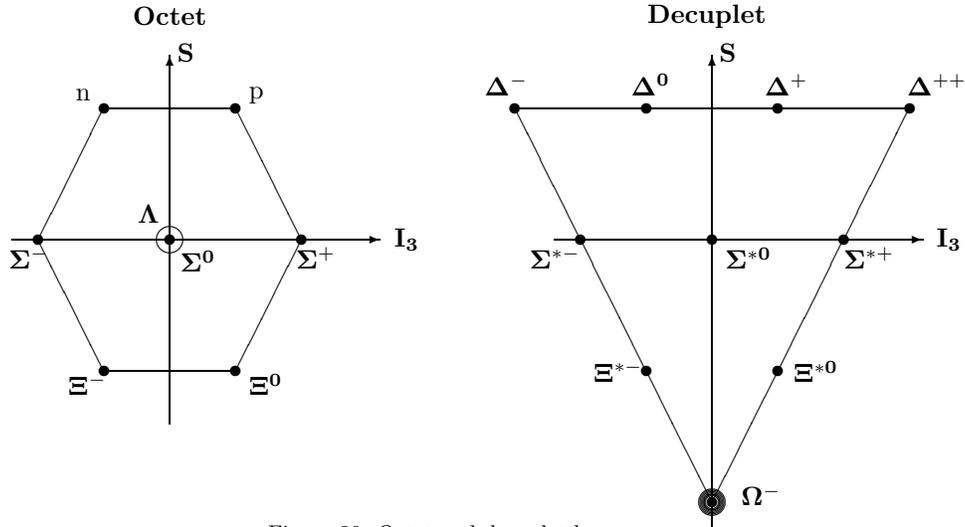

Octet and decuplet are schematically presented in figure~\ref{octet+decu}.
The $\Omega^-$ was predicted~\cite{Gell-Mann:nj}
on the basis of SU(3) by Gell-Mann.
Its experimental discovery~\cite{Barnes:64} 
was a striking confirmation of SU(3)
and of the quark model.
\par
The spin-flavor wave functions can also have mixed
symmetry. The 70-plet can be written as
\begin{equation}
70 = {^2}10\ \oplus\ {^4}8\ \oplus\ {^2}8\ \oplus\ {^2}1.
\end{equation}
Decuplet baryons, e.g. $\Delta^*$, in the 70-plet
have intrinsic spin
1/2; octet baryons like excited nucleons can have spin 1/2 or
3/2. Singlet baryons with J=1/2, the $\Lambda_1$ resonances only exist
for spin-flavor wave functions of mixed symmetry. The ground
state (with no orbital excitation) has no $\Lambda_1$.
\par
The 20-plet is completely antisymmetric and requires an antisymmetric
spatial wave function. It is decomposed into an octet with spin 1/2
and a singlet with spin 3/2:
\begin{equation}
20 = {^2}8\ \oplus\ {^4}1.
\end{equation}
\par
\vspace*{-6mm}
\subsubsection{Regge trajectories}
\begin{figure}[h!]
\bc
\hspace*{5mm}\epsfig{file=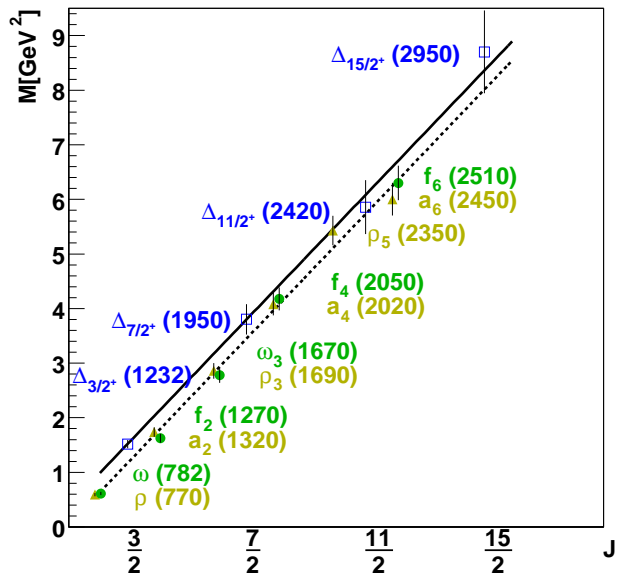,width=0.7\textwidth}
\ec
\vspace*{-5mm}
\caption{
Regge trajectory for $\Delta^*$ resonances in comparison
to the meson trajectory.
}
\label{Delta-mesons}
\end{figure}

In figure~\ref{Delta-mesons} we compare the Regge trajectory
for $\Delta^*$ resonances having $J=L+3/2$ with the meson
trajectory.
The offset is given by the $\Delta (1232)$ mass but the slope
is the same for both trajectories. Mesons
and baryons have the same Regge slope. The QCD forces between
quarks and antiquarks are the same as
those between quark and diquark.
This is an important observation which will be
taken up again in section~5.

\vspace*{-12mm}

%% file: Chapter2_proc.tex
\section{\label{section2}Particle decays and partial wave analysis}
The aim of an analysis is
to determine masses and widths of resonances,
their spins, parities and flavor structure.
\par
In the simplest case, a resonance is described by a
single--channel Breit-Wigner amplitude; however, a resonance may
undergo distortions. The opening of a threshold for a second decay
mode reduces the intensity in the channel in study, an effect
which is accounted for by use of the Flatte formula. The
amplitudes for two resonances close by in masses must not be
added; the sum would violate unitarity. Instead, a K--matrix must be
used. The partial decay widths to different final states may
require the use of multichannel analyses. The couplings of
resonances follow SU(2) and SU(3) relations.
\par
Spin and parity of a resonance are reflected in their
decay angular distributions. These can be described in the
non--relativistic Zemach \cite{Zemach} or relativistic 
Rarita-Schwinger \cite{Rarita}
formalism, or using the helicity formalism \cite{Jacob:at}.

\subsection{\label{section2.1}Particle decays}
The transition rate for particle decays are given by Fermi's
golden rule:
$$
 T_{if} = 2\pi |{\cal M}|^2 \rho (E_f)
$$
$T_{if}$ is the transition probability per unit time. With N particles,
the number of decays in the time interval $dt$ is $N T_{if} dt$ or
$$
dN = - N T_{if} dt
$$
and
$$
N = N_0 e^{-T_{if}t} = N_0 e^{-(t/\tau)} = N_0 e^{-\Gamma t}
$$
$$
\Gamma\tau = \hbar = 1
$$
The latter equation is the well--known uncertainty principle.
\par
Now we turn to short--lived states in quantum mechanics. Consider
a state with energy $E_0 =\hbar\omega$; it is characterized by a
wave function $\psi (t) = \psi_0(t)e^{-iE_0t}$. Now we allow it to
decay:

\begin{minipage}[c]{0.50\textwidth}
\epsfig{file=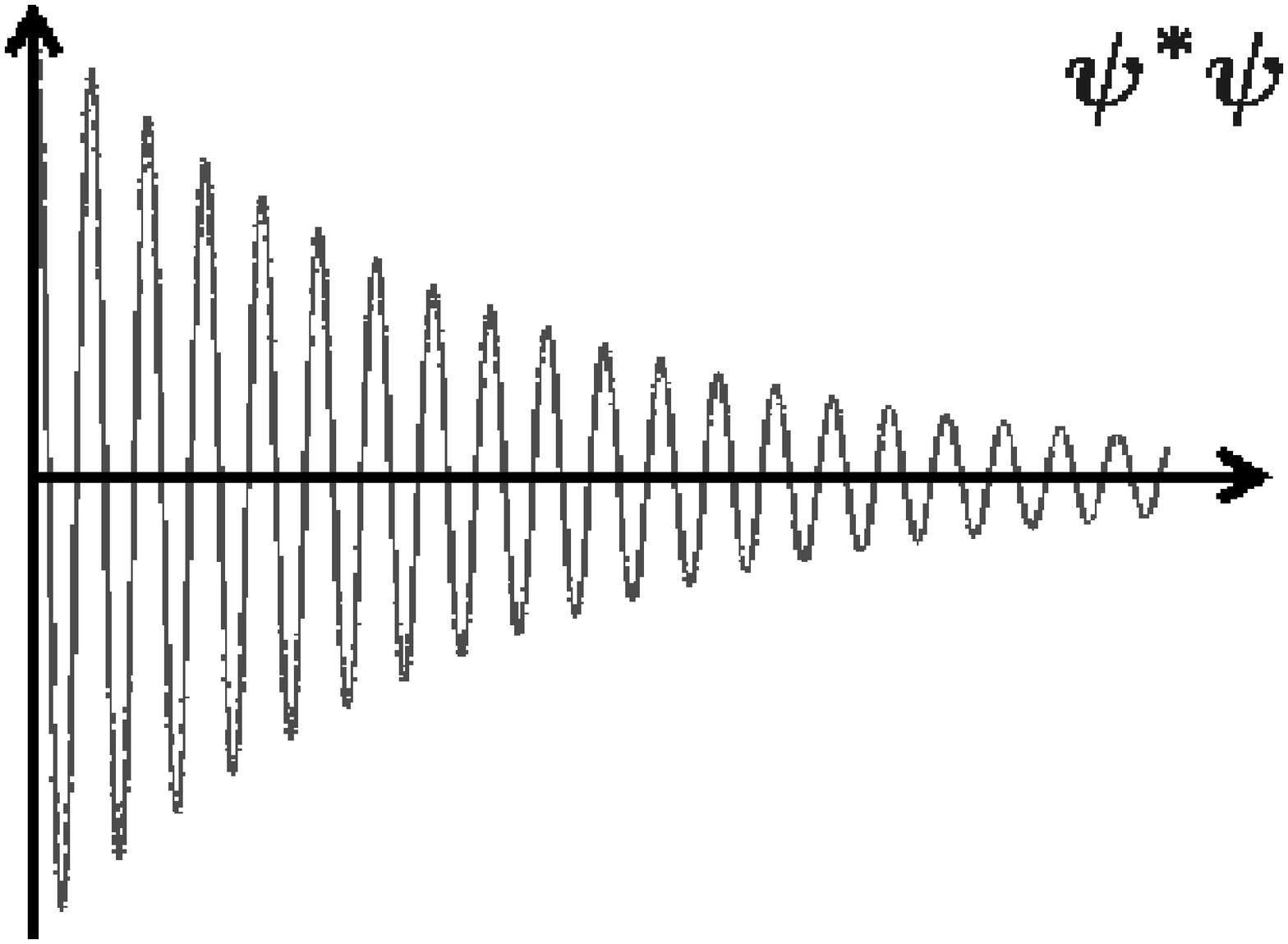,width=\textwidth,clip=}
\end{minipage}
\begin{minipage}[c]{0.48\textwidth}
\vspace*{-8mm} \bc $\psi\psi =  \psi^*_0(t)\psi_0(t)$ $=
\psi^*_0(t=0)\psi_0(t=0)e^{-t/\tau}$. 
\vskip 2mm 
Probability density must decay exponentially. \ec
$$\longrightarrow
\psi (t) = \psi(t=0)e^{-iE_0t}e^{-t/2\tau}
$$
\end{minipage}

A damped oscillation contains more than one frequency. The frequency
distribution can be calculated by the Fourier transformation:
\begin{eqnarray*}
f(\omega) &=& f(E) = \int_0^{\infty}{\psi(t=0)e^{-iE_0t - t/2\tau}}
e^{iEt}dt \\
&=& \int_0^{\infty}\psi(t=0)e^{-i\left((E_0-E) -1/2\tau\right)t}dt
\quad = \quad \frac{\psi(t=0)}{\left(E_0-E\right) - i/\left(2\tau\right)}
\end{eqnarray*}

\begin{figure}[h!]
\bc
\epsfig{file=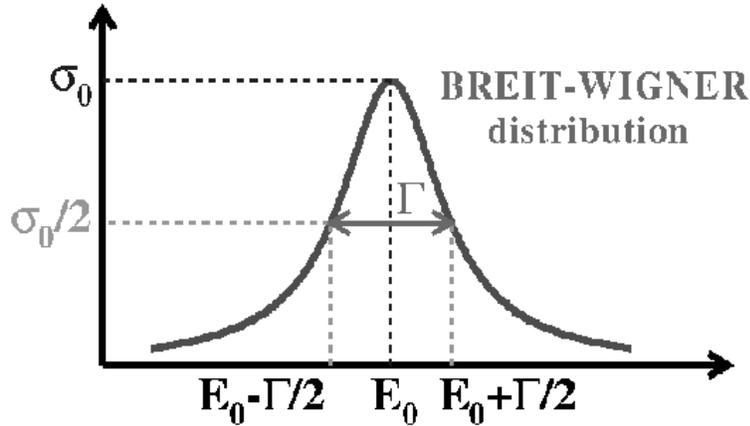,width=0.8\textwidth,clip=}
\ec
\caption{\label{bw}
Breit--Wigner resonance.}
\end{figure}
The probability of finding the  energy $E$ is given by
$$
f^*(E)f(E) =
\frac{|\psi(t=0)|^2}{\left(E_0-E\right)^2 +
1/\left(2\tau\right)^2}
$$
and, replacing $\tau$  with $1/\Gamma$,  
$$
\frac{(\Gamma/2)^2}{\left(E_0-E\right)^2 + (\Gamma/2)^2} 
$$
gives the Breit--Wigner function. However, 
resonances are described by amplitudes:
$$
BW(E) = \frac{\Gamma/2}{\left(E_0-E\right) - i\Gamma/2}
      = \frac{1/2}{\left(E_0-E\right)/\Gamma - i/2}.
$$
With
$$2\left(E_0-E\right)/\Gamma = \cot\delta:
\quad
f(E) = \frac{1}{\cot\delta - i} = e^{i\delta}\sin\delta =
\frac{i}{2}\left( 1 - e^{-2i\delta}\right)
$$
This formula can be derived from $S$ matrix theory; $\delta$ is called phase
shift. The amplitude is zero for $\Gamma/(E-E_0) << 0$ and starts
to be real and positive with a small positive imaginary part. For
$\Gamma/(E-E_0) >> 0$ the amplitude is small, real and positive
with an small negative imaginary part. The amplitude is purely
imaginary $(i)$ for $E=E_0$. The phase $\delta$ goes from 0 to
$\pi/2$ at resonance and to $\pi$ at high energies.

\subsubsection{The Argand circle}
\begin{figure}[h!]
\begin{minipage}[c]{0.6\textwidth}
\epsfig{file=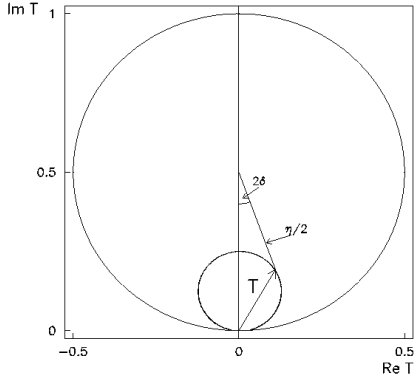,width=0.8\textwidth,clip=}
\end{minipage}
\begin{minipage}[c]{0.33\textwidth}
\caption{\label{circle}
Scattering amplitude
T in case of inelastic
scattering. Definition
of phase $\delta$ and
inelasticity $\eta$.}
\end{minipage}
\end{figure}

The amplitude can be 
represented conveniently in an Argand
diagram (figure~\ref{circle}). 
The scattering amplitude starts when the real and imaginary
part both equal zero. In case of the absence of inelasticities
(only elastic scattering is allowed), the scattering amplitude
makes one complete circle while the energy runs across the
resonance. Inelasticities reduce the amplitude which always stays
inside of the circle.

\begin{figure}[h!]
\begin{minipage}[c]{0.49\textwidth}
\hspace{-2mm}\epsfig{file=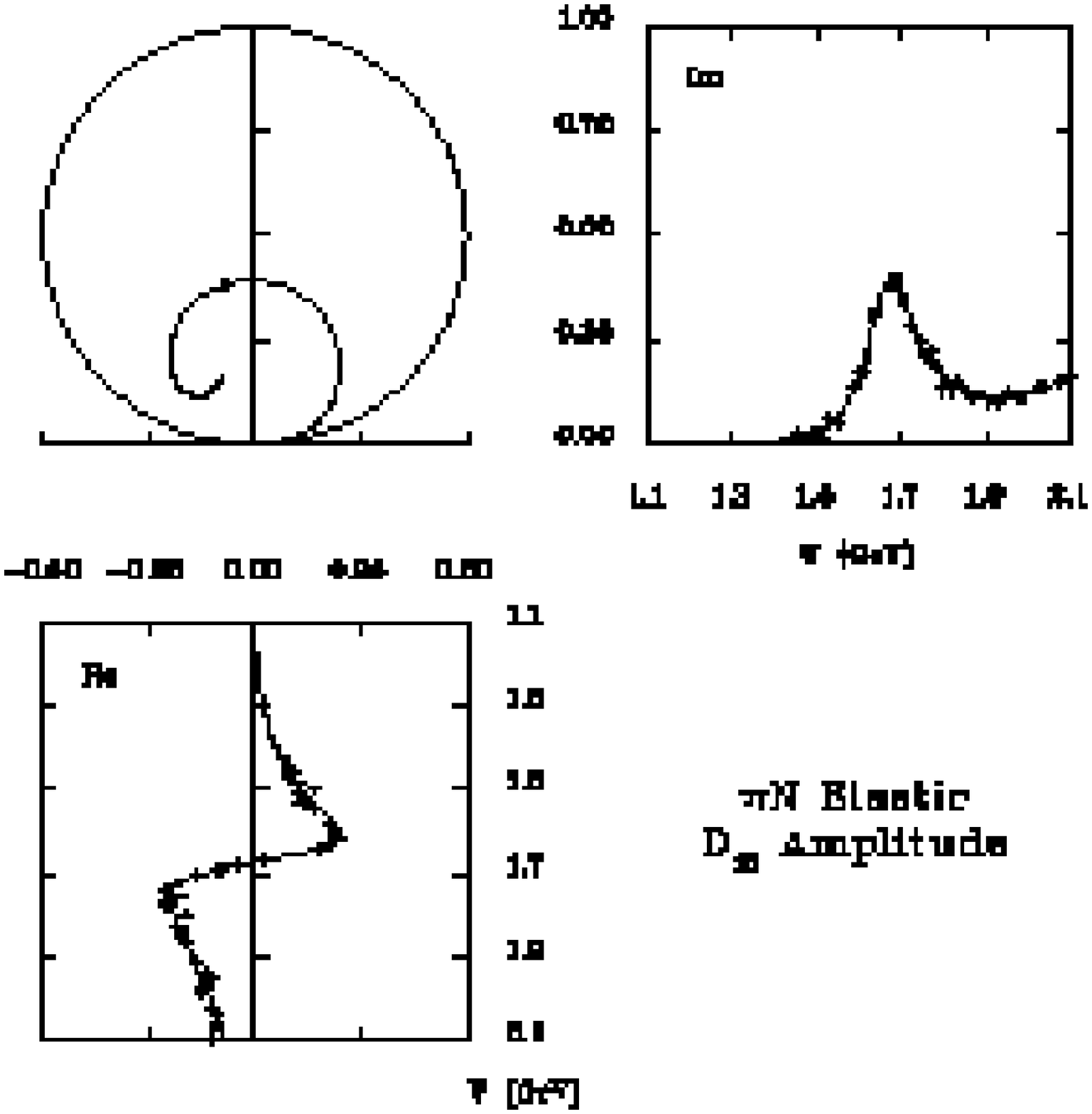,width=0.9\textwidth,clip=}
\end{minipage}
\begin{minipage}[c]{0.49\textwidth}
\hspace{-2mm}\epsfig{file=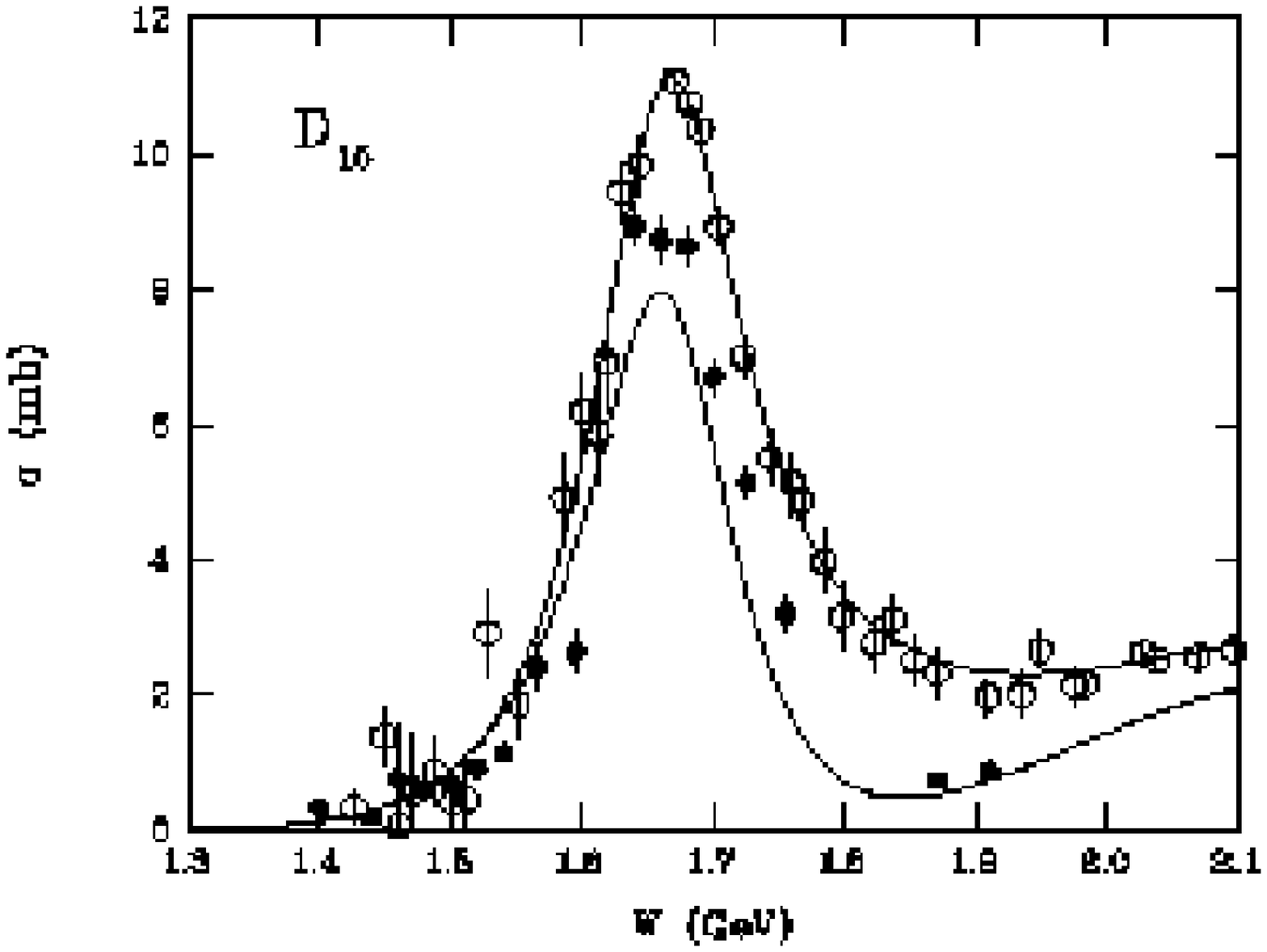,width=1\textwidth,clip=}
\caption{Elastic channel (upper curve), $ \pi p\to \pi\pi p$ (lower
curve), difference $ \pi p\to \eta p$ or $ \pi p\to \Lambda$K;
from~\protect\cite{Manley:fi}.}
\label{argand}
\end{minipage}
\end{figure}

A concrete example~\cite{Manley:fi} is shown in
figure~\ref{argand}: the
Argand diagram and cross section for $ \pi^- p\to \pi^- p$
via formation of the N(1650)$ D_{1,5}$ resonance (L=2).

\subsubsection{The K--matrix}
Consider two-body scattering from the initial state i
to the final state f, $ ab\to cd$. Then
$$
 \frac{d\sigma_{fi}}{d\Omega} =
\frac{1}{(8\pi)^2s}\frac{q_f}{q_i}|{\cal{M}}_{fi}|^2 =
|f_{fi}(\Omega )|^2
$$
where $ s=m^2=$ squared CMS energy; q break-up momenta. In case
of spins, one has to average over initial spin components and
sum over final spin components. The scattering amplitude
can be expanded into partial-wave amplitudes:
$$
 f_{fi}(\Omega ) = \frac{1}{q_i}\sum
(2J+1)T^J_{fi}D^{J*}_{\lambda\mu}(\Phi,\Theta,0)
$$
One may remove the probability that the particles do not interact by
$ S = I + iT$. Probability conservation yields $ SS^{\dagger} = I$ from
which one may define
$$
 K^{-1} = T^{-1} + iI.
$$

From time reversal follows that K is real and symmetric.
Below the lowest inelasticity threshold the $S$--matrix can be written as
$$
 S = e^{2i\delta} \qquad T = e^{i\delta}sin\delta
$$
For a two-channel problem, the S-matrix is a $2\times 2$ matrix with
$$
S_{ik}S^*_{jk} = \delta_{ij}.
$$
$$
S_{11} = \eta e^{2i\delta_1} \qquad
S_{22} = \eta e^{2i\delta_2} \qquad
S_{12} = ie^{i\delta_1+\delta_2}\sqrt{1-\eta^2}.
$$
So far, the T matrix is not relativistically invariant. This
can be achieved by introducing $\hat T$:
$$
T_{ij} = \left(\rho_i^{\frac{1}{2}}\right) \hat T_{ij}
\left(\rho_i)^{\frac{1}{2}}\right)
$$
where $ \rho_n = 2q_n/m$ are phase space factors.
The amplitude now reads
$$
\hat T^J_{fi}(\Omega ) = \frac{1}{q_i}\sum
(2J+1)\hat T^J_{fi}D^{J*}_{\lambda\mu}(\Phi,\Theta,0)
$$
with  \vspace*{-5mm}
$$ \rho_n = 2q_n/m =
\sqrt{\left[1-\left(\frac{m_a+m_b}{m}\right)^2\right]
\left[1-\left(\frac{m_a-m_b}{m}\right)^2\right]}
$$
Now the following relations hold:
$$
\hat T = \frac{1}{\rho}e^{i\delta}sin\delta ; \qquad\ \hat K^{-1} =
\hat T^{-1} +i\rho
$$
$$
\hat T = \frac{1}{1-\rho_1\rho_2\hat D -i\left(\rho_1\hat K_{11}
+ \rho_2\hat K_{22}\right)}
                             \left(
                             \begin{array}{cc} 
                             \hat K_{11} -i\rho_2\hat D & \hat K_{12}\\
                             \hat K_{21} & \hat K_{22} -i\rho_1\hat D
                             \end{array}
                             \right)
$$
and $ \hat D = \hat K_{11}\hat K_{22} - \hat K_{12}^2$.
In case of resonances, we have to introduce poles into the K-matrix:
$$
K_{ij} = \sum_\alpha\frac{g_{\alpha i}(m)g_{\alpha
j}(m)}{m_{\alpha}^2-m^2} + c_{ij}
$$
$$
\hat K_{ij} = \sum_\alpha\frac{g_{\alpha i}(m)g_{\alpha
j}(m)}{\left(m_{\alpha}^2-m^2\right)\sqrt{\rho_i\rho_j}} + \hat c_{ij}
$$
The coupling constants g are related to the partial decay widths.
$$
g_{\alpha i}^2 = m_{\alpha}\Gamma_{\alpha i}(m)\qquad\
\Gamma_{\alpha}(m) = \sum_i \Gamma_{\alpha i}(m)
$$
The partial decay widths and couplings depend on the available phase space,
$$
g_{\alpha i}(m) = g_{\alpha i}(m_{\alpha}) B^l_{\alpha
i}(q,q_{\alpha})\sqrt{\rho_i}
$$
These formulae can be used in the case of several resonances (sum
over $\alpha$) decaying into different final states (i). The
K--matrix preserves unitarity and analyticity. It is a
multi-channel approach.
\par
\begin{figure}[h]
\begin{tabular}{cc}
\hspace*{-5mm}\epsfig{file=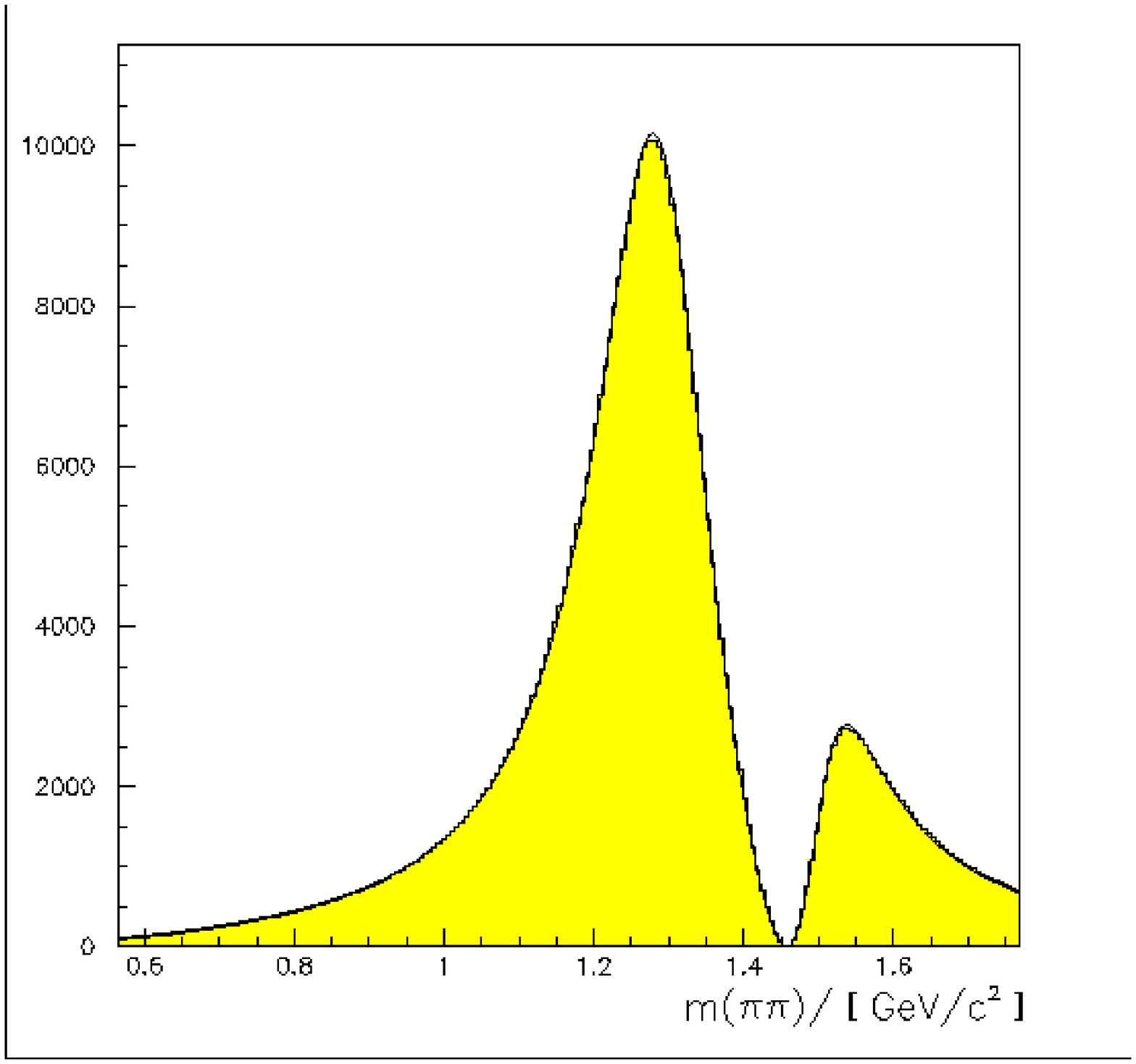,width=0.34\textwidth,clip=}
\hspace*{-5mm}\epsfig{file=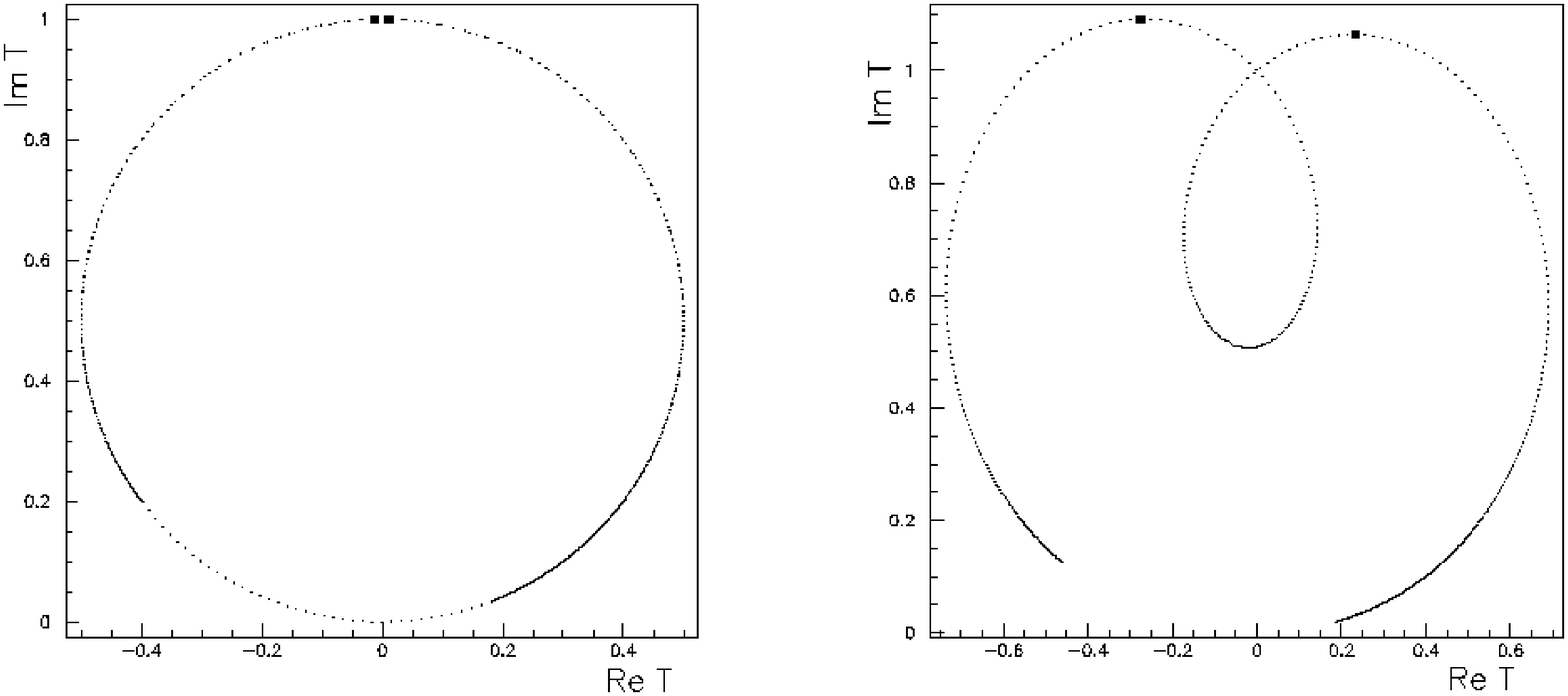,width=0.70\textwidth,clip=}&
\end{tabular}
\caption{\label{km}Two close--by resonances added within a
K--matrix and as sum of two  Breit--Wigner amplitudes
(see~\protect\cite{Chung:dx}.}
\end{figure}

An example for the use of the K--matrix is shown in figure~\ref{km}
where two scalar resonances are added, first within the
K--matrix formalism (left and center) and then, on the right as a sum of 
two Breit--Wigner amplitudes. The latter prescription
violates unitarity. 
A more detailed description can be found 
in~\cite{Chung:dx}.

\subsubsection{Three-body decays}
A particularly important case is that of annihilation into three
final-state particles, $M\to (m_1,m_2,m_3)$. The 3 four--momenta
define 12 dynamical quantities which are constrained by energy and momentum
conservation. The three masses can be determined from a
measurement of the particle momenta,  and from 
dE/dx or time--of--flight measurements. Three arbitrary Euler
angles define the orientation of the three-body system in space.
Hence two variables are needed (and suffice) to be identify the
full dynamics. The two variables used to define the
Dalitz plot are customly chosen as squared
invariant masses $ {m_{12}}^2$ and $ {m_{13}}^2$.
Then the partial
width can be expressed as
$$
\label{dalitz1}
d\Gamma  = \frac{1}{(2\pi)^3}\frac{1}{32M}
        \cdot \overline{|{\cal M}|^2} \cdot dm_{12}^2dm_{13}^2.
$$
Events are uniformly distributed in the ($ m_{12}^2,m_{13}^2$) plane if the
reaction leading to the three particle final state has no internal dynamics.
If particles with spin are produced in flight however, 
the spin may be aligned, the components $ m_j$ can have
a non-statistical distribution and the angular distribution can be distorted.

\subsubsection{The Dalitz plot}
Events are represented in a Dalitz plot by one point
in a plane defined by $m_{12}^2$ in x and $m_{23}^2$ in the y direction.
Since the Dalitz plot represents the phase space, the distribution
is flat in case of absence of any dynamical effects. 
Resonances in $ m_{12}^2$ are given by a vertical line and those in
$ m_{23}^2$ as horizontal lines. Since
$$
m_{13}^2 = (M_p^2 + m_1^2 + m_2^2 + m_3^2) - (m_{12}^2 - m_{23}^2)
$$
particles with defined $ m_{13}^2$ mass are found on the second diagonal.
\par
From the invariant mass of particles 2 and 3
$$
m_{23}^2 = \left(E_2+E_3\right)^2 - \left(\vec p_2+\vec p_3\right)^2
$$
we derive
$$
m_{23}^2 = (m_2^2 + m_3^2 + 2\cdot E_2\cdot E_3) -
              (2\cdot |\vec q_2|\cdot |\vec q_3|) \cdot \cos\theta
$$
with $ \theta$ being the angle between $ \vec q_2$ and $ \vec q_3$.
This can be rewritten as
$$
m_{23}^2 =
\left[\left(m_{23}^2\right)_{max}+\left(m_{23}^2\right)_{min}\right] +
\left[\left(m_{23}^2\right)_{max}-\left(m_{23}^2\right)_{min}\right]
\cdot \cos\theta
$$
For a fixed value of $ m_{1,2}$ the momentum vector $ \vec p_3$
has a $ \cos\theta$ direction w.r.t. the recoil $ \vec p_1$
proportional to $ m_{23}^2$.

\begin{figure}[h!]
\begin{tabular}{cc}
\hspace*{-5mm}\epsfig{file=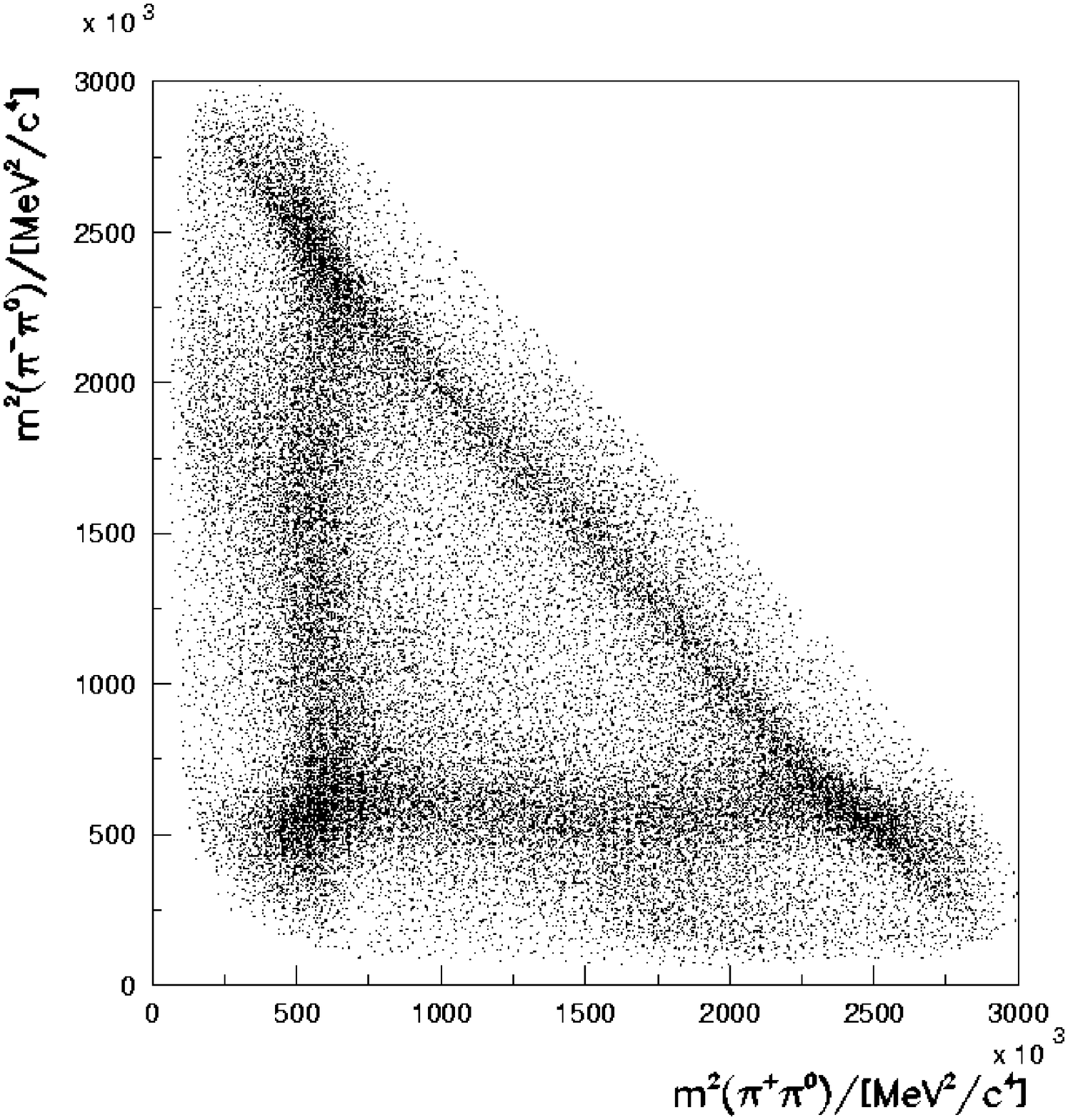,width=0.5\textwidth}&
\hspace*{-13mm}\epsfig{file=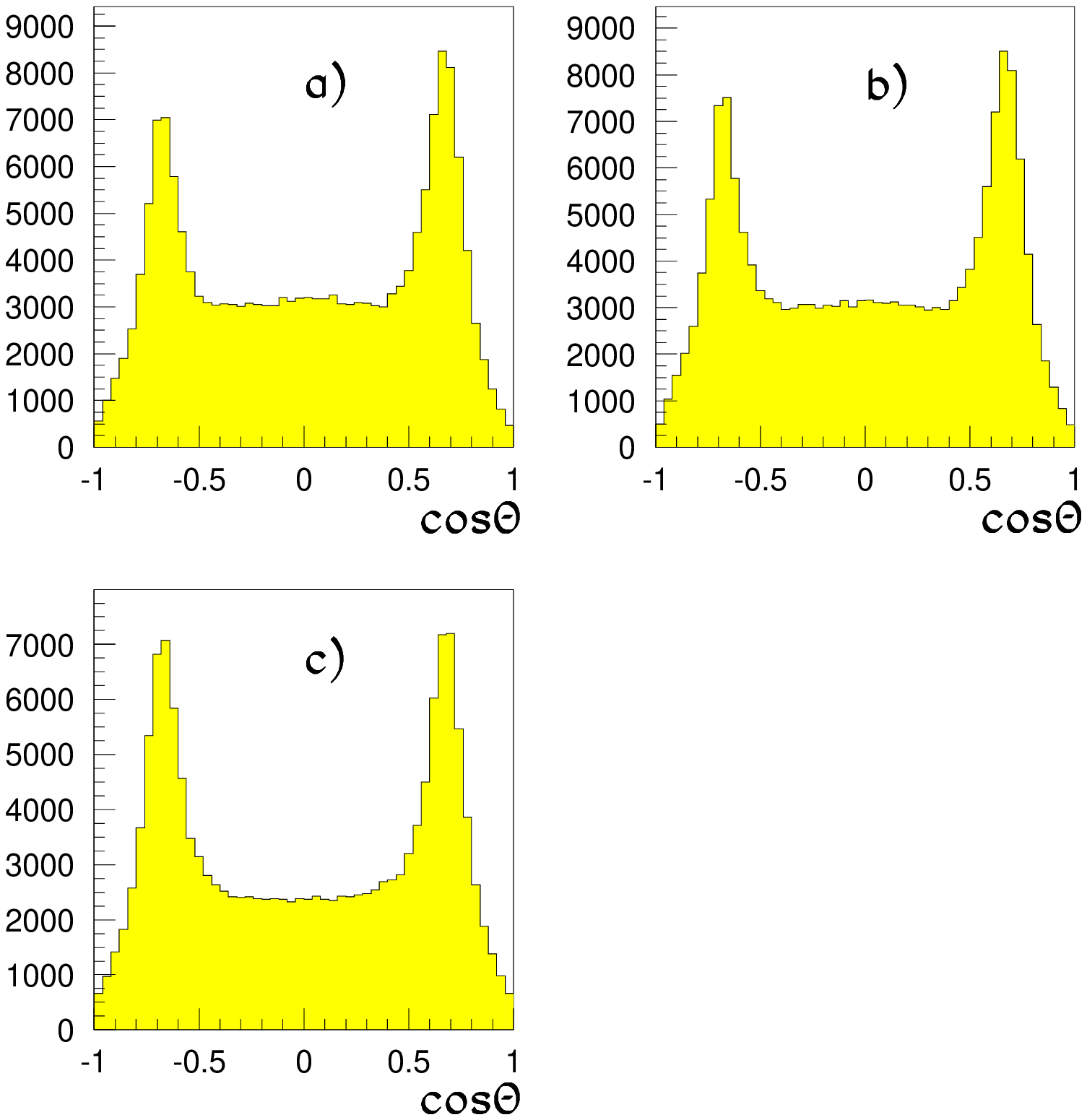,width=0.5\textwidth}
\end{tabular}
\caption{\label{3pidp}
The $ \pi^+\pi^-\pi^0$ Dalitz plot in $ \rm p\bar p$ annihilation
at rest, and $ \rho^+$ (a), $ \rho^-$ (b)
and $ \rho^0$ (c) decay angular distributions~\protect\cite{Abele:ac}.
}
\end{figure}

\par
The Dalitz plot of figure~\ref{3pidp} shows striking evidence for
internal dynamics. High-density
bands are visible at fixed values of $ m_{12}^2, m_{13}^2$, and $ m_{23}^2$.
The three bands correspond to the annihilation modes
\pbp\ $\to$ \rhp\pim , \rhm\pip , and \rhz\piz , respectively. The enhancements
due to $ \rho$ production as intermediate states are described by
dynamical functions $\cal F$.
\par
The three $ \rho$ decay angular distributions exhibit two peaks.
They are generated by choosing a slice
$ m_{12}^2 = m_{\rhp }^2 \pm \Delta m_{\rhp }^2$~
and plotting the number of events
as a function of $ m_{13}^2$.
The origin of the  peaks in the decay angular distributions
is immediately evident from the Dalitz plot.
The three bands due to the three $ \rho$
charged states cross; at the crossing two amplitudes interfere and the
observed intensity increases by a factor of four as one should expect from
quantum mechanics. Apart from the peaks the decay angular distribution
is approximately given by $ \sin^2\theta$, hence $ {\cal A} \sim \sin \theta$.
The three $ \rho$ production amplitudes have obviously the same strength
indicating that the \pbp\ initial state(s) from which $ \rho$ production
occur(s) must have isospin zero.

\subsection{\label{section2.2}Angular distributions}
\subsubsection{Zemach formalism}
Returning to figure~\ref{3pidp}, the right panel presents
decay angular distributions.
The $ \rho$ is emitted with a momentum $\vec{p}_{\rho}$
and then decays in a direction characterized, in the $\rho$ rest frame,
by one angle $\Theta$ and the momentum vector  $\vec p_{\pi}$.
The \pbp\ initial state has $ L=0$;
the parity of the initial state is -1 (in both cases), the
parities of $\pi$ and $\rho$ are -1. Hence
there must be an angular momentum $l_{\rho}=1$ between
$\pi$ and $\rho$. This decay is described by the
vector $\vec{p}_{\rho}$. The $\rho$ decays also with
one unit of angular momentum, with $l_{\pi}=1$.
From the two rank-one tensors (=vectors) we have to construct
the initial state:
$$
\vec J = \vec S = \left\{ \begin{array}{ccccc}
                             J \quad &\quad  ^{2s+1}L_J \quad &\quad  J^{PC} \quad &\quad 
Zemach \quad &\quad \rm  angle \\
                            0 \quad &\quad  ^1S_0      \quad &\quad  0^{-+} \quad &\quad 
\vec{p}_{\rho}\cdot\vec  p_{\pi}\quad &\quad  \cos\Theta \\
                             1 \quad &\quad  ^3S_1      \quad &\quad  1^{--} \quad &\quad 
\vec{p}_{\rho}\times\vec p_{\pi}\quad &\quad  \sin\Theta
                          \end{array}\right.
$$
For higher spins appropiate operators can be constructed according to
the following rules: the operators are
\vspace*{-1mm}
\bi
\item  traceless: $ trace\ t = 0$ \qquad\
$ \sum t^i = 0; \qquad\ \sum t^{ii} = 0; \qquad\ \sum t^{iii} =
0; \ \ldots $
\vspace*{-1mm}
\item symmetric: \qquad\ $ t^{ij}=t^{ji}; \qquad\
t^{ijk}=t^{jik}=t^{ij}=t^{ikj}$
\vspace*{-1mm}
\item can be constructed as products of lower-rank tensors \qquad\\
$ t^it^j\Longrightarrow\frac{1}{2}(t^it^j + t^it^j) -
 \frac{1}{3}t^2\delta^{ij}$
\vspace*{-1mm}
\item To reduce rank, multiply with $\delta^{ij} \delta^{kl}$
or $\epsilon^{ijk}$
\ei
\vspace*{-1mm}
\subsubsection{Helicity formalism}
This section is adapted from an unpublished note written by Ulrike
Thoma~\cite{Thoma}.

The helicity of a particle is defined as the projection of its total
angular momentum $\vec{J}=\vec{l}+\vec{s}$ onto its direction of flight.
$$
\lambda = \vec{J} \cdot \frac{\vec{p}}{|\vec{p}|} = l \cdot \frac{\vec{p}}{|\vec{p}|} + m_s = m_s \,\, ,
$$
Consider a particle A decaying into particles B and C with spins
$s_1$, $s_2$. The particles move along the z-axis (quantization
axis). The final state is described by $(2 s_1 + 1)\cdot(2 s_2 +
1)$ helicity states $\left| p \lambda_1 \lambda_2  \right >$;
$\lambda_i$ are the helicities of the particles and $p$ is their
center of mass momentum.
\par
\begin{minipage}[c]{0.4\textwidth}
The particle B emitted in a arbitrary direction can be described in
spherical coordinates by the angles $\theta$, $\phi$.
\end{minipage}
\begin{minipage}[c]{0.58\textwidth}
\epsfig{file=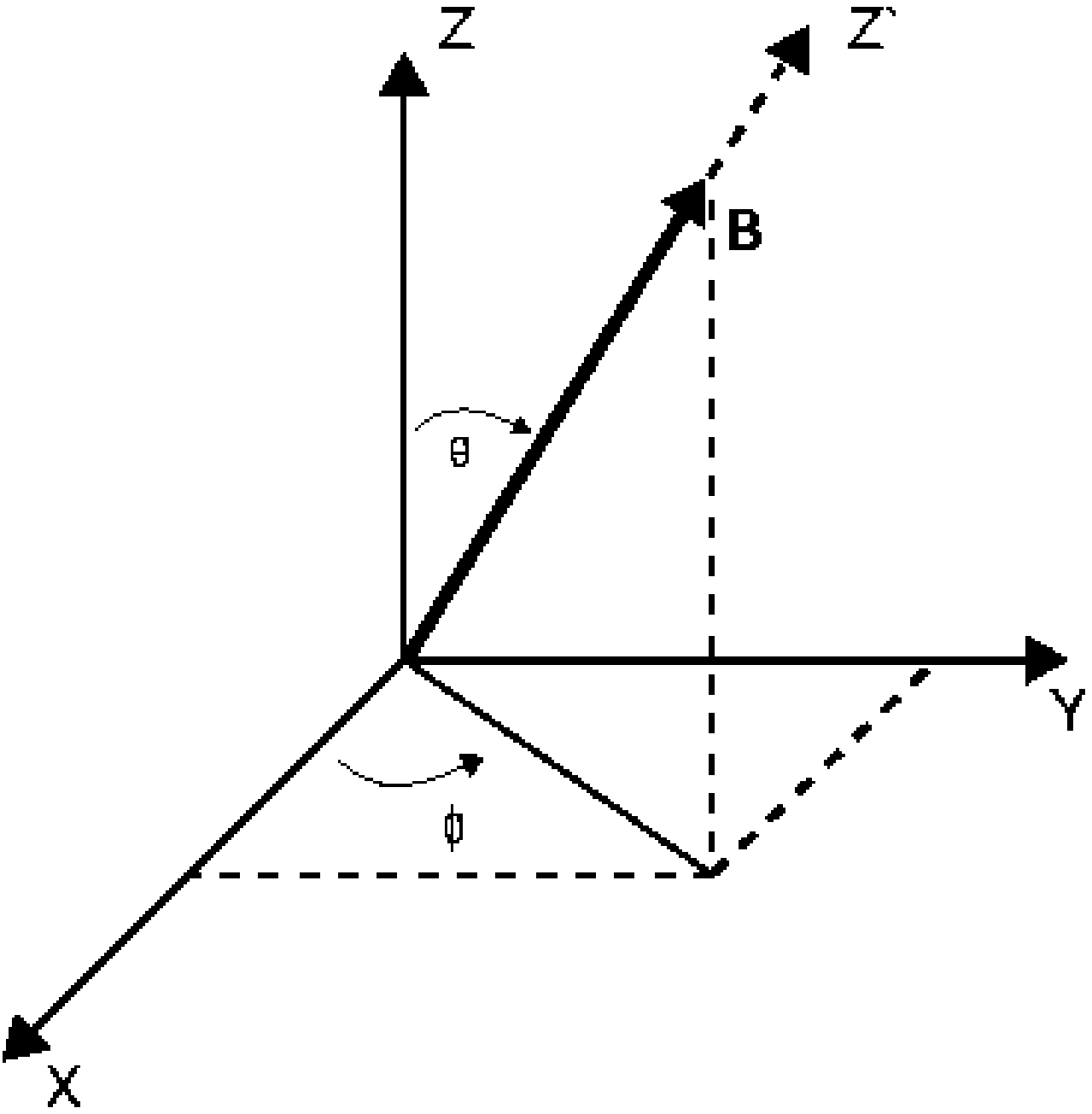,width=0.7\textwidth,clip=}\\
Coordinate system for  $A \to BC$ decays.
\end{minipage}
\vskip 2mm
In this case the helicity states are defined in the coordinate system
$\Sigma_3$ which is produced by a rotation of $\Sigma_1$ into the new
system.
$$
R(\theta, \phi) = \rm{R}_{y2}(\theta)\rm{R}_{z1}(\phi)
$$
Using $d$-functions the rotation can be written as
$$
\label{drehmatrix}
D^J_{mm^{\prime}}(\theta, \phi)=e^{im^{\prime}\phi}
d^J_{mm^{\prime}}(\theta)
$$
The final states in system $\Sigma_1$ can be expressed as
$$
\left| p \theta \phi \lambda_1 \lambda_2 M \right>_1 =
D^J_{M\lambda}(-\theta, -\phi) \cdot \left|p \lambda_1 \lambda_2 \right>_3
$$
with $\lambda = \lambda_1 - \lambda_2$. The transition matrix for the
decay is given by\\

\begin{tabular}{ccccc}
$f_{\lambda_1 \lambda_2,M}(\theta, \phi)$ &=&
$\left<p \theta \phi \lambda_1 \lambda_2 M \right|T
\left|M^{\prime}\right>$
&=& $D^{J\,*}_{M\lambda}(-\theta, -\phi) \left< \lambda_1 \lambda_2
 \right| T \left| M^{\prime}\right>$
\\[1ex]
&=&$e^{i\lambda\Phi}d_{\lambda M}^J(\theta,\phi)\cdot
T_{\lambda_1\lambda_2}$ &=&
$ D^{J'}_{\lambda M}(\theta, \phi)
T_{\lambda_1 \lambda_2} \,\, $
\end{tabular}
\\[1ex]
\par
The interaction is rotation invariant. The transition amplitude is
a matrix with (2$s_1$+1)(2$s_2$+1) rows and (2$J$+1) columns.
$D^{J'}_{\lambda M}$($\theta$, $\phi$) describes the geometry, the
rotation of the system $\Sigma_3$ where the helicity states are
defined, back into the CMS system of the resonance; $T_{\lambda_1
\lambda_2}$ describes the dependence on the spins and the orbital
angular momenta of the different particles in the decay process.
The general form of $T_{\lambda_1 \lambda_2}$ is given by
$$
T_{\lambda_1 \lambda_2}= \sum_{ls}\limits \alpha_{ls}\left<J\lambda|ls0\lambda\right>\left<s\lambda|s_1s_2\lambda_1,-\lambda_2\right>
\label{tallg}
$$
where $\alpha_{ls}$ are unknown fit parameters.

The parameters define the decay
spin and orbital angular momentum configuration.
The brackets are Clebsch-Gordan couplings for
$\vec{J}=\vec{l}+\vec{s}$ and
$\vec{s}=\vec{s}_1+\vec{s}_2$. The sum extends over all
allowed $l$ and $s$. Thus:
$$
\label{wD}
w_D(\theta,\phi) = Tr (\rho_f) = Tr (f \rho_i f^+)
$$
\begin{figure}[h!]
\begin{minipage}[t!]{0.46\textwidth}
where $\rho_f$ is the final state density matrix of the dimension
(2$s_1$+1)(2$s_2$+1) and $\rho_i$ is the initial density matrix of
dimension (2J+1).
\par
Assume that not only A decays into B and C but also B and C decay
further into $ B_1 B_2$ and $ C_1 C_2$.
Figure~\ref{sequence}
 shows a sequential decay of the $\bar p$N
system. Sequential decays are combined to form one common
amplitude. The individual amplitudes are
combined as scalar products if they are
linked by a line, otherwise by a
tensor product. Thus the amplitude is determined to
\end{minipage}
\begin{minipage}[t!]{0.53\textwidth}
\epsfig{file=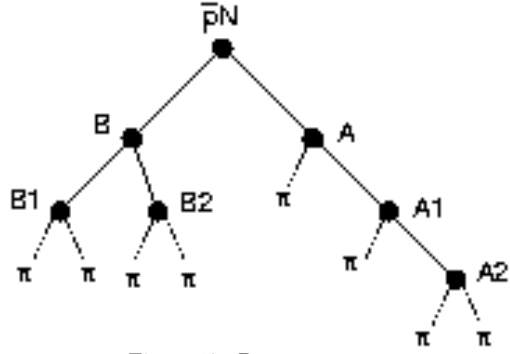,width=\textwidth,clip=}
\vspace*{-10mm}
\caption{\label{sequence} Decay sequence}
\end{minipage}
\vspace*{-15mm}
\end{figure}

$$
f_{tot}=\left[\,\, (\,\, f(A2)f(A1)f(A) \,\,) \otimes (\,\,[\,\,
f(B2)\otimes f(B1)\,\,]\,\, f(B) \,\,) \,\,\right]f(\bar{p}N) \qquad\quad
\label{ftot}
$$
where $\otimes$ represents the tensor product of two matrices.
The total helicity amplitude for a reaction $A \to B C$,  $B \to B_1
B_2$, $C \to C_1 C_2$, has the form: \\

\begin{tabular}{ccc}
$f_{tot}$&=&$\left[f(B) \otimes f(C)\right]f(A)$ \\[1ex]
&=&$\sum_{{\small \lambda(B)\lambda(C)}}\limits
\left[ f_{\lambda(B1)\lambda(B2), \lambda(B)} \otimes
f_{\lambda(C1)\lambda(C2), \lambda(C)} \right]
f_{\lambda(B)\lambda(C), \lambda(A)}$.
\end{tabular}

\subsubsection{The helicity formalism in photo--production processes}
In photo--production, a resonance is produced in the process and
then decays. We first consider a nucleon resonance that is produced and
decays only into two particles, e.g.: $\gamma p \to N^* \to p
\eta$. The $\gamma p$-system defines the z-axis ($\theta,\phi =
0$) and determines the spin density matrix of the $N^*$-resonance.
\par
The reaction $\gamma p \to N^*$ is related to $
N^* \to\gamma p $ (which can be calculated using the formalism
discussed above) by time reversal invariance. Using
$$
f_{\lambda_c \lambda_d , \lambda_a \lambda_b}(\theta) =
(-1)^{\lambda^{\prime}-\lambda} f_{\lambda_a \lambda_b , \lambda_c \lambda_d}(\theta)
 \quad \quad 
$$
valid for $ab\to cd$ with $\lambda^{\prime}=\lambda_c -
\lambda_d$ and $\lambda =\lambda_a - \lambda_d$,
we can write
$$
f(\gamma p \to N^* \to p \eta) = f(N^* \to p \eta) \cdot f^T(N^* \to \gamma p)
$$
Note that
$\lambda^{\prime}=\lambda_{N^*} = \lambda_p -
\lambda_{\gamma} $ always holds.

The photo--production amplitude needs to describe the decay of a
resonance $R$ $f(R\to N\,X)$ into the channel N\,X and its
production $ \gamma p \to R$ calculated using the transposed decay
amplitude $f^T(R \to p \gamma)$. For photo--production of spin 0
mesons this matrix has the form
$$
f^{tot} =
\left(\begin{matrix}
\left[+\frac{1}{2}\,,-1;+\frac{1}{2}\right]&\left[-\frac{1}{2}\,-1;+\frac{1}{2}\right]&
\left[+\frac{1}{2}\,,+1;+\frac{1}{2}\right]&\left[-\frac{1}{2}\,+1;+\frac{1}{2}\right]\cr
\left[+\frac{1}{2}\,,-1;-\frac{1}{2}\right]&\left[-\frac{1}{2}\,-1;-\frac{1}{2}\right]&
\left[+\frac{1}{2}\,,+1;-\frac{1}{2}\right]&\left[-\frac{1}{2}\,+1;-\frac{1}{2}\right]\cr
\end{matrix} \right)
$$
where the numbers in the brackets represent
[$\lambda_{p},\lambda_{\gamma}; \lambda_{p^f}$] with $p,\,p^f$ being
the initial and final state proton.
The angular part of the differential cross section is then given by
$$
\frac{\rm d \sigma}{\rm d \Omega} = \quad \sim \frac{1}{4} \cdot
\sum_{\lambda_p\lambda_{\gamma}\lambda^{\prime}} | T(\lambda_p\lambda_{\gamma}\lambda_{p^{f}})|^2.
\label{cross_section}
$$
We now discuss photo--production of the $\rm S_{11}$ and $\rm
P_{11}$ resonances and their decay into p$\eta$. The photons
are assumed to be unpolarized.
\begin{itemize}
\item \underline{$S_{11} \to p \eta$}
\end{itemize}
To determine the angular distribution of the decay process the
helicity matrix $\rm T_{\lambda_1\lambda_2}$ has to be calculated,
$$
T_{\pm\frac{1}{2}, 0} = \alpha_{0 \,\frac{1}{2}}
\left<\, \frac{1}{2} , \pm \frac{1}{2} \mid 0 , \frac{1}{2} ; 0 , \pm
\frac{1}{2} \right>
\left<\, \frac{1}{2} , \pm \frac{1}{2} \mid  \frac{1}{2} , 0 ; \pm
\frac{1}{2} , 0 \right> = \alpha_{0 \,\frac{1}{2}} = const.
$$
The transition matrix is then given by (the constant is
arbitrarily set to 1):
$$
f_{\pm \frac{1}{2}\,0,\,M}(\theta,\phi) =
 \left( \begin{matrix}
D^{\frac{1}{2}'}_{\frac{1}{2}\,\frac{1}{2}} & D^{\frac{1}{2}'}_{\frac{1}{2}\,-\frac{1}{2}} \cr
D^{\frac{1}{2}'}_{-\frac{1}{2}\,\frac{1}{2}}& D^{\frac{1}{2}'}_{-\frac{1}{2}\,-\frac{1}{2}}\cr
\end{matrix} \right)
$$
where columns represent the spin projections $M_{S_{11}}=\frac{1}{2},\,\,-\frac{1}{2}$
and the rows illustrate $\lambda^{\prime}_p=\frac{1}{2},\,\,-\frac{1}{2}$.
The z-axis corresponds to direction of flight.
The process does not depend on $\phi$. Ee can choose $\phi=0$ and
obtain
$$
f_{S_{11}\to p \eta} =  \left( \begin{matrix}
   \cos(\frac{\theta}{2}) & -\sin(\frac{\theta}{2}) \cr
   \sin(\frac{\theta}{2}) & \cos(\frac{\theta}{2})\cr
\end{matrix} \right).
$$
\begin{itemize}
\item \underline{$\gamma p \to S_{11}$}
\end{itemize}
First we calculate $S_{11} \to p \gamma$, for
$ \frac{1}{2}^- \to \frac{1}{2}^+ 1^- $ and $\ell=0$.
Two proton helicities occur $\lambda_1 = \pm \frac{1}{2}
$; the two photon helicities are $s_2 = 1$, $\lambda_2 =
\pm 1 \,\, $; $\lambda_2 = 0$ is excluded for
real photons.
One finds that
$\rm T_{\frac{1}{2}\,-1} = T_{-\frac{1}{2}\,+1} = 0$,
$T_{-\frac{1}{2}\,-1} = -\sqrt{\frac{2}{3}}$,
$T_{\frac{1}{2}\,1} = +\sqrt{\frac{2}{3}}$.
Setting $\theta$ and $\phi$ to 0 (the $\gamma p$ is
parallel to the z-axis), $D_{\lambda\, M}^J(0,0)$ is non--zero
only for $\lambda = M$.
$$
f_{S_{11}\to  p \gamma } =
    \left( \begin{matrix}
   0 & 0 \cr
-\sqrt{\frac{2}{3}} & 0 \cr
0 & \sqrt{\frac{2}{3}} \cr
0 & 0 \cr
   \end{matrix}
 \right)
$$

The rows correspond to different $\lambda$-values, thus:
$$
f(\gamma p \to S_{11}) = f^T(S_{11}\to p \gamma) =
    \left( \begin{matrix}
   0 & -\sqrt{\frac{2}{3}} & 0 & 0 \cr
   0 & 0 & \sqrt{\frac{2}{3}} & 0\cr
   \end{matrix}
 \right).
$$
\begin{itemize}
\item \underline{$\gamma p \to S_{11} \to p \eta $:} 
\end{itemize}
For the whole reaction we obtain
$$
f_{\gamma p \to S_{11} \to p \eta} =
    \left( \begin{matrix}
   \cos(\frac{\theta}{2}) & -\sin(\frac{\theta}{2}) \cr
   \sin(\frac{\theta}{2}) & \cos(\frac{\theta}{2})  \cr
   \end{matrix} \right)
\cdot
    \left( \begin{matrix}
   0 & -\sqrt{\frac{2}{3}} & 0 & 0 \cr
   0 & 0 & \sqrt{\frac{2}{3}} & 0\cr
   \end{matrix} \right)
$$
$$
\phantom{f_{\gamma p \to S_{11} \to p \eta}} =
    \left( \begin{matrix}
   0 & -\sqrt{\frac{2}{3}} \cos(\frac{\theta}{2})& -\sqrt{\frac{2}{3}}\sin(\frac{\theta}{2}) &
   0 \cr
   0 & -\sqrt{\frac{2}{3}}\sin(\frac{\theta}{2}) & \sqrt{\frac{2}{3}} \cos(\frac{\theta}{2}) &
   0\cr
    \end{matrix} \right)
$$

$$
{\mathrm{With}}\qquad\qquad \frac{\rm d \sigma}{\rm d \Omega} = \quad \sim \frac{1}{4} \cdot
\sum_{\lambda_p\lambda_{\gamma}\lambda^{\prime}} | T(\lambda_p\lambda_{\gamma}\lambda^{\prime})|^2
\label{angular_distribution}
$$
a flat angular distribution is found.
$$
\frac{\rm d \sigma}{\rm d \Omega} \sim |A_{S_{11}}|^2
$$
\begin{itemize}
\item \underline{$\rm \gamma p \to P_{11} \to p \eta$}
\end{itemize}
The helicity matrix for $P_{11} \to p \eta$ involves the
quantum numbers: $ \frac{1}{2}^+ \to \frac{1}{2}^+ 0^- ,\quad \ell=1$
and the matrix elements
$T_{\pm\frac{1}{2}, 0} = \pm \frac{1}{\sqrt{3}} \alpha_{1 \,\frac{1}{2}}$.
The scattering amplitude is now given by
$$
f_{P_{11}\to p \eta} =
    \left( \begin{matrix}
   -\frac{1}{\sqrt{3}} \cos(\frac{\theta}{2}) &  -\frac{1}{\sqrt{3}} (-\sin(\frac{\theta}{2}))
   \cr
   \frac{1}{\sqrt{3}} \sin(\frac{\theta}{2}) &  \frac{1}{\sqrt{3}}\cos(\frac{\theta}{2})\cr
    \end{matrix} \right)
$$
In the next step $P_{11} \to p \gamma$ is calculated
($ \frac{1}{2}^+ \to \frac{1}{2}^+ 1^- ,\quad \ell=1$).
Two values for the spin $s$ are possible: $s=\frac{1}{2}$,
$s=\frac{3}{2}$. The Clebsch Gordan coefficients depend on $s$ leading
to a different $T_{\lambda_1\lambda_2}$ for the two spins.  The D-matrices
depend only on $J, \, \lambda$ and $M$, so that they are the same for both
cases.  One finds \\
$T_{ +\frac{1}{2}\, -1} = T_{ -\frac{1}{2}\, +1} = 0$  and
$T_{ -\frac{1}{2}\, -1} = T_{ +\frac{1}{2}\, +1} =
\alpha_{1\frac{3}{2}}(-\frac{1}{3})+\alpha_{1\frac{1}{2}}(+\frac{\sqrt{2}}{3})=-a$, and 
$$
f_{\gamma p \to P_{11}} = f^T_{P_{11}\to p \gamma} =
    \left( \begin{matrix}
   0 & -a & 0 & 0 \cr
   0 & 0 & -a & 0 \cr
    \end{matrix} \right)
$$
\begin{itemize}
\item $\gamma p \to P_{11} \to p \eta $:
\end{itemize}
For the whole process 
$$
f_{\gamma p \to P_{11} \to p \eta} =
    \left( \begin{matrix}
   0 & +\frac{1}{\sqrt{3}}a\,\cos(\frac{\theta}{2}) &
   -\frac{1}{\sqrt{3}}a\,\sin(\frac{\theta}{2}) & 0\cr
   0 & -\frac{1}{\sqrt{3}}a\,\sin(\frac{\theta}{2}) &
   -\frac{1}{\sqrt{3}}a\,\cos(\frac{\theta}{2}) & 0\cr
    \end{matrix} \right)
$$
which leads again to a flat angular distribution; the differential
cross section does not depend on $\theta$.

Finally we assume that both resonances are produced;
their interference leads to a non-flat angular distribution.
$$
f_{\gamma p \to (S_{11} + P_{11}) \to p \eta} =
    \left( \begin{matrix}
   0 & (-\frac{2}{\sqrt{3}}S_{11} +\frac{a}{\sqrt{3}} P_{11})\cos(\frac{\theta}{2}) &
   (-\frac{2}{\sqrt{3}}S_{11} -\frac{a}{\sqrt{3}} P_{11})\sin(\frac{\theta}{2}) & 0\cr
   0 & (-\frac{2}{\sqrt{3}}S_{11} -\frac{a}{\sqrt{3}} P_{11})\sin(\frac{\theta}{2}) &
   (\frac{2}{\sqrt{3}}S_{11} -\frac{a}{\sqrt{3}} P_{11})\cos(\frac{\theta}{2}) & 0\cr
    \end{matrix} \right).
$$
Fir $s=\frac{2}{\sqrt{3}}S_{11}$ and $p=\frac{a}{\sqrt{3}}
P_{11}$,
$$
\begin{matrix}
\frac{\rm d \sigma}{\rm d \Omega} &\,\sim\,& \frac{1}{4} \cdot 2 \left[
\sin^2(\frac{\theta}{2}) \cdot |s+p|^2 +  \cos^2(\frac{\theta}{2})
\cdot |s-p|^2  \right] \cr
 &\,\sim\,& |s|^2 \,+\,  |p|^2 \,-\, 2Re(s^*p) \cdot \cos(\theta)\cr
    \end{matrix}.
$$
The interference term $2Re(s^*p) \cdot \cos(\theta)$ produces a
non-flat angular distribution.

\subsection{\label{section2.3}Flavor structure of mesons}
\subsubsection{Isoscalar coefficients for meson decays}
Decays of mesons belonging to one SU(3) multiplet are related by
SU(3). The relations are called isoscalar coefficients; they are
generalizations of the Clebsch--Gordan coefficients. A restricted
set is shown below. Note that $\Sigma$ stands for the SU(3)
classification for the isospin triplet system. It can be a $\pi$,
$\rho$ or $a_2(1320)$. The $\Lambda$ is the symbol for both the 
singlet and the octet particle. There are three types of transitions
needed to describe meson decays into two octet mesons:
\par\vskip 2mm
\begin{tabular}{ccc}
$\bf 1 \to 8\otimes 8$&&\\
\end{tabular}

\begin{tabular}{ccc}
$\left( \begin{matrix}
\Lambda
    \end{matrix} \right)$
= &$\left( \begin{matrix}
N\bar K \ & \ \Sigma\pi \ & \ \Lambda\eta \ & \ \Xi K
    \end{matrix} \right)$
= &$\frac{1}{\sqrt 8}\left( \begin{matrix}
2 \ & \ 3 \ & \ -1 \ & \ -2
    \end{matrix} \right)^{1/2}$
\end{tabular}

\par\vskip 2mm
\begin{tabular}{ccc}
$\bf 8_1 \to 8\otimes 8$&&\\
\end{tabular}

\begin{tabular}{ccc}
$\left( \begin{matrix}
 N \ \\ \ \Sigma \ \\ \Lambda \ \\ \ \Xi \ 
    \end{matrix} \right)$
= &\hspace*{-4mm}$\left( \begin{matrix}
\qquad N\pi \ & \ N\eta \ & \ \Sigma K \ & \ \Lambda K \\
\hspace*{-4mm}N\bar K \ & \hspace*{-3mm} \Sigma\pi \ & \hspace*{-2mm} \Lambda\pi \ & \hspace*{-1mm} \Sigma\eta \  &  \Xi K \\
\qquad N\bar K \ & \ \Sigma\pi \ & \ \Lambda\eta \ & \ \Xi K \\
\qquad \Sigma\bar K \ & \ \Lambda\bar K \ & \ \Xi\pi \ & \ \Xi\eta \ \\
    \end{matrix} \right)$
= &\hspace*{-4mm}$\left( \begin{matrix}
\qquad 9 \ & \ -1 \ & \ -9 \ & \ -1 \\
\hspace*{-4mm} -6 \ & \hspace*{-3mm} 0 \ & \hspace*{-2mm} 4 \ & \hspace*{-1mm} 4 \  &  -6 \\
\qquad 2 \ & \ -12 \ & \ -4 \ & \ -2 \\
\qquad 9 \ & \ -1 \ & \ -9 \ & \ -1 \ \\
    \end{matrix} \right)^{1/2}$
\end{tabular}

\par\vskip 2mm
\begin{tabular}{ccc}
$\bf 8_2 \to 8\otimes 8$&&\\
\end{tabular}

\begin{tabular}{ccc}
$\left( \begin{matrix}
 N \ \\ \ \Sigma \ \\ \Lambda \ \\ \ \Xi \ 
    \end{matrix} \right)$
= &\hspace*{-4mm}$\left( \begin{matrix}
\qquad N\pi \ & \ N\eta \ & \ \Sigma K \ & \ \Lambda K \\
\hspace*{-4mm}N\bar K \ & \hspace*{-3mm} \Sigma\pi \ & \hspace*{-2mm} \Lambda\pi \ & \hspace*{-1mm} \Sigma\eta \  &  \Xi K \\
\qquad N\bar K \ & \ \Sigma\pi \ & \ \Lambda\eta \ & \ \Xi K \\
\qquad \Sigma\bar K \ & \ \Lambda\bar K \ & \ \Xi\pi \ & \ \Xi\eta \ \\
    \end{matrix} \right)$
= &\hspace*{-4mm}$\left( \begin{matrix}
\qquad 3 \ & \ 3 \ & \ 3 \ & \ -3 \\
\hspace*{-4mm} 2 \ & \hspace*{-3mm} 8 \ & \hspace*{-2mm} 0 \ & \hspace*{-1mm} 0 \  &  -2 \\
\qquad 6 \ & \ 0 \ & \ 0 \ & \ 6 \\
\qquad 3 \ & \ 3 \ & \ 3 \ & \ -3 \ \\
    \end{matrix} \right)^{1/2}$
\end{tabular}
\par
\vskip 2mm
The $()^{1/2}$ indicates that the square root should be calculated for
each matrix element. There are coupling constants defined
through the relations: 
$$ a_2(1320) \to \etg_1\pi  =  g_1 \qquad
a_2(1320) \to \etg_8\pi  =  \sqrt{\frac{1}{5}}g_8 \qquad
 a_2(1320) \to \etg_{\ssb}\pi = 0 
$$
from which $g_1=-\sqrt{\frac{4}{5}}g_8$ can be derived. 
\par
The singlet component of the $ f_2(1270)$ and $f2(1525)$ decays into 
$\pi\pi, \eta\eta , \eta\eta'$, and $\rm K\bar K$ with ratios 3:1:0:4. A
useful 'rule of thumb' helps to decide if $8_1$ or $8_2$ decays should
be used; $8_1$ is responsible for decays in which the decay of
the neutral member of the primary octet into the neutral members of
the two octets in the final states are allowed. In this case the
product $C'$ of the $C$ parities of the neutral members of the three
involved octets is positive. $8_2$ should be used when $C'$ is negative.
Examples for $8_1$ decays (with $C'=1$) are $ f_2(1270), f2(1525)\to
\pi\pi, \eta\eta , \eta\eta', \rm K\bar K$ or
$ a_2(1320) \to\eta\pi ,\rm K\bar K ;\ K_2^*(1430) \to K\pi, K\eta,
K\eta'$, $ a_2(1320) \to\rho\pi$ is an example  
for  $8_2$ decays (with $C'=-1$).

\begin{table}[h!]
\caption{\label{codd}Isoscalar coefficients for decays of 
mesons with $I=1$ and $I=1/2$.}
\begin{tabular}{@{}lll@{}@{}lll@{}} \hline
decay & sym. & antissym. & \qquad decay & sym. & antisym. \\ \hline
$ \pi\rightarrow\pi\pi$ & 0 & \ssfrac{2}{3} &\qquad
$ K\rightarrow K \pi$ & \sfrac{3}{20} & $ \frac{1}{2}$ \\
$ \pi\rightarrow K\bar{K}$ & \sfrac{1}{6}  & \ssfrac{3}{10} &\qquad
$ K\rightarrow K\eta, K \phi$ & \sfrac{\cf+2\sqrt{2}\sf}{20} & $ \frac{\cf}{2}$ \\
$ \pi\rightarrow\pi\eta, \pi\phi$ & 0 & \sfrac{\cf-\sqrt{2}\sf}{5} &\qquad
$ K\rightarrow K\eta^\prime, K\omega$ &  \sfrac{2\sqrt{2}\cf-\sf}{20} & $ \frac{\sf}{2}$ \\
$ \pi\rightarrow\pi\eta^\prime, \pi\omega$ & 0 & \sfrac{\sf+\sqrt{2}\cf}{5} &&&\\
\hline
\end{tabular}
\end{table}

The matrix elements become a bit tedious when final state mesons with
mixing angles are involved. With $i$ and $f$ denoting the initial and
final state nonet mixing angle, respectively,  the
SU(3) couplings for $ \eta^\prime$--, $f_2(1270)$-- and $ \omega$
(dominant \uubar{} and \ddbar{})--like mesons with nonet mixing are
given as

\begin{table}[h]
\caption{\label{ceven1}Isoscalar coefficients for decays of isoscalar
($\omega$--type) mesons.}
\bc
\begin{tabular}{@{}lll@{}} \hline
decay & symmetric & antisymmetric \\ \hline
$ f\rightarrow \pi\pi$ & $ \sqrt{3}\sfrac{\sqrt{2}\ci+\si}{10}$ & 0 \\
$ f\rightarrow K\bar{K}$ & $ \sfrac{\si-2\sqrt{2}\ci}{10}$ & $ \frac{\si}{2}$ \\
$ f\rightarrow \eta\eta, \phi\phi$ & $ \sfrac{-\cf^2\si-\sqrt{2}(2\si\cf\sf-\ci)}{5}$ & 0 \\
$ f\rightarrow \eta^\prime\eta^\prime, \omega\omega$ & $ \sfrac{\sf^2\si+\sqrt{2}(2\si\cf\sf+\ci)}{5}$ & 0 \\
$ f\rightarrow \eta\eta^\prime, \phi\omega$ & $ \sfrac{\si(\sqrt{2}(\cf^2-\sf^2)-\cf\sf}{5}$ & 0 \\
\hline
\end{tabular}
\ec
\end{table}

while those for $ \eta$--, $ f^\prime_2(1525)$-- and $ \phi$ (dominant
\ssbar{})--like mesons read as follows:

\begin{table}[h]
\caption{\label{ceven2}Isoscalar coefficients for decays of isoscalar
($\Phi$--type) mesons.}
\bc
\begin{tabular}{@{}lll@{}} \hline
decay & symmetric & antisymmetric \\ \hline
$ f^\prime\rightarrow \pi\pi$ & $ \sqrt{3}\sfrac{\sqrt{2}\si-\ci}{10}$ & 0 \\
$ f^\prime\rightarrow K\bar{K}$ & $ \sfrac{\ci+2\sqrt{2}\si}{10}$ & $ \frac{\ci}{2}$ \\
$ f^\prime\rightarrow \eta\eta, \phi\phi$ & $ \sfrac{-\cf^2\ci-\sqrt{2}(2\ci\cf\sf+\si)}{5}$ & 0 \\
$ f^\prime\rightarrow \eta^\prime\eta^\prime, \omega\omega$ & $ \sfrac{\sf^2\si+\sqrt{2}(2\si\cf\sf+\ci)}{5}$ & 0 \\
$ f^\prime\rightarrow \eta\eta^\prime, \phi\omega$ & $ \sfrac{\ci(\sqrt{2}(\cf^2-\sf^2)-\cf\sf}{5}$ & 0 \\
\hline
\end{tabular}
\ec
\end{table}
The names in the tables are generic, i.e. $f'$ stands for $\Phi_3$,
$f_4'$ and so on.
\subsubsection{Fits}
We now ask if the SU(3) isoscalar coefficients are as useful as the
Clebsch--Gordan coefficients proved to be. For this purpose we apply
the matrix elements to relate tensor decays into two pseudoscalar
mesons and decays into a vector and a pseudoscalar meson. The former 
transitions are of type $C'=1$, the latter ones of  type $C'=-1$.
\par
The matrix element $ \cal M$
$$
\label{Width}
d\Gamma = \frac{1}{32\pi^2}|{\cal M}|^2 \frac{q}{m^2} d\Omega
$$
contains a coupling constant, $ C_{T\ra PS+PS}$ or
$ C_{T\ra V+PS}$ (which is calculable in dynamical models), the SU(3)
amplitudes $ c_{isoscalar}$ and a dynamical function $ F(q)$
with $ q$ being the breakup momentum.
$$
\label{Blatt}
B_2(qR)  =  \sqrt{\frac{13(qR)^2}{9+3(qR)+9(qR)^2}}
$$
$$
\label{SBW}
BW(m)  =  \frac{m_0 \Gamma_0}{m^2-m_0^2- i m_0 \Gamma_0}\
$$
\begin{small}
\begin{minipage}[c]{0.50\textwidth}
\begin{center}
\begin{tabular}{|l|r@{.}l@{$ \pm$}r@{.}l|r@{.}l|r@{.}l|}
\hline
 Decay &  \multicolumn{4}{|c}{Data} &   \multicolumn{2}{|c}{Fit}
 & \multicolumn{2}{|c|}{$ \chi^2$}   \\
 \cline{2-5}
 & \multicolumn{2}{|c}{$ \Gamma$} & \multicolumn{2}{|c|}{$ \sigma_{\Gamma}$}
 & \multicolumn{2}{c|}{$ \Gamma$} & \multicolumn{2}{c|}{\ }  \\
 \hline
 $ a_2\ra\pi\eta$      & 15 & 95 &  1 & 32 & 24 & 8 & 2 & 99  \\
 $ a_2\ra\pi\eta'$     &  0 & 63 &  0 & 12 &  1 & 2 & 4 & 39  \\
 $ a_2\ra K\bar{K}$    &  5 & 39 &  0 & 88 &  5 & 2 & 0 & 01  \\
 $ f_2\ra\pi\pi$       & 157 & 0 &  5 &  0 &117 & 1 & 2 & 77  \\
 $ f_2\ra K\bar{K}$    &  8 &  5 &  1 &  0 &  8 & 0 & 0 & 08  \\
 $ f_2\ra \eta\eta$    &  0 &  8 &  1 &  0 &  1 & 5 & 0 & 44  \\
 $ f'_2\ra\pi\pi$      &  4 &  2 &  1 &  9 &  3 & 7 & 0 & 07  \\
 $ f'_2\ra K\bar{K}$   & 55 &  7 &  5 &  0 & 48 & 6 & 0 & 43  \\
 $ f'_2\ra \eta\eta$   &  6 &  1 &  1 &  9 &  5 & 3 & 0 & 12  \\
 $ f'_2\ra \eta\eta'$  &  0 &  0 &  0 &  8 &  0 & 7 & 0 & 77  \\
 $ K_2\ra K\pi$        & 48 &  9 &  1 &  7 & 61 & 1 & 0 & 99  \\
 $ K_2\ra K\eta$       &  0 & 14 &  0 & 28 &  0 & 2 & 0 & 02  \\
 \hline
 \end{tabular}
 \end{center}
\end{minipage}
\begin{minipage}[c]{0.5\textwidth}
 \begin{center}\begin{tabular}{|l|r@{.}l@{$ \pm$}r@{.}l|r@{.}l|r@{.}l|}
 \hline
 Decay &  \multicolumn{4}{|c}{Data} &   \multicolumn{2}{|c}{Fit}
 & \multicolumn{2}{|c|}{$ \chi^2$}   \\
 \cline{2-5}
 & \multicolumn{2}{|c}{$ \Gamma$} & \multicolumn{2}{|c|}{$ \sigma_{\Gamma}$}
 & \multicolumn{2}{c|}{$ \Gamma$} & \multicolumn{2}{c|}{\ }  \\
\hline
 $ a_2\ra\pi\rho$      & 77 &  1 &  3 &  5 & 66 & 0 & 0 & 67  \\
 $ f_2\ra K^*\bar{K}$  &  0 &  0 &  1 &  8 &  0 & 2 & 0 & 01  \\
 $ f'_2\ra K^*\bar{K}$ &  10 & 0 & 10 & 0  & 11 & 8 & 0 & 03  \\
 $ K_2\ra K\rho$       &  8 &  7 &  0 &  8 & 11 & 5 & 1 & 29  \\
 $ K_2\ra K\omega$     &  2 &  7 &  0 &  8 & 1 & 0 & 0 & 00   \\
 $ K_2\ra K^*\pi$      & 24 &  8 &  1 &  7 & 24 & 1 & 0 & 02  \\
 $ K_2\ra K^*\eta$     &  0 &  0 &  1 &  0 & 0 & 9 & 0 & 81   \\
 \hline
 \end{tabular}
 \label{results}

\vskip 5mm
Results of the final fit. The $ \chi^2$ values\\
include 20\% SU(3) symmetry breaking.\end{center}
\end{minipage}
\end{small}
\par
\vskip 2mm
Obviously tensor meson decays are nearly compatible with SU(3). One
has to assume 20\% symmetry breaking to achieve a fit with
$\chi^2/N_F\sim 1$. From the fit nonet mixing angles can be determined.
They are not inconsistent with the values obtained from the
Gell-Mann--Okubo mass formula. 
$$
\begin{matrix}
\Theta_{ps\ }  &=& -(14.4 \pm\ 2.9)^{\circ} &\qquad\
& R &=&  0.2 \pm\ 0.04\,fm \cr
\Theta_{vec}\  &=& +(37.5 \pm\ 8.0)^{\circ}&&C_{T \ra PS + PS} &=& 1.11 \pm 0.05 \cr
\Theta_{ten}\ &=&  +(28.3 \pm\ 1.6)^{\circ} &&C_{T \ra PS + V\ } &=& 2.07 \pm 0.13 \cr
\lambda\   &=&  0.77 \pm\ 0.10   &&&&\cr
\end{matrix}
$$
SU(3) is broken; the chance of producing an $\bar ss$ pair
out of the vacuum is reduced by $0.77 \pm\ 0.10$ compared to
the chance of producing a light $\bar qq$ pair. The matrix elements and
fits are taken from~\cite{Peters:jv}.

%% file: Chapter3_proc.tex
\section{\label{section3}Particles and their interaction}
\subsection{\label{section3.1}The particles: quark and leptons}
Quarks and leptons are the basic building blocks of matter.
These particles have spin 1/2 and are fermions fulfilling
the Pauli principle which states: the wave functions of two identical
fermions must be antisymmetric with respect to their
exchange. Fermions interact via exchange of bosons, with spin 
0 (e.g. pions in nuclear physics), spin 1 (photons, gluons,
vector mesons, weak interaction bosons) or spin 2 (gravitons, tensor 
mesons).
\par
\subsubsection{Leptons}
Table~\ref{leptonqn} lists charged ($e^-, \mu^-, \tau^-$)
and neutral ($v_e, \nu_{\mu}, \nu_{\tau}$) leptons. All 
leptons (and all quarks) have their own antiparticles.
Fermions and antifermions have two spin components but weak 
interaction couples only to left--hand currents of fermions 
and to the right--hand currents of antifermions.
The separate conservation of the 3 lepton numbers 
is deduced from, e.g., the absence of electrons in
a beam of high energy neutrinos originating from
$\pi^-\to\mu^-\nu$ decays, or from the non-observation
of $\tau^- \to e^- \gamma$ decays. We now know that the 3 generations
are mixed, i.e. that the mass eigenstates are not identical with the
weak--interaction eigenstates. 

\vspace*{-3mm}

\begin{table}[h!]
\caption{Leptons and their quantum numbers.}
\renewcommand{\arraystretch}{1.3}
\begin{center}
\begin{tabular}{lccc}
\hline\hline
Classification        & $e^-/\nu_e$ &$\mu^-/\nu_{\mu}$ & $\tau^-/\nu_{\tau}$ \\
e-lepton number       &      1    &      0         &         0         \\
$\mu$-lepton number   &      0    &      1         &         0         \\
$\tau$-lepton number  &      0    &      0         &         1         \\
\hline\hline
\end{tabular}
\end{center}
\label{leptonqn}
\renewcommand{\arraystretch}{1.0}
\end{table}

\vspace*{-3mm}

\subsubsection{Quarks and their quantum numbers}
Quarks have charges of 2/3 or -1/3 and not 1 (in units
of the positron charge $e$). Additionally quarks carry a new type
of charge, called color, in 3 variants 
defined to be red, blue, and green. Antiquarks have
the complementary colors anti-red, anti-blue, and anti-green.
Mesons composed of a quark and an antiquark can be
written as superposition (in the quantum mechanical sense)
$q_{red}\overline{q_{red}} + q_{blue}\overline{q_{blue}} +
q_{green}\overline{q_{green}}$, and baryons as
$q_{red}q_{blue}q_{green}$. Color and anti-color neutralize
and so do three colors or three anti-colors. 
\begin{table}[h!]
\caption{\label{quarkqn}
Quarks and their quantum numbers}
\renewcommand{\arraystretch}{1.3}
\begin{center}
\begin{tabular}{lcccccc}
\hline\hline
Classification        & $d$ & $u$ & $s$ & $c$ & $b$ & $t$ \\
Charge                &-1/3 & 2/3 &-1/3 & 2/3 &-1/3 & 2/3 \\
Isospin  $I$          & 1/2 & 1/2 &  0  &  0  &  0  &  0  \\
$I_3$                 &-1/2 & 1/2 &  0  &  0  &  0  &  0  \\
Strangeness  $s$      &  0  &  0  & { -1}  &  0  &  0  &  0  \\
Charm        $c$      &  0  &  0  &  0  & {1}  &  0  &  0  \\
Beauty (bottom) $b$   &  0  &  0  &  0  &  0  &{ -1}   &  0  \\
Truth (top)     $t$   &  0  &  0  &  0  &  0  &  0  & {1}   \\
\hline\hline
\end{tabular}
\end{center}
\renewcommand{\arraystretch}{1.0}
\vspace*{-2mm}
\end{table}
All quantities like strangeness $s$ or topness $t$ 
except the isospin $I$ change sign when
a particle is replaced by its antiparticle.  
The sign of the flavor in Table~\ref{quarkqn} is given by the sign of 
the meson charge. Examples: 
\vskip 2mm
\begin{tabular}{clc}
Charge(K$^+$) = Charge($u{\bar s}$) = +1 &\qquad\ Strangeness of 
$\bar s$:& $S=1$\\
Charge(D$^+$) = Charge(${c}\bar d$) = +1 &\qquad\ Charm of $c$:& $C=1$\\
Charge(B$^-$) = Charge(${ b}\bar u$) = -1 &\qquad\ Beauty of $b$:& $B=-1$
\end{tabular}
\vskip 2mm
\par
The masses of quarks are much more difficult to determine.
Even the concept of a quark mass is difficult to understand. 
No free separated quark has ever been observed; quarks
are {\it confined} and we cannot make a quark mass
measurement. What we can do is construct a model 
for mesons and baryons as being composed of two or
three {\it constituent} quarks. We may guess an interaction
and then hope that for a good choice of quark masses
there is approximate agreement between model and experiment.
In this way we determine constituent quark masses. 
Or we may try to solve the theory of strong
interactions (to be outlined below). For the full theory,
we have no chance except for very high momentum transfer
(perturbation theory) or in the framework of effective field theory
at very low momenta (chiral perturbation theory).
The quark masses enter these
calculations as parameters which can then be determined 
by comparison of the computational results with data.
In this case, we solve the equations of strong interactions
and the resulting quark masses are called {\it current} quark
masses. In table~\ref{quarkmass}, mean values~\cite{Hagiwara:fs} are
given. 

\vspace*{-3mm}

\begin{table}[h!]
\caption{Constituent and current quark masses}
\begin{center}
\renewcommand{\arraystretch}{1.3}
\begin{tabular}{lcccc|cccc}
\hline\hline
Quark masses:         &&&&&&\\
Classification        & $d$ & $u$ & $s$&& $c$ & $b$ & $t$ & \\
Current mass          & $\sim$6  &$\sim$ 3 & $\sim$115 &MeV& $\sim$1.2 &  $\sim$4.2  & $\sim$174 &GeV  \\
Constituent mass      & $\sim$340 & $\sim$340 & $\sim$510 &MeV& $\sim$1.2 &  $\sim$4.2  & $\sim$174 &GeV  \\
\hline\hline
\end{tabular}
\end{center}
\renewcommand{\arraystretch}{1.0}
\label{quarkmass}
\vspace*{-2mm}
\end{table}

\subsection{\label{section3.2}Quarks and leptons and their interactions}
\subsubsection{The Standard Model and QCD}
\begin{figure}
\epsfig{file={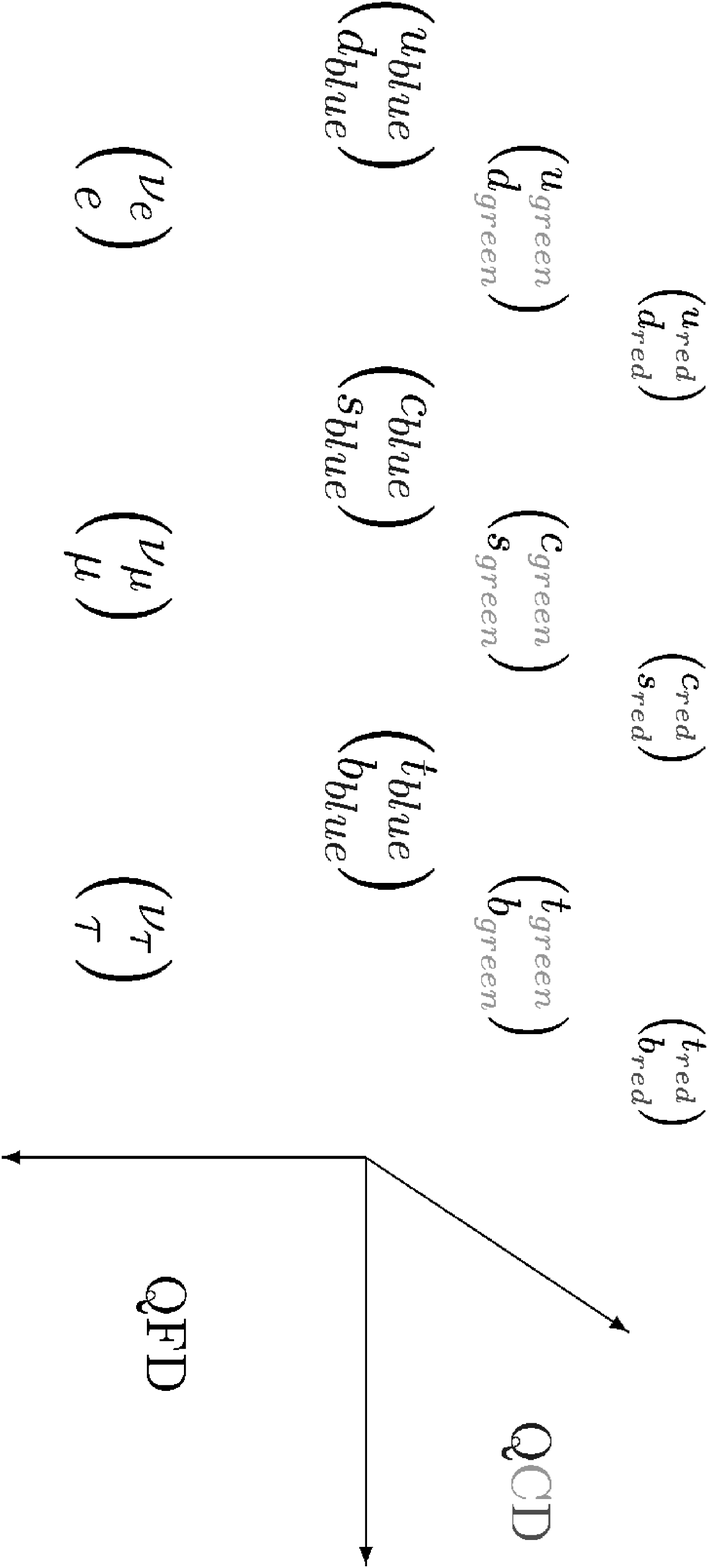},width=0.50\textwidth,angle=90}
\vspace*{-6mm}
\caption{Particles and interactions in the Standard Model.}
\label{SM}
\vspace*{-3mm}
\end{figure}

Within the Standard Model we have 6 leptons and 
6 quarks with different flavors (see figure~\ref{SM}). 
They interact via exchange
of photons or of the 3 weak interaction bosons $W^{\pm}, Z^0$.
This part of the interaction is called quantum flavor dynamics; it
unifies quantum electrodynamics and weak interactions. The four
vector bosons ($\gamma , W^{\pm}, Z^0$) couple to the electric and weak
charges but not to color. Quarks carry a further charge, color,
and interact in addition via the exchange of gluons. Color is triple valued;
all objects directly observable in experiments are color--neutral. 
Gluons can be thought to carry one color and one anticolor in a 
color octet configuration; the completely
symmetric configuration $1/\sqrt 3(\bar rr+\bar bb+\bar gg)$
is a color singlet and excluded, hence there exist 8 gluons.
The gauge group of strong interactions is thus $SU_C(3)$. Likewise
the gauge group of weak interaction is $SU(2)$ and $U(1)$
the gauge group of QED. 

The Standard Model can be broken into its components:
\bc
\begin{tabular}{ccccc}
         $SU_C(3)$    & 
        $\otimes$     & 
\quad $SU(2)$ \quad   & 
\quad $\otimes$ \quad & 
\quad $U(1)$ \quad   \\
$\Downarrow$ && $\Downarrow$ && $\Downarrow$ \\
8 gluons && $W^{\pm},Z^0$ && photon \\
strong interactions &&
\multicolumn{3}{c}{electro-weak interactions}
\end{tabular}
\ec
Gluons are massless particles like photons. It is unclear if 
the notion of a constituent gluon mass (from an effective
parameterization of gluon self--interactions)
is meaningful. Sometimes a constituent mass of 700\,MeV
is assigned to them. Gluons carry the same quantum numbers as
photons, $J^{PC}=1^{--}$. Unlike the electrically neutral photons, 
they carry color. Not only quarks and gluons interact;  
gluon--gluon interactions are possible as well with three--
and four--point vertices.
\bc
\label{fig:1}
\epsfig{file={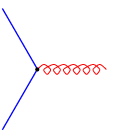},width=0.2\textwidth,origin=c}
\epsfig{file={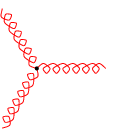},width=0.2\textwidth,origin=c}
\epsfig{file={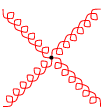},width=0.2\textwidth,origin=c}
\ec
A theory of strong interactions based on the exchange of
colored gluons between colored quarks can be constructed in
analogously to quantum electro dynamics.
It was a great success that the resulting theory, quantum
chromo dynamics, can be shown to be renormalizable. Like
in QED, some expressions give infinite contributions, but
the renormalization scheme allows one to control all divergencies.
The color-electromagnetic fields 
\begin{equation}
F_{\mu\nu}^{a} =
\delta_{\mu}A_{\nu}^{a} -
\delta_{\nu}A_{\mu}^{a} +
g_s f_{abc}
A_{\mu}^{b} A_{\nu}^{c}
\nonumber
\end{equation}
resemble QED very much,  except for the color indices $a,b$, and $c$
	and the third term describing gluon--gluon interactions.
\par
\begin{figure}
\centerline{\epsfig{file=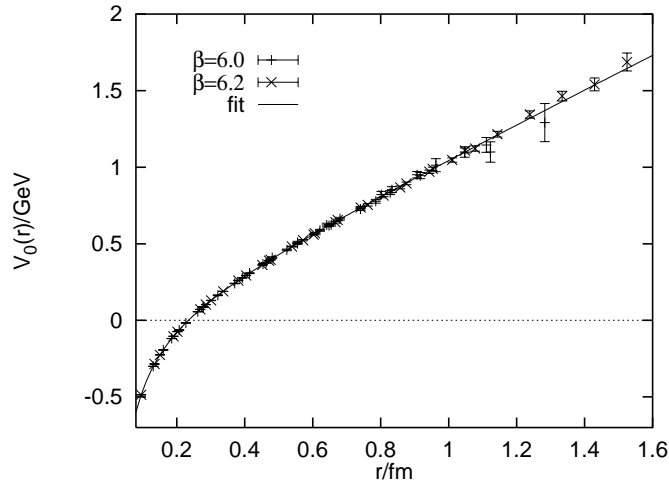,width=9.2cm,clip=}}
\vspace*{-5mm}
\caption{\label{fig:potential}
The potential energy between two heavy quarks (with fixed
positions) as a function of their separation from lattice QCD.
The solid line represents a potential in the form
$V(r) = \frac{4}{3}\frac{\alpha_s}{r} + b\cdot r$;
from ~\protect\cite{Bali:2000gf}.}
\vspace*{-5mm}
\end{figure}

The beauty of QCD as a theory of strong interactions has some ugly
spots. The coupling constant $\alpha_s$ increases dramatically 
with decreasing momentum transfer, and QCD predictions
in the low--energy regime are (mostly) not possible. Only
at high momentum transfer is $\alpha_s$  small and does QCD
become a testable and useful theory. Numerically, there is
progress to calculate QCD quantities on a discrete space--time
lattice. Figure~\ref{fig:potential} shows as example the static $Q\bar{Q}$ 
potential as a function of separation~\cite{Bali:2000gf}. It is the potential
energy between two heavy quarks; the possibility that 
virtual $\bar qq$ pairs can be created is
neglected. The line represents a superposition of a $1/r$
potential as expected from one--gluon exchange between quarks
and a linearly rising part reflecting confinement.  

\subsubsection{From large energies to large distances }
\begin{figure}[h!]
\setlength{\unitlength}{1.2mm}
\bc
\begin{picture}(150.00,90.00)
\put(5,15){\epsfig{file=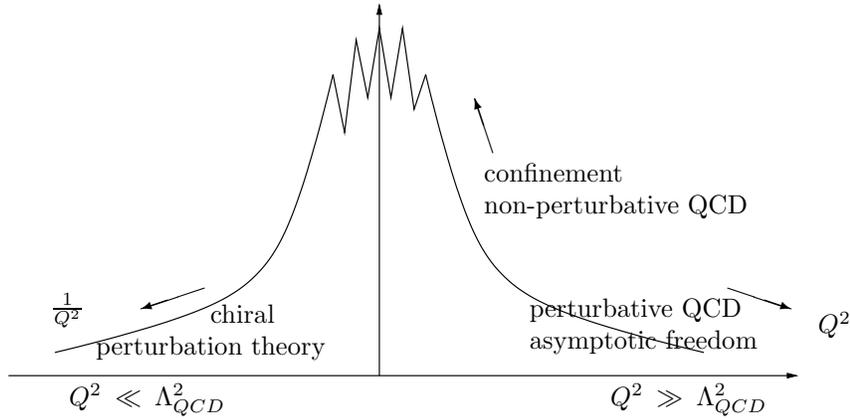,height=5cm}}
\put(10,50){\begin{minipage}[h]{5cm}

          \end{minipage}}
\put(58,35){\begin{minipage}[h]{4cm}
               { confinement\\ non-perturbative QCD}
          \end{minipage}}
\put(95,20){$Q^2$}
\put(10,22){$\frac{1}{Q^2}$}
\put(72,12){$Q^2\:\gg\:\Lambda^2_{QCD}$}
\put(12,12){$Q^2\:\ll\:\Lambda^2_{QCD}$}
\put(15,36){\begin{minipage}[h]{7cm}\vspace*{4cm}
               {  \hspace{14mm} chiral\\ perturbation theory}
          \end{minipage}}
\put(63,20){\begin{minipage}[h]{6cm}
               { perturbative QCD\\
                             asymptotic freedom}
          \end{minipage}}
\put(27,25){{\vector(-3,-1){7}}}
\put(85,25){{\vector(3,-1){7}}}
\put(58,40){{ \vector(-1,3){2}}}
\end{picture}
\ec
\vspace*{-15mm}
\caption{QCD observables as a function of $Q^2$.}
\label{strongQCD}
\vspace*{-5mm}
\end{figure}

Figure~\ref{strongQCD} sketches the situation. At very large energies
QCD can be treated perturbatively. The strong interaction constant $\alpha_s$
decreases and particles behave asymptotically as if they were
free. For lower $Q^2$ confinement becomes the most important
aspect of strong interactions. This is the realm of non--perturbative
QCD or of {\it strong QCD}. At very small $Q^2$, in the
{\it chiral limit}, observables can be expanded in powers of 
masses and momenta and chiral perturbation theory leads to reliable
predictions~\cite{Meissner:2003hr}. 
In an extremely hot and dense environment we
expect quarks to become free; a phase transition 
to the quark--gluon plasma is expected~\cite{Satz:2000bn}
and has likely been observed~\cite{Abreu:2000ni}.
\par
The region of interest here is the one where QCD is really
strong, where perturbative QCD and chiral perturbation theory 
both fail. This is the region most relevant to our daily
life; in  this region protons and neutrons and their
excitations exist. For momentum scales given by typical hadron masses,
not only $\alpha_s$ changes but 
also the relevant degrees of freedom change from current quarks
and gluons to constituent quarks, instantons and vacuum condensates.
To understand this transition is one of the most challenging
intellectual problems. 
\par
\vspace*{-42mm}

\subsubsection{Basic questions in strong QCD}
A clarification of the following central issues is needed:
\begin{itemize}
\vspace{-2mm}\item What are the relevant degrees of freedom that govern
hadronic phenomena\,? 
The relevant degrees of freedom
in superconductors are Cooper pairs and not electrons. 
In meson and baryon spectroscopy, the constituent quarks seem to
play an important role, but how are they formed and
what is their interaction\,?\hspace*{-3mm}
\vspace{-2mm}\item What is the relation
between partonic degrees of freedom in the infinite momentum frame
and the structure of hadrons in the rest frame\,? 
At large
momentum transfers we know that about 50\% of the momentum of
a proton is carried by gluons and not by quarks. In deep inelastic
scattering structure functions reveal the importance of sea
quarks. Do these participate in the dynamics of mesons and
baryons and, is so, how\,?
\vspace{-2mm}\item What are the mechanisms for confinement and for
chiral symmetry breaking\,? Are deconfinement and chiral symmetry restoration
linked and can precursor phenomena be seen in nuclear physics\,?
At very large densities and temperatures, quarks can no longer
be assigned to a particular proton; quarks can be exchanged frequently
and propagate freely in this dense material, conserving
chirality. This is a phase 
transition from the regime of broken chiral symmetry to a
regime were it is restored; and from the hadronic phase to the 
quark--gluon plasma. It is unknown whether these two phase
transitions are identical, occurring under the same conditions. 
\end{itemize}
\vspace*{-3mm}

\subsubsection{Modeling strong QCD}

The answers to these questions will not be the direct result of 
experiments. Models are needed to link observables
to these fundamental questions. Significant observables are
the nucleon excitation spectrum and their electromagnetic couplings
including their off-shell behavior and the response of hadronic
properties to the exposure by a nuclear environment.

For many physicists the ultimate hope of 'solving' QCD is
performing numerical calculations on a space--time lattice. 
At least for the
years to come we have to rely on models. These models try to 
shape what we know or believe about QCD; they are called 
{\it QCD inspired models}. 
\vspace*{-3mm}

\subsubsection{Gluon exchange and the flux tube model}
A very popular version
introduces a linear confinement potential and a kind
of 'effective' one-gluon exchange (with $\alpha_s$ chosen
arbitrarily and neglecting higher orders of an expansion in
$\alpha_s$). Flux tube models concentrate the gluon field 
connecting a quark and an antiquark in a tube
of constant energy density. The flux tube introduces a new
degree of freedom into hadrons; while the orbital angular
momentum along the direction between $e^+$ and $e^-$ in
positronium vanishes, the flux tube can rotate around this axis. This
dynamical enrichment leads to a richer spectrum. The
additional states are called hybrids. 
\vspace*{-3mm}

\subsubsection{Chiral symmetry and instanton--induced interactions}
 
The $u,d$,and $s$ quarks are nearly massless; the so--called ``current'' 
quark masses (.i.e. those which appear in the fundamental QCD
Lagrangian) are in the mass range, respectively, from 1.5 to 5\,MeV;
3 to 9\,MeV, or 60 to 170\,MeV~\cite{Hagiwara:fs}. 
In the chiral limit QCD possesses
a large symmetry, and quarks with vanishing mass 
preserve their handiness, their chirality. 
If chiral symmetry were unbroken, all
baryons would appear as parity doublets. Obviously this is not
the case since the masses of proton and its first orbital--angular
momentum excitation N(1535)S$_{11}$ are very different. 
Hence the (approximate) chiral symmetry is broken spontaneously. As a 
result the eight pseudoscalar mesons $\rm\pi, K$, and $\eta$ are light
Goldstone bosons. They are light but not massless
(as the Goldstone therorem~ \cite{Goldstone:eq} would require)
because the current quarks have small (but finite) masses.
\par
\begin{figure}[t!]
\vspace*{-3mm}
\begin{tabular}{cc}
\epsfig{file=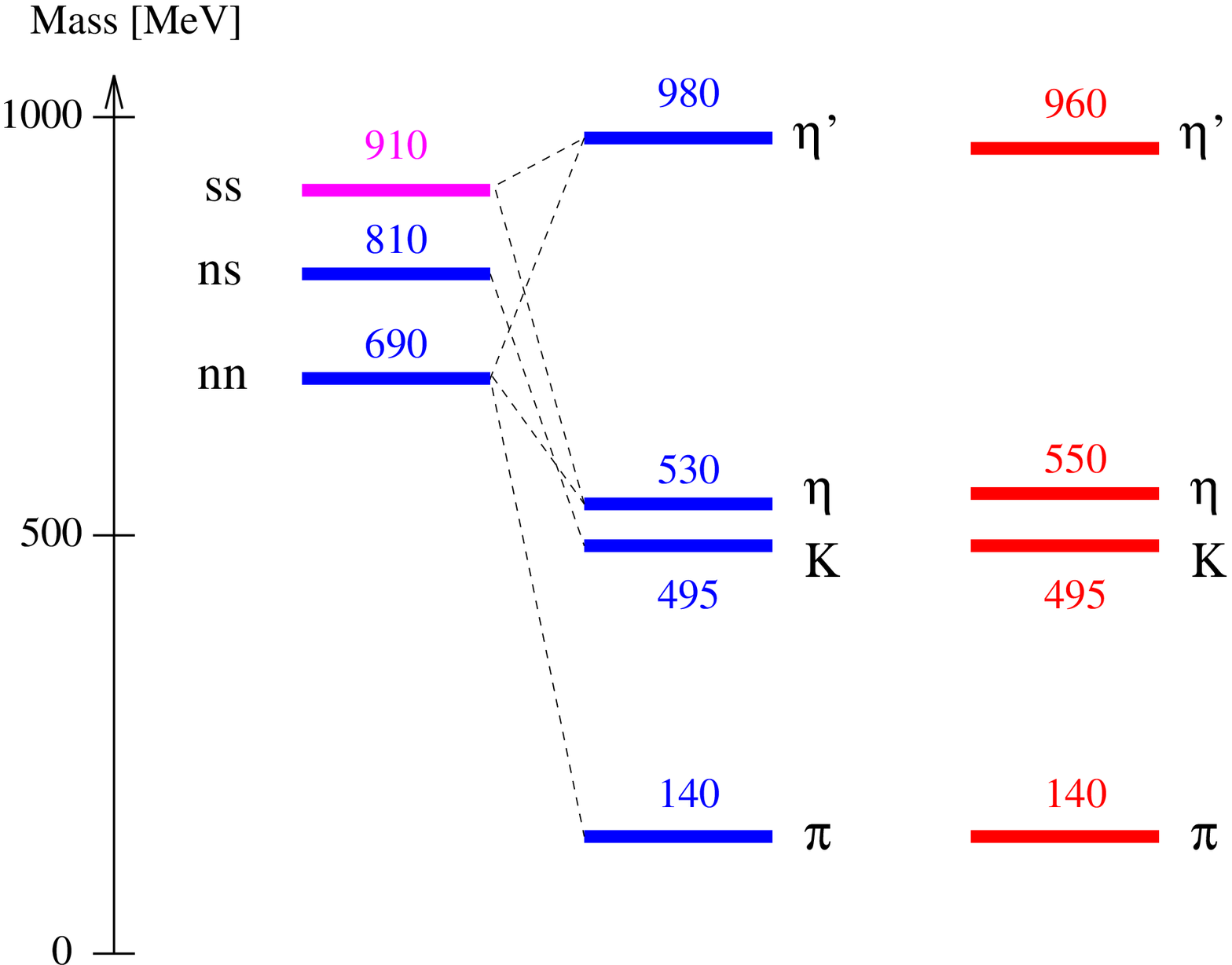,width=0.45\textwidth}&
\epsfig{file=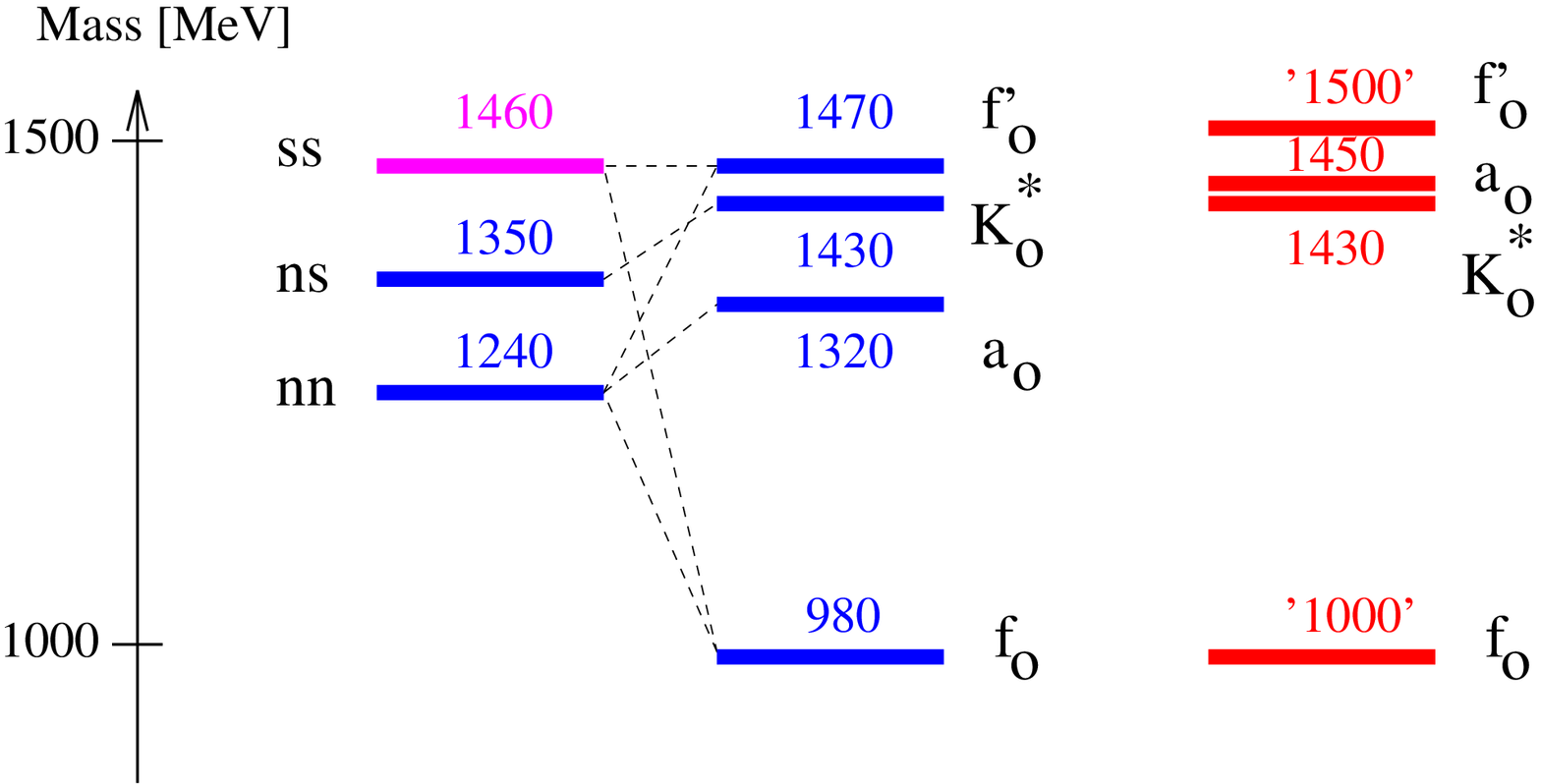,width=0.45\textwidth}
\end{tabular}
\caption{The action of instantons in the mass spectrum
of  pseudoscalar (left) and scalar (right) mesons. Shown are
the spectra as calculated using only a confinement potential 
and the mass shifts resulting from instantons 
(from~\protect\cite{Klempt:1995ku}).}
\label{sps}
\vspace*{-3mm}
\end{figure}

Compared to the proton or $\Lambda$, the 8 pseudoscalar mesons 
$\pi , \eta$ and K have small masses; their masses
are remnants of the Goldstone theorem. Now we have a problem.
The $\eta^{\prime}$ mass is close to the proton mass, but due
to  flavor $SU(3)$, or $SU_F(3)$, it should have a small mass too
but it has not. As an $SU_F(3)$ singlet it couples directly to the gluon 
fields. This gives rise to an additional interaction introduced
by 't Hooft~\cite{'tHooft:1986nc}. It originates 
from the spontaneously broken chiral symmetry and the 
occurance of instantons in the QCD 
vacuum~\cite{Diakonov:vw,Diakonov:1987ty,Diakonov:2002fq}. 
Their action on the masses of pseudoscalar and scalar mesons
can be seen in figure~\ref{sps}~\cite{Klempt:1995ku}).
\par
Instanton--induced interactions originate from vacuum fluctuations
of the gluon fields. Already in QED there are vacuum fluctuations
of the electromagnetic fields. 
Unlike QED, QCD allows solutions to have ``topological charge'' or
a ``winding number'' (I like to interprete theses as field vortices
since they can flip quark spins; technically, QCD
vortices are different objects and may have non--integer winding numbers). 
Quarks can flip helicity when scattering on instanton 
fluctuations. Instantons 
change quark helicity from right to
left, anti--instantons from left to right. Therefore, when quarks
scatter on many instantons and anti--instantons they acquire
a dynamical mass signaling the spontaneous breaking of chiral 
symmetry. 
\par
These 
induced spin--flips are the origin of the instanton--induced
interactions. The same interaction acts independently
on $u,d$, and $s$ quarks so that each of them flip helicity. Averaging
over the positions of the instanton fluctuation induces a correlation
between the $u,d$, and $s$ quarks, which can be written conventiently in the
form of the 't Hooft interaction.
Instanton--induced interactions violate the OZI rule. In mesons,
a quark can flip its spin only when the antiquark flips
its spin simultaneously since the total spin is conserved. 
$J$ is conserved, too. Thus $J$ must vanish, and
instanton--induced interactions contribute only to pseudoscalar
and scalar mesons. In baryons, spin and flavor flips can only occur
when the two--quark wave function is antisymmetric 
in spin and in flavor~\cite{Shuryak:1981ff} when the two 
quarks are exchanged. 
In baryons with a total quark spin 1/2, the $(qq)$--spin vanishes 
for one component of the baryonic wave function, 
and this component is antisymmetric. In octet baryons one
$(qq)$ is antisymmetric in flavor. In singlet baryons all three 
$(qq)$ pairs are antisymmetric w.r.t. their exchange.
\par
Due to spontaneous symmetry breaking, the
isovector $q\bar q$ pairs in $^3S_1$ acquire the $\rho$
mass; the pion remains massless. However  
there is a second kind how chiral symmetry is broken. The
massless quarks couple, like leptons, to the Higgs field
which generates the current quark masses. 
The current quark mass then gives a finite mass to the pion.
Chiral symmetry leads to constituent 
quarks and is responsible for the largest 
fraction of the proton mass. 
\par
The situation can be compared to the
more familiar magnetism. Individual Fe atoms have a
magnetic moment and their directions are arbritrary. 
Many Fe atoms cluster to the Weiss districts with a
macroscopic magnetization in a fixed direction. 
The direction is random; even though the atoms within
the Weiss district have no 'reason', they decide
spontaneously to magnetically point in a specific direction.
An external magnetic field may induce a preferred
direction; this is an induced (external) breaking
of rotational symmetry. 
\subsubsection{The chiral soliton model}
The concept of a nucleon composed of 3 constituent
quarks is certainly oversimplified and the hadronic properties
of nucleons cannot be understood or, at least, are not understood
in terms of quarks and their interactions. Skyrme studied the
pion field and discovered that by adding a non--linear ``$\sigma$
term'' to the pion field equation, stable solutions can 
result~\cite{Skyrme61}.
These solutions have half integer spin and a winding number 
identified by Witten~\cite{Witten83}   
as the baryon number.
These stable solutions of the pion field
equation are called soliton solutions. 
\par
Of course the Skyrme model does not imply that there are no quarks. 
Again we compare the situation with magnetic interactions.
The theory of ferromagnetism does not imply that Weiss districts
are elementary physics. The Skyrme view can be used to understand
aspects of baryons from a different point of view. In a modification of
the Skyrme model, in the chiral soliton model~\cite{Diakonov:2000pa},
the Skyrme solitons tur out to be the self--consistent field that
binds quarks inside  a baryon. 
\par
\vspace*{-10mm}
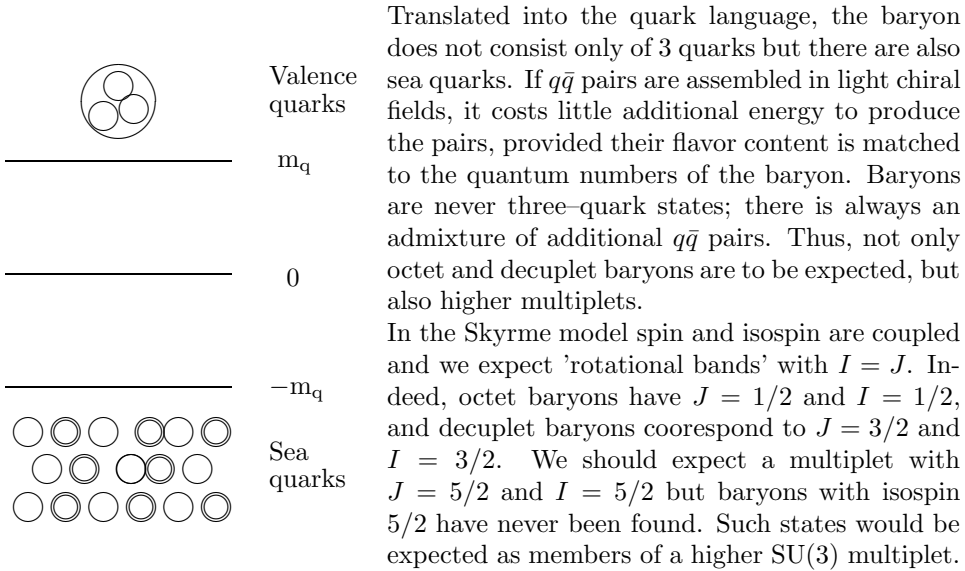
\begin{figure}[h!]
\setlength{\unitlength}{1mm}
\begin{minipage}[c]{0.39\textwidth} 
\begin{picture}(150.00,90.00) 
\put(13.00,73.00){\bf \circle{4.00}} 
\put(15.00,77.00){\bf \circle{4.00}} 
\put(17.00,74.00){\bf\circle{4.00}} 
\put(35.00,76.00){\makebox(12.50,5.00)[l]{Valence }} 
\put(35.00,72.00){\makebox(12.50,5.00)[l]{quarks}} 
\put(15.00,75.00){\circle{10.00}} 
\put(0.00,67.00){\line(1,0){30.00}} 
\put(35.00,64.00){\makebox(12.50,5.00)[l]{$\ \bf\rm m_q$}} 
\put(0.00,52.00){\line(1,0){30.00}} 
\put(35.00,49.00){\makebox(12.50,5.00)[l]{$\ \ \bf\rm 0$}} 
\put(0.00,37.00){\line(1,0){30.00}} 
\put(35.00,34.00){\makebox(12.50,5.00)[l]{$\bf\rm -m_q$}} 
\put(03.00,31.00){\circle{4.00}} 
\put(08.00,31.00){\circle{4.00}} 
\put(08.00,31.00){\circle{3.00}} 
\put(13.00,31.00){\circle{4.00}} 
\put(18.00,31.00){ \circle{4.00}} 
\put(18.00,31.00){ \circle{3.00}} 
\put(23.00,31.00){\circle{4.00}} 
\put(28.00,31.00){\circle{4.00}} 
\put(28.00,31.00){\circle{3.00}} 
\put(05.50,26.00){\circle{4.00}} 
\put(10.50,26.00){\circle{4.00}} 
\put(10.50,26.00){\circle{3.00}} 
\put(15.50,26.00){ \circle{4.00}} 
\put(15.50,26.00){ \circle{4.00}} 
\put(20.50,26.00){\circle{3.00}} 
\put(20.50,26.00){\circle{4.00}} 
\put(25.50,26.00){\circle{4.00}} 
\put(03.00,21.00){\circle{4.00}} 
\put(08.00,21.00){\circle{4.00}} 
\put(08.00,21.00){\circle{3.00}} 
\put(13.00,21.00){\circle{4.00}} 
\put(18.00,21.00){\circle{4.00}} 
\put(18.00,21.00){\circle{3.00}} 
\put(23.00,21.00){\circle{4.00}} 
\put(28.00,21.00){\circle{4.00}} 
\put(28.00,21.00){\circle{3.00}} 
\put(35.00,26.00){\makebox(12.50,5.00)[l]{Sea}} 
\put(35.00,22.00){\makebox(12.50,5.00)[l]{quarks}} 
\end{picture} 
\vspace*{-20mm}
\caption{Quarks and sea quarks are dynamically coupled. The equations
of motion support soliton solutions which can be organized into
multiplets. The lowest lying multiplets are ${\bf 8}$
and ${\bf 10}$ and $\bf\overline{10}$. }
\end{minipage} 
\begin{minipage}[c]{0.60\textwidth}
\vspace*{7mm}
Translated into the
quark language, the baryon does not consist only of 3 quarks but there
are also sea quarks. If $q\bar q$ pairs are assembled in light
chiral fields, it costs little additional energy to produce the
pairs, provided their flavor content is matched to the quantum numbers
of the baryon. Baryons are never three--quark states; there is always
an admixture of additional $q\bar q$ pairs. 
Thus, not only octet and decuplet baryons are to be
expected, but also higher multiplets.
\par
In the Skyrme model  spin and isospin are coupled and we expect 
'rotational bands' with $I=J$. Indeed, octet baryons have $J=1/2$ and
$I=1/2$, and decuplet baryons coorespond to
$J=3/2$ and $I=3/2$. We should expect a
multiplet with  $J=5/2$ and $I=5/2$ but 
baryons with isospin 5/2 have never been found. 
Such states would be
expected as members of a higher SU(3) multiplet.
An excuse may be that these baryons could be very broad. 
\par
The chiral soliton model predicts the existence 
of an antidecuplet~\cite{Chemtob85,Walliser92} 
shown in figure~\ref{anti10}. The
flavor wave function in the minimum quark model
configuration is given by $\Theta^+ = uud d\bar s$; it is called 
\end{minipage}
\end{figure}
\vspace*{-3mm}
\noindent
a pentaquark~\cite{Diakonov97}. The strange quark fraction
increases from 1 to 2 units in steps of 1/3 additional $s$ quark.
The increase in mass per unit of strangeness is 
is 540\,MeV, instead of 120\,MeV
when the $\rho$ or $\omega$ mass is compared to the K$^*$ mass.
The splitting is related to the so--called $\sigma_{\pi N}$ term
in low--energy $\pi$\,N scattering. Its precise value is difficult
to determine and undergone a major revision. The splitting
is now expected to be on the order of 110\,MeV for an
additional 1/3 $s$ quark~\cite{Diakonov:2003jj}.
Note that the three corner states have quantum numbers which cannot
be constructed out of 3 quarks.

\begin{figure}[h!]
\vspace*{-12mm}
\begin{minipage}[c]{0.48\textwidth}
\large
\vskip 16mm
\renewcommand{\arraystretch}{1.5}
\hspace*{-5mm}
\begin{tabular}{c}
$\rm  uud d\bar s$ \\
\\
 $\rm  uud(1/\sqrt{3} d\bar{d} + \sqrt{2/3}  s \bar{s})\phantom{rrrrr}$\\
\\
$\rm  uu{ s}(1\sqrt{3} { s\bar{s}} + 
\sqrt{2/3} d \bar{d})$\phantom{rrrrrrrr}\\
\\
$\rm  uu{ ss} \bar{d}$\phantom{rrrrrrrrr}\\
\end{tabular} 
\renewcommand{\arraystretch}{1}
\end{minipage}
\begin{minipage}[c]{0.48\textwidth}
\begin{center}
\setlength{\unitlength}{0.6mm}
\hspace*{-15mm}
\begin{picture}(150.00,90.00)
\put(10.00,20.00){ \vector(1,0){70.00}}
\put(45.00,-10.00){\vector(0,1){90.00}}
\put(82.50,20.00){\makebox(5.00,5.00){$I_3$}}
\put(47.50,82.00){\makebox(5.00,5.00){ Antidecuplet }}
\put(07.50,-5.00){\circle*{2.00}}
\put(7.50,-5.00){  \circle{4.80}}
\put(7.50,-5.00){  \circle{4.40}}
\put(7.50,-5.00){  \circle{4.00}}
\put(7.50,-5.00){  \circle{3.60}}
\put(7.50,-5.00){  \circle{3.20}}
\put(7.50,-5.00){  \circle{2.80}}
\put(7.50,-5.00){  \circle{2.40}}
\put(32.50,-5.00){\circle*{2.00}}
\put(57.50,-5.00){\circle*{2.00}}
\put(82.50,-5.00){  \circle{4.80}}
\put(82.50,-5.00){  \circle{4.40}}
\put(82.50,-5.00){  \circle{4.00}}
\put(82.50,-5.00){  \circle{3.60}}
\put(82.50,-5.00){  \circle{3.20}}
\put(82.50,-5.00){  \circle{2.80}}
\put(82.50,-5.00){  \circle{2.40}}
\put(82.50,-5.00){\circle*{2.00}}
\put(45.00,20.00){\circle*{2.00}}
\put(70.00,20.00){\circle*{2.00}}
\put(20.00,20.00){\circle*{2.00}}
\put(32.50,45.00){\circle*{2.00}}
\put(57.50,45.00){\circle*{2.00}}
\put(45.00,70.00){  \circle{4.80}}
\put(45.00,70.00){  \circle{4.40}}
\put(45.00,70.00){  \circle{4.00}}
\put(45.00,70.00){  \circle{3.60}}
\put(45.00,70.00){  \circle{3.20}}
\put(45.00,70.00){  \circle{2.80}}
\put(45.00,70.00){  \circle{2.40}}
\put(45.00,70.00){\circle*{2.00}}
\put(45.00,70.00){\circle*{2.00}}
\put(7.50,-5.00){\line(1,0){75.00}}
\put(82.50,-5.00){\line(-1,2){37.50}}
\put(7.50,-5.00){\line(1,2){37.50}}
\put(83.00,67.50){\makebox(15.00,5.00)[r]{$\mathbf \Theta^+(1530)$\ \  S=+1}}
\put(17.00,42.50){\makebox(13.25,5.00)[r]{$  N^0{(1710)}$}}
\put(97.00,42.50){\makebox(13.25,5.00)[r]{$  N^+{(1710)}$\ \  S=+0}}
\put(-11.00,12.50){\makebox(12.50,5.00)[l]{$  \Sigma^-(1890)$}}
\put(35.00,12.50){\makebox(12.50,5.00)[l]{$  \Sigma^0(1890)$}}
\put(80.00,12.50){\makebox(12.50,5.00)[l]{$  \Sigma^+(1890)$\quad  S=-1}}
\put(-11.00,-15.00){\makebox(12.50,5.00)[l]{$  \bf \Xi^{--}(2070)$}}
\put(22.00,-15.00){\makebox(12.50,5.00)[l]{$  \Xi^{-}(2070)$}}
\put(55.00,-15.00){\makebox(12.50,5.00)[l]{$  \Xi^{0}(2070)$}}
\put(85.50,-15.00){\makebox(12.50,5.00)[l]{$\mathbf\Xi^+(2070)$\quad  S=-2}}
\end{picture}
\end{center}
\end{minipage}
\vskip 5mm
\caption{The antidecuplet and its quark model decomposition. 
The antidecuplet predicted by the chiral soliton model
describes nucleons in terms of the pion field and not
by the number of quarks~\protect\cite{Diakonov97}}
\label{anti10}
\vspace*{-5mm}
\end{figure}
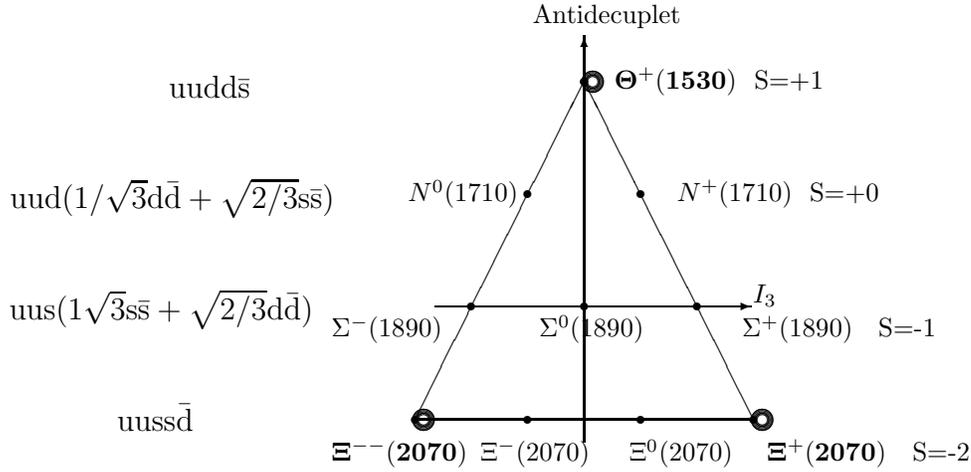

The recent discovery of the $\Theta^+(1540)$ (the experimental
evidence for it will be discussed in section~\ref{section5.4}) 
with properties 
as predicted in the chiral soliton model has ignited a considerable
excitation about this new spectroscopy and its interpretation. In the
chiral soliton model, the members of the antidecuplet all have
$J^P=1/2^+$. This must be been tested experimentally. A principle
concern is the lack of predictions in the Skyrme model of baryons 
with negative 
parity. I do not know if this is a limitation of the model or if
this fact just reflects the limited interest and scope of the physicists
working on the Skyrme model.

\subsubsection{Confinement}
The formation of constituent quarks and their confinement
is a central issue of theoretical developments. These questions
are beyond the scope of these lectures. We refer to two recent
papers~\cite{Alkofer:2000wg,Szczepaniak:2001rg}.
\subsection{\label{section3.3}Quark models for mesons} 
Explicit quark models start from a confining  potential, mostly in the
form  $V = V_0 + a\,r$ where $a$ is the string constant,
$a = 0.2$\,GeV$^2$. 
At small distances, a Coulomb--like potential
$V = -\frac{4}{3}\frac{\alpha_s}{r}$ due to one-gluon exchange 
is added. The (constituent) mass of the quarks is a
parameter of the model. A central question now is
how the {\it effective interaction} between 
constituent quarks should be described. Three suggestions are presently
discussed: 
\vspace*{-2mm}
\begin{enumerate}
\item Is there an effective one-gluon exchange\,?
\vspace*{-2mm}
\item Do quarks in baryons exchange 
Goldstone bosons, i.e.  pseudoscalar mesons\,?\hspace*{-3mm}
\vspace*{-2mm}
\item Or is the interaction
best described by instanton-induced interactions\,?
\vspace*{-2mm}
\end{enumerate}
\subsubsection{The  Godfrey-Isgur  model}

The first unified constituent quark model for all $q\bar q$-mesons was
developed by Godfrey and Isgur~\cite{Godfrey:xj}.
The model starts from a Hamiltonian

\begin{displaymath}
	H \Psi = (H_0 + V) \Psi = E \Psi,
\qquad
H_0 = \sqrt{{m_q}^2+|\vec p|^2} +
\sqrt{{m_{\bar q}}^2+|\vec p|^2}\,,
\end{displaymath}

with $\vec p$ the relative momentum in the CM-frame, and
an interaction part 
\begin{displaymath}
	V = H^{c} + H^{h\!f} + H^{SS} + H^{A}
\end{displaymath}
which contains the central potential (linear confinement 
$br + c$ and Coulomb potential), the spin--spin and tensor interaction and
an annihilation contribution for flavor--neutral mesons.

The potential is generated by a vector (gluon) exchange
\vspace*{-1mm}
$$
	G(Q^2) = -\frac{4}{3}\,\alpha_s(Q^2)\frac{4\pi}{Q^2}
$$
where $\alpha_s(Q^2)=\sum_k\,\alpha_k\,\ e^{-\frac{Q^2}{4\gamma_k^2}}$
is a parameterization of the running coupling, with
${\alpha_s(0)}$\,finite,
and a long-range confining potential $ S(Q^2)$,
$S(r)={b}r+{c}$.
Here, $\vec Q = \vec p' -\vec p$.
These potentials are ``smeared out''
to avoid singularities at the origin. 
Relativistic effects are partly taken into account,
but spin--orbit forces are suppressed; {\it 
there are no spin-orbit forces in the Hamiltonian}.
The excuse for this suppression is the experimental
observation that these are weak or absent in the data. From the
theoretical side, spin--orbit forces are at least partly
compensated by the so--called Thomas precession, a relativistic
generalisation of Coriolis forces. 
Within a fully relativistic treatment, the Thomas precession can be
calculated but it fails to cancel the
spin--orbit forces at the level required by data~\cite{metschpc}. 
\par
Annihilation is taken into account by parameterizing the 
annihilation amplitude, one for non-pseudoscalar flavor-neutral
mesons and a different one for pseudoscalar mesons. 
All mesons are assumed to be ``ideally mixed'', except the
pseudoscalar mesons.
Finally, it may be useful to give (table~\ref{GS})
the list of parameters which were
tuned to arrive at the meson spectra.
\vspace*{-3mm}
{\small
\begin{table}[h!]
\caption{Parameters of the \textsc{Godfrey-Isgur} model. 
The quark masses are constituent masses, $b$ and $c$ describe the
confinement potential, $\alpha(0)$ is the coupling constant
as defined in the text, the $\epsilon$ are various correction factors,
the $A$'s give mass contributions from virtual annihilationand $M_0$
is a mean meson mass~\protect\cite{Godfrey:xj}.}
\begin{center}
\begin{tabular}{ccrlcrl}
\hline
\hline
masses 		& $m_n$ 	& 220 	& MeV
		& $m_s$ 	& 419 	& MeV \\
confinement 	& b 		& 910 	& MeV/fm
		& c 		&-253 	& MeV \\
OGE 		& $\alpha_s(0)$ & 0.60 	&
		& $\Lambda$ 	& 200 	& MeV \\
    		& $\epsilon_{SS}$ & -0.168 &
		& $\epsilon_T$ 	& 0.025 & \\
    		& $\epsilon_{LS}^C$ & -0.035 &
		& $\epsilon_{LS}^S$ & 0.055 & \\
``smearing'' 	& $\sigma_0$ 	& 0.11 & fm
		& $s$ 		& 1.55 & \\
annihilation 	& $A(^3 S_1)$ 	& 2.5 &
		& $A(^3 P_2)$ 	& -0.8 & \\
             	& $A_0$		& 0.5 (0.55) &
		& $M_0$		& 550(1170)	& MeV\\
\hline\hline
\end{tabular}
\end{center}
\label{GS}
\vspace*{-5mm}
\end{table}
}

\vspace*{-3mm}
\subsubsection{Meson exchange between quarks\,?}
\vspace*{-1mm}
The absence of strong spin--orbit forces in the meson and baryon 
spectrum sheds some doubts on one--gluon exchange as a leading
mechanism in hadron spectroscopy. Also the low mass of the 
N(1440)$P_{11}$ (Roper) resonance is a point of concern.
As radial excitation it belongs to the second excitation band,
but its mass is lower than the N(1535)$S_{11}$  and N(1520)$D_{13}$.
Riska and Glozman~\cite{Glozman:1995fu} suggested that
constituent quarks may interact via exchange of Goldstone bosons 
(i.e. of pseudoscalar mesons). The masses of low--lying N$^*$ and $\Delta^*$ 
resonances have been reproduced very well; no attempt was made to
calculate the full mass spectrum or to address questions like
missing resonances. Glozman emphasized that in the high--mass spectrum
new phenomena may occur. He suggested that chiral symmetry could be
restored at large excitation energies and that the mass spectrum
organizes into parity doublets~\cite{Glozman:2002jf}. 
A large number of parity doublets is indeed observed but the 
doublets can also be explained~\cite{Klempt:2002tt} 
by assuming that radial excitation
energy (per $N$) and orbital excitation energy (per $L$) are the same. 
This is approximately true in meson spectroscopy (see figure~\ref{va}).

\vspace*{-2mm}
\subsubsection{The Bonn model}
\vspace*{-2mm}
An ambitious program was started in Bonn many years ago. The
aim is to calculate meson~\cite{Koll:2000ke} and 
baryon~\cite{Metsch} resonances and their
properties starting from field theory, which amounts
to solving the homogeneous, instantaneous
Bethe--Salpeter equation. 
The interactions used in the model differ from
the models described above in two important aspects. First, as
a relativistic theory, the confinement is described by a linear
potential  only in the rest frame. The potential can be boosted
into any other system and then develops a time component. The
potential has a Lorentz structure; the relativistic
transformation properties are chosen to minimize the (unwanted)
spin--orbit coupling. Two variants of the Lorentz structure
are used, defining model $\cal A$ and $\cal B$. The two Lorentz
structures are given in table~\ref{BM} which also lists the parameters
of the model. Second, not one--gluon exchange is used but 
instanton-induced interactions are used instead. 
The strengths of the interaction for
$u\bar u\to d\bar d$ and for $u\bar u\to s\bar s$ transitions
are fit to describe the ground state pseudoscalar mesons.
\par
\begin{figure}
\begin{minipage}[b]{0.74\textwidth}
\vspace*{-6mm}
\epsfig{file={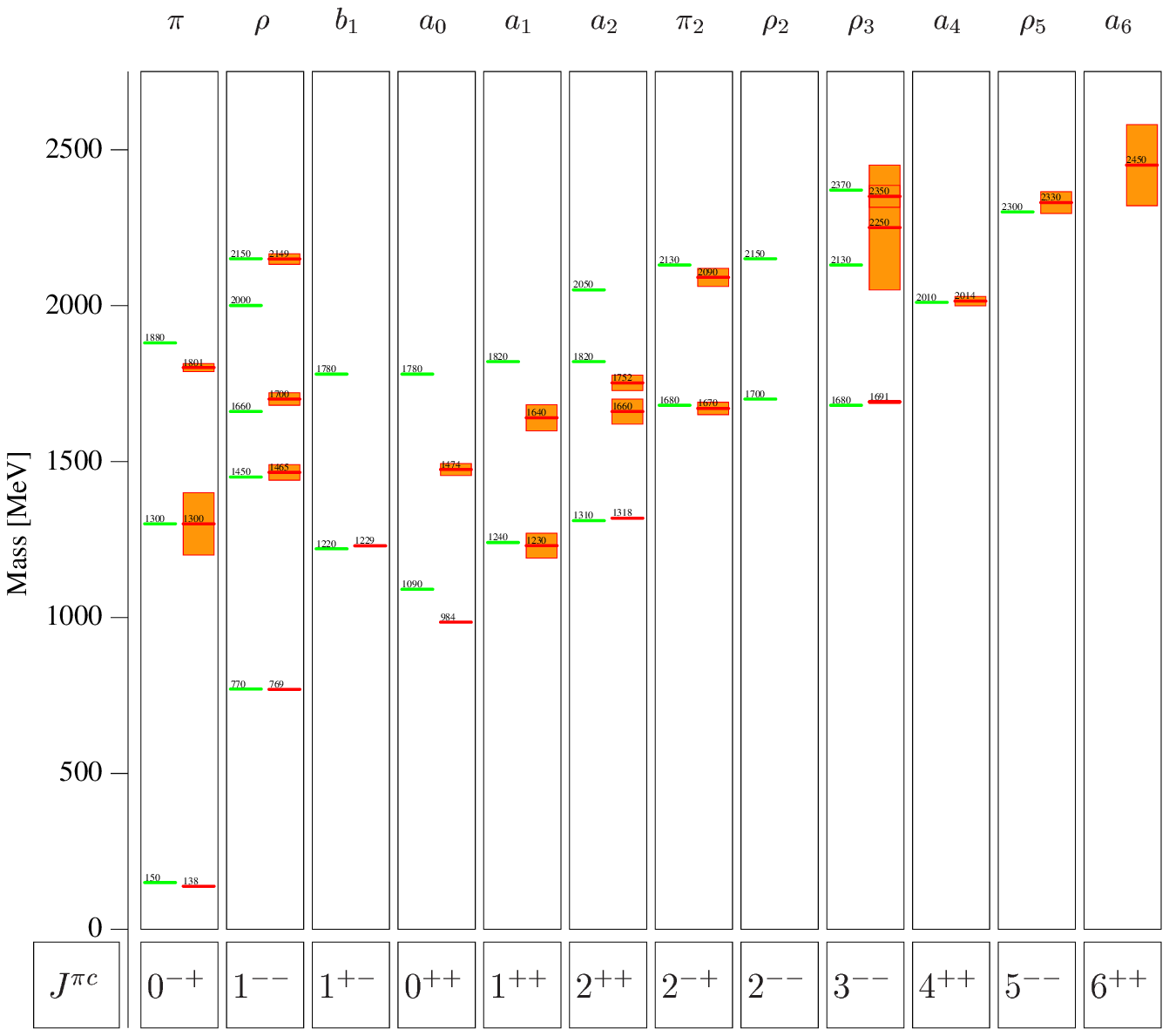},width=\textwidth,height=0.35\textheight,origin=c}\\
\vspace*{-4mm}
\epsfig{file={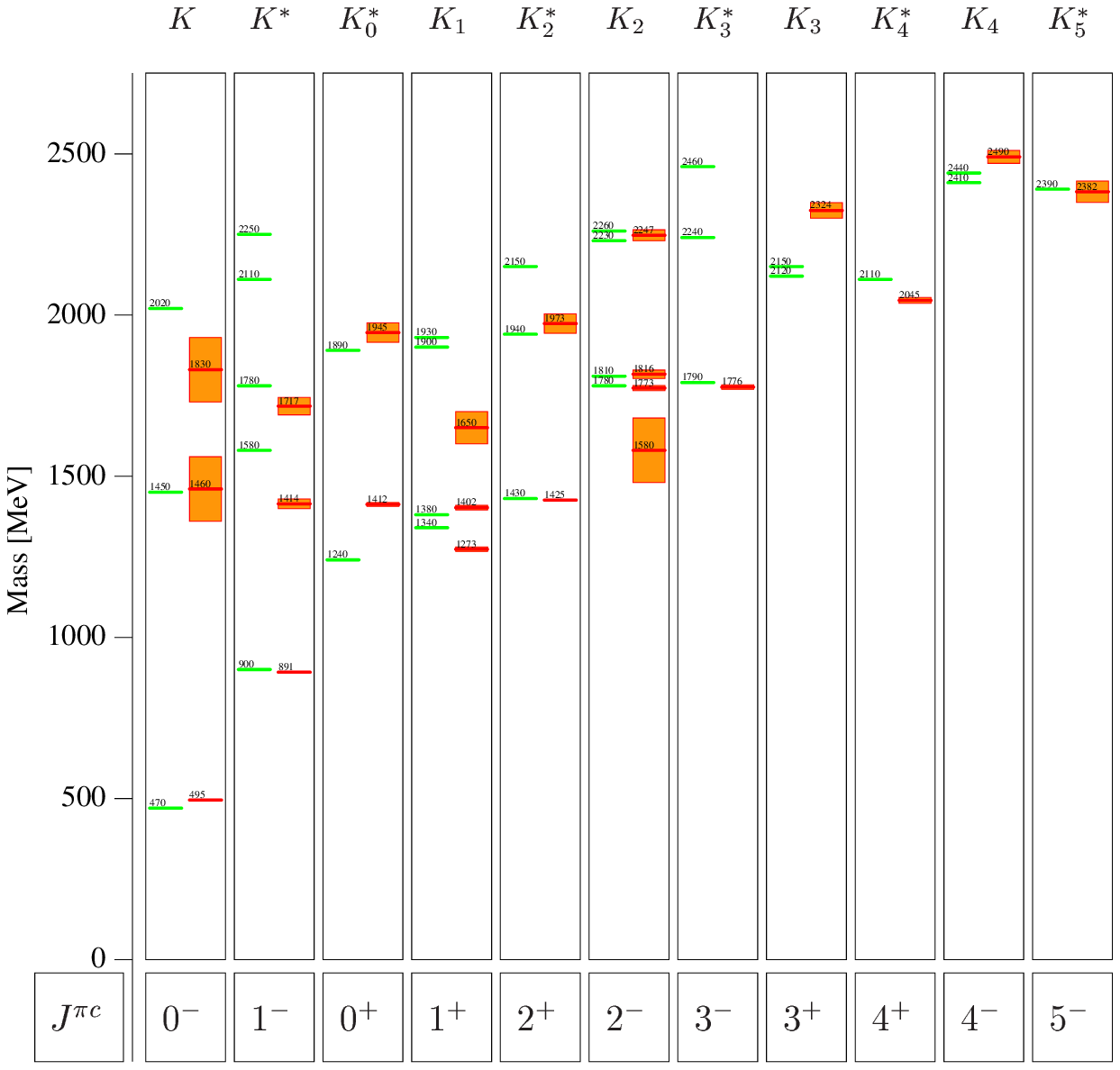},width=\textwidth,height=0.35\textheight,origin=c}\\
\vspace*{-4mm}
\epsfig{file={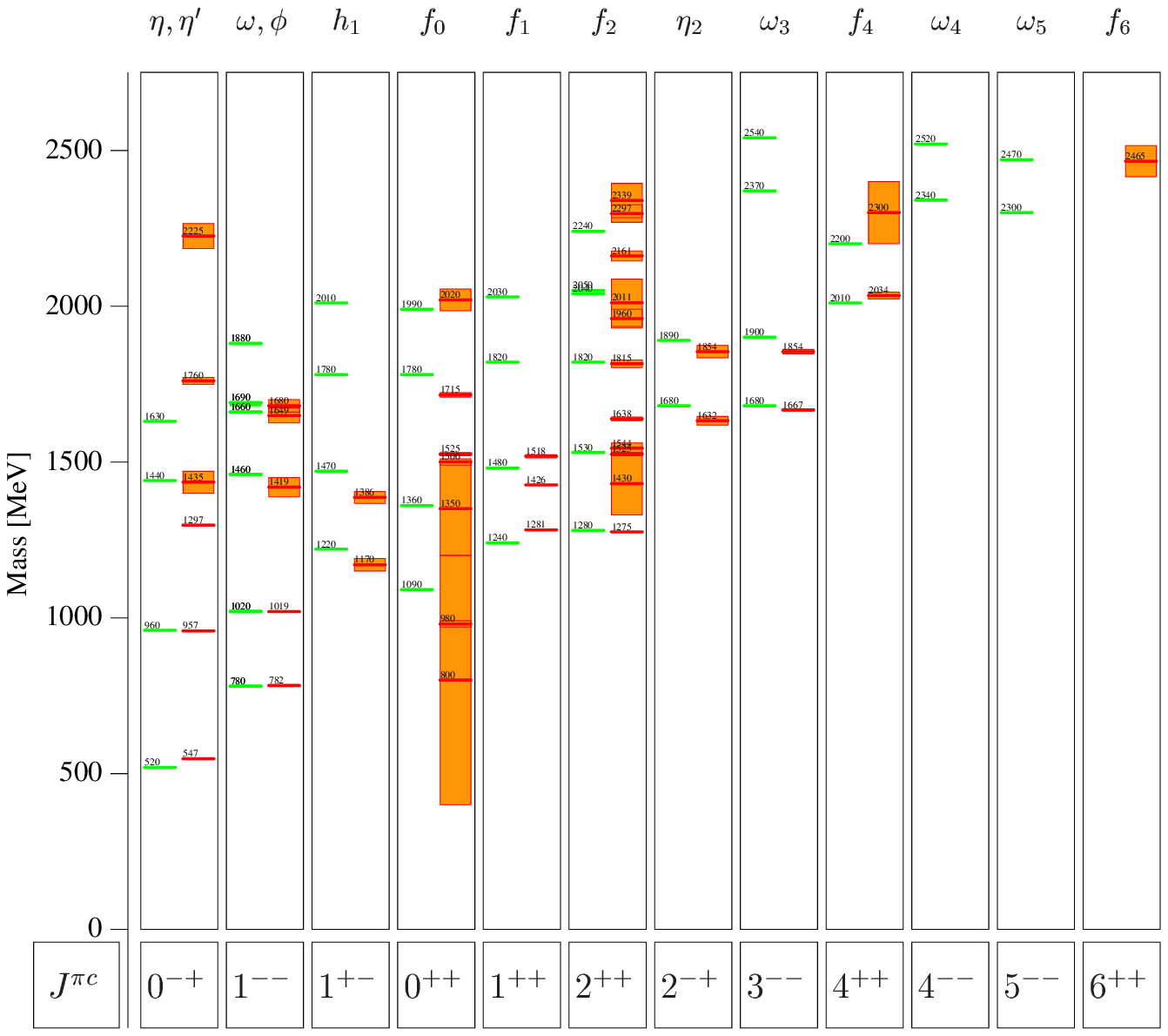},width=\textwidth,height=0.35\textheight,origin=c}
\end{minipage} 
\begin{minipage}[t]{0.25\textwidth}
\vspace*{-40mm}
\caption{\label{meson-isgur}
Mass spectra of light mesons for isovector (top), \qquad\
isodoublet (center) and \qquad\ isoscalar (bottom) \qquad\
mesons. 
Experimental results (on the right, with error bars) 
are compared with the Godfrey--Isgur model
(left)~\protect\cite{Godfrey:xj}.} 
\end{minipage} 
\end{figure}
\begin{figure}
\begin{minipage}[b]{0.74\textwidth}
\vspace*{-6mm}
\epsfig{file={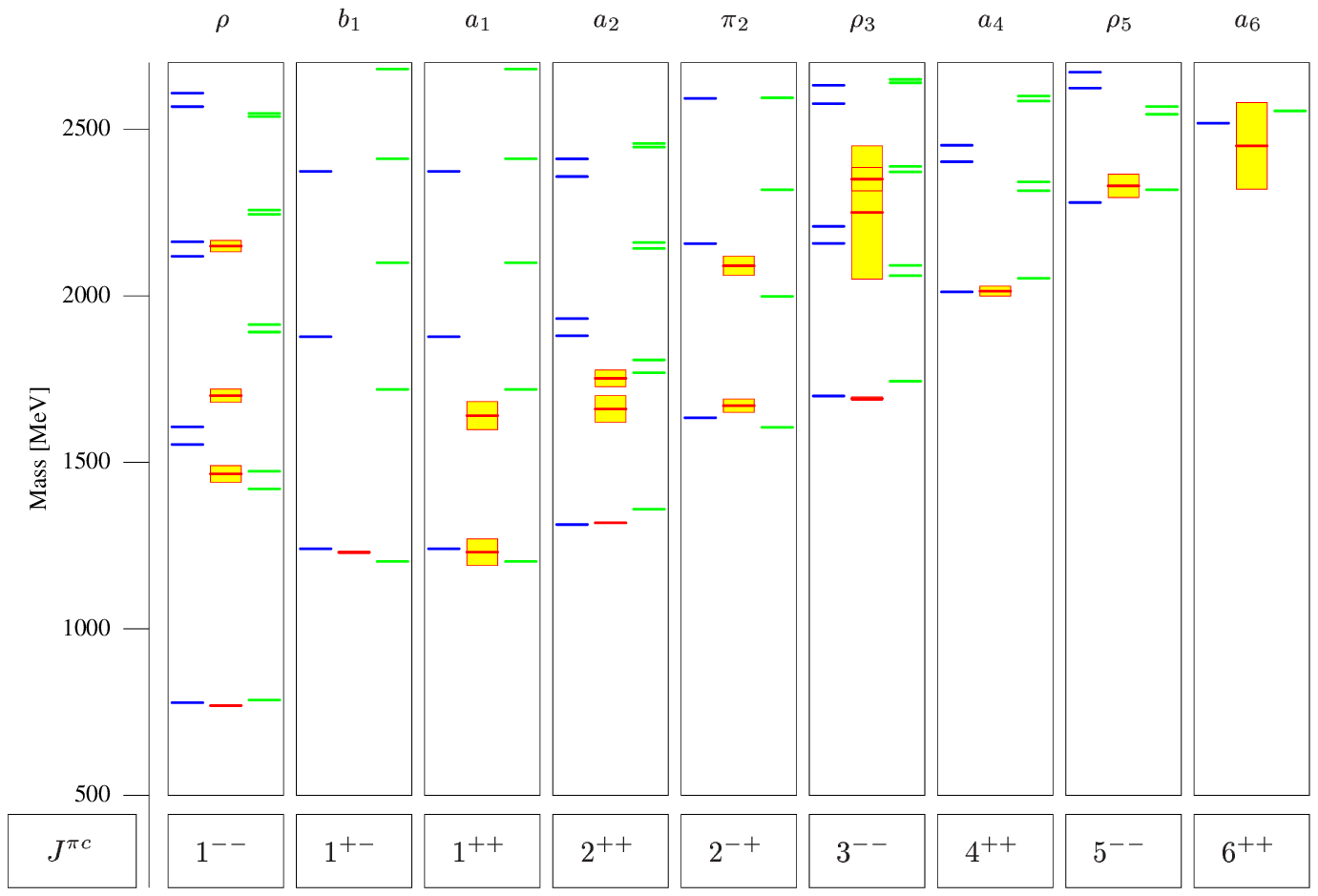},width=\textwidth,origin=c}\\
\vspace*{-4mm}
\epsfig{file={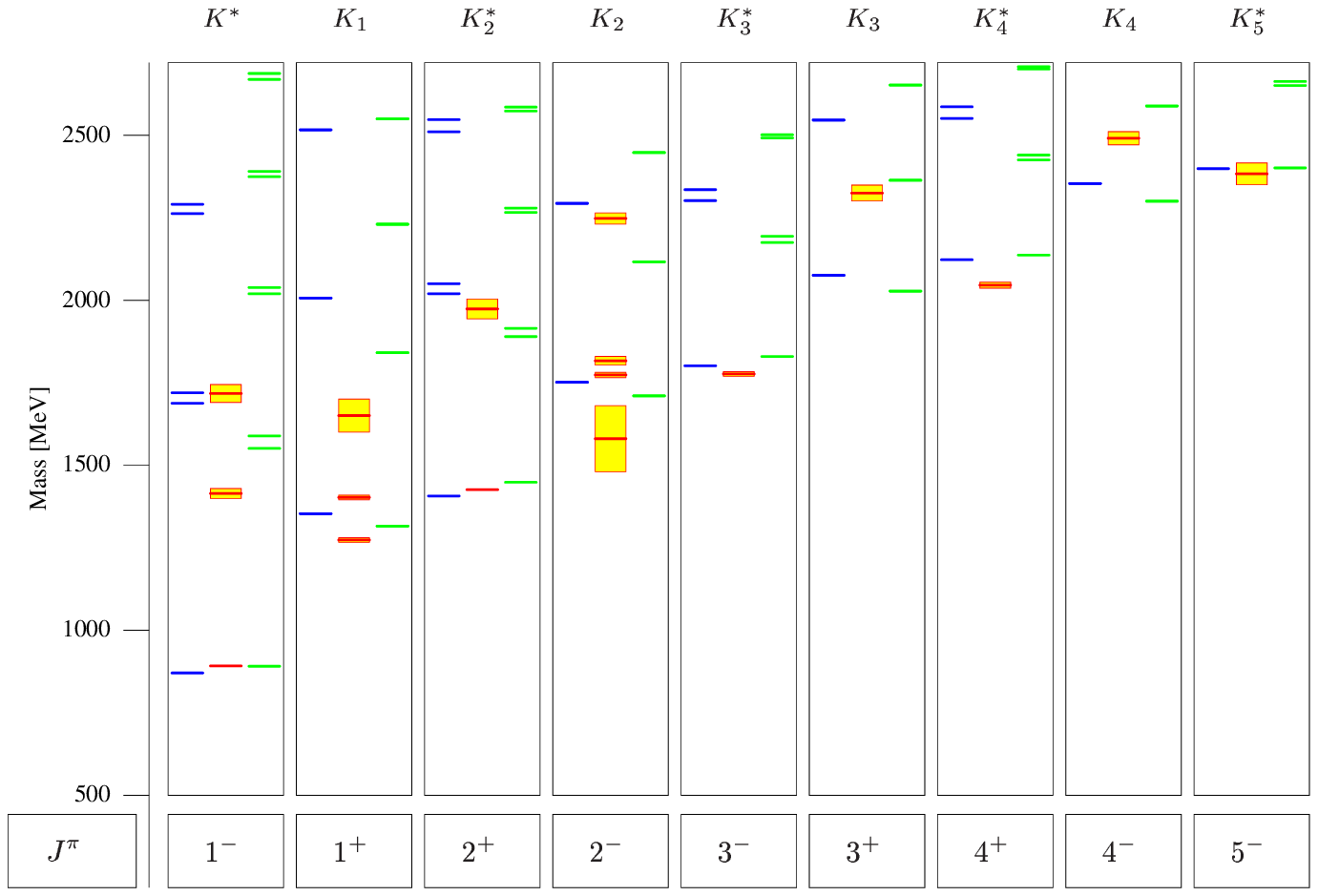},width=\textwidth,origin=c}\\
\vspace*{-4mm}
\epsfig{file={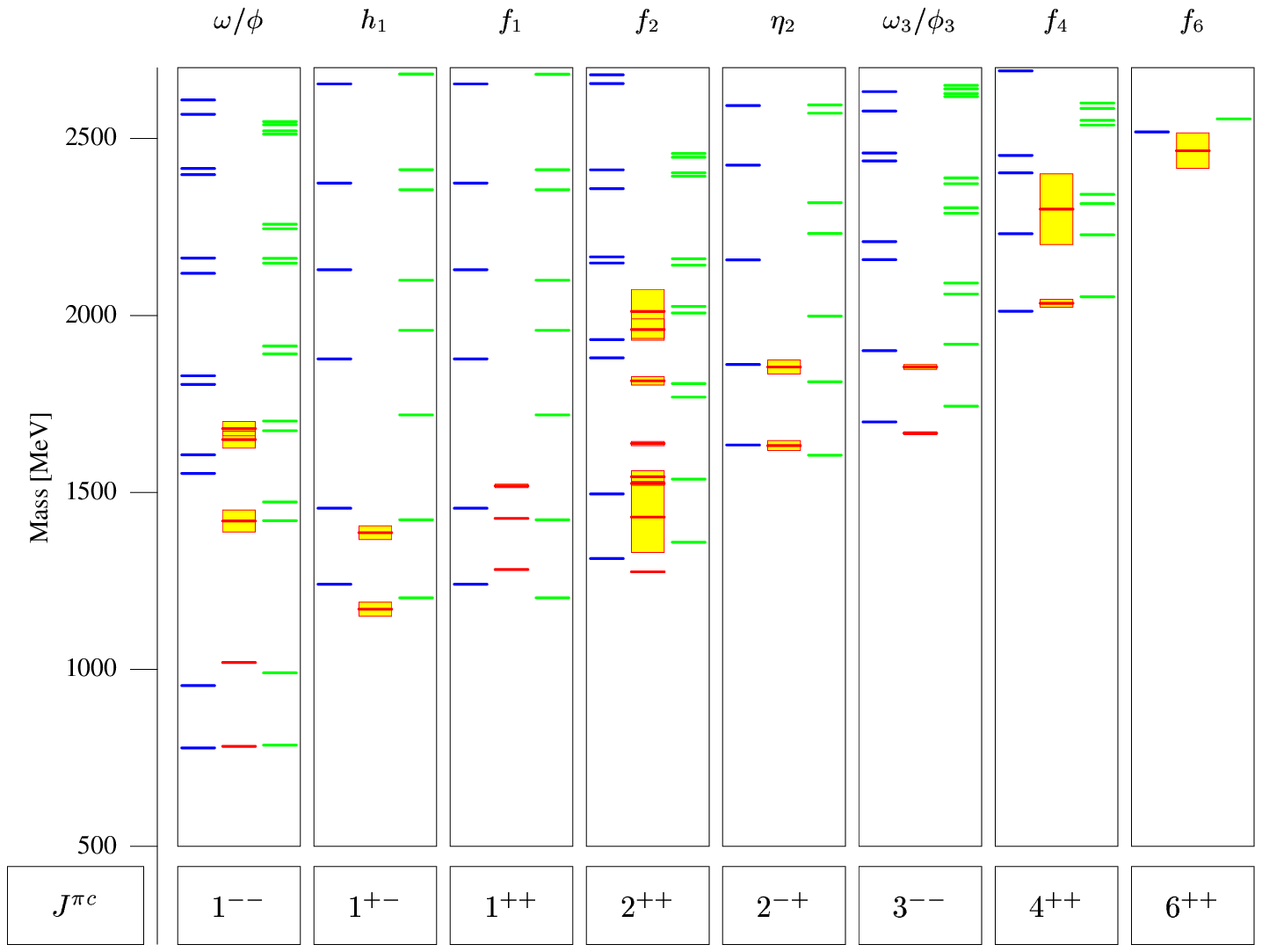},width=\textwidth,origin=c}
\end{minipage} 
\begin{minipage}[t]{0.25\textwidth}
\vspace*{-40mm}
\caption{\label{meson-metsch}
Mass spectra of light mesons for isovector (top),
isodoublet (center) and isoscalar (bottom) mesons.  
Experimental results (for each partial wave in the center)
are compared to the Bonn model, model A (left) and model B 
(right)~\protect\cite{Koll:2000ke}.} 
\end{minipage} 
\end{figure}

Figs.~\ref{meson-isgur}-\ref{meson-metsch} 
compare the experimental 
meson mass spectrum with model calculations. 
Notice the
good agreement in the number of states and their approximate
positions except for four cases.
\vfill
{\small
\begin{table}[b!]
\vspace*{-9mm}
\caption{Parameters used in the Bonn model. The confinement potential
is given by a constant $a_C$ and a slope $b_C$; the strength of 
instanton--induced interactions is $g$ for $u\bar u\to d\bar d$ and
$g'$ for  $n\bar n\to s\bar s$. The interaction is regularized by
by a cut--off $\lambda$. The confinement potential is defined 
including its change in relativistic boosts, written symbolically
as $\Gamma\cdot\Gamma$~\protect\cite{Koll:2000ke}.} 
\begin{center}
\begin{tabular}{|lc|rl|rl|}
\hline
		& & {Model ${\cal A}$} &&
		{Model ${\cal B}$} &\\
\hline
masses 		& $m_n$ 	& 306 	& MeV & 419 	& MeV \\
		& $m_s$ 	& 503 	& MeV & 550 	& MeV \\
confinement 	& $a_C$ 		& -1751	& MeV & -1135 	& MeV
		\\
		& $b_C$		& 2076	& MeV/fm & 1300    & MeV/fm \\
		 & $\Gamma\cdot\Gamma$ &
		\multicolumn{2}{c|}{
		$\frac{1}{2}(\Id\cdot\Id-\gamma_0\cdot\gamma_0)$} &
		\multicolumn{2}{c|}{
		$\frac{1}{2}(\Id\cdot\Id
		-\gamma_5\cdot\gamma_5
		-\gamma^\mu\cdot\gamma_\mu)$}\\
\hline
instanton	& $g$		& 1.73	&GeV$^{-2}$ & 1.63 &
		GeV$^{-2}$ \\
induced		& $g'$		& 1.54 &GeV$^{-2}$ & 1.35 &
		GeV$^{-2}$ \\
interaction	& $\lambda$	& 0.30 & fm & 0.42 & fm \\
\hline
\end{tabular}
\end{center}
\label{BM}
\end{table}
}
\clearpage
\bi
\item There is one extra isoscalar pseudoscalar state, 
the \etg (1295), which is not expected in quark models.
\vspace*{-1mm}
\item There is one extra isoscalar pseudovector state, 
the $f_1(1420)$, which is not expected in quark models.
\vspace*{-1mm}
\item There is an abundacy of scalar and tensor states.
Figs.~\ref{meson-isgur}-\ref{meson-metsch} 
do not support an easy identification
of quark--antiquark mesons and intruders (glueballs, hybrids
or multiquark states).
\ei

\subsection{\label{section3.4}Quark models for baryons} 
\subsubsection{The spatial wave function}
\par
The three-particle motion can be
decomposed in Jacobian coordinates into two relative motions and the
center-of-mass motion. The two internal oscillations can be assigned
to an oscillation of two quarks in a diquark and of the third quark
against the diquark. These two oscillators (usually called $\rho$--
and $\lambda$ oscillators) 
support rotational and vibrational excitations and lead to a large
number of expected resonances. The spatial wave functions of mesons
can be classified in the 3-dimensional rotational group O(3); the
three-body motion requires O(6).
\par
The quark dynamics can be approximated by two harmonic oscillators.
To first order, harmonic-oscillator wave-functions can be used.
The rotational group O(6) can be expanded into
O(6)$\rightarrow$O(3)$\otimes$O(2)~\cite{Hey:1982aj}.
Table \ref{tab:hey} gives
the expected multiplet structure in an O(6)$\otimes$SU(6)
classification scheme for the four lowest excitation
quantum numbers $N$. With increasing $N$, an increasing number of
multiplets develop. The decomposition of the orbital wave-functions
results in a complicated multiplet structure of harmonic-oscillator
wave-functions. It should be mentioned that some of these multiplets need two
quark excitations. In the lowest 20-plet, at $N=2$, two quarks are
excited, each carrying one unit of orbital angular momentum;
the two orbital angular momenta add to a total orbital angular
momentum 1 with positive (!) parity.
\par
\begin{table}[h!]
\caption{\label{tab:hey}
Multiplet-structure of harmonic oscillator wave 
functions~\protect\cite{Hey:1982aj}.}
\bc
\renewcommand{\arraystretch}{1.3}
\begin{tabular}{|cccc|}
\hline
\ N \ &$\ O(6)$ & $\ O(3)\otimes O(2)$  &  $\ (D,L^P_N)$    \\
\hline
0  &   1   & $\ 1\otimes 1$        &  $\ (56,0^+_0)$  \\
1  &   6   & $\ 3\otimes 2_1$      &  $\ (70,1^-_1)$  \\
2  &  20   & $\ (5+1)\otimes 2_2$  &  $\ (70,2^+_2)$, $\ (70,0^+_2)$    \\
   &       & $\ 5\otimes 1$        &  $\ (56,2^+_2)$  \\      
   &       & $\ 3\otimes 1$        &  $\ (20,1^+_2)$  \\
   &   1   & $\ 1\otimes 1$        &  $\ (56,0^+_2)$  \\
3  &  50   & $\ (7+3)\otimes 2_3$  &  $\ (56,3^-_3)$, $\ (20,3^-_3)$, 
$\ (56,1^-_3)$, $\ (20,1^-_3)$ \\
   &       & $\ (7+5+3)\otimes 2_1$&  $\ (70,3^-_3)$, $\ (70,2^-_3)$, $\ (70,1^-_3)$ \\
   &   6   & $\ 3\otimes 2_1$      &  $\ (70,1^-_3)$  \\
4  & 105   & $\ (9+5+1)\otimes 2_4$&  $\ (70,4^+_4)$, $\ (70,2^+_4)$, $\ (70,0^+_4)$ \\
   &       & $\ (9+7+5+3)\otimes 2_2$&  $\ (70,4^+_4)$, $\ (70,3^+_4)$, 
$\ (70,2^+_4)$, $\ (70,1^+_4)$ \\
   &       & $\ (9+5+1)\otimes 1$  &  $\ (56,4^+_4)$, $\ (56,2^+_4)$, $\ (56,0^+_4)$ \\ 
   &       & $\ (7+5)\otimes 1$    &  $\ (20,3^+_4)$, $\ (20,2^+_4)$  \\  
   &  20   & $\ (5+1)\otimes 1$    &  $\ (70,2^+_4)$, $\ (70,0^+_4)$  \\
   &       & $\ 3\otimes 1$        &  $\ (20,1^+_4)$                \\
   &       & $\ 5\otimes 1$        &  $\ (56,2^+_4)$                \\
   &   1   & $\ 1\otimes 1$        &  $\ (56,0^+_4)$                \\
\hline
\end{tabular}
\renewcommand{\arraystretch}{1.0}
\ec
\end{table}
The ground state $N=0$ is readily identified with the well-known octet
and decuplet baryons. The first excitation band ($N=1$) has internal
orbital angular momentum  L=1 ; both oscillators are excited
coherently and there is one coherent excitation mode of the two
oscillators. This information is comprised in the notation
$3\otimes 2_1$. The next excitation band involves several dynamical
realizations. The intrinsic orbital angular momentum L can be
associated with
two different quarks; the vector sum of the two l$_i$ can be 0, 1 or 2,
giving rise to the series  $5\otimes 1$ to $1\otimes 1$ where the
$3\otimes 1$ is antisymmetric w.r.t. quark exchange, and the other
two are symmetric.
Two linearly independent coherent two-oscillator excitations exist
having mixed--symmetry spatial wave functions. The baryon is
excited radially where the three quarks oscillate against their
common center of mass. This mode is represented by $1\otimes 1$.
\par
With increasing $N$ the number of multiplets increases strongly;
multiplets belonging to bands of up to 12 were 
calculated~\cite{Dalitz:me}.
So far, the multitude of predicted resonances have escaped 
experimental observation. This is the  so--called  missing-resonance
problem and the basis for experimental searches for new states
~\cite{Napolitano:1993kf,Thoma-prop}.
\subsubsection{Quark model predictions}
\begin{figure}[h!]
\bc
\epsfig{file={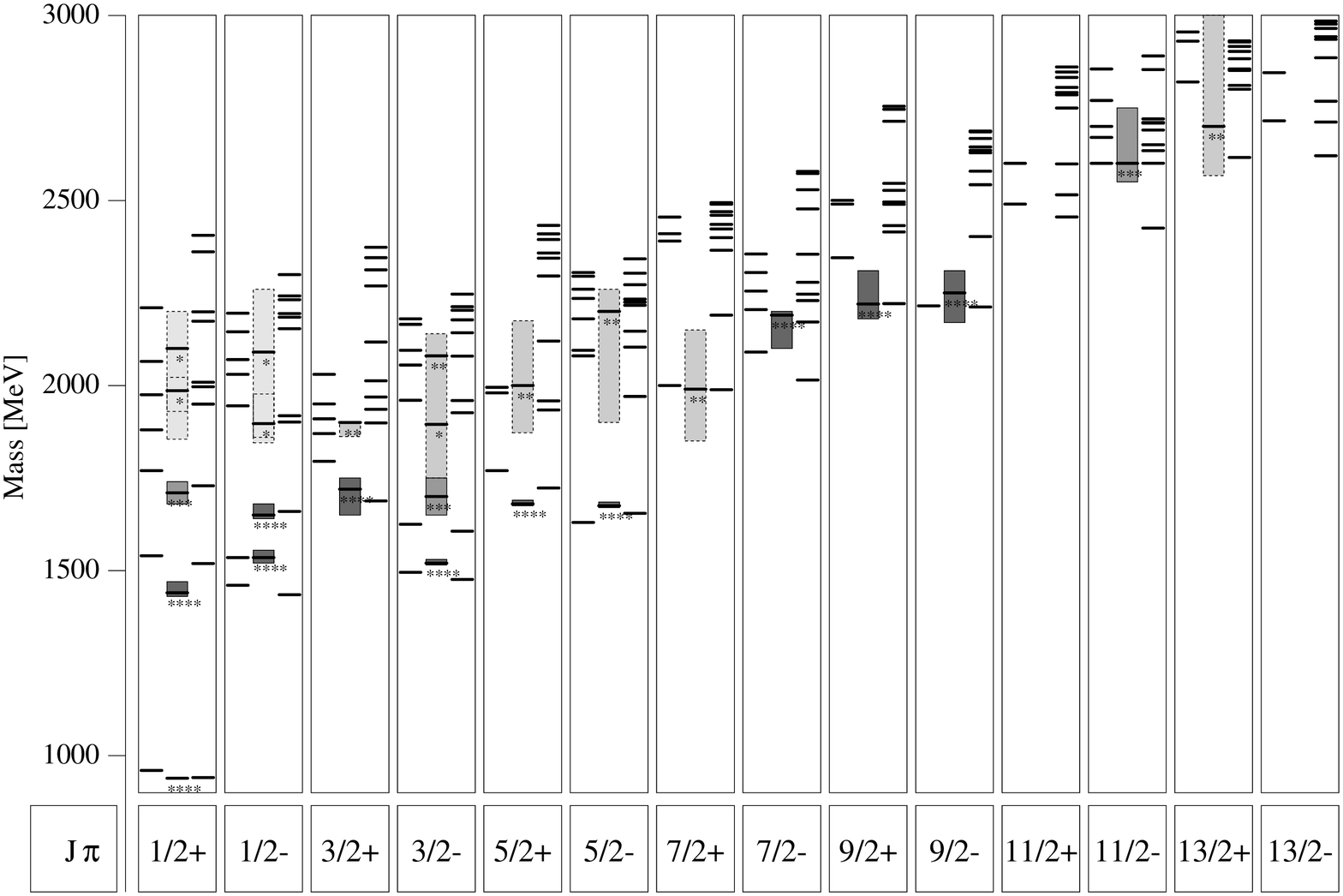},width=0.95\textwidth}
\ec
\caption{\label{bar1}
The spectrum of N$^*$ resonances. For each partial
wave, data (in the center row) are compared to predictions from
one--gluon exchange~\protect\cite{Godfrey:xj} (left) and from
instanton--induced interactions~\protect\cite{Metsch} (right). 
}
\bc
\epsfig{file={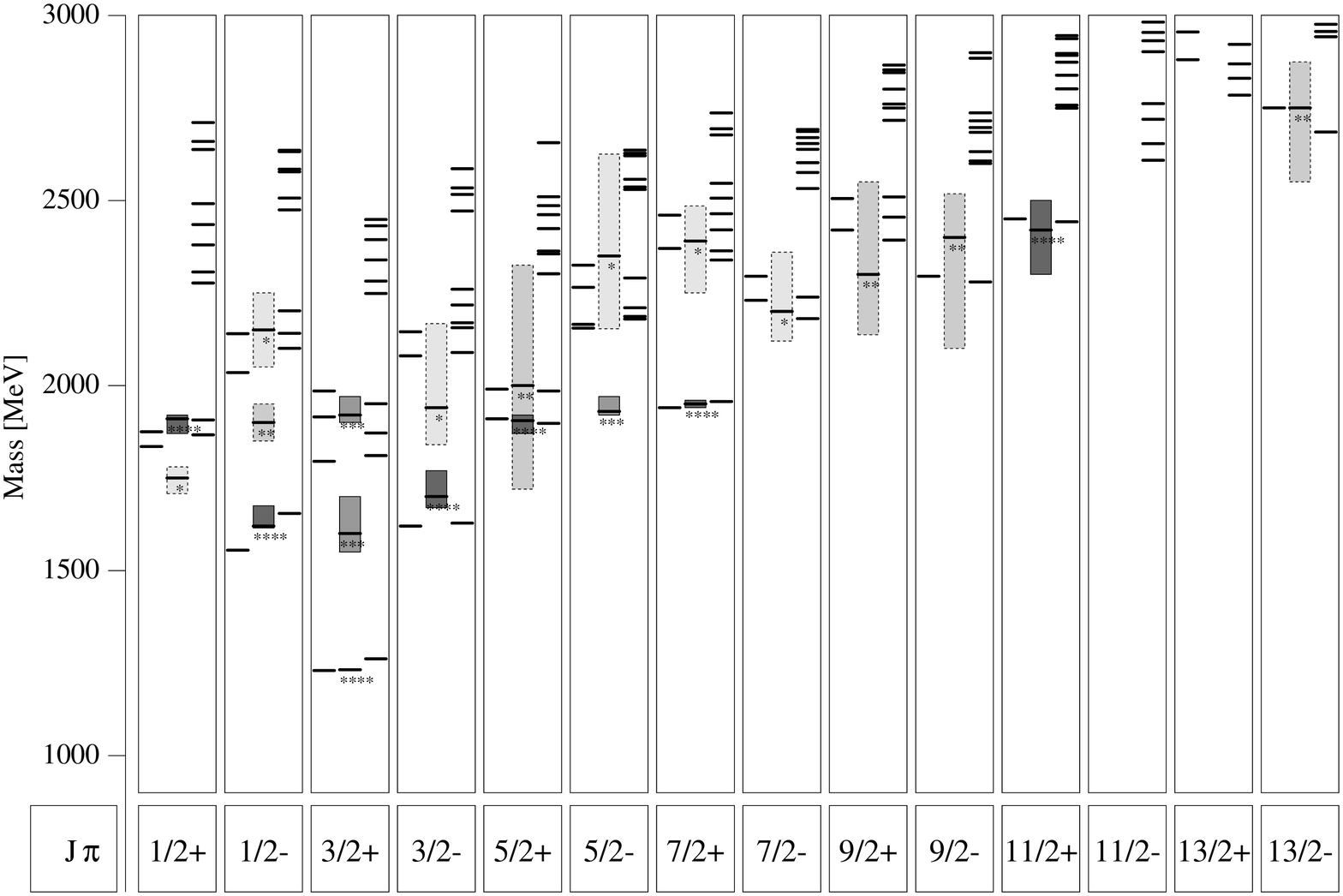},width=0.9\textwidth}
\ec
\caption{\label{bar2}
The spectrum of $\Delta^*$ resonances. For each partial
wave, data (in the center row) are compared to predictions from
one--gluon exchange~\protect\cite{Capstick:bm} (left) and from
instanton--induced interactions~\protect\cite{Metsch} (right). 
}
\end{figure}

Figs.~\ref{bar1}-\ref{bar2} show the mass spectra of N$^*$
and $\Delta^*$ resonances using one--gluon exchange or 
instanton--induced interactions
plus a linear confinement potential; the spectra for baryons with
strangeness were also calculated in these models but are not
reproduced here. 

\subsection{\label{section3.5}Conclusions}
QCD inspired models are well suited to describe hadron mass spectra.
In meson spectroscopy, wildly discrepant interpretations
are found in the pseudoscalar and scalar sector. This
is very exciting, since QCD allows not only the existence of
$\bar qq$ mesons and $qqq$ baryons but also
other forms of hadronic matter, like glueballs, hybrids, and 
multiquark states. In baryon physics, the most exciting issue
is the possible discovery of the $\Theta^+(1540)$. If confirmed,
the $\Theta^+(1540)$ would open a new spectroscopy, and it is still
completely open what we will learn from it about QCD. 
Even if the $\Theta^+(1540)$ should not survive,  
I am convinced that
baryon spectroscopy, due to the richness of the 3--particle dynamics, 
is the best testing ground to determine the degrees of freedom
relevant to understand low--energy QCD. These questions will be
discussed in the second half of the lecture series.

%% file: Chapter4_proc.tex
\section{\label{section4}The quest for glueballs}
\subsection{\label{section4.1}Glueballs in ``gluon-rich'' processes}

\begin{figure}[h!]
\begin{tabular}{ccc}
\epsfig{file=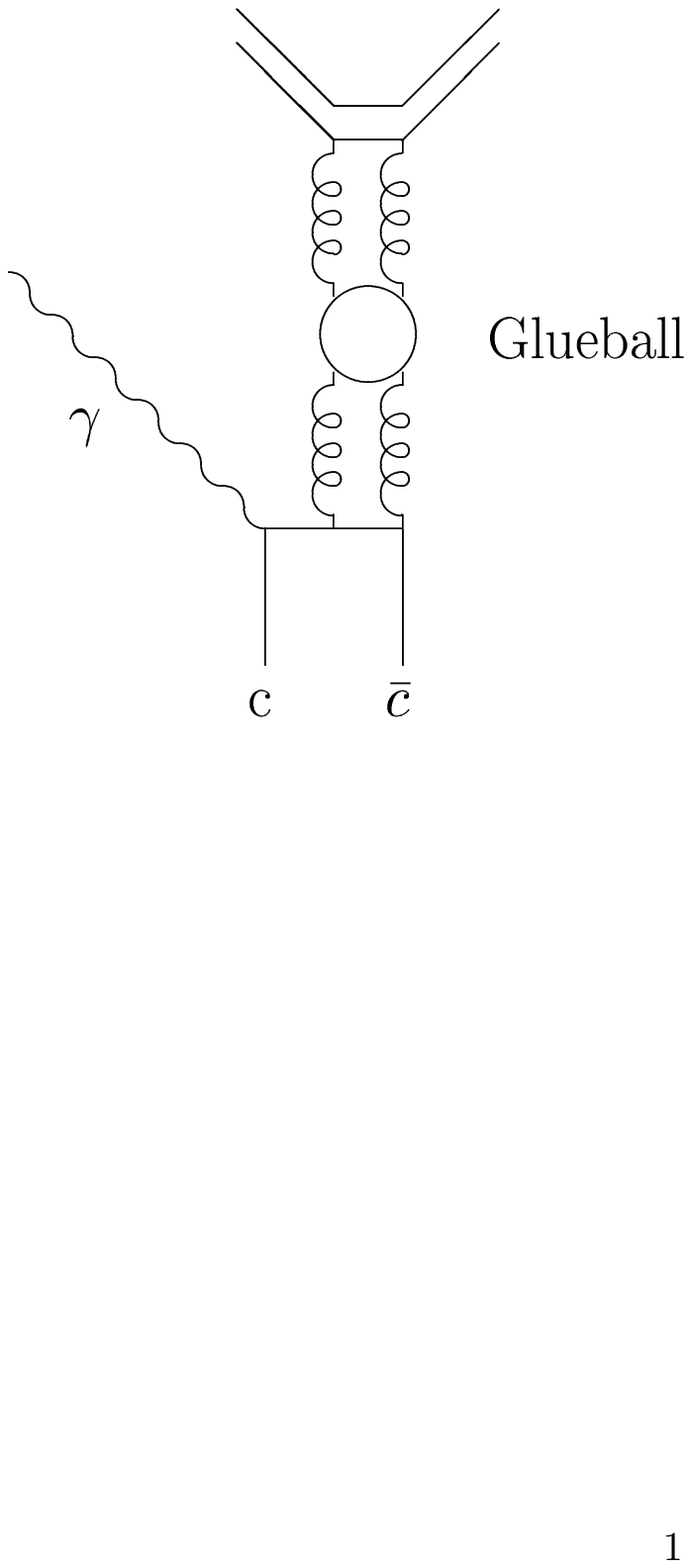,width=40mm,clip=,bbllx=120,bblly=360,bburx=340,bbury=600}&
\epsfig{file=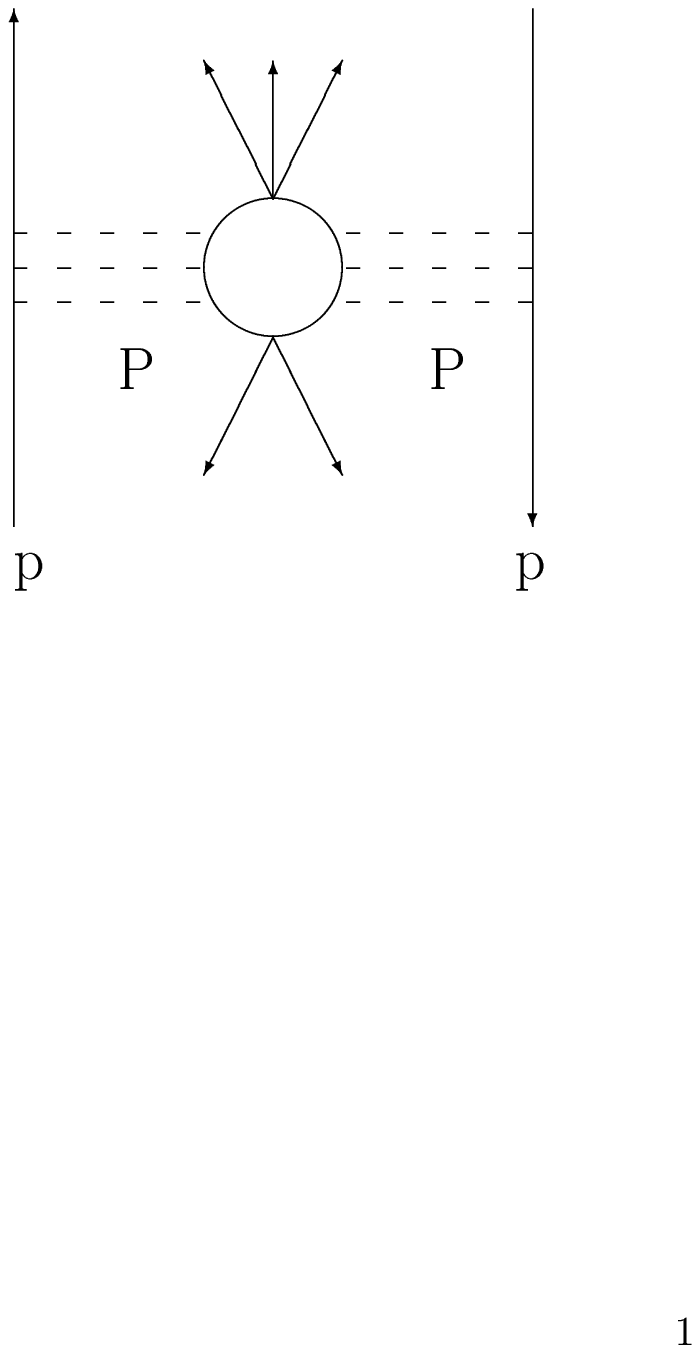,width=40mm,clip=,bbllx=110,bblly=250,bburx=340,bbury=600}&
\epsfig{file=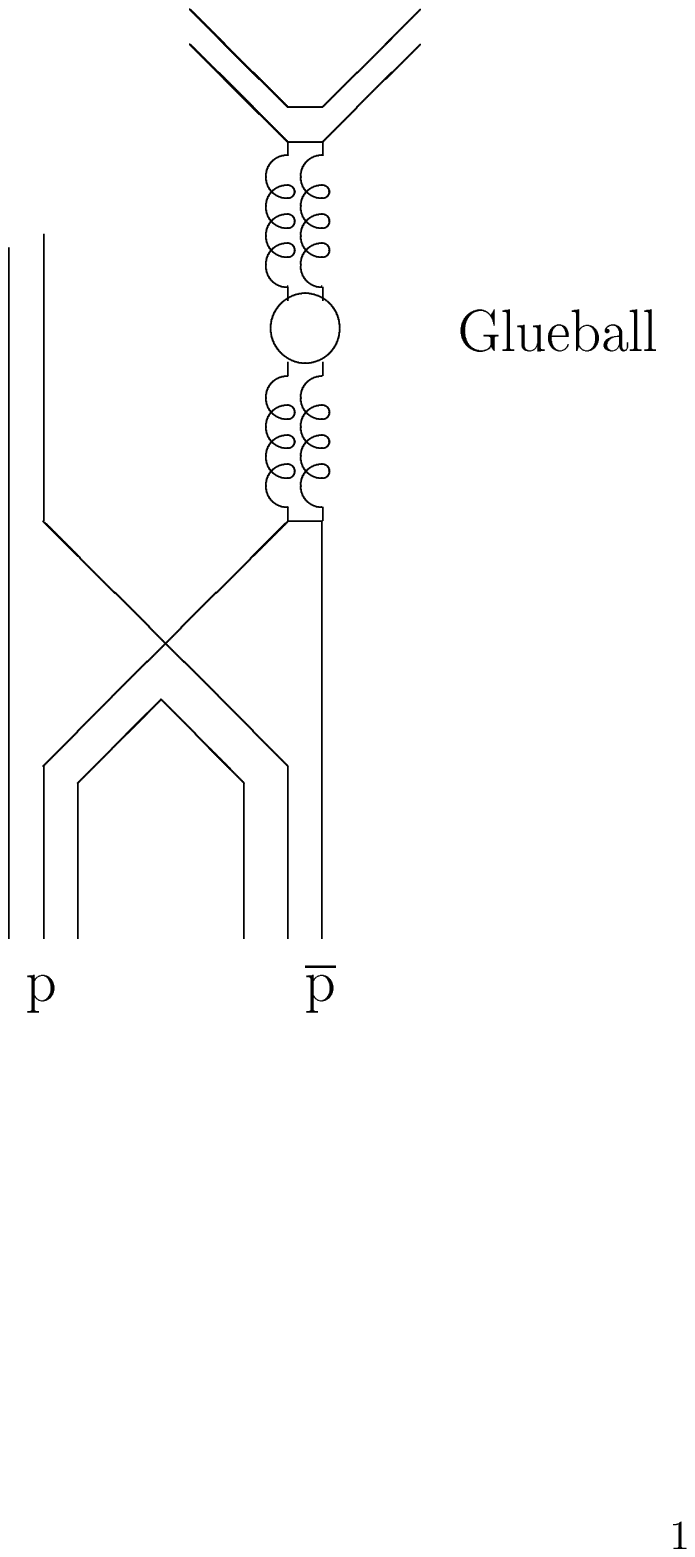,width=40mm,clip=,bbllx=110,bblly=290,bburx=340,bbury=600}
\end{tabular}
\vspace*{-5mm}
\caption{\label{gluemod}
Diagrams possibly leading to the formation of glueballs:
radiative J/$\psi$ decays, Pomeron-Pomeron collisions in hadron
hadron diffractive scattering, and in $\rm p\bar p$ annihilation.}
\end{figure}

Glueballs, bound states of gluons with no constituent quarks,
are predicted to exist and have been searched for in numerous
experiments. There is an extensive folklore on how to hunt for
glueballs~\cite{Robson:1977pm}
and which distinctive features should identify them as 
non--$\bar qq$ mesons. Glueballs should, e.g.,  
be produced preferentially in so--called
gluon--rich processes; some are depicted in
Fig.~\ref{gluemod}.
\par
\vspace*{-13mm}

\par
The most suggestive process is the radiative
J/$\psi$ decay. The J/$\psi$ is narrow; the $D\bar D$ threshold is above
the mass of the J/$\psi$ and the OZI
rule suppresses decays of the $c\bar c$ system into light
quarks. In most decays, the J/$\psi$ undergoes a transition
into 3 gluons which then convert into hadrons. But the
J/$\Psi$ can also decay into 2 gluons and a photon. The photon
can be detected, the two
gluons interact and must form glueballs - if they exist.
\par
Central production is another process in which glueballs should be
produced abundantly. In central production two hadrons pass by
each other `nearly untouched'
and are scattered diffractively in forward direction. No valence
quarks are exchanged. The process is often called
Pomeron-Pomeron scattering. The absence of valence quarks
in the production process makes central production a good place
to search for glueballs.
\par
In \pbp\ annihilation, quark-antiquark
pairs annihilate into gluons, they interact and may form
glueballs. Glueballs decay into hadrons and hence
hadro--production of glueballs is always possible. 
\par
Production of glueballs should be suppressed in $\gamma\gamma$
collisions since photons couple to the intrinsic charges. So we
should expect a glueball to be strongly produced in radiative
J/$\psi$ decays but not in  $\gamma\gamma$ fusion. Radial
excitations might be visible only weakly in J/$\psi$ decays but
they should couple to $\gamma\gamma$.
\par
Further distinctive features can be derived from their
decays (glueballs are flavor singlets). Decays
to $\eta\eta^{\prime}$ identify a flavor octet; radiative
decays of glueballs are forbidden. All these arguments have to
be taken with a grain of salt: mixing of a glueball with mesons
having the same quantum numbers can occur and would dilute
any selection rule.

\subsection{\label{section4.2}E/$\iota$ saga}
The $\eta(1440)$ was the first glueball candidate
and is still topic of a controversial discussion.
It is instructive to outline its history.
\subsubsection{Short history of the $\eta (1440)$}
The E/$\iota$ was discovered 1967 in $p\bar p$
annihilation at rest into $(K\bar K\pi)\pi^+\pi^-$.
It was the first meson found in a European experiment,  
and was called E-meson~\cite{Baillon67}. 
Mass and width
were determined to $M = 1425 \pm 7, \Gamma = 80\pm 10$\, MeV; the
quantum numbers to $J^{PC} = 0^{-+}$.
Also seen, 1967, was a state with
$M = 1420 \pm 20, \Gamma = 60\pm 20$\,MeV but
$J^{PC} = 1^{++}$, now in the charge exchange reaction
$\pi^- p \to n\rm K\bar K\pi$ using a
1.5 to 4.2\,GeV/c pion beam~\cite{Dahl:ad}. Even though the quantum
numbers had changed, it was still called E-meson.
\par
\begin{figure}[htb]
\begin{minipage}[c]{0.65\textwidth}
\bc
\epsfig{file=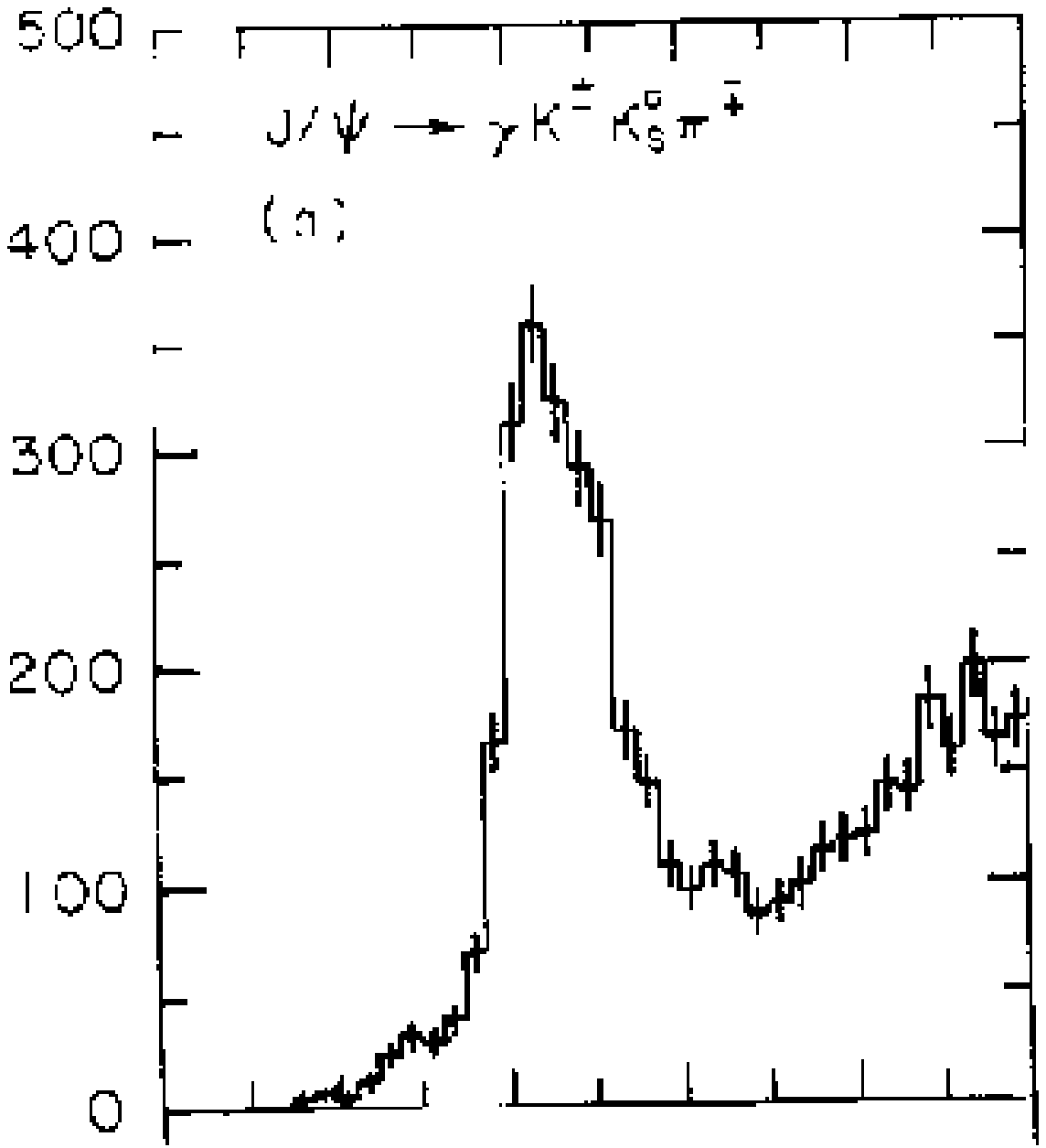,height=7cm}\\
\ec
\hspace{16.5mm}
{1295 MeV} 
$\bf \Uparrow$ \qquad\ $\bf \Uparrow$ \ {1440 MeV}
\end{minipage}
\begin{minipage}[c]{0.34\textwidth}
\vspace*{-10mm}
\quad\ Further studies showed that the $\iota (1440)$
is split into two components, a
$\eta_L\to a_0(980)\pi$ with $M = 1405 \pm 5, \Gamma = 56\pm 6$\,MeV
and a $\eta_H\to \rm K^*\bar K +\bar K^*K$ with $M = 1475 \pm 5,
\Gamma = 81\pm 11$\,MeV. Hence, there seem to be 3 $\eta$ states
in the mass range from 1280 to 1480\, MeV.
\vspace*{10mm}
\caption{\label{jradglue}
The $\iota (1440)$ is a strong signal in radiative
J/$\psi$ decay. It cannot be
described by a single Breit-Wigner resonance.
There is no evidence for the $\eta (1295)$
from radiative J/$\psi$ decay~\protect\cite{Kopke:1988cs}.
}
\end{minipage}
\end{figure}

In 1979, there was a claim for a $\eta (1295)$ which was later
confirmed in several experiments~\cite{Stanton:ya}.
The E--meson was observed in 1980 
in radiative J/$\psi$ decays~\cite{Scharre:1980zh} into
$(K\bar K\pi)$ with $M = 1440 \pm 20, \Gamma = 50\pm 30$\,MeV;
and quantum numbers `rediscovered'~\cite{Edwards:1982nc} 
to be $J^{PC} = 0^{-+}$. It was now called $\iota (1440)$ to
underline the claim that it was the $\iota^{\rm st}$ glueball.
The $\iota (1440)$ is a very strong signal, one of the strongest in
radiative J/$\psi$ decays (see Fig.~\ref{jradglue}). 
The radial excitation $\eta (1295)$
is not seen in this reaction; hence the  $\iota (1440)$
must have a different nature. At that time it was
proposed (and often still is) to be a glueball. 
\par
The $\eta (1295)$ is then likely the radial excitation of the
$\eta$. It is mass degenerate with the $\pi (1300)$, hence
the pseudoscalar radial excitations seem to be ideally mixed\,!
Then, the $\bar ss$ partner is expected to have a mass
240 MeV higher. The $\eta_H$ could play this role.
The $\eta_L$ finds no slot in the spectrum of $\bar qq$ mesons;
the low mass part of the $\iota (1440)$ could be a glueball.
This conjecture is consistent with the observed decays.
A pure flavor octet $\eta (xxx)$ state decays into $\rm K^*K$ but
not into $a_0(980)\pi$. In turn, a pure flavor singlet $\eta
(xxx)$ state decays into $a_0(980)\pi$ but not into $\rm K^*K$.
(Both, $(\bar uu + \bar dd )$ and $\bar ss$ states, may decay
into $\rm K^*K$ and $a_0(980)\pi$. For $\sigma\eta$ decays there
are no flavor restrictions.) The $\eta_H$, with a large coupling 
to $\rm K^*K$, cannot 
possibly be a glueball, while the $\eta_L$ with its
$a_0(980)\pi$ decay mode can be.
\par
Two quantitative tests have been proposed to
test if a particular meson is glueball--like:
the stickiness and the gluiness.
The stickiness of a resonance R with mass $m_{\rm R}$ and two--photon width $\Gamma _{{\rm R} \ra \gamma\gamma}$ 
is defined as:
$$
S_{\rm R} = N_l \left(\frac{m_{\rm R}}{K_{{\rm J}\to\gamma {\rm R}}}\right)^{2l+1}
\frac{\Gamma _{{\rm J}\to\gamma {\rm R}}}{\Gamma _{{\rm R} \ra \gamma\gamma}} \ ,
$$
where $K_{{\rm J}\to\gamma {\rm R}}$ is the energy of the photon  in the J rest frame, 
$l$ is the orbital angular momentum of the two initial photons or gluons ($l=1$ for $0^-$), 
$\Gamma _{{\rm J}\to\gamma {\rm R}}$ is the J radiative decay width for R, 
and $N_l$ is a normalization factor chosen to give $S_{\eta} = 1$.
The L3 collaboration determined~\cite{Acciarri:2000ev} this 
parameter to $S_{\eta(1440)}=79\pm 26$.

The gluiness ($G$) was introduced~\cite{Close:1996yc,Paar:pr}
to quantify the ratio of the two--gluon and two--photon coupling of a particle,
it is defined as:
$$ 
G = \frac{9\,e^4_q}{2}\,\biggl(\frac{\alpha}{\alpha _s}\biggr)^2 \,
\frac{\Gamma _{{\rm R} \ra {\rm gg}}}{\Gamma _{{\rm R} \ra \gamma\gamma}} \ ,
$$
where $e_q$ is the relevant quark charge, 
calculated assuming equal amplitudes for 
\uubar{} and \ddbar{} and zero amplitude for 
\ssbar{}. $\Gamma _{{\rm R} \ra {\rm gg}}$ is 
the two--gluon width of the resonance {\rm R}, 
calculated from equation (3.4) of Reference~\cite{Close:1996yc}.
Whereas stickiness is a relative measure, 
the gluiness is a normalised quantity and is expected to be near 
unity for a \qqbar{} meson.
The L3 collaboration determined~\cite{Acciarri:2000ev} this quantity to 
$G_{\eta(1440)}=41\pm 14$.
\par
These numbers can be compared to those for the $\eta '$ for which 
$S_{\eta '} = 3.6 \pm 0.3$ and $G_{\eta '} = 5.2 \pm 0.8$ 
for $\alpha_s(958 MeV)=0.56\pm0.07$ is determined. 
Hence also the $\eta'$ is `gluish', but much more the $\eta_L$.
The $\eta_L$ is the first glueball\,!

\par
We should not stop here, instead we should also collect
arguments which speak against this interpretation. 
Is the $\eta_H$ a $\bar ss$ state, and is the $\eta(1295)$
the radial excitation of the $\eta$\,? 

\begin{minipage}[t]{.48\textwidth}
\bc
As $\bar ss$ state, the $\eta_H$
should be produced in
$\rm K^- \ p \to\ \Lambda\ \ \eta_H$.
\ec
\end{minipage}
\begin{minipage}[t]{.48\textwidth}
\bc
As $\bar ss$ state, the $\eta_H$
should not be produced in
$\rm\pi^- \ p \to\ n \ \eta_H$.
\ec
\end{minipage}
\vspace*{-5mm}
\begin{figure}[!htpc]
\begin{center}
\psset{xunit=.5cm,yunit=.5cm,linearc=.2}
\pspicture(-1,-.1)(19,5)
\psline[linewidth=1pt,linecolor=blue](0,4.5)(6,4.5)
\psline[linewidth=1pt,linecolor=blue](0,4)(6,4)
\psline[linewidth=1.5pt,linecolor=red](0,0)(6,0)
\psline[linewidth=1.5pt,linecolor=red](6,0.5)(3.25,0.5)(3.25,3.5)(6,3.5)
\psline[linewidth=1pt,linecolor=blue](0,0.5)(2.75,0.5)(2.75,3.5)(0,3.5)
\rput[cl](6.8,4.0){$\mathrm{\Lambda}$}
\rput[cl](6.8,.25){$s\bar s$}
\rput[cr](-.6,4){$\mathrm{P}$}
\rput[cr](-0.6,0.25){$\mathrm{K^-}$}
\rput(12,0){
\psline[linewidth=1pt,linecolor=blue](0,4.5)(6,4.5)
\psline[linewidth=1pt,linecolor=blue](0,4)(6,4)
\psline[linewidth=1pt,linecolor=blue](0,0)(6,0)
\psline[linewidth=1pt,linecolor=blue](6,0.5)(3.25,0.5)(3.25,3.5)(6,3.5)
\psline[linewidth=1pt,linecolor=blue](0,0.5)(2.75,0.5)(2.75,3.5)(0,3.5)
\rput[cl](6.8,4.0){$\mathrm{N}$}
\rput[cl](6.8,.25){$n\bar n$}
\rput[cr](-.6,4){$\mathrm{P}$}
\rput[cr](-0.6,0.25){$\mathrm{\pi^-}$}
}
\endpspicture
\end{center}
\end{figure}

\vspace*{-5mm}
\begin{minipage}[t]{.48\textwidth}
\bc
\underline{It is not\,!}
\ec
\end{minipage}
\begin{minipage}[t]{.48\textwidth}
\bc
\underline{But it is\,!}
\ec
\end{minipage}
\vskip 5mm
In the diagram, the thicker lines represent a strange quarks.
The $\eta (1440)$ region does not contain a large $s\bar s$ component.
The $\eta_H =\eta_{\ssb}$ is not a $s\bar s$ state\,! 
\par
Fig.~\ref{jradglue}, with the strong $\eta(1440)$ signal, shows
no sign of the hypothetical radial excitation of the
$\eta$, of the $\eta(1295)$. Is there evidence for this state
in other reactions\,?
\subsubsection{The $\eta (1295)$ and the $\eta (1440)$ in $\gamma\gamma$ at LEP}
Photons couple to charges; in $\gamma\gamma$ fusion
a radial excitation is hence expected to
be produced more frequently than a
glueball. In $\gamma\gamma$ fusion, both electron and positron scatter
by emitting a photon. If the momentum transfer to
the photons is small, the $e^+$ and $e^-$ are scattered into
forward angles (passing undetected through the beam pipe), thus
\begin{figure}[b!]
\begin{minipage}[c]{0.55\textwidth}
\epsfig{file=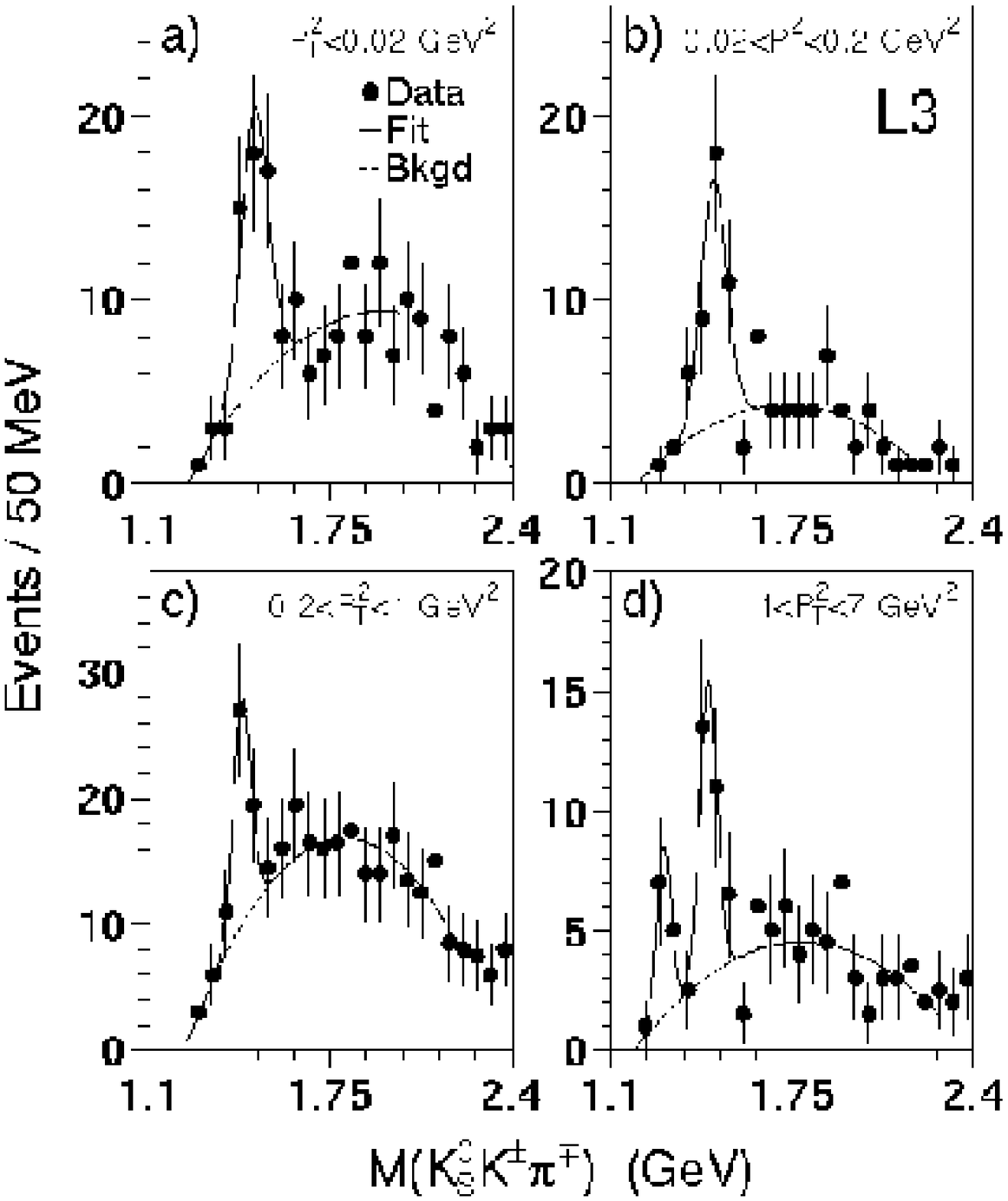,width=0.9\textwidth}
\end{minipage}
\begin{minipage}[c]{0.44\textwidth}
\bc
$\gamma\gamma\to \eta(1295)$: \\
CG${(1,0)+(1,0)\to (0,0)}\ne 0$\\
\vskip 2mm
$\gamma\gamma\not\to f_1(1285)$: \\
CG$_{(1,0)+(1,0)\to (1,0)}=0$\\
\vskip 2mm
$\gamma^*\gamma\to f_1(1285)$: \\
CG$\_{(1,M)+(1,0)\to (1,M)}\ne 0$
\ec
\caption{\label{ggfusion}
$\rm\gamma\gamma^*\to K^0_sK^{\pm}\pi^{\mp}$ from L3.
At low $q^2$, a peak at 1440\,MeV is seen, it requires high
$q^2$ to produce a peak at 1285\,MeV. A pseudoscalar
state is produced also at vanishing $q^2$ while $J^{PC}=1^{++}$
is forbidden for $q^2\to 0$. Hence the structure at 1285\,MeV
is due to the $f_1(1285)$ and not due to $\eta(1295)$. There is
no evidence for the $\eta (1295)$ from $\gamma\gamma$ fusion.
The stronger peak contains contributions from the   
$\eta (1440)$ and  $f_1(1420)$~\protect\cite{Acciarri:2000ev}.}
\end{minipage}
\end{figure}
the two photons are nearly real. If the $e^+$ or $e^-$
has a large momentum transfer, the photon acquires mass, and we
call the process $\gamma\gamma^*$ collision.
\par
Fig.~\ref{ggfusion} shows data from the L3 
experiment~\cite{Acciarri:2000ev}. Selection
rules for production of $\eta$ and $f_1$ mesons are also given.
The peak at low mass and high $Q^2$ must be the $f_1(1285)$, since
it is not produced for low $Q^2$. The higher mass peak can have
contributions from both, from the $f_1(1420)$ and the $\eta
(1440)$. These contributions can be separated due to their
different dependence on $Q^2$ or $P_T^2$. As a result we can state
that the $\eta (1440)$ is definitely produced in $\gamma\gamma$
collisions. There is no sign of the \etg (1295). The coupling of
the \etg (1440) to photons is stronger than that of the \etg
(1295): the assumption that the \etg (1295) is a $(u\bar u+d\bar d)$
radial excitation must be wrong\,!
The mass of the pseudoscalar resonance in
$\gamma\gamma$ fusion is about 1460\,MeV, and it
decays mainly into $\rm K^*K$. Hence we identify the state
with the $\eta_H$.
\vspace*{-3mm}

\subsubsection{The $\eta (1295)$ and $\eta (1440)$ in $p\bar p$ annihilation}
\begin{figure}[h!]
\begin{minipage}[c]{0.49\textwidth}
\begin{center}
\epsfig{file=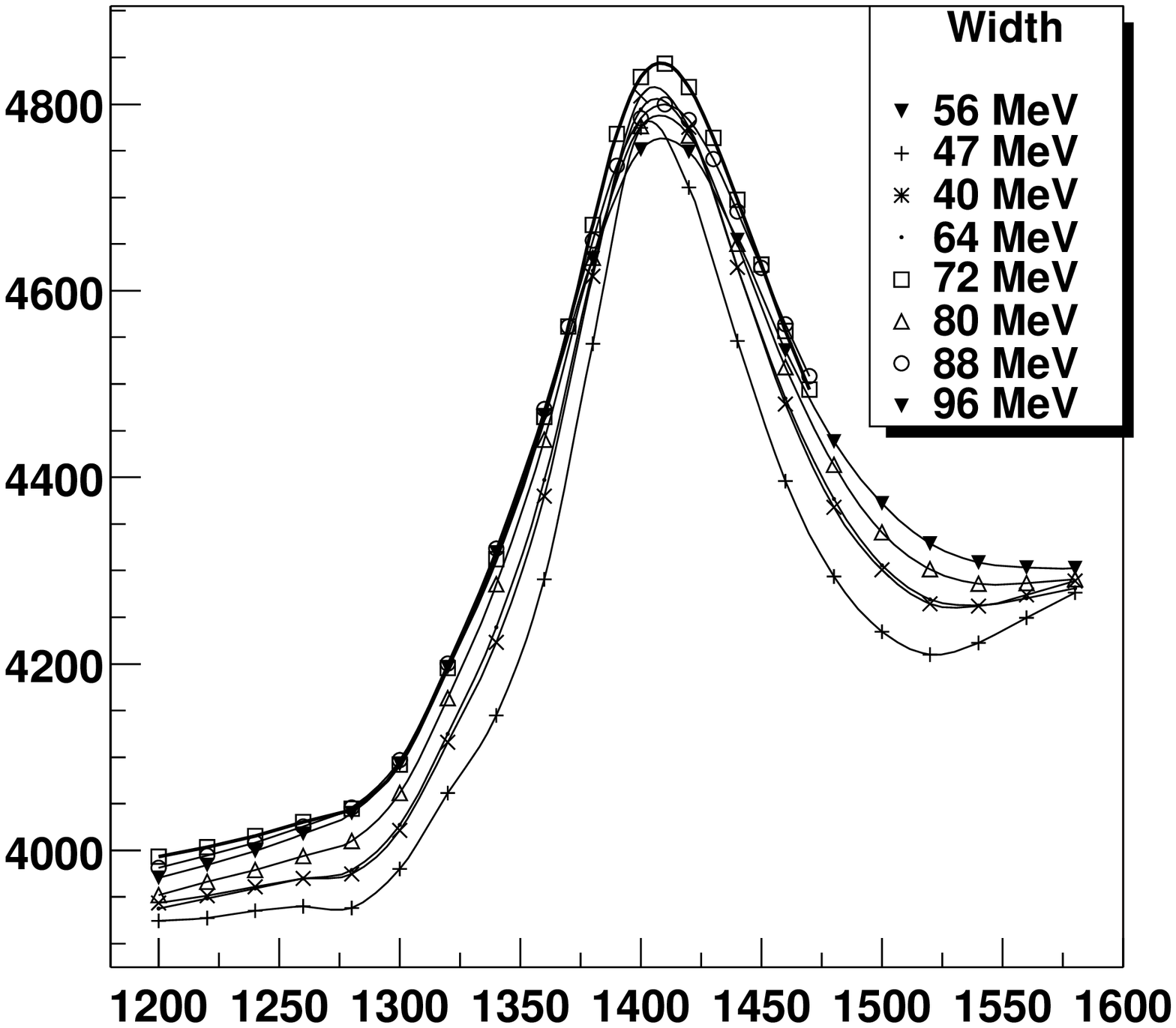,width=\textwidth}
\end{center}
\end{minipage}
\begin{minipage}[c]{0.49\textwidth}
\epsfig{file=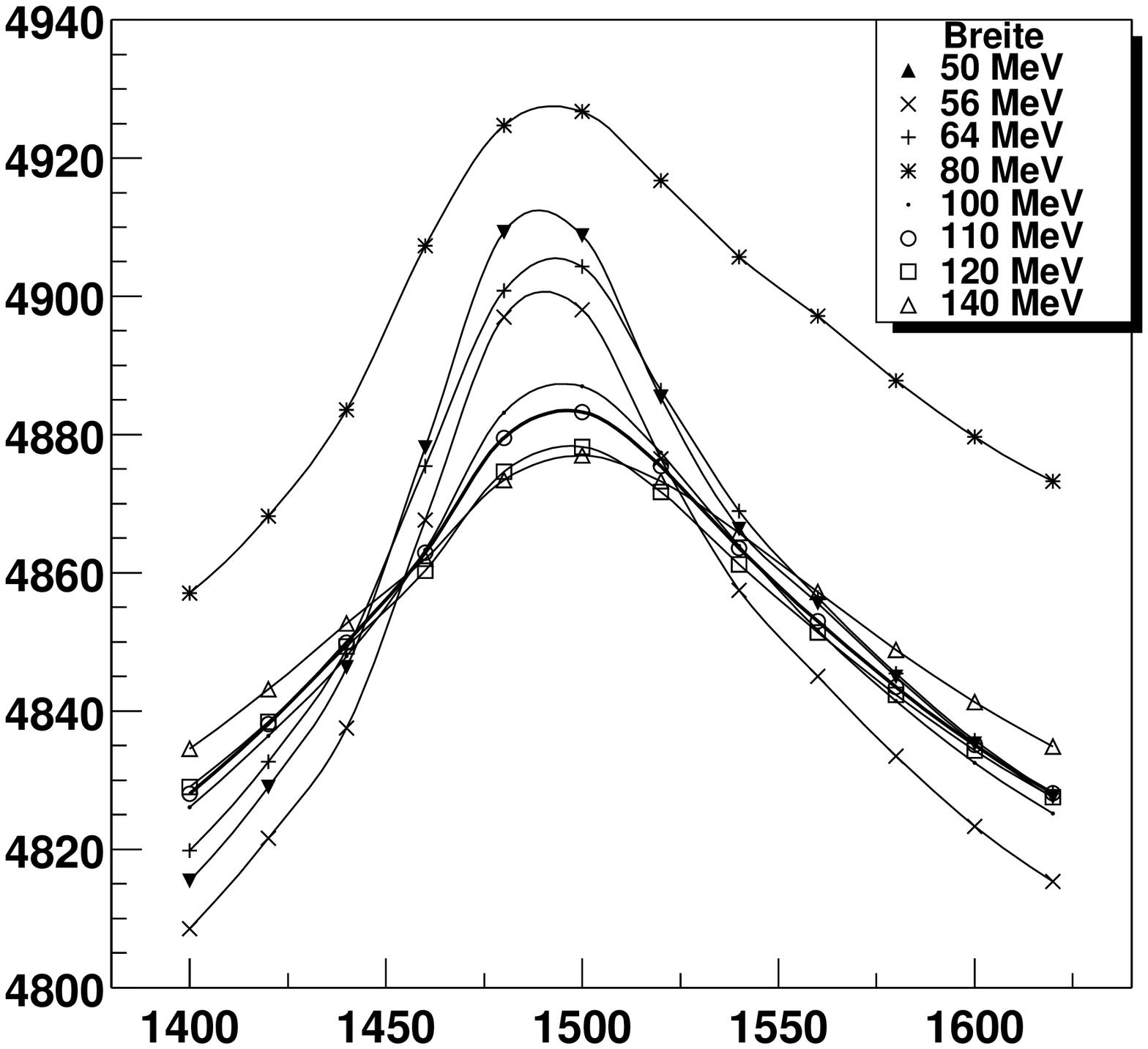,width=\textwidth}
\end{minipage}
\caption{\label{escan}
Scan for a $0^+ 0^{- +}$ resonance with
 different widths. The likelihood optimizes for
$M=1407\pm 5, \Gamma = 57\pm 9$\,MeV. The resonance 
is identified with the $\eta_L$.  
A search for a second pseudoscalar resonance (right panel)
gives evidence for the $\eta_H$ with 
$M=1490\pm 15, \Gamma = 74\pm 10$\,MeV.
From~\protect\cite{Reinnarth}.}
\vspace*{-3mm}
\end{figure}

The $\eta (1295)$ and $\eta (1440)$ can be searched for in
the reaction $p\bar p\to\pi^+\pi^-\eta (xxx)$,  
$\eta (xxx)\to\eta\pi^+\pi^-$. The search is done by assuming the presence
of a pseudoscalar state of given mass and width, mass and width
are varied and the likelihood of the fit is plotted.
Fig.~\ref{escan} shows such a plot~\cite{Reinnarth}.
A clear pseudoscalar resonance signal is seen 
at 1405\,MeV. Two decay modes are observed, $a_0(980)\pi$ and
$\eta\sigma$\footnote{\footnotesize We use the notation $\sigma (600)$
for a particle discussed in section \ref{section4.4} and $\sigma$ for the
full $\pi\pi$ S--wave.} with a ratio $0.6\pm0.1$.
\par
A scan for an additional $0^+ 0^{- +}$ resonance 
provides no evidence for the $\eta (1295)$ but for a second
resonance at 1480\,MeV, see Fig.~\ref{escan}, with $M=1490\pm 15
,\Gamma=74\pm 10$. This is the $\eta_H$. It decays to
$a_0(980)\pi$ and $\eta\sigma$ with a ratio $0.16\pm0.10$. 

\vspace*{-3mm}
\vspace*{-3mm}
\subsubsection{$E/\iota$ decays in the $^3P_0$ model}

The phenomena observed in the pseudoscalar sector are confusing:
The $\eta (1295)$, the assumed radial excitation of the $\eta$, is
only seen in $\pi^- p\to n (\eta\pi\pi)$, not in $p\bar p$
annihilation, nor in radiative J/$\psi$ decay, nor in
$\gamma\gamma$ fusion. In all these reactions, except perhaps in
radiative J/$\psi$ decays, it should have been observed. There is
no reason for it not being produced if it is a $\bar qq$ state. On
the other hand, we do not expect glueballs, hybrids or multiquark
states so low in mass. In the 70's, the properties of the
$a_1(1260)$ were obscured by the so--called Deck effect
($\rho$--$\pi$ rescattering in the final state). Possibly,
$a_0(980)\pi$ rescattering fakes a resonant--like behavior but
the $\eta(1295)$ is too narrow to make this possibility realistic.
Of course, there is the possibility that the $\eta(1295)$ is
mimicked by feed--through from the $f_1(1285)$. In any case, we
exclude the $\eta (1295)$ from the further discussion.
\par
The next puzzling state is the $\eta (1440)$. It is  not
produced as $\bar ss$ state but decays with a large fraction
into $\rm K\bar K\pi$ and it is split into two components.
We suggest that the origin of all these anomalies are due
to a node in the wave function of the $\eta(1440)$\,!
The node has an impact on the decay matrix element which
were calculated by~\cite{Barnes:1996ff} within the $^3P_0$ model.
\par
\begin{figure}[b!] 
\vspace*{-3mm}
\begin{tabular}{ccc} 
\hspace*{-6mm}\includegraphics[width=0.38\textwidth,height=6cm]{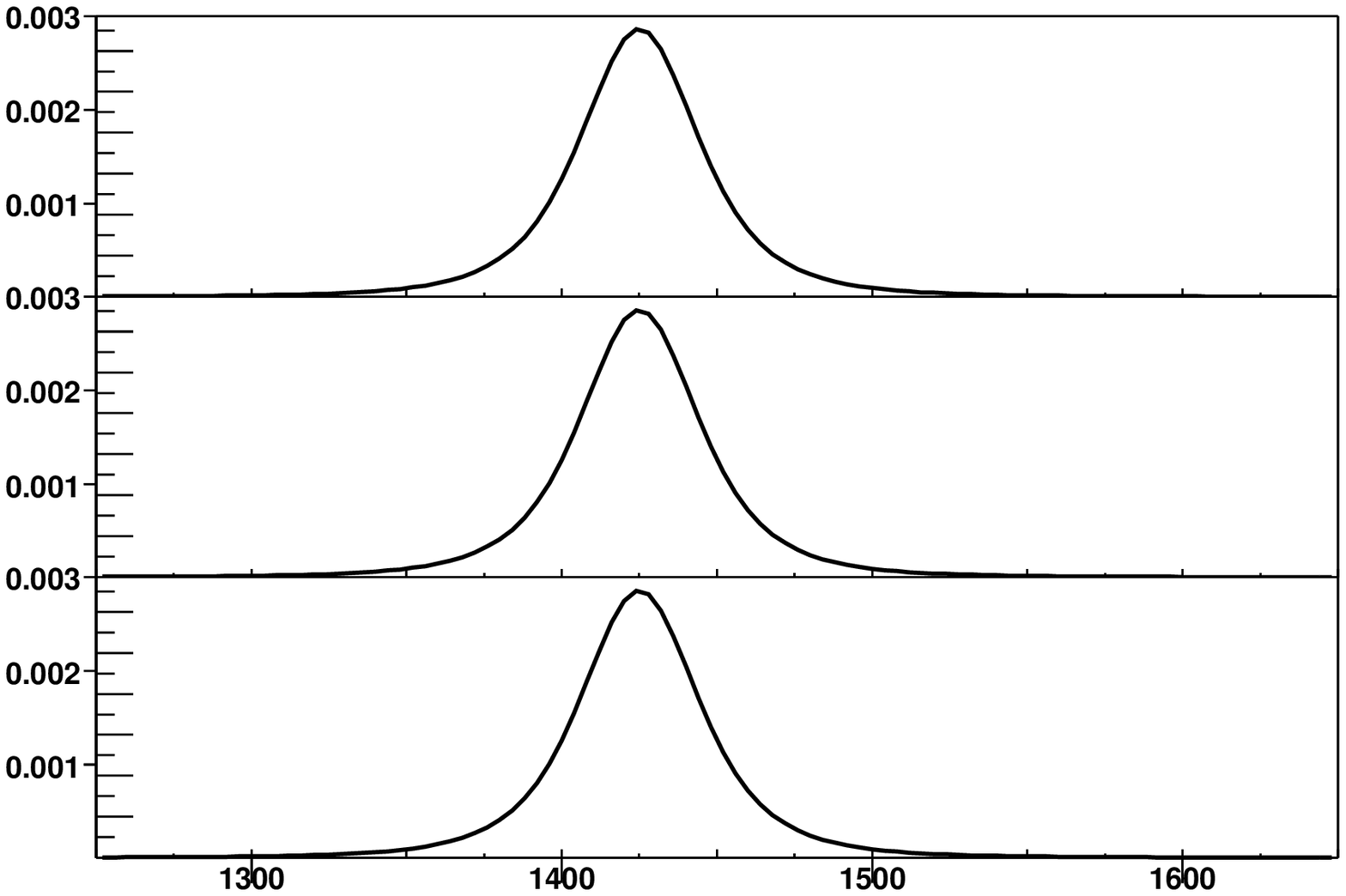}&
\hspace*{-9mm}\includegraphics[width=0.38\textwidth,height=6cm]{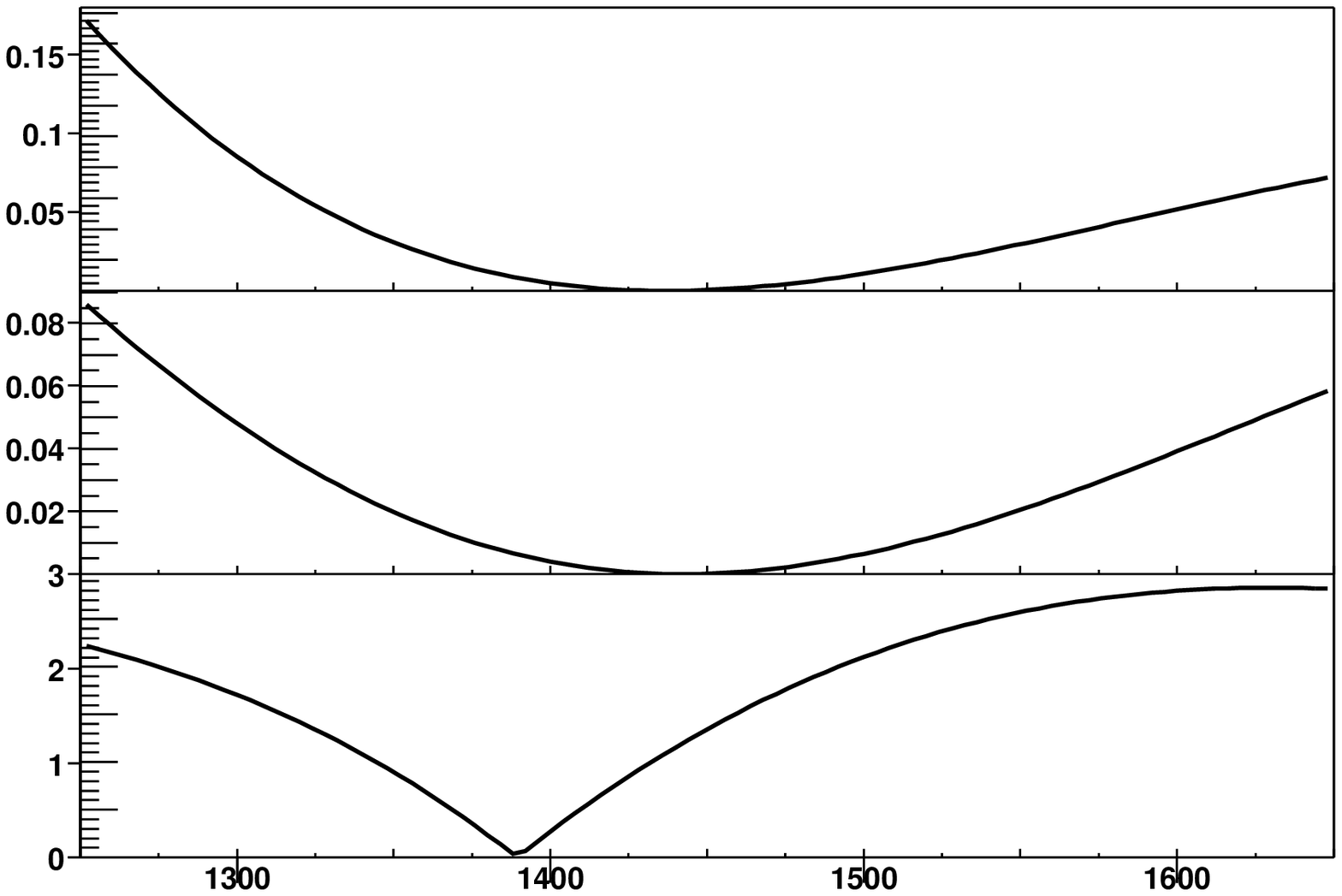}&
\hspace*{-8.5mm}\includegraphics[width=0.38\textwidth,height=6cm]{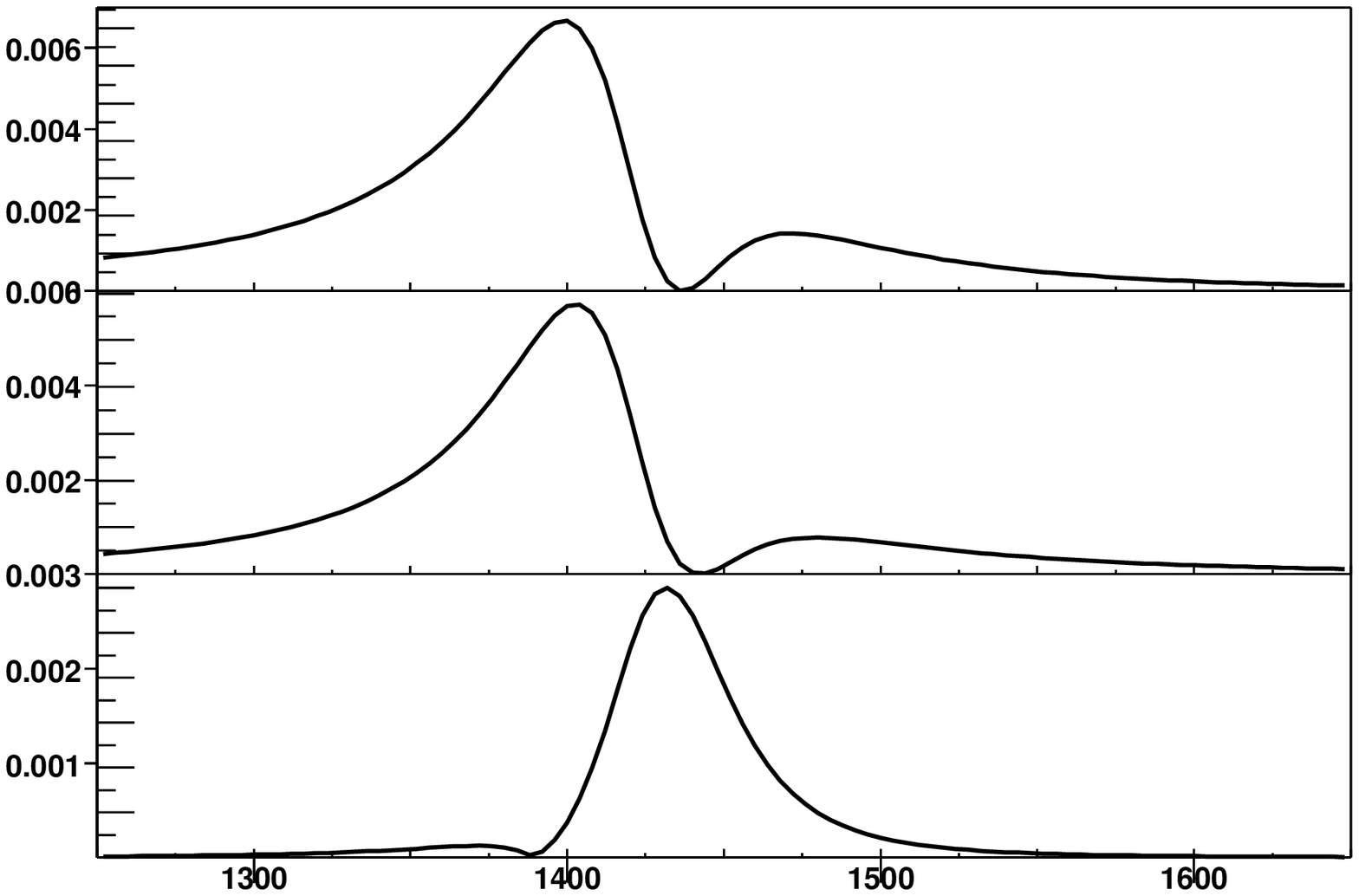}
\end{tabular}
\vspace*{-5mm}
\caption{\label{node}
Amplitudes for $\eta(1440)$ decays to $a_0\pi$ (first row), 
$\sigma\eta$ (second row):  $\rm K^*\bar K$ (third row)
the Breit-Wigner functions are shown on the left, then the squared 
decay amplitudes~\protect\cite{Barnes:1996ff} and, on the right, 
the resulting squared transition matrix element.}
\end{figure}
\par
The matrix elements for decays of the $\eta (1440)$ as a
radial excitation (=$\eta_R$) depend on spins, parities and decay
momenta of the final state mesons. For
$\eta_R$ decays to $\rm K^*K$, the matrix element is given by
$$
f_P = \frac{2^{9/2}\cdot 5}{3^{9/2}}\cdot x\left(1-\frac{2}{15}x^2 \right).
$$
In this expression, $x$ is the decay momentum in units of 400\,MeV/c,
the scale is determined from comparisons of measured partial widths
to model predictions. The matrix element vanishes for $x=0$ and
$x^2 = 15/2$, or $p=1$\,GeV/c. These zeros have little effect on the
shape of the resonance.
\par
The matrix element for $\eta_R$ decays to $a_0(980)\pi$ or
$\sigma\eta$ has the form
$$
f_S = \frac{2^{4}}{3^{4}}\cdot \left(1-\frac{7}{9}x^2 +
\frac{2}{27}x^2 \right)
$$
and vanishes for  $p=0.45$\,GeV/c. So, if
$\eta_R = \eta (1440)$, the decay to $a_0(980)\pi$
vanishes at the mass 1440\,MeV. This does have a decisive impact
on the shape, as seen in Figure~\ref{node}. Shown are 
the transition matrix elements as given by Barnes et al.~\cite{Barnes:1996ff}
and the product of the squared matrix elements and a Breit--Wigner
distribution with mass 1420\,MeV and a width of 60\,MeV.

We note that $\eta(1440)\to a_0(980)\pi$ and
$\to\rm K^*K$ have different peak positions;  at approximately
the $\eta_L$ and $\eta_H$ masses. Hence there is no need to introduce
the $\eta_L$ and $\eta_H$ as two independent states. One
$\eta(1420)$ and the assumption that it is a radial excitation
describes the data.
\par
\begin{figure}[h!]
\begin{minipage}[r]{0.35\textwidth}
\includegraphics[width=\textwidth]{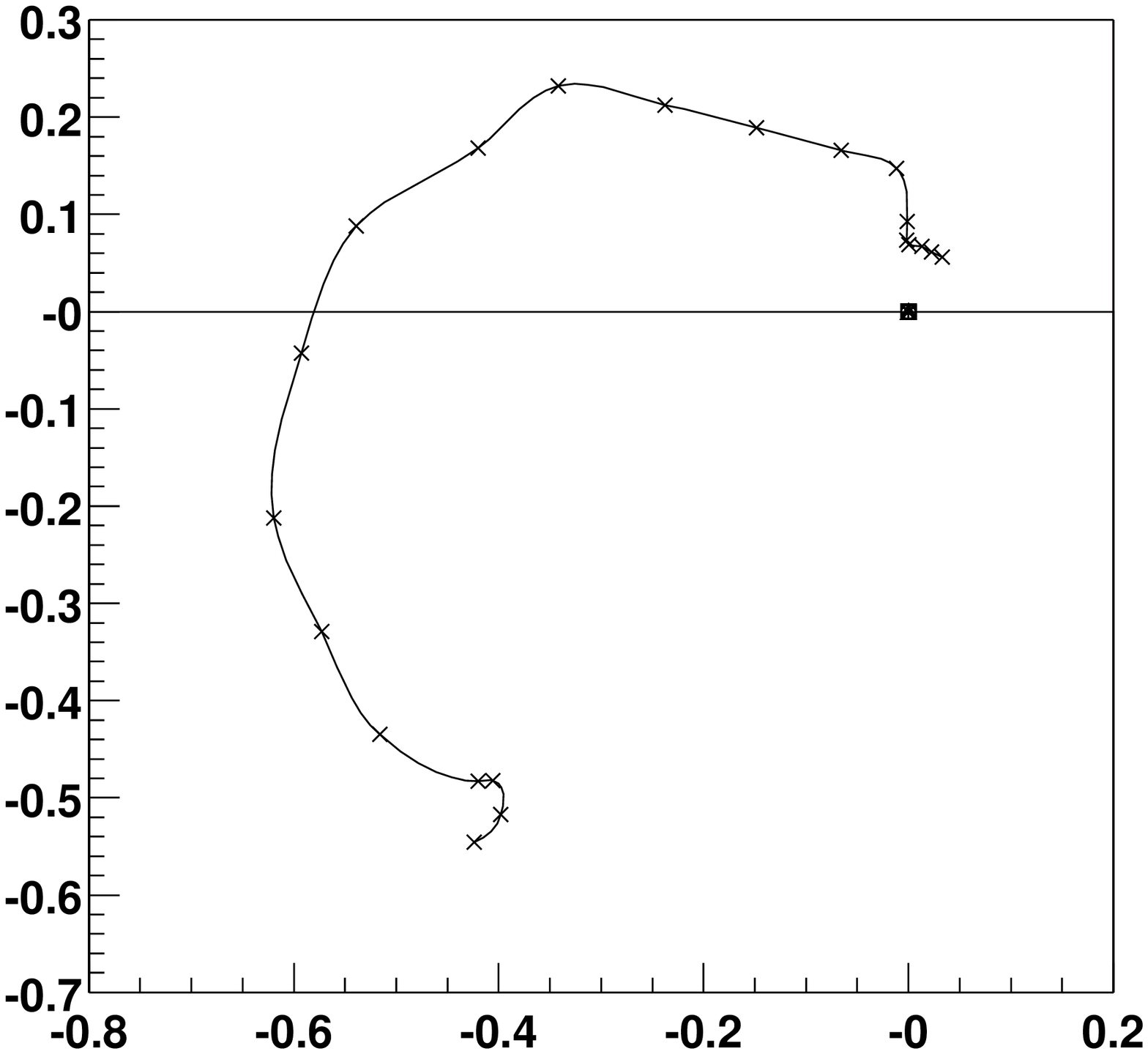}
\end{minipage}
\begin{minipage}[r]{0.35\textwidth}
\includegraphics[width=\textwidth]{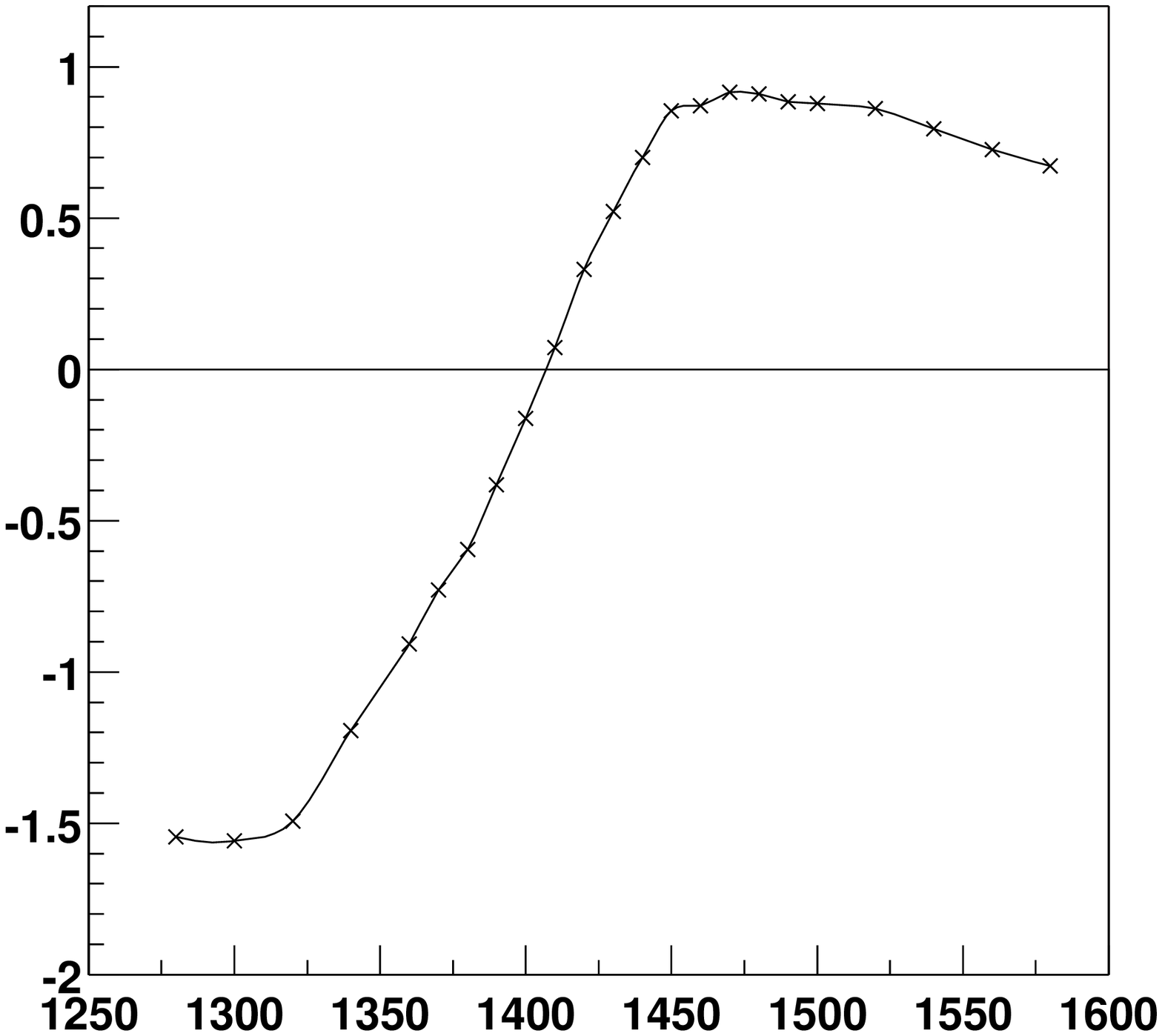}
\end{minipage}
\begin{minipage}[c]{0.28\textwidth}
\caption{\label{phase}
Complex amplitude and phase motion of the 
$a_0(980)\pi$ isobars
in $\rm p\bar p$ annihilation into $4\pi\eta$. In the mass range
from 1300 to 1500\,MeV the phase
varies by $\pi$ indicating that there is only one resonance in the
mass interval. The $\sigma\eta$ (not shown) exhibits 
the same behavior~\protect\cite{Reinnarth}.
}
\end{minipage}
\end{figure}

This can be
further tested by following the phase motion of the
$a_0(980)\pi$ or $\sigma\eta$ isobar~\cite{Reinnarth}. 
The phase changes by $\pi$ and
not by 2$\pi$, see Fig.~\ref{phase}.
\par
\subsubsection{Conclusions}
We summarize the results for the radial excitations of
pseudoscalar mesons.
\bi
\vspace*{-2mm}\item The $\eta(1295)$ is not a $q\bar q$ meson.
\vspace*{-2mm}\item The $\eta (1440)$ wave function
has a node leading to two appearantly different states $\eta_L$
and $\eta_H$. 
\vspace*{-2mm}\item The node suppresses OZI allowed decays into
$a_0(980)\pi$ and allows $K^*K$ decays.
\vspace*{-2mm}\item There is only one $\eta$ state, the $\eta(1420)$
in the mass range from 1200 to 1500 MeV and not 3\,!
\vspace*{-2mm}\item
The $\eta(1440)$ is the radial excitation
of the $\eta$.
\vspace*{-2mm}\item  The radial excitation of the $\eta'$ is expected
at about 1800\,MeV; it might be the $\eta (1760)$.
\ei

The following states are most likely the pseudoscalar
ground states and radial excitations:

\bc
\renewcommand{\arraystretch}{1.3}
\begin{tabular}{lcccc}
\hline\hline
$1^1S_0$& $\pi$ & $\eta^{\prime}$ & $\eta$ & K  \\
$2^1S_0$& $\pi(1300)$ & $\eta(1760)$ & $\eta(1440)$ & K(1460) \\
\hline\hline
\end{tabular}
\renewcommand{\arraystretch}{1.3}
\ec
\vskip 2mm
{\bf
\bc
Warning lesson from the $\bf \iota (1440)$: \\
\bf You can build up a case, convince the community, yet still be wrong\,!
\ec
}
\subsection{\label{section4.3}Glueball masses from the lattice}
\begin{figure}[h!]
\begin{minipage}[b]{0.38\textwidth}
\caption{\label{glueballs}
The glueball spectrum from an
anisotropic lattice study~\protect\cite{Morningstar:1999rf}.
A pseudoscalar glueball should have a mass of about 2.5 GeV\,!
Obviously, the $\eta (1440)$ cannot be a glueball.
The scalar glueball is expected at $1.7$\,GeV.}
\end{minipage}
\begin{minipage}[t]{0.55\textwidth}
\hspace*{5mm}\epsfig{file=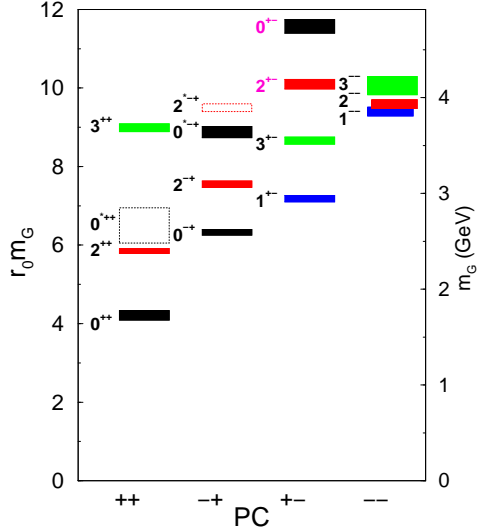,width=0.9\textwidth,angle=0}
\end{minipage}
\end{figure}

At the time when the $\eta(1440)$ was claimed to be the first
glueball, mass estimates were not yet reliable and required
normalization. Often, the mass of the $\eta(1440)$ was  used
as input to define the scale of glueball masses. Today, the best
estimates come from lattice gauge calculations. In
figure~\ref{glueballs}, we show the results obtained from an
anisotropic lattice (where the (Eukledian) time grid extends
over more grid points than the spatial grid).

\subsection{\label{section4.4}The enigmatic scalar mesons}

The lowest--mass glueball has scalar quantum numbers. Its 
predicted mass ($\sim 1700$\,MeV)
falls into a region in which one may hope to get a consistent
picture of the mass spectrum of all scalar mesons.
Table~\ref{scalar-data} lists the spectrum of scalar
mesons as given by the Particle Data Group. 
Let us recall what we expect: scalar mesons
have intrinsic orbital angular momentum $L=1$ and
quark spin $S=1$ which couple to $J=0$. Since spin--orbit
interactions are not very large (the $a_1(1260)$ and
$a_2(1320)$ masses are not very different), we might expect
an $a_0(1300)$, a K$^*_0(1430)$, and two $f_0$ with a
mass difference of about 250\,MeV (like the
$f_2(1270)$ and $f_2(1525)$ mass splitting) or about
400\,MeV which is the mass difference between the
$\eta$ and the $\eta^{\prime}$. In the case of baryons,
we have seen that radial excitations have a gap in mass square
to the ground states in the order of
1.2\,GeV$^2$. Thus we expect radial excitations to have masses
of about 1700\,MeV
and above. Including radial excitations, there should be
two  $a_0$, two K$^*_0$ (which we find) and four $f_0$'s.
But there are 7  $f_0$'s. (At high masses we combine separate
candidates in one entry. We will omit the $f_0(2200,2330)$
from our discussion.) Hence we need
to identify the four $\bar qq$ states and discuss what
the nature of the remaining states might be.

\begin{table}[h!]
\begin{minipage}[b]{.65\textwidth}
\bc
\renewcommand{\arraystretch}{1.3}
\begin{tabular}{|c|c|cc|}
\hline
\quad   I = 1/2 \quad\  & \quad  I = 1  \quad\   &
\quad   I = 0   \quad\           &      \\  %
\hline
             &           & $f_0(600)$           &      \\  
             &           &                      &      \\  
             & {$a_0(980)$} &
{$f_0(980)$} &      \\  
             &           &                      &      \\  
             &           &   {$f_0(1370)$}          &\\  
{$K_0^*(1430)$}   & {$a_0(1490)$} &
            {$f_0(1500)$}           & \\ 
            &           &                      &      \\  
            &           & {$f_0(1710)$}           &      \\  
$K_0^*(1950)$   &           &                      &      \\  
            &           & {$f_0(2020,2100)$}           &      \\  
            &           & {$f_0(2200,2330)$}           &      \\  
\hline
\end{tabular}
\renewcommand{\arraystretch}{1.0}
\ec
\end{minipage}
\begin{minipage}[t]{.30\textwidth}
\caption{\label{scalar-data}
The Particle Data Group lists
12 scalar mesons. Within the quark model
we expect 4 ground state
mesons and 4 radial excitations. }
\vskip 20mm
\end{minipage}
\end{table}
\subsubsection{Scalar mesons below 1\,GeV}
The $f_0(600)$, the lowest mass scalar meson often called $\sigma (600)$, 
has rather ill-defined properties. The Particle Data Group
assigns to it a mass range from 400 to 1200\,MeV. In 
partial wave analyses, it is seen as a pole at about 470\,MeV.
However, the phase reaches $90^{\circ}$ only at $\sim 780$\,MeV.
Its nature is hotly debated: a very attractive conjecture 
assigns the  $\sigma (600)$ to  a nonet 
($a_0(980), \sigma (600), f_0(980)$, $\kappa (900)$) where $\kappa
(900)$ represents 
the K$\pi$ S--wave which may have a pole at about 900\,MeV.
As a nonet of `normal' $q\bar q$ mesons, their mass seem to be too low,
but Jaffe~\cite{Jaffe:1976ig} showed that the nonet may be composed of
$qq\bar q\bar q$ 
states where the $a_0(980)$ and $f_0(980)$ carry an additional
$s\bar s$ pair (explaining their large coupling to $\rm K\bar K$).
Or, they may be relativistic S-wave $q\bar q$ states
(`chiralon')s~\cite{Ishida:eg}. In
the limit of chiral symmetry, we expect scalar partners of the
pseudoscalar nonet, and these 9 scalar resonances are 
identified as scalar companions of the pseudoscalar mesons. 
\par
Even if all four particles ($\sigma (600), \kappa (900),
a_0(980)$ and $f_0(980)$ exist, there is
no proof that they form one nonet~\cite{ochs_hadron03}. 
Other scenarios are
feasible where the dynamical origin of the $\sigma (600), \kappa
(900)$ and of the $a_0(980)$ and $f_0(980)$ are different. The 
$a_0(980)$ and $f_0(980)$ are often considered
as $\rm K\bar K$ molecular--like bound states. 
Their masses are close to the
$\rm\bar KK$ threshold, hence K and $\rm\bar K$ could be weakly
bound, forming two resonances in isospin $I=0$ and $I=1$,
as suggested 
by Isgur and Weinstein~\cite{Weinstein:gu}, by Speth and 
collaborators~\cite{Janssen:1994wn} or by Markushin
and Locher~\cite{Locher:1997gr}. The $\sigma(600)$ and 
$\kappa(900)$ are both very wide objects; they might be due
to attractive $\pi\pi$ or $\rm K\pi$ interactions, generated 
dynamically, (or by `left--hand cuts' in a technical language).
Practically, the $\sigma(600)$ and 
$\kappa(900)$ do not play a role in the discussion of glueballs,
and the reader is referred to a recent review~\cite{Tuan:2003bu}. 
\par
\subsubsection{Scalar mesons above 1\,GeV}
\begin{figure}[h!]
\begin{tabular}{cc}
\vspace*{-5mm}
\hspace*{-1mm}\includegraphics[width=0.43\textwidth]{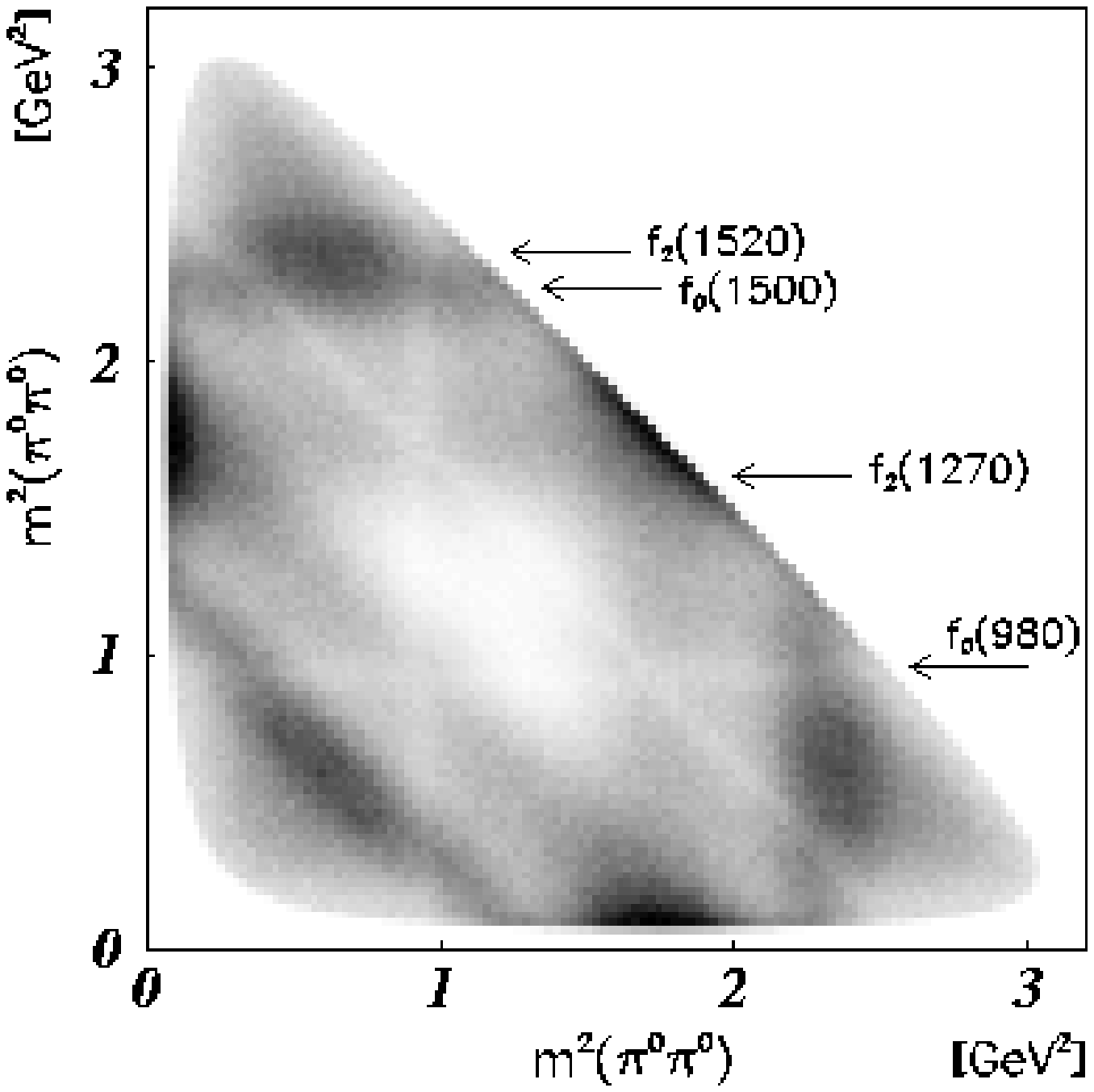}&
\hspace*{-3mm}\includegraphics[width=0.44\textwidth]{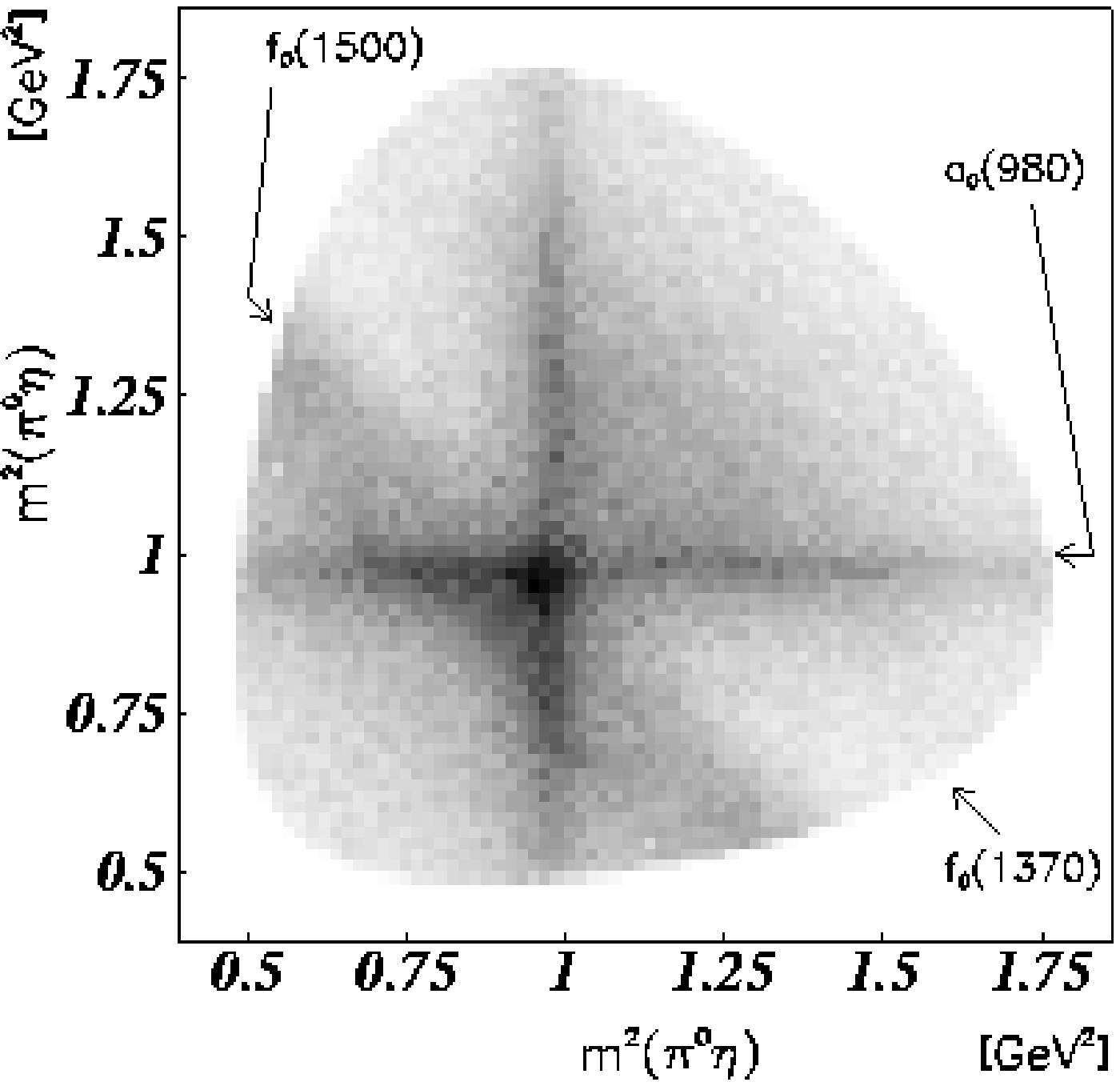}\\
\vspace*{-2mm}
\hspace*{-1mm}\includegraphics[width=0.46\textwidth]{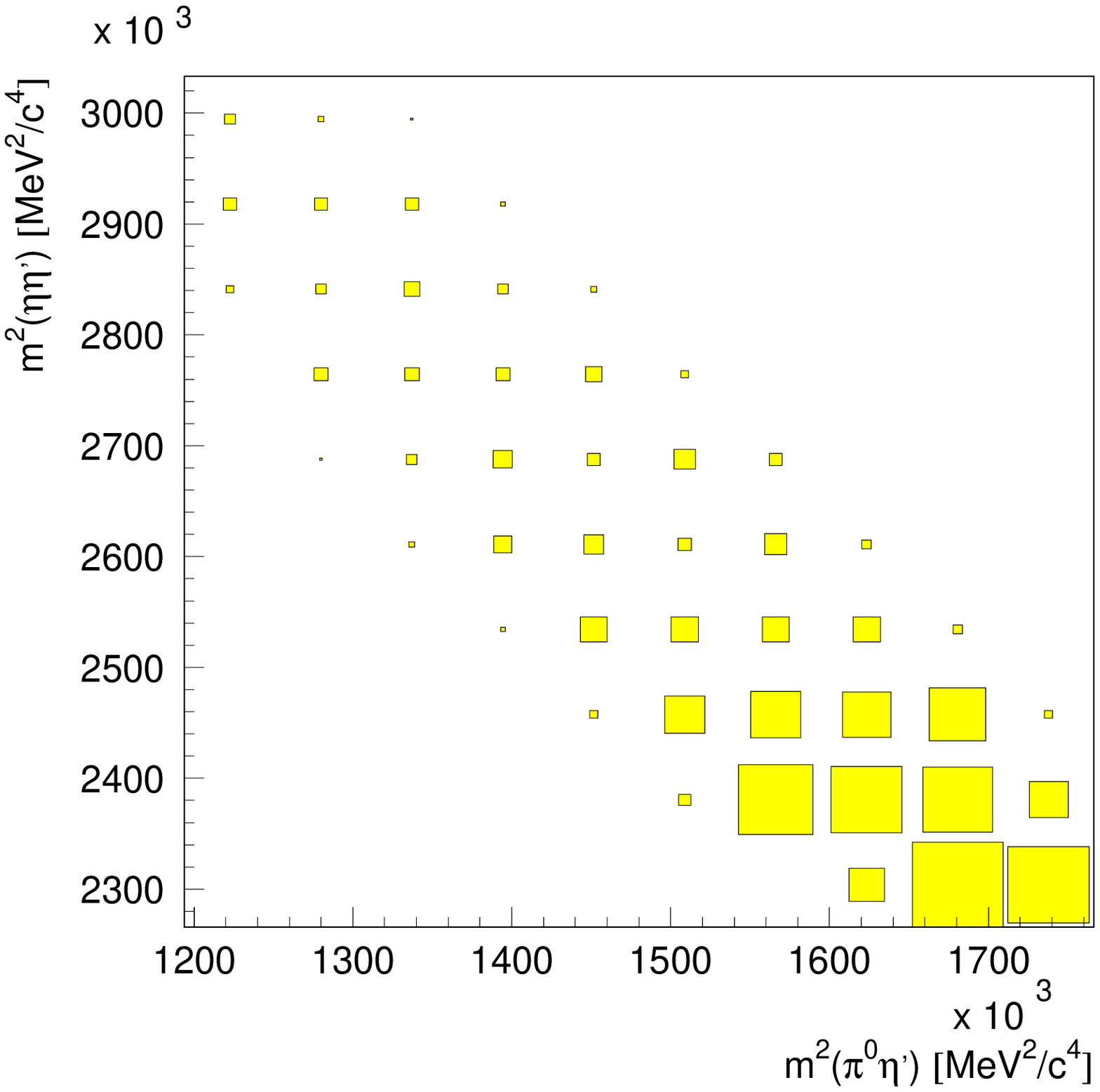}&
\hspace*{-3mm}\includegraphics[width=0.43\textwidth]{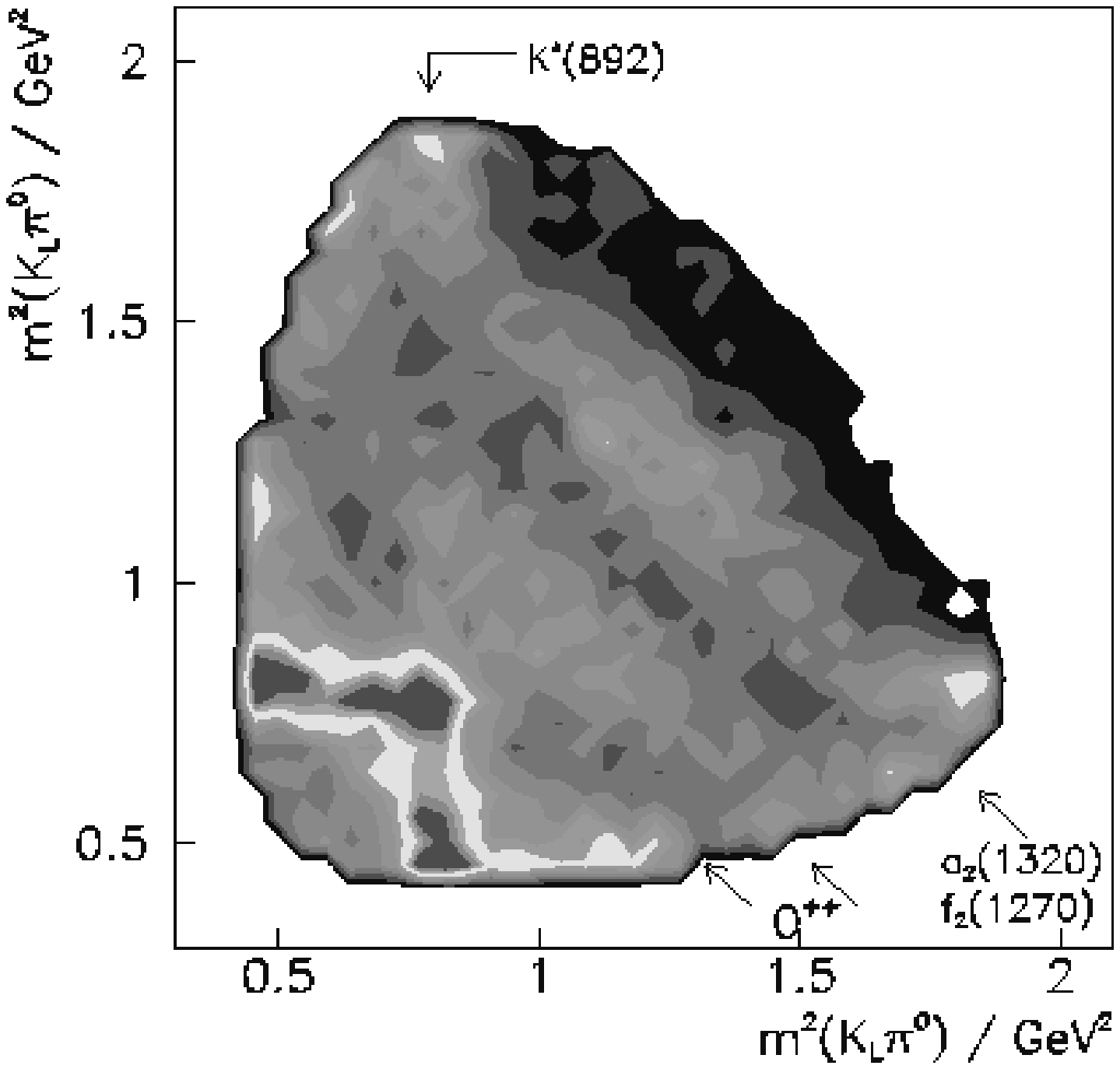}\\
\vspace*{-2mm}
\hspace*{-1mm}\includegraphics[width=0.44\textwidth]{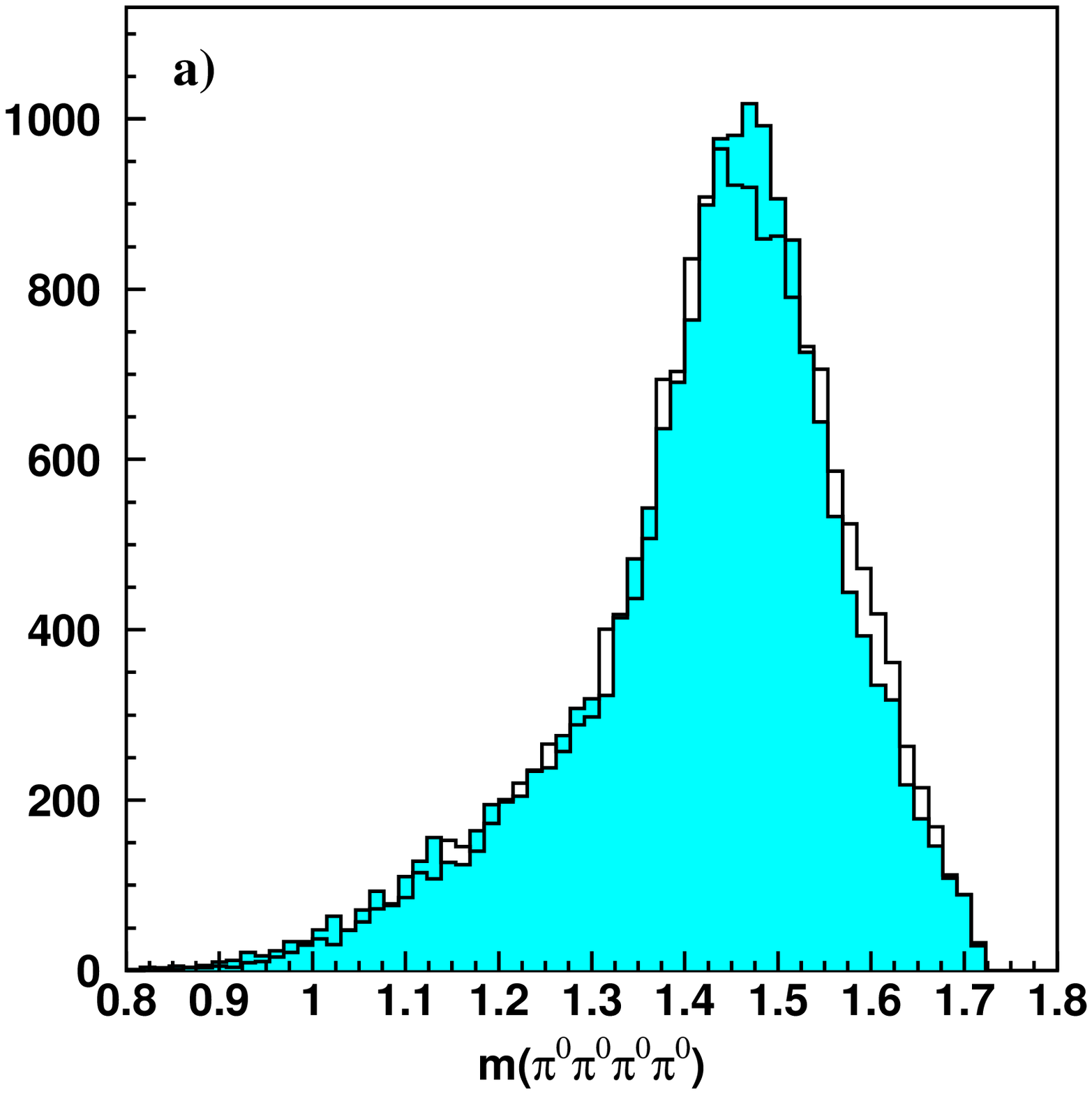}&
\hspace*{-3mm}\includegraphics[width=0.44\textwidth]{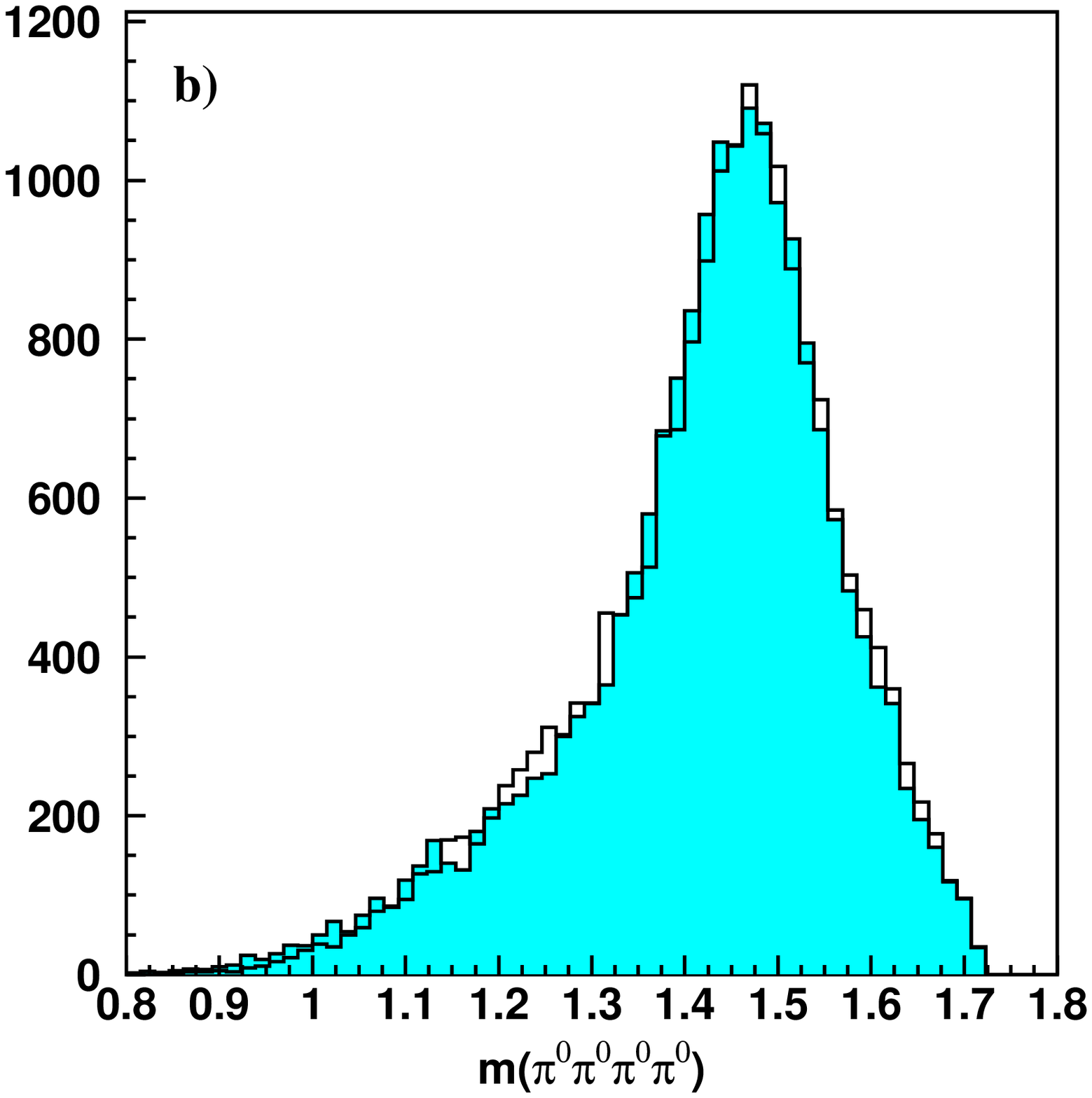}
\end{tabular}
\caption{\label{four-dp}
Dalitz plots for $p\bar p$ annihilation at rest into
3$\pi^0$ (upper left), $\pi^0 2\eta$ (upper right),
$\pi^0\eta\eta^{\prime}$ (lower left), $\rm K_lK_l\pi^0$ (lower
right). The $f_0(1370)$ contributes to (a,b,d), the $f_0(1500)$ to
all 4 reactions. The  $\rm K_lK_l\pi^0$ is difficult to interpret
in the black-and-white version;  the colored Dalitz plot can be
found on the web. The data are from
\protect\cite{Amsler:1995gf,Amsler:1995bz,Amsler:1994ah,Abele:1996nn}. 
The two lowest plots show the 4$\pi^0$ invariant mass in the reaction
$\rm\bar pn\to\pi^- 4\pi^0$. A fit (including other amplitudes)
with one scalar state fails; two scalar resonances at 1370 and
1500 MeV give a good fit. Note that the full 8-dimensional phase
space is fitted and not just the mass projection shown 
here~{\protect\cite{Abele:js}}.
} 
\end{figure}

The Crystal Barrel collaboration proposed the existence of two scalar
isoscalar mesons, the $f_0(1370)$ and $f_0(1500)$. Their main
properties were derived from four Dalitz plots 
~\cite{Amsler:1995gf,Amsler:1995bz,Amsler:1994ah,Abele:1996nn}, 
shown in figure~\ref{four-dp}, and from the
analysis of different 5 pion final 
states~\cite{Amsler:1994rv,Abele:1996fr,Abele:js,Abele:pv}.
In the 3\piz\ (upper left) and
the \piz 2\etg\ (upper right) Dalitz plots the $f_0(1500)$ is
clearly seen as band structure. In \piz\etg\etp\
a strong threshold enhancement in the \etg\etp\ invariant
mass is seen (lower left); the final state K$_l$K$_l$\piz\
has prominent K$^*$ bands; their interference with the
$f_0(1500)$ makes the intensity so large in the left corner
of the Dalitz plot (lower right).
The reactions $p\bar p\to \pi^+\pi^- 3\pi^0$~\cite{Amsler:1994rv},
$p\bar p\to 5\pi^0$~\cite{Abele:1996fr}, $p\bar n\to\pi^- 4\pi^0$
~\cite{Abele:js} and  $p\bar n\to 2\pi^- 2\pi^0\pi^+$ 
~\cite{Abele:pv}
were studied to determine decays into 4 pions.
\par
\par
We have seen that decays of mesons are constrained by SU(3)
relations.  So there is hope that the glueball nature
of a state can be unraveled by inspecting the
coupling to various final states.
\par
Table \ref{f0decay} lists partial widths of
the $f_0(1370)$ and $f_0(1500)$ as derived from the
Crystal Barrel Collaboration.
Neither the $f_0(1370)$ nor the $f_0(1500)$, 
has a large coupling to
$\rm K\bar K$, so none of them carries a large $s\bar s$ fraction.
For an interpretation as ground state plus radial excitation,
their mass difference is too small. Hence one of them might be the
scalar glueball.
The partial decay widths for the decays into \etg\etg\ and \etg\etp\
and the smallness of the $\rm K\bar K$ coupling of the $f_0(1500)$
show that the  $f_0(1500)$ cannot be a pure glueball:
For a glueball or any other isosinglet meson we expect ratios for
$\pi\pi$\,:\,\etg\etg\,:\,\etg\etp\,:\,$\rm K\bar K$ of 3\,:1\,:\,0\,:\,4,
after removal of phase space. Since the coupling to \etg\etp\ is
large, the $f_0(1500)$ 

\begin{table}[h!]
\vspace*{-7mm}
\begin{minipage}[c]{.5\textwidth}
\caption{\label{f0decay}
Partial decay widths of the \protect\phantom{$f_0(1370)$}
\protect\phantom{$f_0(1370)$rrr} $f_0(1370)$ and $f_0(1500)$.}
\renewcommand{\arraystretch}{1.2}
\bc
\begin{tabular}{lcc}
\hline\hline
                        & $f_0(1370)$    &  $f_0(1500)$ \\
\hline
$\Gamma_{tot}$
                        &  $\sim 350$    & $\sim 109$   \\
$\Gamma_{\pi\pi}$
                        & $\sim 90$  & $\sim 32$ \\
$\Gamma_{\eta\eta}$
                        & $\sim 1$   & $\sim 6$   \\
$\Gamma_{\eta\eta'}$
                        &                 & $\sim 3$   \\
$\Gamma_{\bar{K}K}$     &$\sim 50$
                                          & $\sim 6$   \\
$\Gamma_{4\pi}$
                        & $\sim 210$  & $\sim 62$ \\
$\Gamma_{\sigma\sigma}$
                        & $\sim 106$  & $\sim 20$ \\
$\Gamma_{\rho\rho}$
                        &  $\sim 55$  & $\sim 10$ \\
$\Gamma_{\pi^*\pi}$
                        & $\sim 36$     & $\sim 25$ \\
$\Gamma_{a_1\pi}$
                        & $\sim 13$   & $\sim 7$ \\

\hline\hline
\end{tabular}
\ec
\renewcommand{\arraystretch}{1.0}
\end{minipage}
\vspace*{-5mm}
\begin{minipage}[c]{.49\textwidth}
cannot be a pure glueball, it must mix with nearby
states. The $f_0(1370)$ has important couplings to
two pairs of \piz-mesons, to $\sigma\sigma$. This is evident from
the two plots at the bottom of figure~\ref{four-dp}.
\par
Three striking peaks  were observed in the
\etg\etg\ invariant mass spectrum produced 
in \pbp\ annihilation in flight into \piz\etg\etg~\cite{E760},
$1500, 1750$ and $2100$\,MeV. 
The data were not decomposed into partial waves
in a partial wave analysis, so the peaks could have 
J$^{\rm P\rm C}=0^{++}$, $2^{++}$, or higher. If the states would have
J$^{\rm P\rm C}=2^{++}$, their decay into \etg\etg\ would 
be suppressed by the angular momentum barrier. 
The peaks are seen very clearly suggesting 
$0^{++}$ quantum numbers. 
\end{minipage}
\end{table}
\vspace*{3mm}
\par
\begin{figure}[h]
\vspace*{-5mm}
\epsfig{file=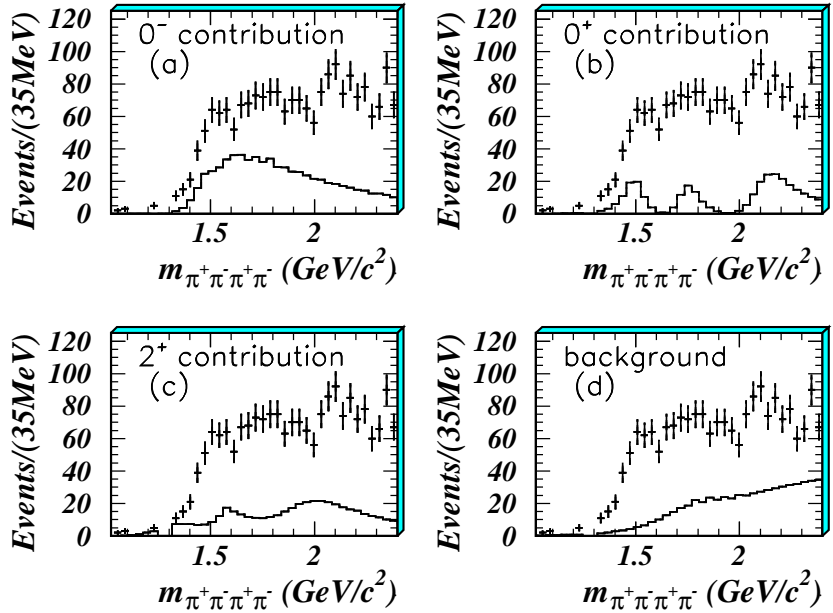,width=0.90\textwidth}
\vspace*{-5mm}
\caption{\label{jto4pi} 
Partial wave decomposition of radiative J/$\psi$ 
decays into 2\pip 2\pim\protect\cite{bes4p}.
}
\vspace*{-3mm}
  \end{figure}

The same pattern of states 
was seen at BES in radiative J/$\psi$ decays~\cite{bes4p}
into 2\pip 2\pim . The results of a partial wave analysis 
in figure~\ref{jto4pi}
show a slowly rising instrumental background and 
3 important contributions with scalar, pseudoscalar and 
tensor quantum numbers. 
The scalar part contains three resonances, at 1500, 1710
and 2100 MeV. This pattern of states was already suggested in 
a reanalysis of MARKIII data~\cite{tokibugg}. 
The $f_0(1500), f_0(1710)$ and the $f_0(2100)$ have a similar 
production and decay pattern. Neither a  $f_0(1370)$ nor a
`background' intensity is assigned to the scalar isoscalar 
partial wave. 
\par 
\subsection{\label{section4.5}Scalar mesons: interpretation}
There is no agreement how to interpret the scalar spectrum, there
are numerous experimental and theoretical contributions to this
field. Here we discuss three typical scenarios.
\subsubsection{The `narrow' glueball}
The first interpretation~\cite{Amsler:1995td}, 
also adopted by the Particle Data Group,
identifies the $a_0(980)$ and  $f_0(980)$ as non--$q\bar q$ states.
They might form, together with the $\sigma (600)$ and $\kappa (900)$, a nonet
of four--quark states, or they could form a nonet of `chiralons', but
they are left out for further discussion. 
In the mass region where the scalar $1^3P_0$
mesons are expected, there are now 10 states
while the quark model predicts only 9
(3 $a_01450)$, 4K$^*_0(1430)$, and 2 $f_0$'s). One of
the states,  $f_0(1370)$ or $f_0(1500)$ or $f_0(1710)$,
must be the scalar glueball\,!
\par
However, the  $f_0(1500)$ couples strongly to $\eta\eta^{\prime}$;
these are two SU(3) orthogonal states and cannot come from a singlet.
The  $f_0(1500)$ must hence have a strong flavor--octet component, 
it cannot be a pure glueball. The $f_0(1370)$ and $f_0(1500)$
decay strongly to $2\pi$ and into  $4\pi$ and weakly to
$\rm\bar KK$, they both cannot carry a large $\bar ss$ component.
The  $f_0(1370)$ is, perhaps, too light to be the scalar
glueball. So, none of the three states 'smells' like a glueball.
A way out is mixing; the two scalar $\bar qq$ states and the scalar
glueball have the same quantum numbers, they mix and 
form the three observed states. Table~\ref{inter1} summarizes
this interpretation.
\par
\begin{table}[h!]
\vspace*{-3mm}
\caption{\label{inter1}
Possible interpretation of the scalar mesons. The three
states $f_0(1370)$, $f_0(1500)$ and $f_0(1710)$ originate from
2 $q \bar q$ states and a glueball.}
\renewcommand{\arraystretch}{1.3}
\bc
\begin{tabular}{ccccc}
\hline\hline
\quad   I = 1/2 \quad\ & \quad  I = 1  \quad\  & \quad   I = 0  \quad\    &         & \\  %
\hline
              &             & $f_0(600)$  &        &  $\sigma (600)$ meson                 \\   
              &             &             &        &  chiral partner of the $\pi$ \\  
              & $a_0(980)$  & $f_0(980)$  &        & $\rm K\bar K$ molecules   \\
              &             &             &        & \\                 
\hline
              &             & $f_0(1370)$ &        &$q\bar q$ state\\  
$K_0^*(1430)$ & $a_0(1490)$ & $f_0(1500)$ &        &2 $q\bar q$ states, glueball \\ 
              &             &             &        &  \\    
              &             & $f_0(1710)$ &        &$q\bar q$ state  \\    
\hline
$K_0^*(1950)$ &             &             &        &$q\bar q$ state  \\  
              &             & $f_0(2100)$ &        &$q\bar q$ state  \\  
              &             & $f_0(2200,2330)$ &   &$q\bar q$ state  \\  
\hline\hline
\end{tabular}
\renewcommand{\arraystretch}{1.0}
\ec
\vspace*{-3mm}
\end{table}
Several mixing scenarios have been
suggested~\cite{Amsler:1995td,Lee:1999kv,Li:2000yn,Close:2000yk,Celenza:uk,Strohmeier-Presicek:1999yv,Amsler:2002ey}
and some of them are capable of reproducing the decay
pattern. So, in these scenarios two scalar states plus an
intruder, the scalar glueball, mix. The lattice gauge predictions
for the existence of a glueball and the mass estimates are
beautifully confirmed, and there is only the need to confirm
some further glueball predictions.

\subsubsection{The `narrow' glueball scrutinized}

An important ingredient of 
the `narrow--glueball' is the interpretation of the
$f_0(980)$ and $a_0(980)$ as alien objects, unrelated 
to the spectroscopy of \qqbar\ mesons. Several experiments
were directed to determine the structure of these two mesons,
like two-photon production~\cite{Boglione:1999rw},
or $\Phi$ radiative decay rate into 
$f_0(980)$~\cite{Achasov:2000ym,Aloisio:2002bt}
and into $a_0(980)$~\cite{Achasov:2000ku,Aloisio:2002bs}.   
The conclusions drawn from these results are, however, 
ambiguous.

At LEP, the fragmentation of quark- and gluon jets has been
studied intensively~\cite{bohrer}. In particular the inclusive
production of the $f_0(980)$ and $a_0(980)$ provides insight
into their internal structure. Some total inclusive rates are 
listed in Table \ref{Z0decay}. The rates depend on the
meson mass and on the spin multiplicity.
The three mesons \etp , $f_0(980)$
and $a_0(980)$ - which have very similar masses - have 
production rates which are nearly identical (the two charge 
modes of the $a_0(980)^{\pm}$ need to be taken into
account). Hence there is primary evidence that the 
three mesons have the same internal structure and that they are all
three $q\bar q$ states. This conclusion was substantiated
by further studies~\cite{opal} of the production characteristics 
of the $f_0(980)$ 
as compared to those of f$_2(1270)$ and $\Phi(1020)$ mesons,
and with the Lund string model of hadronization 
within which the $f_0(980)$  is treated as a conventional meson.
No difference is observed in any of these comparisons between
the $f_0(980)$ and the f$_2(1270)$ and $\Phi(1020)$. 

\begin{table}[h!]
\begin{minipage}[h]{.65\textwidth}
\renewcommand{\arraystretch}{1.3}
\bc 
\begin{tabular}{|lc|}
\hline
\piz          & $ 9.55\pm 0.06\pm 0.75$ \\
\etg          & $ 0.97\pm 0.03\pm 0.11$ \\
\etp          & $ 0.14\pm 0.01\pm 0.02$ \\
$a_{0}^{\pm}(980)$& $ 0.27\pm 0.04\pm 0.10$\\
$f_0(980)$    & $ 0.141 \pm 0.007 \pm 0.011$\\
$\Phi (1020)$ & $ 0.091\pm0.002\pm0.003$\\
$f_2(1270)$   & $ 0.155\pm0.011\pm0.018$ \\
\hline
\end{tabular}
\ec
\end{minipage}
\begin{minipage}[b]{.33\textwidth}
\caption{\label{Z0decay}
Yield of light mesons per hadronic $Z^0$
decay~\protect\cite{bohrer,opal}.
}
\renewcommand{\arraystretch}{1.0}
\end{minipage}
\end{table}   

Now, what is the nature of the $f_0(980)$ and $a_0(980)$\,?
Presumably, their wave function has a complex 
mass and momentum dependence. 
Likely, the outer part of the wave
function contains a large $\rm K\bar K$ component, in particular
close to the $\rm K\bar K$ threshold.
The fragmentation process couples to the \qqbar\ core. 
If this is true, the $f_0(980)$ and $a_0(980)$ mesons cannot be
disregarded when the spectrum of scalar \qqbar\ mesons is
discussed.

\subsubsection{Evidence for a very wide glueball}
In meson--meson scattering in relative S--wave, 
coupled channel effect play a decisive role. The
opening of thresholds attracts pole positions and the resonances
found experimentally do not need to agree with masses as
calculated in quark models. Under normal circumstances,
$K$--matrix poles, poles of the scattering matrix $T$ and
positions of observed peaks agree apprximately, and the interpretation is
unambiguous. In S--waves, the situation is more complicated.
\par
\begin{figure}[h!]
\vspace*{-4mm}
\begin{minipage}[c]{.60\textwidth}
\epsfig{file=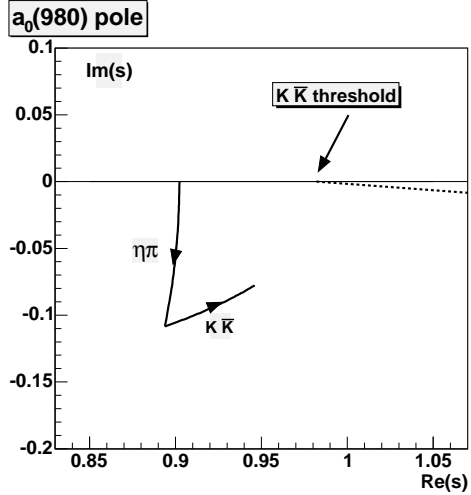,width=0.9\textwidth}
\vspace*{-2mm}
\caption{\label{pole-pos}
Pole positions of a $T$-matrix describing
$\pi\eta$ S--wave interactions as a function of the
couplings to $\eta\pi$ and $\rm\bar KK$.
}
\end{minipage}
\begin{minipage}[c]{.35\textwidth}
Figure~\ref{pole-pos} shows e.g. the $T$-matrix pole position of 
a $a_0(980)$ resonance coupling to $\eta\pi$ and 
$\rm\bar KK$. When the coupling to the final state is set to zero,
the pole position falls onto the real axis, and coincides with the
$K$--matrix pole, chosen at $900$\,MeV. When the coupling
to $\eta\pi$ is turned on, the resonances acquires width and the
pole moves into the complex plane $\sqrt s$ plane. Then the coupling
to $\rm\bar KK$ is turned on; the resonance gets narrower and
the pole approaches the $\rm\bar KK$ threshold. 
\end{minipage}
\vspace*{-5mm}
\end{figure}
\par
The mass of the resonance as quoted by experiments is the
\begin{table}[b!]
\vspace*{-5mm}
\caption{\label{aas}
The K--matrix poles of \protect\cite{andrey} show a remarkable
agreement with the results of the Bonn model~\protect\cite{Koll:2000ke}, 
version B. There is an
additional pole at $1400\pm200$\,MeV, 
far from the real axis (i.e. $\sim 1000$\,MeV broad)
which is a flavor singlet and could be the glueball. 
}
\renewcommand{\arraystretch}{1.3}
\begin{tabular}{ccc|ccc}
\multicolumn{3}{c}{K-matrix poles}&
\multicolumn{3}{c}{Bonn model, B}\\
\hline\hline
             &$a_0(980\pm30)$
             &$f_0(680\pm50)$ &
             &$a_0(1057)$
             &$f_0(665)$       \\  
             &&&&&\\  
             &           &           &        
             &           &                \\  
             $\rm K_0^*(1230\pm40)$
             &$a_0(1630\pm 40)$
             &$f_0(1260\pm 30)$
             &$\rm K_0^*(1187)$
             &$a_0(1665)$
             &$f_0(1262)$ \\  
             &&$f_0(1400\pm 200)$
             &&&\\  
             &&$f_0(1600)$
             &&&$f_0(1554)$\\  
             &&&&&\\  
             $\rm K_0^*(1885^{+50}_{-100})$ &&&
             $\rm K_0^*(1788)$ &&\\
             &&$f_0(1810\pm 50)$
             &&&$f_0(1870)$     \\
\hline\hline
\end{tabular}
\renewcommand{\arraystretch}{1.3}
\end{table}
$T$ matrix pole.
Quark models usually do not take into account the couplings
to the final state. So we might need to compare
the $K$--matrix poles with quark model results.
\par
This comparison is made in Table~\ref{aas}. The $K$--matrix poles come
from a series of coupled--channel
analyses~\cite{Anisovich:1996qj,Anisovich:1997qp,Anisovich:2002ij},
mean values and errors are estimates provided by one
of the authors~\cite{andrey}. 
The quark model states are from the Bonn model~\cite{Koll:2000ke}, 
with the Lorentz structure B of the confinement potential.
\par
There is excellent agreement. The two lowest scalar nonets are
identified, and there is one additional state, the 
$f_0(1400\pm 200)$. Its coupling to two pseudoscalar mesons
are flavor--blind, it is an isoscalar state. So it can be
identified as a scalar glueball. Problematic is the width:
it exceeds 2\,GeV. In the next section we ask if we
can identify a pole of such an enormous width as
a resonance. An excellent review of this approach can be found 
in~\cite{vvanis}.

\subsubsection{The `wide' glueball scrutinized}

\begin{figure}[h!]
\begin{minipage}[t]{.60\textwidth}
\epsfig{file=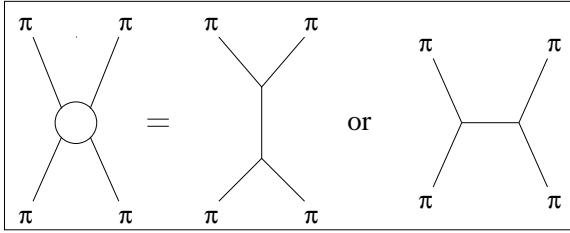,width=\textwidth}
\end{minipage}
\begin{minipage}[b]{.35\textwidth}
\caption{\label{pipiscatt}
Scattering of two pions (left) via $s$-channel
resonances (center) and $t$-channel exchange.
}
\end{minipage}
\end{figure}

Before we continue the discussion we have to introduce a further
concept: {\it $s$--channel resonances and $t$--channel exchanges.}
There are two processes which may contribute to the $\pi\pi$
scattering amplitude: formation of $s$-channel resonances and
scattering via $t$-channel exchanges. They are schematically drawn
in Figure~\ref{pipiscatt}. Scattering processes can be represented 
by a sum of $s$-channel
resonances or by $t$-channel exchanges; in Regge theory this is
called duality and is the basis for the Veneziano model. So you
may analyze a data set and describe the data by a sum over
$s$-channel resonances and get a very good description with a
finite number of complex poles in the $\pi\pi$ S-wave scattering
amplitude. You could also analyze the data by a summation over
$t$-channel exchange amplitudes and also get a good fit. If you
add amplitudes for both processes, you run the risk of double
counting. 
\par
\par
There is a common belief that the interpretation of
a pole in the complex scattering energy plane as
originating from $s$-, $t$- or $u$-channel phenomenon
is a matter of convenience.
Indeed, $t$--channel
exchanges can lead to resonances; the exchanges represent forces
and can lead to binding, can create poles. The $f_0(400-1200)$,
often called $\sigma (600)$ meson, is certainly present in
scattering data with a
pole in the complex energy plane. And you may choose
to describe this pole as $s$-channel resonance even
if its true origin might be $t$-channel exchange.
But it is hard to believe
that the $\omega$ can be created by $t$--channel forces
in the $\rho\pi$ channel.
Hence particles may exist which are
not created by $t$--channel exchanges. Also the reverse statement is true:
not all poles created by exchanges forces need to have particle
properties. So, how does one decide if a
particular pole in the scattering plane is due to a $s$-channel
resonance or to $t$-channel exchanges\,?
\par
$s$-channel resonances always have 
the same ratio of
couplings to different final states. The partial widths of the
$f_0(1500)$ must not depend on the way in which it was produced. This
is different for poles generated by $t$-channel exchanges.
If properties of a pole depend 
on the production process, then the pole is not
a particle.
\par

\begin{figure}[h!]
\begin{minipage}[c]{.65\textwidth}
\epsfig{file=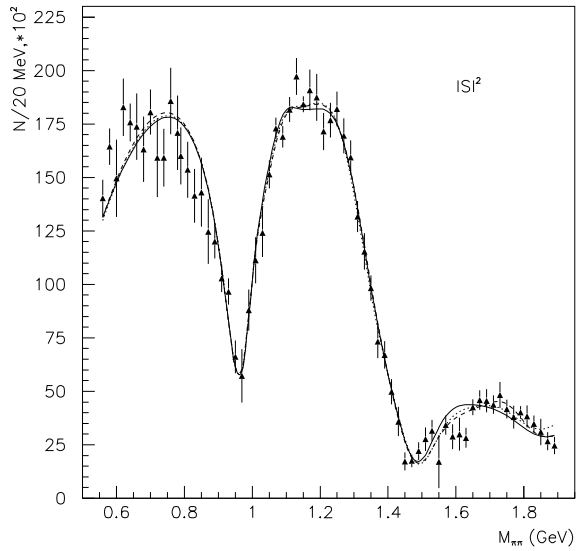,width=0.92\textwidth}
\end{minipage}
\begin{minipage}[c]{.34\textwidth}
\caption{\label{GAMS}
The $\pi\pi$ scattering amplitude measured
in the GAMS experiment. From~\protect\cite{Anisovich:1996qj}.
}
\par
\vskip 5mm
The $\pi\pi$ scattering amplitude exhibits a continuously and slowly
rising phase and a sudden phase increase at 980 MeV. The rapid phase
motion is easily identified with the $f_0(980)$, the slowly rising
phase can be associated with an $s$-channel resonance which was
called $f_0(1000)$ by Morgan and Pennington~\cite{Morgan:td}
and the {\it Red Dragon} by Minkowski and Ochs~\cite{Minkowski:1998mf}.
It
\end{minipage}
\vspace*{-4mm}
 \end{figure}

Figure~\ref{GAMS} shows the $\pi\pi$ scattering
amplitude as seen in the GAMS experiment.
The modulus of the amplitude shows two dips, at the mass of the
$f_0(980)$ and $f_0(1500)$: intensity
is taken from $\pi\pi$ scattering to inelastic
channels. The first peak in the scattering amplitude at
low energy is the $f_0(400-1200)$ and often called
$\sigma$-meson; the second
bump at 1300 MeV was called $\epsilon (1300)$.

\noindent
extends at least up to 1400 MeV. It has been 
suggested~\cite{Minkowski:1998mf} that this broad enhancement 
is the scalar glueball. 
It seems to agree with the broad glueball discussed in the
last section. 
This broad background amplitude - including
the monotonously rising phase - can however well be reproduced
by an amplitude for \rh\ exchange in the $t$-channel. From
a fit to the $\pi\pi$ S-wave scattering data even mass and width
of the \rh\ exchanged in the $t$-channel
can be determined. So this background amplitude
\begin{figure}[t!]
\vspace*{-10mm}
\includegraphics[width=0.9\textwidth]{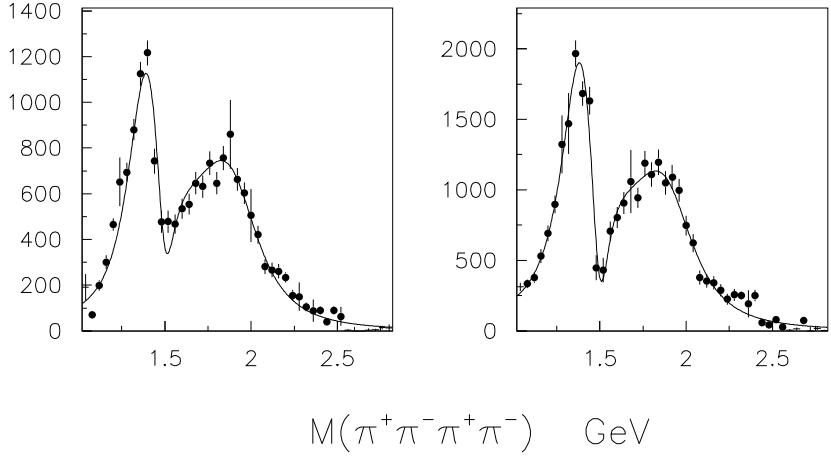}
\vspace*{-0.7cm}

\caption{\label{wa_dip}
4$\pi$ invariant mass spectrum produced by two protons in
central collisions. A cut is made on
the angle in the transverse direction between the two
outgoing protons. Left, 90-135$^{\circ}$. Right,
135-180$^{\circ}$.
The latter setting corresponds to the so-called
glueball filter. The $f_0(1500)$ shows up as a dip
just like the  $f_0(980)$ in $\pi\pi$ 
scattering\protect\cite{Barberis:2000em}.
}
\vspace*{-4mm}
\end{figure}
is likely not a  $f_0(1000)$ $q\bar q$ state, nor two mesons,
$\sigma (600)$ and $\epsilon (1300)$; it is caused
by \rh\ (and possibly other less important) exchanges
in the $t$-channel.
\par
Now, consider figure~\ref{wa_dip}. On the right side,
a selection is made for small momentum transfers to the
4$\pi$ system. At small momentum transfer, the $f_0(1500)$ is
seen as a dip. This resembles very much the data of the
GAMS collaboration on $\pi\pi$ scattering (Figure~\ref{GAMS}).
So the question arises if the enhancement seen in the
left part of figure~\ref{wa_dip} is
a $q\bar q$ resonance. Or can it be traced to
\rh\ and other exchanges in the $t$-channel\,? We now
argue that the latter is indeed the case.
\par
\begin{figure}[h!]
\vspace*{-2mm}
\begin{minipage}[t]{.60\textwidth}
\epsfig{file=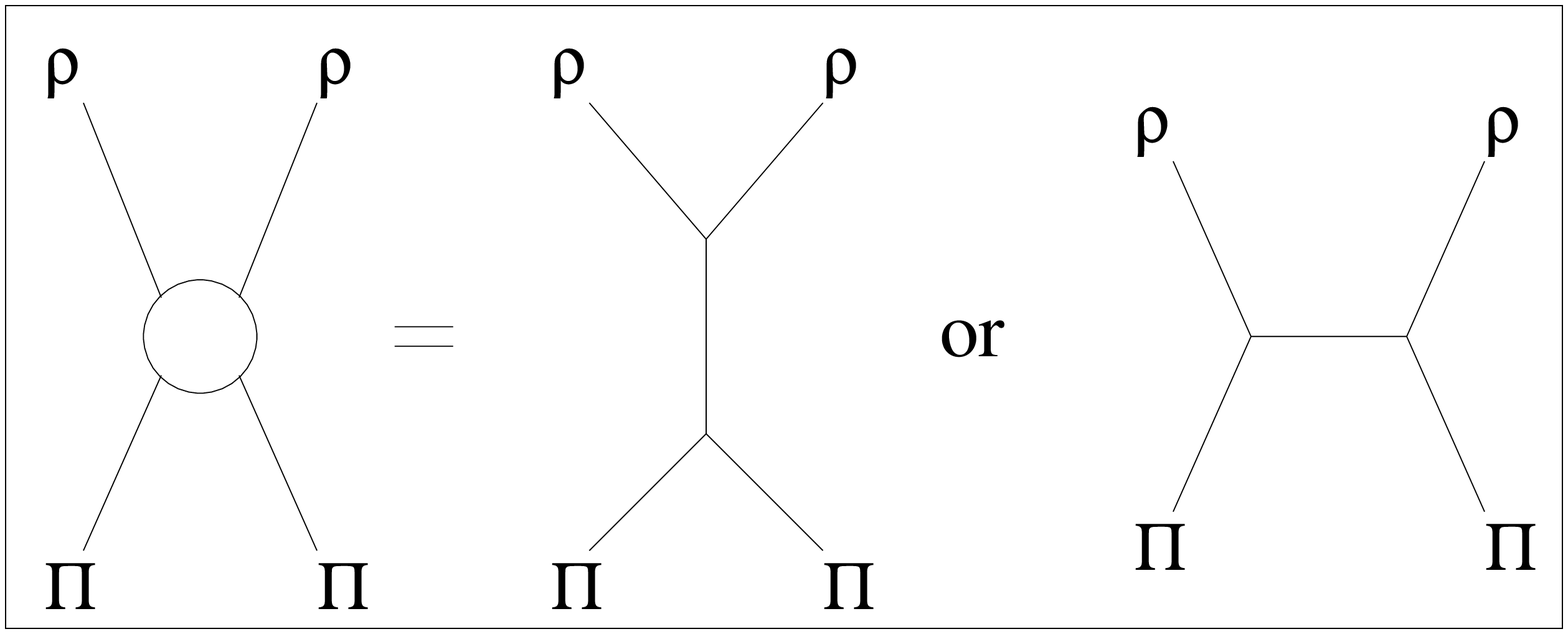,width=0.85\textwidth}\\
\epsfig{file=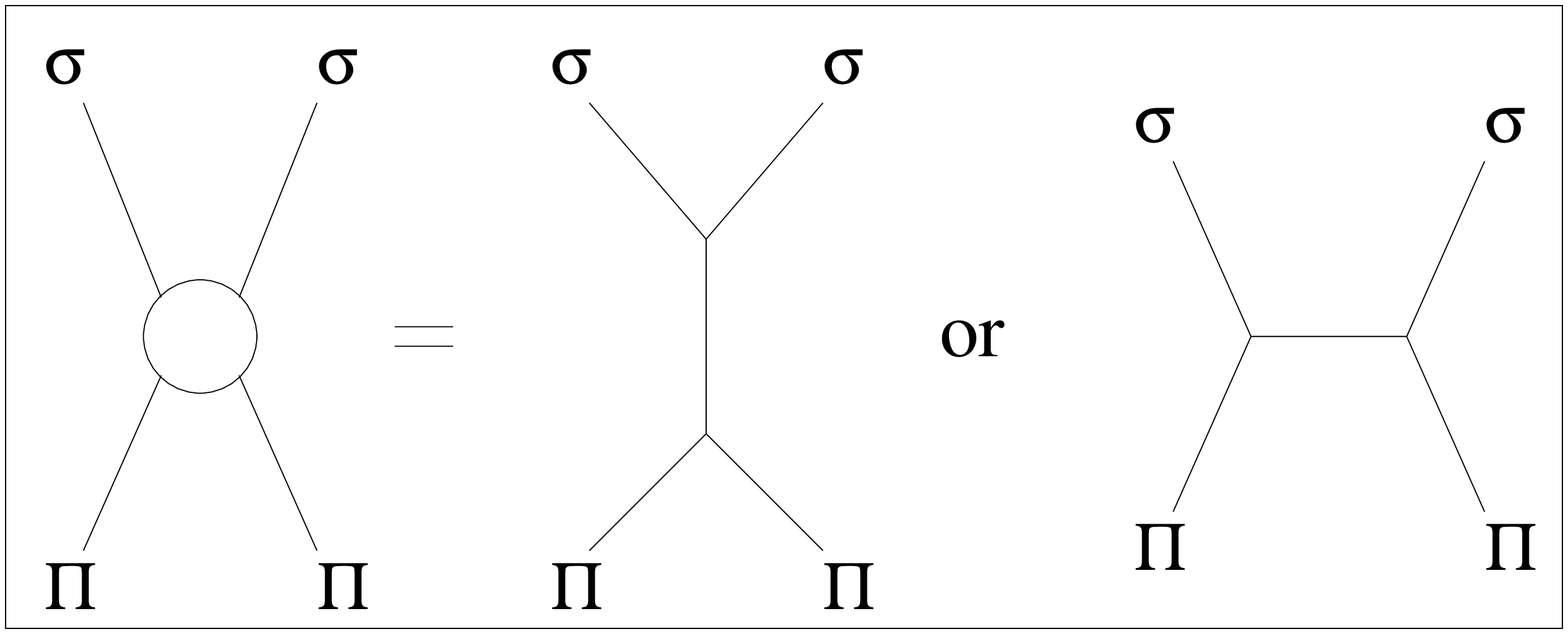,width=0.85\textwidth}
\end{minipage}
\begin{minipage}[b]{.35\textwidth}
\caption{\label{pompomscatt}
Scattering of two Pomerons via $s$-channel
resonances and $t$-channel exchange into $\rho\rho$
and into $\sigma\sigma$. Production of $\sigma\sigma$
via $t$--channel exchange seems to be suppressed.
}
\end{minipage}
\vspace*{-2mm}
\end{figure}

We now assume that Pomeron-Pomeron scattering can also proceed via
$\rho$ exchange in the $t$-channel (figure~\ref{pompomscatt}). 
This $t$-channel amplitude
then interferes with the production of the $q\bar q$ state
$f_0(1500)$ producing a dip, very much alike the dip seen at 980
and 1500\,MeV in $\pi\pi$ scattering. This conjecture leads to
measurable consequences.
\par
In Pomeron-Pomeron scattering, $\rho$
exchange in the $t$-channel may occur leading to production
of two $\rho$ mesons. Isospin conservation 
does not allow $\sigma\sigma$ production via $\rho$
exchange. Hence we may expect the 4$\pi$ background
amplitude not to couple to  $\sigma\sigma$.
\par
\begin{figure}[h!]
\includegraphics[width=\textwidth]{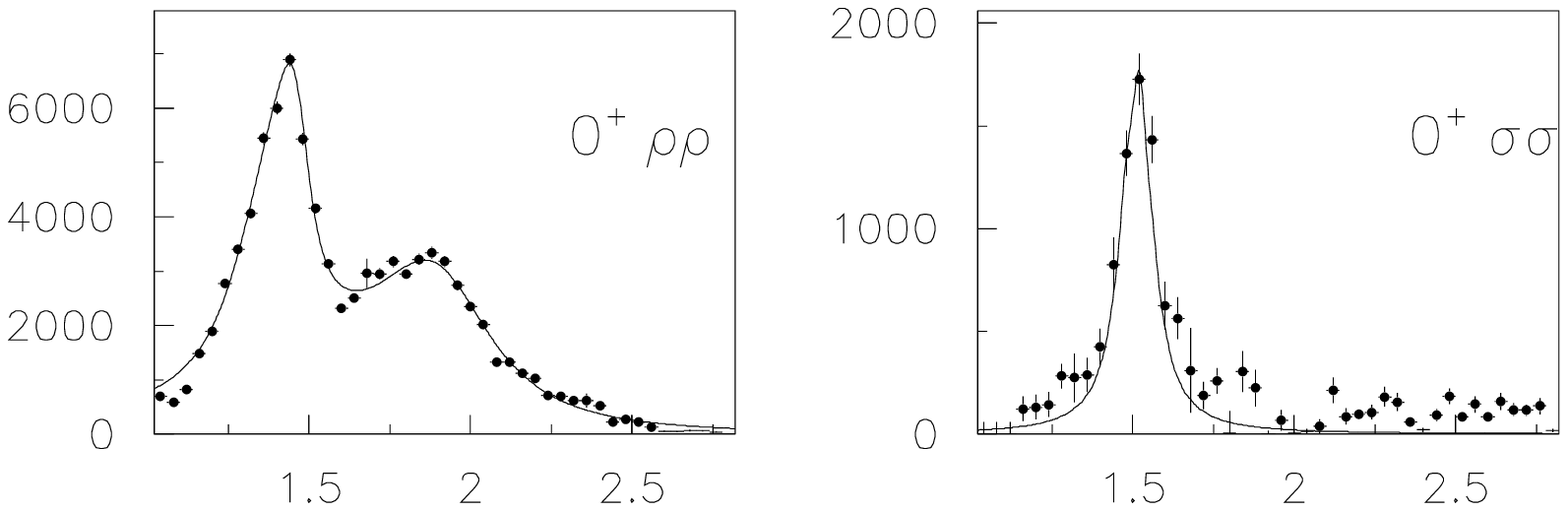}\\
\vspace*{-2cm}

\includegraphics[width=\textwidth]{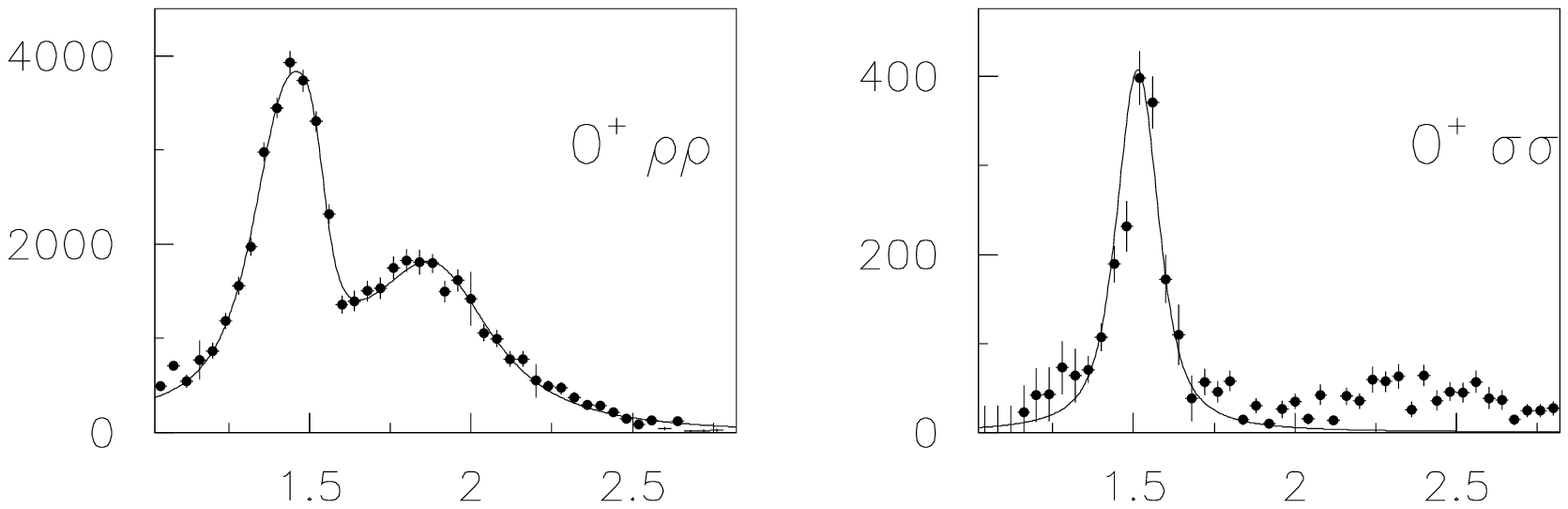}\\
\vspace*{-2.7cm}

\begin{minipage}[t][4cm][t]{0.65\textwidth}
\hspace*{-1.4cm}
\includegraphics[width=\textwidth]{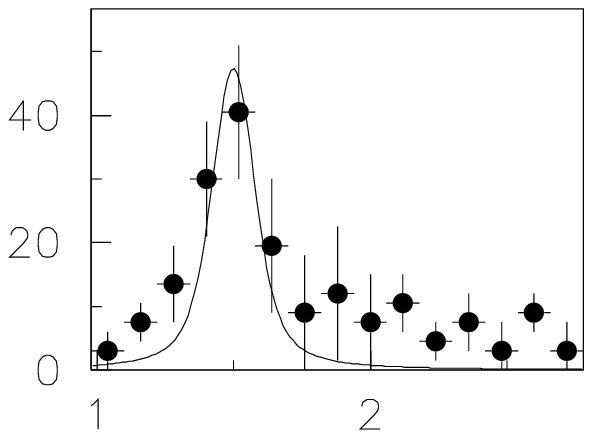}
\end{minipage}

\vspace*{-6cm}

\hspace*{-16cm}\hfill
\begin{minipage}[t][4cm][t]{.4\textwidth}
\vspace*{15mm}
\caption{\label{wa4pi}
4$\pi$ invariant mass (in GeV) spectra from central
production. First row: 2\pip 2\pim ; second row \pip\pim 2\piz ;
left: $\rho\rho$ S-wave; right: $\sigma\sigma$ S-wave.
Third row: $\sigma\sigma$ S-wave in
4\piz~\protect\cite{Barberis:2000em}.
}
\end{minipage}
\vspace*{-3mm}
  \end{figure}

Figure~\ref{wa4pi} shows 4$\pi$ invariant mass spectra from
the WA102 experiment~\cite{Barberis:2000em}.
A large peak at 1370 MeV is seen followed by a dip in the
1500 MeV region and a further (asymmetric) bump.

The 4$\pi^0$ invariant mass spectrum shows the
$f_0(1500)$ but nearly no background\,!
The partial wave analysis confirms these findings:
it determines contributions from
several scalar
resonances, the $f_0(1370), f_0(1500)$ and $f_0(1750)$ and
a new $f_0(1900)$. The partial wave analysis
finds $f_0(1370)$ decays into $\rho\rho$ but not
into $\sigma\sigma$ while the $f_0(1500)$ shows both decay
modes. In the Crystal Barrel experiment the $f_0(1370)$
decays into $\rho\rho$ and into $\sigma\sigma$ with similar
strength, see Table \ref{f0decay}.
\par
Here we have made an important step. We now understand why the
left-hand spectra of figure~\ref{wa4pi} differ so much from
the right-hand spectra. The $\sigma\sigma$ final state can
be reached only via $s$-channel resonances and there is only
one, the $f_0(1500)$. The \rh\rh\ final state is produced
by $t$-channel exchanges; they generate the broad enhancement
extending over the full accessible mass range. It rises
at threshold for 4$\pi$ production and falls off because
of the kinematics of central production. High mass
systems are suppressed with 1/M$^2$.
\par
Notice the similarity of the  $f_0(1370)$ and the old
$\epsilon (1300)$. The relation between these two phenomena
is not well understood. The reason that the old
$\epsilon (1300)$ was not identified with the $f_0(1370)$
lies just here. The $\epsilon (1300)$ was seen in $\pi\pi$
scattering with a small inelasticity, i.e. small coupling
to 4$\pi$ while the $f_0(1370)$ has small coupling to $\pi\pi$
and a large one to 4$\pi$. This is naturally explained
when the 1300 MeV region interacts via $t$-channel exchange.
Then $\pi\pi$ goes to $\pi\pi$ , $\rm K\bar K$ to $\rm K\bar K$,
Pomeron-Pomeron to $\pi\pi$ by pion
exchange, to $\rho\rho$ via $\rho$ exchange, etc.
\par
\subsubsection{Is there no glueball\,?}
A broad enhancement is observed in the isoscalar S--wave, in $\pi\pi$
and $\rho\rho$ interactions. The enhancement can be interpreted as a scalar
glueball~\cite{Minkowski:1998mf}. 
However, at least a large fraction of it must be
generated by $t$--channel exchanges, by left--hand cuts. 
The identification of a glueball component, beneath the intensity
generated by $t$--channel exchanges, seems to be hopeless, at least
at present. We conclude that there is no `narrow' glueball; a
scalar glueball with a width of $\sim 2$\,GeV may exist but
the experimental methods are not adequate to identify such a
broad object as genuine resonance.
\par
\begin{figure}[h!]
\begin{tabular}{ccc}
\includegraphics[width=0.35\textwidth]{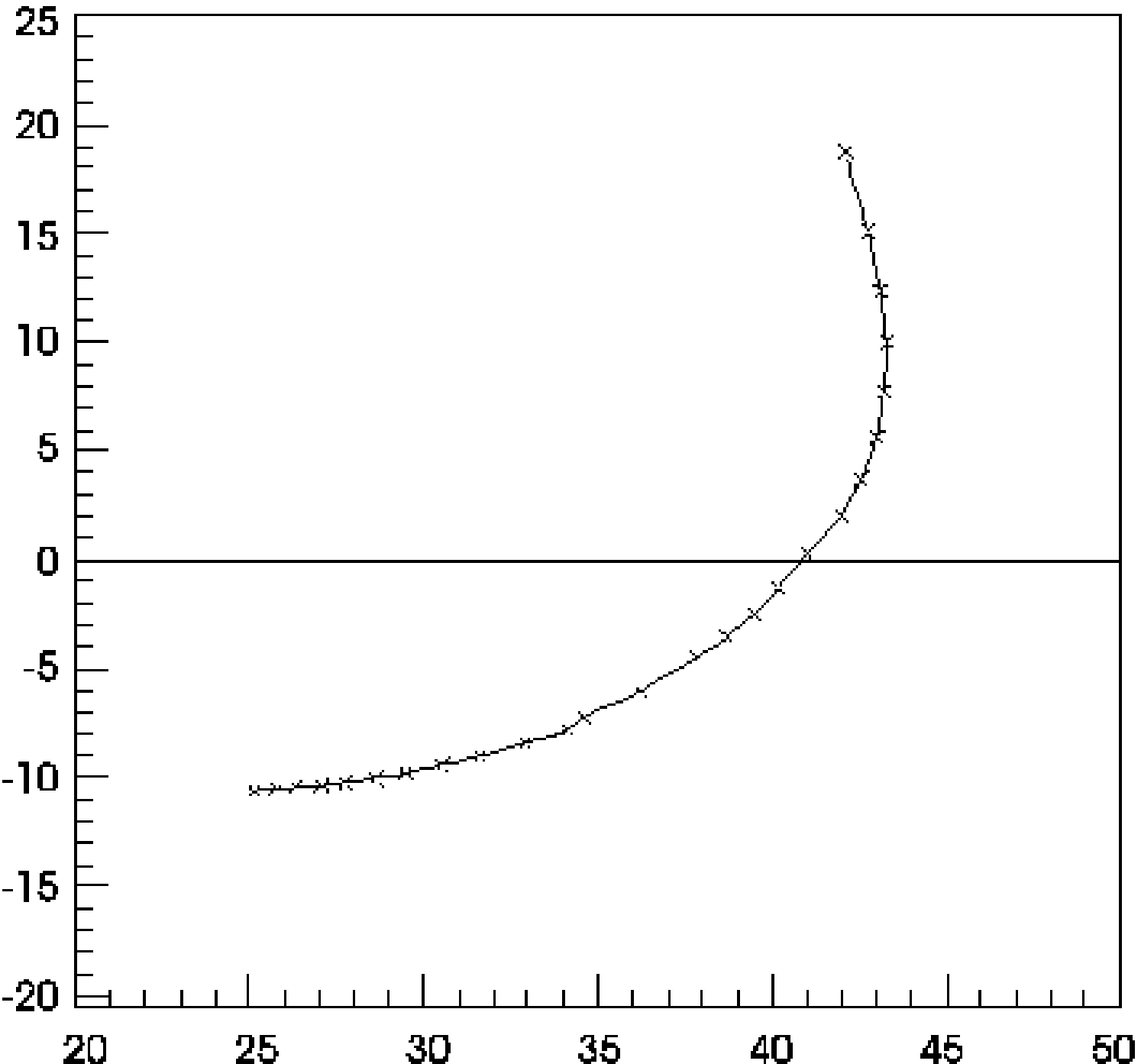}&\quad&
\includegraphics[width=0.35\textwidth]{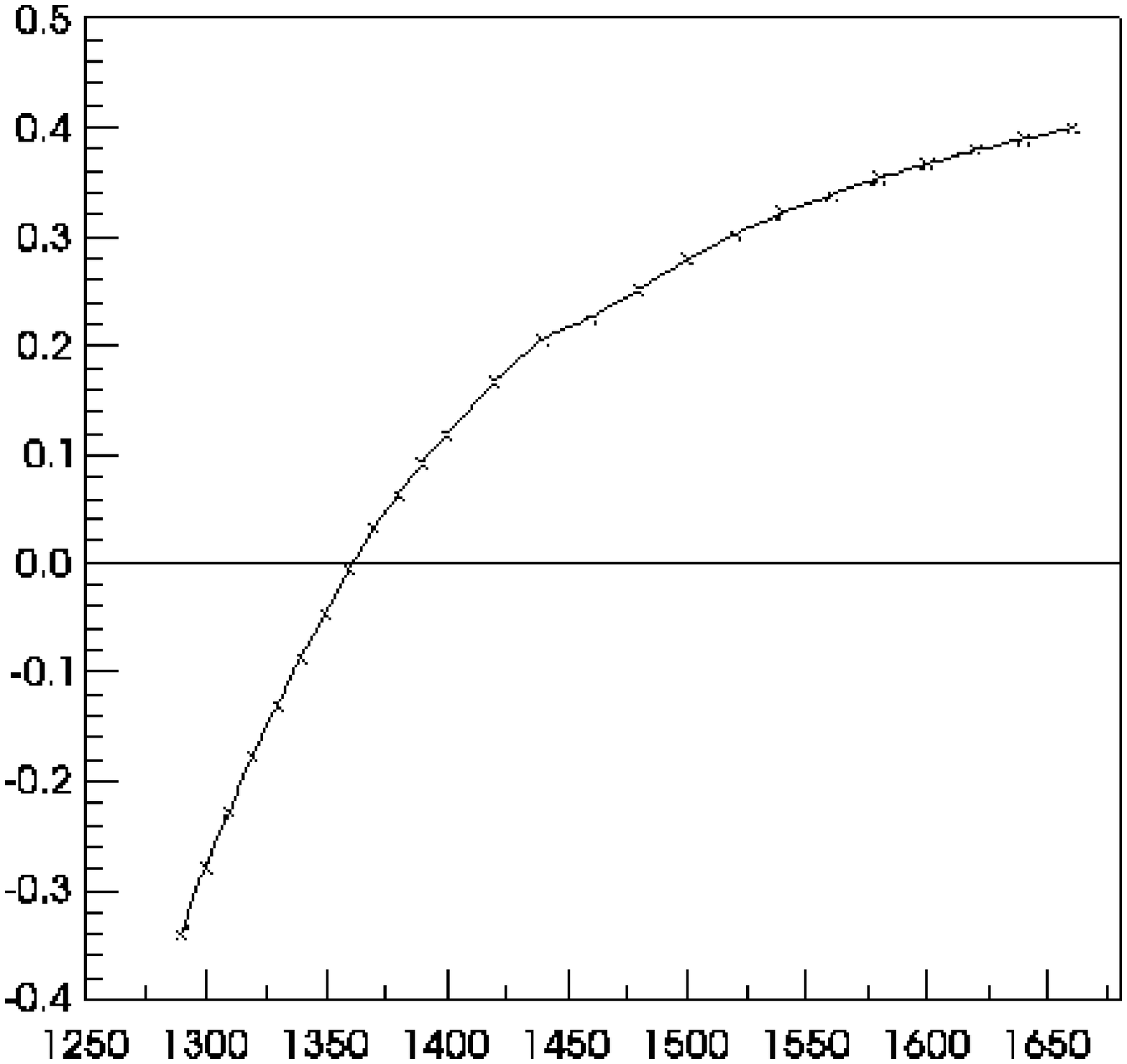}
\end{tabular}
\caption{\label{phase_1370}
Complex amplitude and phase motion of the $\rho\rho$ scalar isoscalar 
isobar in $\rm p\bar p$ annihilation into $4\pi\eta$. In the mass range
from 1300 to 1500\,MeV the phase varies by $<0.6$ 
questioning (but not excluding) the $f_0(1370)$ as genuine resonance.
The $\sigma\sigma$ (not shown) exhibits the same 
behavior~\protect\cite{Reinnarth}.}
\vspace*{-3mm}
\end{figure}

 
Finally, we come back to the question wether  
the $f_0(1370)$ is a genuine resonance. The reaction
$\bar pp\to\eta 2\pi^+ 2\pi^-$ was studied~\cite{Anisovich:jb}
and it was shown that a large fraction of the
final state is reached via the $f_0(1370)\eta$ isobar. Due to
phase space limitations, no influence by the 
 $f_0(980)$ or  $f_0(1500)$ must be expected. Figure~\ref{phase_1370}
shows amplitude and phase of the $\rho\rho$ system (the 
$\sigma\sigma$ system shows the same behavior). Amplitude and phase
are determined by fitting the data with a $f_0(1370)$ of variable
width (scanning the mass). For each mass, the fits returns a best
complex amplitude. The 4$\pi$
decays give the largest contributions to its width; hence there should
be a phase variation by $\pi$. On the contrary, there is not. 
The $f_0(1370)$, the
cornerstone of most interpretations of the $f_0(1500)$ as scalar 
glueball, does not behave like a genuine resonance. On the other
hand the determination of the phase motion is indirect. The least one
can say is that the results in figure~\ref{phase_1370} do not
support the interpretation of the $f_0(1370)$ as a normal resonance.
\par
Assuming that the $f_0(1370)$ is not a genuine $q\bar q$ resonance but
generated by $t$--channel exchange,   
there are left four distinct scalar
isoscalar resonances at 980, 1500, 1710 and 2100\,MeV. The
complicated structure of the resonances with distinct $T$
and $K$-matrix poles may originate from the production and 
dynamics of hadrons. In ``simple'' situations like
$p\bar p$ annihilation in flight into
$\pi\eta\eta$~\cite{E760}
or radiative J/$\psi$ decays into $4\pi$~\cite{bes4p}
only poles at 1500, 1710, and 2100\,MeV show up. These are
the ``true'' quarkonium states. A scalar glueball is not needed to
understand the mass spectrum of scalar mesons.

\subsection{\label{section4.6}Hybrids}
\par
Hybrids, mesons with an intrinsic gluonic excitation, were first
predicted  shortly after the development of the bag model
~\cite{Chodos:1974pn}. At that time, hybrids were thought of as
$q\bar q$ pair in color octet neutralized in color by a
constituent gluon~\cite{Horn:1978rq,Barnes:1982tx}. More
recent autors
expect hybrids as excitations of the gluon fields providing the
binding forces between quark and antiquark, as excitations of the
color flux tube linking quark and antiquark~\cite{Isgur:1985bm}.
The QCD sum rule approach is not as model dependent and  
finds the lowest $J^{PC} = 1^{-+}$ excitation at about 
1400\,MeV~\cite{Balitsky:ps}.
Due to its inclusive approach, sum rules do
not predict if an exotic meson should have a large
``$q\bar q$ + gluon field'' contribution, or if it is dominantly
a multiquark state.

The flux tube can have a non--zero 
orbital--angular--momentum component $\Lambda$ 
along the $q\bar q$ axis, and the following quantum numbers are
now possible:
\begin{eqnarray*}
S=0 \Rightarrow &J^{PC} = 1^{++}, 1^{--}\\
S=1 \Rightarrow &J^{PC} = (0,1,2)^{-+}, (0,1,2)^{+-}
\end{eqnarray*}
The quantum numbers $J^{PC}=1^{-+}, 0^{+-}, 2^{+-}, \cdots$ are of
particular interest; 
they are {\it exotic}. In the quark model $q\bar q$ states 
cannot be formed with these quantum numbers. 
Hybrids are expected at masses around 2 GeV
and higher~\cite{Michael:2003ai}
and to decay into two mesons with one of them
having one unit of orbital angular momentum~\cite{Isgur:vy}.

\subsubsection{The  $\pi_1(1370)$}

Indeed an exotic meson
has been seen to decay into a $p$-wave $\eta\pi$ system.
The quantum numbers in this partial wave are
$I^G(J^{PC})=1^-(1^{-+})$. These are not quantum numbers which
are accessible to the $q\bar q$ system; they are exotic.
\par
A meson with quantum numbers $I^G(J^{PC})=1^-(0^{-+})$
is called a $\pi$ and one with $I^G(J^{PC})=1^-(2^{-+})$
is called $\pi_2$. These latter two mesons are well established
$q\bar q$ mesons. A meson with quantum numbers $I^G(J^{PC})=1^-(1^{-+})$
is called  $\pi_1$. Its mass is added to the name
in the form  $\pi_1(1370)$ to identify
the meson because  there could be (and there are) more than one
resonance in this partial wave.
\par
A meson with exotic quantum numbers like the  $\pi_1(1370)$ cannot
be a regular $q\bar q$ meson. It must have a more complex structure.
It could be a hybrid but it might also be a four-quark
($qq\bar q\bar q$) resonance. The quantum numbers give no hint which
of the two possibilities is realized in nature. Before
we discuss arguments in favor of a  four-quark assignment
let us first have a look at the experimental findings.
\par
\begin{figure}[h!]
\begin{minipage}[c]{0.70\textwidth}
\epsfig{file=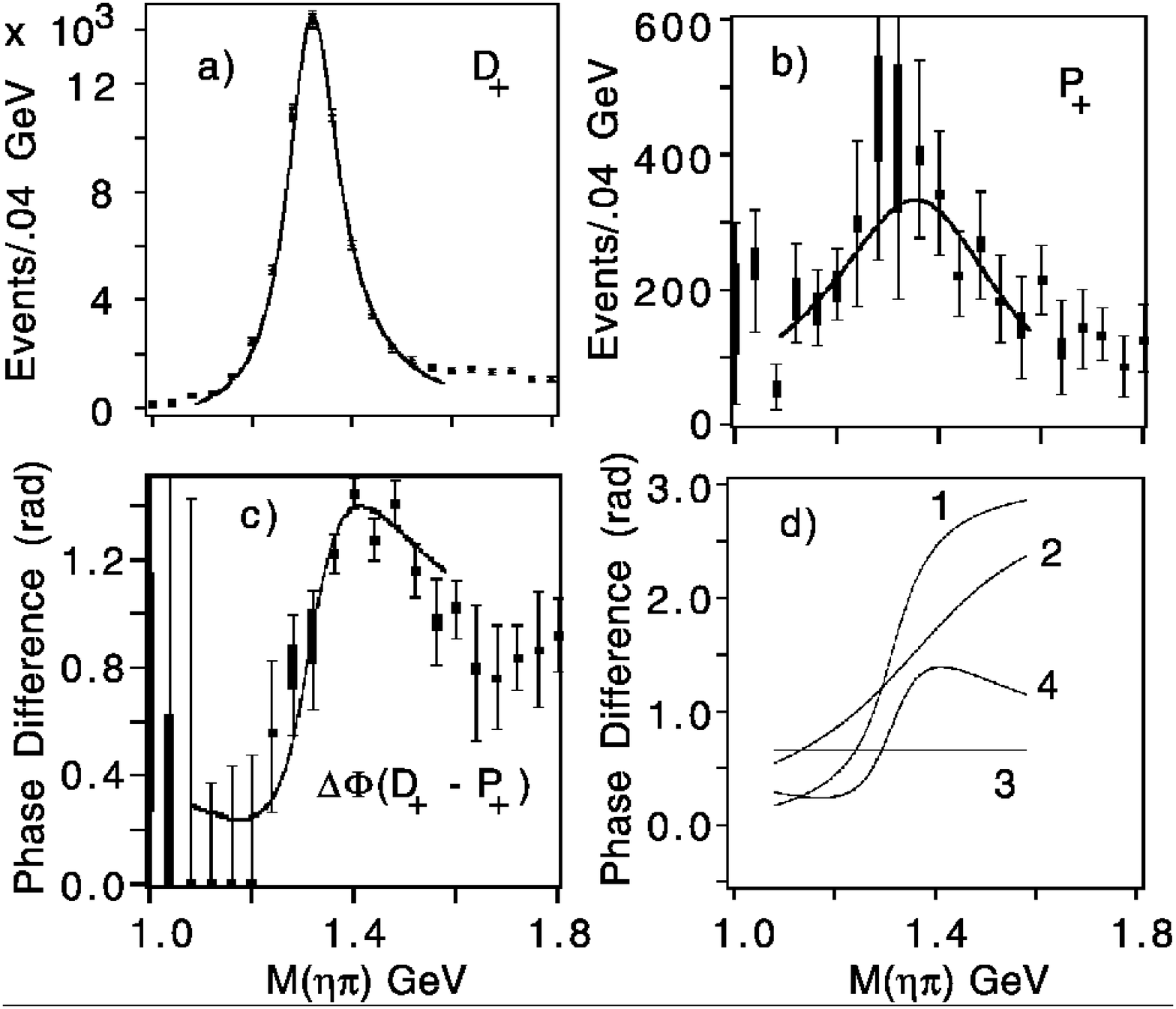,width=0.95\textwidth}
\end{minipage}
\begin{minipage}[c]{0.29\textwidth}
\caption{\label{bnl}
The squared scattering amplitude for the $D^+$ (a)
and $P^+$ (b) waves. The +sign indicates natural parity exchange.
The relative phase between the two waves is shown
in (c). The lines
correspond to the expectation for two Breit-Wigner amplitudes.
In (d) the (fitted) phases for the D- (1) and P-wave (2) are
shown. The $P$- and $D$-production phases are free parameters
in the fits; their difference is plotted as line 3, and -
with a different scale - as 4~\protect\cite{Chung:1999we}.}
\end{minipage}
\vspace*{-2mm}
\end{figure}

At BNL, the reaction
$$
\pim\ p \ra\ \pim\etg\ p
$$
was studied at 18 GeV/c~\cite{Thompson:1997bs,Chung:1999we}.
The data shows a large asymmetry
in the angular distribution evidencing interference
between even and odd angular momentum contributions.
Figure~\ref{bnl} shows data and the results of the partial wave
analysis. In a scattering process, the $\pi\eta$ system can be produced
\begin{figure}[thb]
\begin{tabular}{cc}
\hspace*{-5mm}
\includegraphics[width=0.48\textwidth]{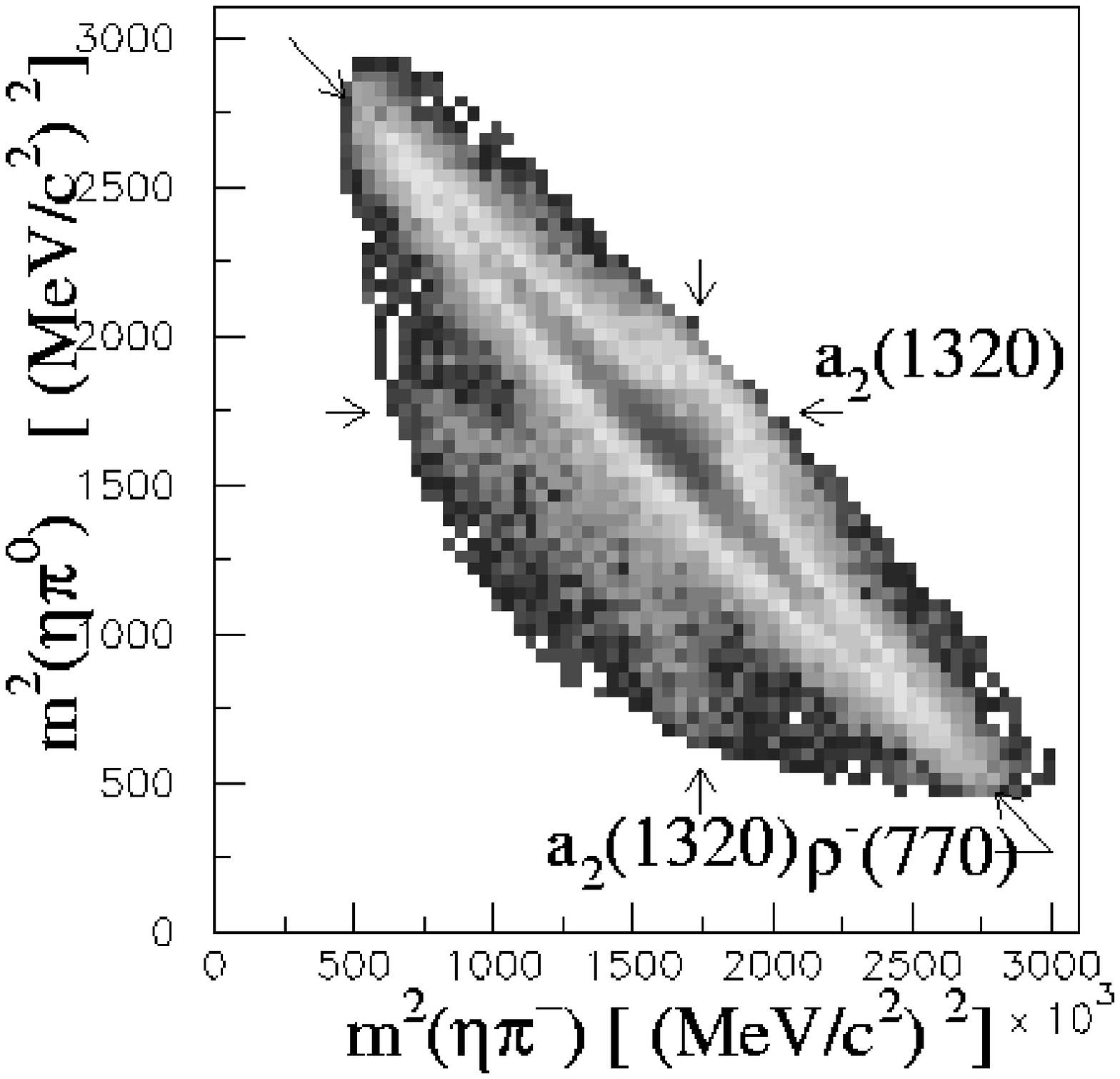}&
\includegraphics[width=0.48\textwidth]{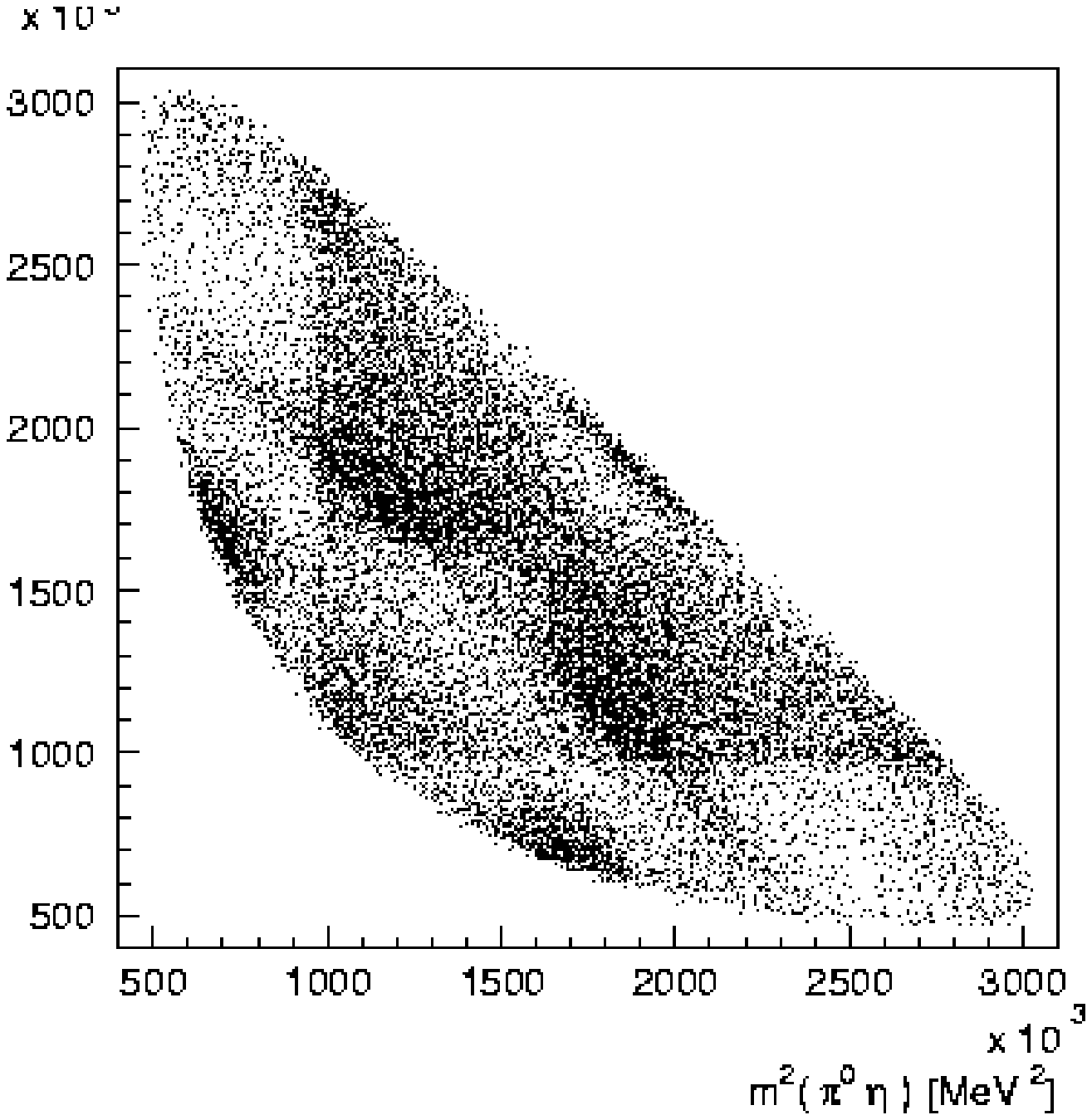}
\end{tabular}
\vspace*{-3mm}
\caption{\label{wd}
Dalitz plot for the reaction $\bar pn\to$
\pim\piz\etg\ for antiproton annihilation at rest
in liquid D$_2$. Annihilation on quasi-free neutrons is
enforced by a cut in the proton momentum
($p_{proton}\leq 100$\,MeV/c. The data requires
contributions from the
$I^G(J^{PC})=1^-(1^{-+})$ partial wave in the
$\eta\pi$ system~\protect\cite{Abele:1998gn,Abele:1999tf}.}
\vspace*{-5mm}
\end{figure}
in different partial waves ($S, P, D$ waves). In the $t$-channel
quantum numbers are exchanged corresponding to natural
($0^{++},1^{--},2^{++}$) or unnatural ($0^{-+},1^{+-},2^{-+}$)
parity. The naturality is a good quantum number for a given partial wave
and is added as suffix, + for natural and  - for unnatural exchange.
\par
The data is fully compatible with the existence
of a resonance in the
$I^G(J^{PC})=1^-(1^{-+})$ partial wave produced
via natural parity exchange.
Mass and width are fit to values given in Table \ref{exotics}.
Since the spin in the
final state is one, the exchanged particle cannot
have scalar quantum numbers. The resonance is not observed
in the charge exchange reaction~\cite{Dzierba:2003fw}
(with \piz\etg\ in the final state), hence the exchanged particle
cannot be the \rh . The particle exchanged
is the $f_2(1270)$ (or the tensor part of a Pomeron).
The authors of ref.~\cite{Dzierba:2003fw} deduce from the
absence of the $\pi_1(1370)$ in  \piz\etg\ that the
 $\pi_1(1370)$ might not exist, but what
follows is only that it is not produced
by $\rho$ exchange. This is further discussed
in ref.~\cite{Chung:2002fz,Chung:2003qy}.  
\par
The VES collaboration used $\pi^-Be$ interactions
at momenta of about 25\,GeV/c. They
observed a very similar amplitude 
and phase, in their data from 1993~\cite{Beladidze:km} 
and in their more recent data~\cite{Dorofeev:2001xu}. 
They cautiously pointed out that background amplitudes
can be constructed leading to an acceptable fit
to the data in figure~\ref{bnl}. Thus they do not see
the mandatory need to introduce a new resonance in an
exotic partial wave. 
\par
The Crystal Barrel Collaboration studied the
reaction $\bar pn\to\pi^-\pi^0\eta$ . Figure~\ref{wd} shows
the \pim\piz\etg\ Dalitz plot. Clearly visible
are $\rh^-\eta$ , $a_2(1320)\pi$ with $a_2(1320)
\to\eta\pi$ (in two charge modes) as intermediate states.
A fit with only conventional mesons gives
a bad description. The difference between data and
predicted Dalitz plot shows a pattern very
similar to the contributions expected from the interference
of the $\pi_1(1370)$ with the amplitudes for production of conventional
mesons~\cite{Abele:1998gn}.
Introducing the exotic partial wave, the fit optimizes for
values listed in Table \ref{exotics}. Selection rules
(and the PWA) attribute the production of the exotic partial wave
to the \pbp\ ($^3S_1$) initial atomic state.
\par
A similar analysis on the reaction \pbp\ra 2\piz\etg\ was
carried out. In this case the $\pi_1(1370)$ can only be
produced from the $^1S_0$ state; its production is considerably
reduced in this
situation. The small contribution could only be unraveled
when data taken by stopping antiprotons in liquid and gaseous
H$_2$ was analyzed.
In these two data sets the fraction of annihilation contributions from
atomic S and P states is different (their ratio is known
from cascade models). Thus S and P wave contributions are
constrained. It is only under these conditions that
positive evidence for the small contribution from the exotic
partial wave could be found~\cite{Abele:1999tf}, see
however ref~\cite{AS}.
\par
In a partial wave analysis of $\bar{\textrm{p}}\textrm{n}\to
\pi^-3\pi^0$ a $\pi_1(1370)$ is seen at slightly higher mass, 
but with a production characteristic distinctively different
from the $\pi_1(1370)$ seen in $\pi\eta$~\cite{WD}. Hence
it is listed as separate exotic meson in Table~\ref{exotics}.
\par
{\small
\begin{table}[htb]
\vspace*{-5mm}
\caption{\label{exotics}
Evidence for  J$^{\textrm{PC}}$ = $1^{-+}$ exotics. States
supposed to be distinct are separated by double--lines.
The six entries in the 1600 to 1700\,MeV range might be
one or two states.
}
\vspace*{-5mm}
\begin{center}
\begin{tabular}[t]{c l l c c}
\hline
\hline
Experiment & mass (MeV/c$^2$) & width (MeV/c$^2$) &decay mode& reaction\\
\hline
BNL~\cite{Thompson:1997bs}
& 1370 $\pm$ 16 $^{+\,\,\,\,\,\,50}_{-\,\,\,\,\,\,30}$& 385 $\pm$ 40 $^{+\,\,\,65}_{-105}$& $\eta \pi$&$\pi ^-
\textrm{p}\to \eta \pi ^- \textrm{p} $\\
BNL~\cite{Chung:1999we}
& 1359 $\,^{+\,\,\,\,\,16}_{-\,\,\,\,\,14}$ $\,^{+\,\,\,\,\,10}_{-\,\,\,\,\,24}$ & 314 $\,^{+31}_{-29}$ $\,\,\,\,^{+\,\,9}_{-66}$ & $\eta \pi$&$\pi ^-
\textrm{p}\to \eta \pi ^- \textrm{p} $\\
CBar~\cite{Abele:1998gn}& 1400 $\pm$ 20 $\pm$ 20& 310 $\pm$ 50
$^{+50}_{-30}$ &$\eta \pi$&
$\bar{\textrm{p}}\textrm{n}\to \pi ^- \pi ^0\eta$\\
CBar~\cite{Abele:1999tf}& 1360 $\pm$ 25 & 220 $\pm$ 90 &$\eta
\pi$&$\bar{\textrm{p}}\textrm{p}\to \pi ^0\pi ^0\eta$ \\
\hline
\hline
CBar~\cite{WD}& $\sim$1440 & $\sim$400 & 
$\rho\pi$&$\bar{\textrm{p}}\textrm{n}\to \pi^-3\pi^0$\\
\hline
\hline
BNL~\cite{Adams:1998ff}& 1593 $\pm$ 8 $^{+29}_{-47}$ & 168 $\pm$ 20
$^{+150}_{-\,\,\,12}$ & $\rho \pi$&$\pi ^- \textrm{p}\to \pi
^+\pi ^-\pi ^-\textrm{p}$\\
BNL~\cite{Ivanov:2001rv} & 1596 $\pm$ 8 & 387 $\pm$ 23 &$\eta' \pi$&$\pi ^-
\textrm{p}\to
\pi^-\eta'\textrm{p}$\\
VES~\cite{Khokhlov:2000tk}& 1610 $\pm$ 20 & 290 $\pm$ 30 & $\rho \pi , \eta '\pi$&$\pi^-$N$\to\pi^-\eta'$N\\
BNL~\cite{Kuhn:2004en}& 1709 $\pm$24$\pm$41 & 403$\pm$80 $\pm$115 & 
$f_1(1285)\pi$&$\pi ^- \textrm{p}\to \eta\pi^+\pi^-\pi^-\textrm{p}$\\
BNL~\cite{Lu}& 1664 $\pm$8$\pm$4 & 185 $\pm$ 25 $\pm$12 & 
$b_1(1235)\pi$&$\pi ^- \textrm{p}\to \omega\pi^0\pi^-\textrm{p}$\\
CBar~\cite{Baker:jh}& 1590$\pm$50 & 280 $\pm$75 & 
$b_1(1235)\pi$&$\bar{\textrm{p}}\textrm{p}\to \pi^+\pi^-\pi^0\omega$\\
\hline
\hline
BNL~\cite{Kuhn:2004en} & $\sim$2003$\pm$88$\pm$148 &306$\pm$132$\pm$121  & 
$f_1(1285)\pi$&$\pi ^- \textrm{p}\to \eta\pi^+\pi^-\pi^-\textrm{p}$\\
BNL~\cite{Lu}& 2000 $\pm$20$\pm$10 & 230$\pm$32$\pm$15 & 
$\omega\pi^0\pi^-$&$\pi ^- \textrm{p}\to \omega\pi^0\pi^-\textrm{p}$\\
\hline
\hline
\end{tabular}
\end{center}
\end{table}
}
\subsubsection{The $\pi_1(1625)$}
The $\pi_1(1370)$ 
is not the only resonance observed in this partial wave.
At BNL, the $\eta^{\prime}\pi$ is also observed to exhibit a
resonant behavior~\cite{Ivanov:2001rv} at about 1600 MeV. A partial
wave analysis of the $\rho\pi$ system~\cite{Adams:1998ff} reveals
an exotic meson  with mass and width given in Table \ref{exotics}.
\par
At Protvino, the $\pi\eta^{\prime}$ , $\rho\pi$ and the $b_1(1235)\pi$
systems  were studied in a 40\,GeV/c \pim\ beam. In all three systems
a resonant contribution in the exotic  $I^G(J^{PC})=1^-(1^{-+})$
partial wave was found. A combined fit found a mass of
$\sim$\,1600 MeV and a width of $\sim$\,300 MeV~\cite{Khokhlov:2000tk}.
Possibly these are three different 
decay modes of one resonance. 
\par
\begin{figure}[h!]
\includegraphics[width=\textwidth,]{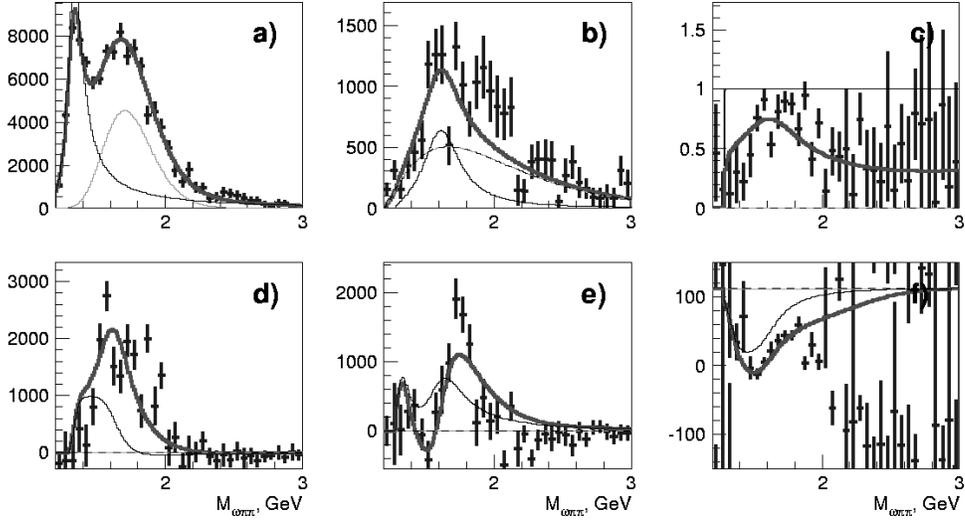}
\caption{\label{fig:9}\qquad\qquad
a) $2^+(\omega\rho)$ intensity.
b) $1^-(b_1\pi)$ intensity.
c) Coherence parameter. 
The real d) and imaginary 
e) parts of their non-diagonal $\rho$-matrix element.
f) The $1^-$ phase relative to $2^+$. 
The smooth curves are fit results.}
\end{figure}

The VES collaboration has carried out a partial wave analysis of 
data on $\omega\pi\pi$ production. The partial waves 
$\omega\rho$, $b_1(1235)\pi$, $\rho_3(1690)\pi$, and $\rho_1(1450)\pi$
are included in the wave set. 
The $2^+(\omega\rho)$ wave, shown in figure~\ref{fig:9}a
was found to be a dominant wave  with a clear
$a_2(1320)$ peak and a broad bump at 1.7\,GeV. 
A significant $1^-(b_1\pi)$ wave shown in figure~\ref{fig:9}b
is observed with a broad bump at 1.6\,GeV.
In figure~\ref{fig:9}f an $80^{\circ}$ phase rise of the $1^-$-wave 
phase relative to the $2^+$-wave phase is observed,
which may be attributed to a $1^-$ resonance.   
\par
A fit describes the interference pattern satisfactorily
as seen in figure~\ref{fig:9}. 
The data is consistent with the resonant description of the 
$1^-1^+S1(b_1\pi)$ with the mass 1.6\,GeV and width 0.33\,GeV.
Other hypotheses, including a partially 
coherent background,  
did not result in better fits.

\subsubsection{Higher-mass exotics}
\begin{figure}[h!]
\begin{minipage}[c]{0.76\textwidth}
\includegraphics[height=9.5cm,angle=-90]{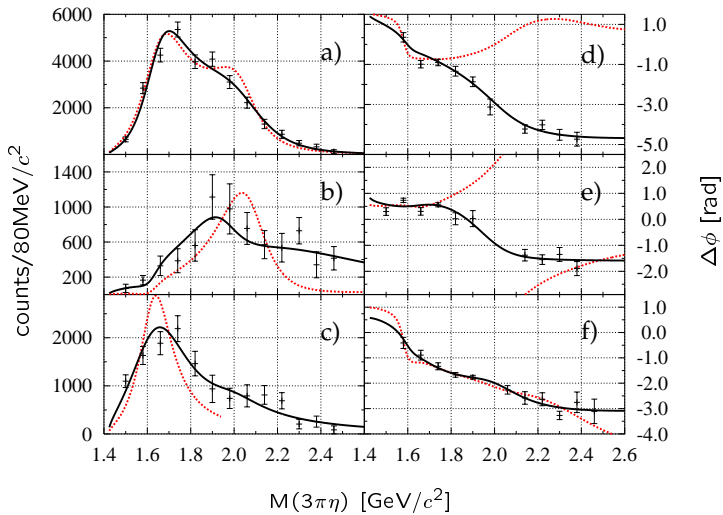}
\end{minipage}
\begin{minipage}[c]{0.22\textwidth}
\caption{\label{fig:PWAresults}
PWA results: $f_{1}(1285)\pi^{-}$ intensity distributions \quad
(a) $1^{++}0^{+}f_{1}\pi^{-}P$,
(b) $2^{-+}0^{+}f_{1}\pi^{-}D$, 
(c) $1^{-+}1^{+}f_{1}\pi^{-}S$ 
and phase difference distributions 
(d) $\phi(1^{-+})-\phi(2^{-+})$, 
(e) $\phi(1^{-+})-\phi(1^{++})$, 
(f) $\phi(1^{++})-\phi(2^{-+})$. 
The results from a least squares fit are
overlaid as the solid line (two poles in the $1^{-+}f_{1}\pi$ wave)
and the dashed line (one pole in the $1^{-+}f_{1}\pi$ wave).
From~\protect\cite{Lu}.} 
\end{minipage}
\vspace*{-3mm}
\end{figure}

Hybrid mesons are expected to have masses of about 2\,GeV
and to decay into a P--wave and a S--wave meson. Due to
its narrow width, hybrid decays into $\pi^0f_1(1285)$ 
are particularly well suited for a search.
The two mesons are produced in part with
zero orbital angular momentum between them, and this leads to
the exotic $I^G(J^{PC})=1^-(1^{-+})$ partial wave. 
The amplitude and phase motion in this partial wave
(and two other partial waves) are shown in figure~\ref{fig:PWAresults}.
A fit with one pole in the $\pi^0f_1(1285)$ $1^{-+}$
partial wave is obviously not consistent with the data (a fit
with no pole is not shown); two poles are required. The results
for mass and width are given in Table~\ref{exotics}. 

The exotic $1^{-+} f_{1}\pi^{-}$ contribution is only observed in
positive reflectivity waves, indicating that the process is mediated
by exchange of natural parity Reggeons, most likely $\rho(770)$ or
$f_{2}(1270)$/Pomeron.  
An unpublished thesis~\cite{Lu} reports observation of
two  $1^{-+}$ exotic mesons decaying into $b_1(1235)\pi$. 
The results are included in Table~\ref{exotics}. 
\par
Why are so many exotic resonances in this one partial wave\,? The number 
of resonances seems to be three or four: $\pi_1(1370)$, $\pi_1(1440)$,
$\pi_1(1625)$, and $\pi_1(2000)$. 
The large number of states in one partial and
in such a narrow mass interval is certainly surprising.
All have exotic quantum numbers, so they cannot
possibly be $q\bar q$ states.
We now discuss whether they are likely four-quark states or the
searched--for hybrid mesons.

\subsubsection{The Fock-space expansion}

The majority of established mesons can be interpreted as $q\bar q$
bound states. This can be an approximation only; the $\rho$-meson
e.g. with its large coupling to $\pi\pi$ must have a four-quark
component and could as well have contributions from
gluonic excitations. The Fock space of the $\rho$
must be more complicated than just $q\bar q$. We may
write
\begin{equation}
\rho\ = \alpha q\bar q +  \beta_1b\bar qq\bar qq + ... +
\gamma_1q\bar qg + ...
\label{fock}
\end{equation}
where we have used $q\bar qg$ as short--hand for a gluonic excitation.
The orthogonal states may be shifted into the $\pi\pi$ continuum.
Now one might ask, ``are the higher-order terms important and what is
the relative importance of the $\beta$ and $\gamma$ series\,?''
\par
Possibly this question can be answered by truncating the
$\alpha$-term. Exotic mesons do not contain a $q\bar q$ component
and they are rare. Naively we may expect the production of
exotics in hadronic reactions to be suppressed by a factor 10 when
one of the coefficients,
$\alpha_1$ or $\beta_1$, is of the order 0.3.
We thus expect additional states
having exotic quantum numbers, quantum numbers which are not accessible
to the $q\bar q$ system. Their production rate should be
suppressed compared to those for regular $q\bar q$ mesons. In
non-exotic waves the four-quark and hybrid configurations are
likely subsumed into the Fock expansion. If we can decide what
kind of exotic mesons we observe, four-quarks, hybrids or both, we
can say what the most important contributions in (\ref{fock}) are.
\par
\subsubsection{SU(3) relations}
\begin{figure}[h]
\begin{minipage}[c]{0.70\textwidth}
\epsfig{file=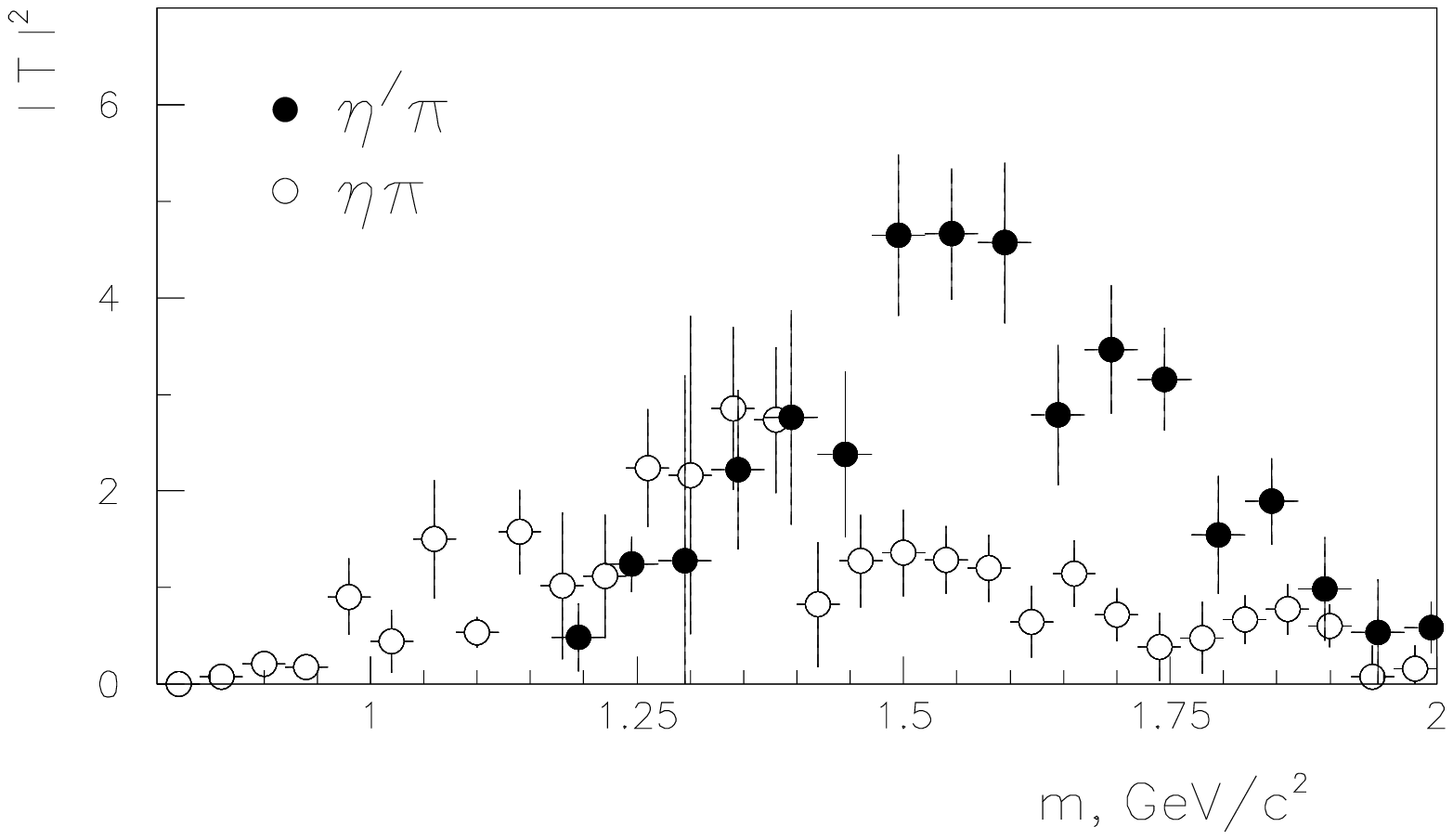,width=\textwidth}
\end{minipage}
\begin{minipage}[c]{0.29\textwidth}
\caption{\label{etapietp}
The squared scattering amplitude in the
$I^G(J^{PC})=1^-(1^{-+})$ partial wave for the
$\pi\eta$ and $\pi\eta^{\prime}$
systems~\protect\cite{Khokhlov:2000tk}.
}
\end{minipage}
\vspace*{-5mm}
  \end{figure}

The  $\pi_1(1370)$ decays strongly into $\pi\eta$ and the
$\pi_1(1625)$ into $\pi\eta^{\prime}$; decays of the  $\pi_1(1370)$ into
$\pi\eta^{\prime}$ and of the $\pi_1(1625)$ into $\pi\eta$ were not observed
or reported. Figure~\ref{etapietp} shows the exotic wave
for the $\pi\eta$ and $\pi\eta^{\prime}$ systems as a function of their mass;
the $\pi\eta$ intensity is concentrated around 1400 MeV, the
$\pi\eta^{\prime}$ intensity at 1600 MeV. A resonance decaying
into $\pi\eta^{\prime}$  should also decay into $\pi\eta$ , and a $\pi\eta$
resonance should also have a sizable coupling to $\pi\eta^{\prime}$.
Why is there such  a strange decay pattern\,?
\par
We first consider the limit of flavor symmetry. The $\eta$ is
supposed to belong to the pseudoscalar octet and the
\etp\ is considered to be a pure singlet SU(3) state. Mixing is neglected.
The $\pi_1$ states, having isospin one, cannot be isoscalar
states. Now I claim that a meson belonging to an octet
with exotic quantum numbers  $J^{PC}=1^{-+}$
cannot decay into two octet pseudoscalar mesons.
\par
The argument goes as follows: decays of particles belonging to an octet
of states into two other octet mesons, decays
of the type
$8 \ra 8 \otimes 8$, may have symmetric or antisymmetric
couplings. The two octets can be combined using
symmetric structure constants $d_{ijk}$ or antisymmetric
structure constants $f_{ijk}$. The decay $\pi_1(1370)$ into
two pseudoscalar mesons is governed by the symmetric
couplings. SU(3) demands the decay amplitude
for $\pi_1$ decays into two pseudoscalar
mesons not to change sign when the two mesons are
exchanged. The orbital angular momentum $l=1$
between the two mesons requires the opposite. The
two mesons must be in a state $\pi\eta -\eta\pi$.
Both requirements cannot be fulfilled at the same time: the
decay of a $\pi_1$ which belongs to an SU(3) octet
into two octet pseudoscalar mesons is 
forbidden~\cite{Chung:2002fz}.
\par
There are immediate consequences. Let us begin
with the $\pi_1(1625)$ and assume that it belongs to
an octet of states. Then it must decay
into $\pi\eta^{\prime}$  while the decay into $\pi\eta$ is forbidden.
This is precisely what we see. But what happens
in case of the $\pi_1(1370)$\,? It does decay into
$\pi\eta$, why\,? As we have seen, it cannot belong to
an SU(3) octet; it must belong to a SU(3) decuplet $10$
or $\bar{10}$. As member of a decuplet, it cannot
decay into $\pi\eta^{\prime}$, into an octet and a singlet meson,
and it cannot possibly be a hybrid since gluonic excitations do
not contribute to the flavor. Mesonic hybrids can only
be SU(3) singlets or octets.
The strange phenomenon that the  $\pi_1(1625)$ does
not decay into $\pi\eta$ thus provides the clue for the
interpretation of the  $\pi_1(1370)$ as a four--quark (decuplet) state.
\par
The above arguments hold in the limit of flavor symmetry.
Due to  $\eta -\eta^{\prime}$ mixing, the exotic $\pi_1(1625)$
could decay into $\eta\pi$ via the small singlet component of the
$\eta$. A small coupling of the  $\pi_1(1370)$
to $\eta^{\prime}\pi$ is also possible.
\par
\subsubsection{Four-quark states in SU(3)}
In the limit of SU(3) symmetry, the $\pi_1(1370)$ with its large
$\pi\eta$ decay rate must belong to a decuplet and must hence be a
four-quark state. The $\pi_1(1625)$ must belong to an octet of
states and could thus be a hybrid. There is no rigid argument
against this conjecture. However, the mass difference between the
$\pi_1(1370)$ and $\pi_1(1625)$ is typical for SU(3) multiplet
splitting. It has the same order of magnitude as the
octet-decuplet splitting in baryon spectroscopy, only the mass
ordering is reversed.
\par
Let us discuss how we can construct a decuplet of states from
two quarks and two antiquarks. Two quarks in flavor 3 combine to
$3 \otimes 3 = \bar 3 + 6$, two antiquarks to $3 + \bar 6$. Now we
construct
$$
(\bar 3 + 6) \otimes (3 + \bar 6) = \bar 3 \otimes 3 +
\bar 3 \otimes \bar 6 +  6 \otimes 3  + 6 \otimes \bar 6
$$
$$
= 1 + 8 + 8 + 10 + 8 + \bar{10} + 1 + 8 + 27
$$
We can construct $10+\bar{10}$ and $10-\bar{10}$ multiplets and four
different octets. A large
number of different states with the same quantum numbers should be expected
from four-quark states.
\subsection{\label{section4.7}Exotic(s) summary}
Several exotic mesons are observed, all in one partial wave
$I^G(J^{PC})=1^-(1^{-+})$. The
decay pattern of the two resonances at 1400 and 1600 MeV
suggests that at least the $\pi_1(1370)$ should be a four-quark
resonance belonging to a decuplet of states. Then further
resonances with quantum numbers $I^G(J^{PC})=1^-(1^{-+})$
are to be expected and may have been found.
\par
Even though there is no argument against the hypothesis
that one of the other observed resonances could be a hybrid,
there is at present no experimental support for
this hypothesis. Once Pandora's box of four-quark
states has been opened, it is very hard to establish experimentally
that mesons with gluonic excitations be found
in an experiment.

Finally, do glueballs exist\,? This is still an unresolved issue\,!
We do find natural explanations for the spectrum without
request for the presence of a glueball while all
interpretations of the spectrum of scalar resonances (which
necessitates a scalar glueball) contradict some experimental results,
or make questionable assumptions. Of course it is very difficult 
or even impossible to
prove that glueballs and hybrids do not exist. There is, however, 
no positive
evidence that they are needed to understand low--energy hadron physics.

%% file: Chapter5_proc.tex
\section{\label{section5}Baryon spectroscopy}
Baryon resonances have been studied since the early
50's and a wealth of information is available. Still
there are many open questions. 
The Particle Data Group lists more than 100
baryon resonances; 85 of them have known spin parity
but only $\sim 50$ of these are well established,
with 3$^*$ and 4$^*$ ratings. The breakdown of
baryon resonances according to their SU(3) classification is
given in table~\ref{pdg_baryons}.

\begin{table}[h!]
\begin{minipage}[t]{0.40\textwidth}
\caption{Status of baryon resonances according to
the The Particle Data Group.}
\end{minipage}
\begin{minipage}[b]{0.58\textwidth}
\renewcommand{\arraystretch}{1.4}
\begin{tabular}{|ccccccc|}
\hline
Octet   & \ N \ &          & \ $\Sigma$ \ & \ $\Lambda$ \ & \ $\Xi$ \ & \\
Decuplet&       & $\Delta$ &  $\Sigma$    &               &$\Xi$ & $\Omega$\\
Singlet &       &             &                 & $\Lambda$ &         & \\
\hline
****    & 11    &      7      &      6      &       9      &   2     &     1      \\
***  & 3  &    3      &      4      &       5      &   4     &     1      \\
**   & 6  &    6      &      8      &       1      &   2     &     2      \\
*    & 2  &    6      &      8      &       3      &   3     &     0      \\
\hline
No J & -  &    -      &      5      &       -      &   8     &     4      \\
Total& 22 &   22      &     26      &      18      &  11     &     4      \\
\hline
\end{tabular}
\renewcommand{\arraystretch}{1.0}
\end{minipage}
\label{pdg_baryons}
\end{table}
\subsection{\label{section5.1}N$^*$ and $\Delta^*$ resonances}
\subsubsection{Spin--orbit forces and the multiplet structure}
A severe problem of quark models using one--gluon exchange as
residual interaction is posed by the absence of spin--orbit
interactions. (The residual interaction was defined by the
additional interaction between constituent quarks once confinement
is taken into account by a linear potential and chiral symmetry
breaking by giving quarks an effective mass.) If the $\Delta$--N
mass difference is assigned to the magnetic hyperfine interaction,
then large spin--orbit splittings are expected, which is 
in contrast to
experimental findings. The conjecture that the Thomas precession
may counterbalance spin--orbit forces is an excuse; model
calculations do not support the idea~\cite{metschpc}. In this
section we show that the systematics of baryon masses suggest that
the residual interactions are induced by instantons.
\par
The smallness of spin--orbit forces provides a distinctive benefit
when the baryon resonances are to be grouped into supermultiplets.
In absence of strong spin--orbit splitting, there are multiplets
of approximate equal masses -- four resonances with
$J=L\pm 3/2,L\pm 1/2$ and two resonances with  $J=L\pm1/2$. Often
not all of them are experimentally established, so some
interpretation of the data is needed. An assignment of all known
N$^*$ and $\Delta^*$ states is suggested in Table~\ref{next}.
\par
\begin{table}[h!]
\bc
\renewcommand{\arraystretch}{1.3}
\hspace*{-2mm}\begin{tabular}{|cc|cccc|c|}
\hline
\hspace*{-1mm}56\hspace*{-2mm}&S=1/2;L=0;N=0&&\hspace*{-4mm}N$_{1/2^+}$(939)
&&&939 MeV  \\
&S=3/2;L=0;N=0&&&\hspace*{-5mm}$\Delta_{3/2^+}$(1232)&& 1232 MeV \\
\hline
\hline
\hspace*{-1mm}70\hspace*{-2mm}&S=1/2;L=1;N=0&&\hspace*{-4mm}
N$_{1/2^-}(1535)$      &\hspace*{-5mm} N$_{3/2^-}(1520)$      &
&1530 MeV  \\
  &S=3/2;L=1;N=0&&\hspace*{-4mm} N$_{1/2^-}$(1650)
&\hspace*{-5mm}N$_{3/2^-}(1700)$
&\hspace*{-5mm}N$_{5/2^-}(1675)$&1631 MeV\\
  &S=1/2;L=1;N=0&&\hspace*{-4mm} $\Delta_{1/2^-}(1620)$
&\hspace*{-5mm} $\Delta_{3/2^-}(1700)$ &                 &1631 MeV \\
\hline
\hspace*{-1mm}70\hspace*{-2mm}&S=1/2;L=1;N=2&&\hspace*{-4mm}
N$_{1/2^-}(2090)$      &\hspace*{-5mm} N$_{3/2^-}(2080)$      &
&2151 MeV  \\
  &S=3/2;L=1;N=2&&\hspace*{-4mm} N$_{1/2^-}$
&\hspace*{-5mm} N$_{3/2^-}$            &\hspace*{-5mm}N$_{5/2^-}$
&2223 MeV\\
  &S=1/2;L=1;N=2&&\hspace*{-4mm} $\Delta_{1/2^-}(2150)$
&\hspace*{-5mm} $\Delta_{3/2^-}$       &                 &2223 MeV \\
\hline
\hspace*{-1mm}56\hspace*{-2mm}&S=1/2;L=1;N=1&&\hspace*{-4mm}
N$_{1/2^-}$            &\hspace*{-5mm} N$_{3/2^-}$            &
&1779 MeV  \\
  &S=3/2;L=1;N=1&&\hspace*{-4mm}
$\Delta_{1/2^-}(1900)$&\hspace*{-5mm}$\Delta_{3/2^-}(1940)$&
\hspace*{-5mm}$\Delta_{5/2^-}(1930)$&1950 MeV \\
\hline
\hline
\hspace*{-1mm}56\hspace*{-2mm}&S=1/2;L=2;N=0&&\hspace*{-4mm}
N$_{3/2^+}(1720)$      &\hspace*{-5mm}N$_{5/2^+}(1620)$       &
&1779 MeV  \\
  &S=3/2;L=2;N=0&\hspace*{-1mm}$\Delta_{1/2^+}(1910)$
&\hspace*{-5mm}$\Delta_{3/2^+}(1920)$&\hspace*{-5mm}
$\Delta_{5/2^+}(1905)$&\hspace*{-5mm}$\Delta_{7/2^+}(1950)$&1950 MeV \\
\hline
\hspace*{-1mm}70\hspace*{-2mm}&S=1/2;L=2;N=0&&\hspace*{-4mm}
N$_{3/2^+}$            &\hspace*{-5mm}N$_{5/2^+}$             &
&1866 MeV  \\
  &S=3/2;L=2;N=0&\hspace*{-4mm}N$_{1/2^+}$&\hspace*{-5mm} N$_{3/2^+}(1900)$
&\hspace*{-5mm}N$_{5/2^+}(2000)$&\hspace*{-5mm} N$_{7/2^+}(1990)$&1950 MeV\\
  &S=1/2;L=2;N=0&&\hspace*{-4mm} $\Delta_{3/2^+}$ &
  \hspace*{-5mm} $\Delta_{5/2^+}$      &                 &1950 MeV \\
\hline
\hline
\hspace*{-1mm}70\hspace*{-2mm}&S=1/2;L=3;N=0&&\hspace*{-4mm}
N$_{5/2^-}$             &\hspace*{-5mm}N$_{7/2^-}$ &&2151 MeV  \\
  &S=3/2;L=3;N=0&\hspace*{-4mm}N$_{3/2^-}$&\hspace*{-5mm}
  N$_{5/2^-}(2200)$& \hspace*{-5mm}N$_{7/2^-}(2190)$  & \hspace*{-5mm}
  N$_{9/2^-}(2250)$&2223 MeV  \\
  &S=1/2;L=3;N=0&&\hspace*{-4mm}$\Delta_{5/2^-}$
  &\hspace*{-5mm}$\Delta_{7/2^-}(2200)$  &       &2223 MeV  \\
\hline
\hspace*{-1mm}56\hspace*{-2mm}&S=1/2;L=3;N=1&&\hspace*{-4mm}N$_{5/2^-}$
  &\hspace*{-5mm}N$_{7/2^-}$             &                 &2334 MeV  \\
  &S=3/2;L=3;N=1&\hspace*{-4mm}$\Delta_{3/2^-}$&\hspace*{-5mm}
  $\Delta_{5/2^-}(2350)$&\hspace*{-5mm}$\Delta_{7/2^-}$
  &\hspace*{-5mm}$\Delta_{9/2^-}(2400)$&2467 MeV   \\
\hline
\hline
\hspace*{-1mm}56\hspace*{-2mm}&S=1/2;L=4;N=0&&\hspace*{-4mm}N$_{7/2^+}$
  &\hspace*{-5mm}N$_{9/2^+}(2220)$       &                 &2334 MeV   \\
  &S=3/2;L=4;N=0&\hspace*{-4mm}$\Delta_{5/2^+}$&\hspace*{-5mm}
$\Delta_{7/2^+}(2390)$&\hspace*{-5mm}$\Delta_{9/2^+}(2300)$
&\hspace*{-5mm}$\Delta_{11/2^+}(2420)$&2467 MeV \\
\hline
\hline
\hspace*{-1mm}70\hspace*{-2mm}&S=1/2;L=5;N=0&&\hspace*{-4mm}
 N$_{9/2^-}$&\hspace*{-5mm}N$_{11/2^-}(2600)$&     & 2629 MeV  \\
\hspace*{-1mm}56\hspace*{-2mm}&S=3/2;L=5;N=1&\hspace*{-4mm}
$\Delta_{7/2^-}$ &\hspace*{-5mm}$\Delta_{9/2^-}$ &\hspace*{-5mm}
$\Delta_{11/2^-}$ &\hspace*{-5mm} $\Delta_{13/2^-}(2750)$ & 2893 MeV \\
\hline
\hline
\hspace*{-1mm}56\hspace*{-2mm}&S=1/2;L=6;N=0&&\hspace*{-4mm}
N$_{11/2^+}$&\hspace*{-5mm}N$_{13/2^+}(2700)$ &    & 2781 MeV  \\
&S=3/2;L=6;N=0&\hspace*{-4mm}$\Delta_{9/2^+}$&\hspace*{-5mm}
$\Delta_{11/2^+}$ & $\hspace*{-5mm}\Delta_{13/2^+}$ &\hspace*{-5mm}
$\Delta_{15/2^+}(2950)$ & 2893 MeV \\
\hline
\hline
\hspace*{-1mm}70\hspace*{-2mm}&S=1/2;L=7;N=0&&\hspace*{-4mm}
N$_{13/2^-}$&\hspace*{-5mm}N$_{15/2^-}$&     & 3033 MeV  \\
\hspace*{-1mm}56\hspace*{-2mm}&S=3/2;L=7;N=1&\hspace*{-4mm}
$\Delta_{11/2^-}$ &\hspace*{-5mm}$\Delta_{13/2^-}$ &\hspace*{-5mm}
$\Delta_{15/2^-}$ &\hspace*{-5mm} $\Delta_{17/2^-}$ & 3264 MeV \\
\hline
\hline
\hspace*{-1mm}56\hspace*{-2mm}&S=1/2;L=8;N=0&&\hspace*{-4mm}
N$_{15/2^+}$& \hspace*{-5mm}N$_{17/2^+}$ &    & 3165 MeV  \\
&S=3/2;L=8;N=0&\hspace*{-4mm}$\Delta_{13/2^+}$& \hspace*{-5mm}
$\Delta_{15/2^+}$ & \hspace*{-5mm}$\Delta_{17/2^+}$ & \hspace*{-5mm}
$\Delta_{19/2^+}$ & 3264 MeV \\
\hline
\end{tabular}
\renewcommand{\arraystretch}{1.0}
\caption{Multiplet structure of nucleon and $\Delta$ resonances. The
table contains all known resonances except radial excitations of the
N$_{1/2^+}$(939) and $\Delta_{3/2^+}$(1232).
}
\label{next}
\ec
\end{table}

\subsubsection{Regge trajectories}
\begin{figure}[h!]
\begin{minipage}[c]{0.7\textwidth}
\epsfig{file=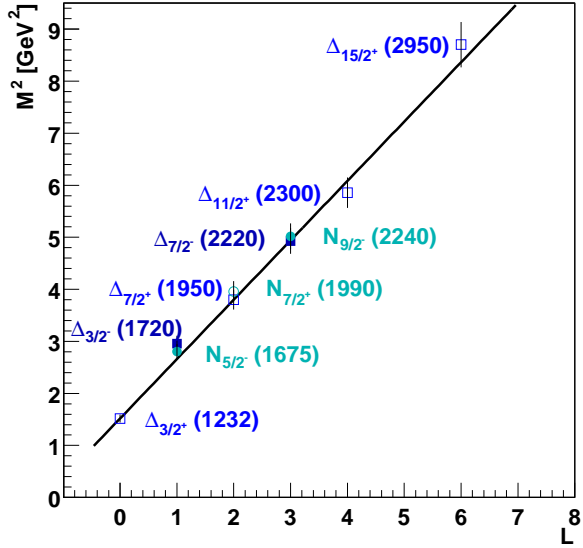,width=8.5cm}
\end{minipage}
\begin{minipage}[c]{0.28\textwidth}
\caption{\label{N-Delta}
Regge trajectory for $\Delta^*$ resonances
with with intrinsic spin $S=1/2$ and $3/2$, and for N$^*$'s with spin 
$S=3/2$. They all fall onto the same
trajectory.}
\end{minipage}
\end{figure}

The masses of meson- and $\Delta^*$ resonances for which all
intrinsic angular momenta are aligned
($J=L+1$ and $J=L+3/2$, respectively)
fall onto Regge trajectories as we had seen
in figure~\ref{Delta-mesons}.
Figure~\ref{N-Delta} differs from the standard Regge trajectory
by plotting the mass square
against the orbital angular momentum L instead of J. Since
spin-orbit forces are small, the orbital angular momentum is well
defined. We choose L to be one variable because this allows us to combine
baryons of positive and negative parity. The figure
includes $\Delta^*$ resonances with intrinsic spin 3/2
and 1/2 (the latter ones must have negative parity\,!), and N$^*$
intrinsic quark spin 3/2.
\par
\vspace*{-5mm}
\par
\begin{figure}
\caption{\label{instant}Mass square shift (in GeV$^2$) 
of spin-1/2 baryons w.r.t. the Regge trajectory 
$\rm  M^2 = M_{\Delta}^2 + {n_s}/3\cdot M_s^2  + a\,(L + N)$
defined by
baryons with S=3/2 (hyperfine splitting).  
The mass shifts scale as 1\,:\,1/2\,:\,3/2\,:\,0 times
$\rm M_{\Delta}^2 - M_{\rm N}^2$ as we proposed in
mass formula (\ref{mass})~\protect\cite{Klempt:2002vp}.} 
\includegraphics[width=\textwidth]{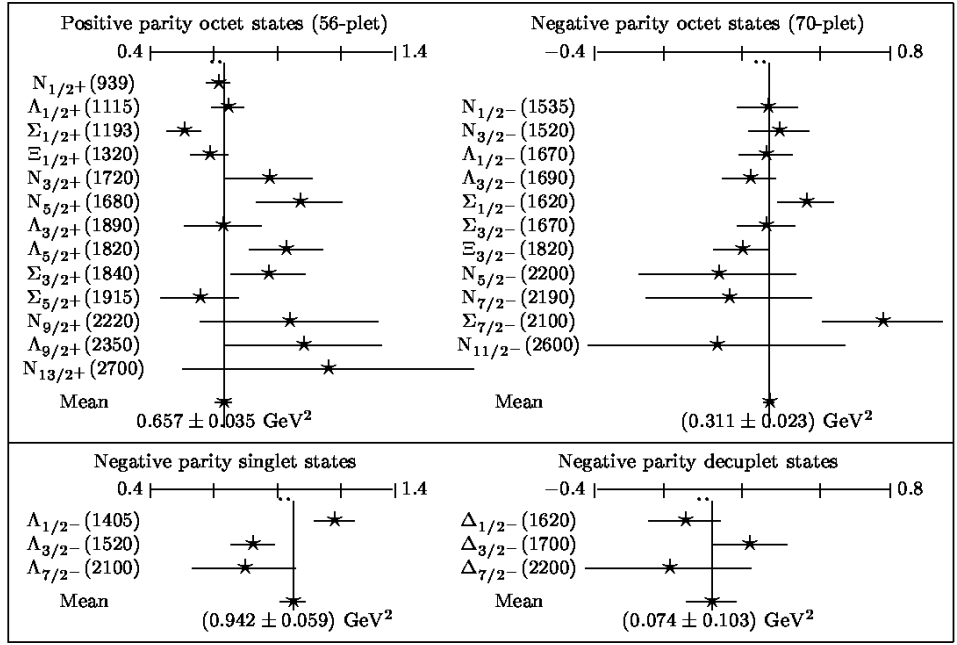}
\end{figure}

Not included in figure~\ref{N-Delta} are the nucleon and nucleon
resonances with spin 1/2. In figure~\ref{instant} we compare
the squared masses of positive- and negative-parity baryon
resonances having intrinsic spin $S=1/2$
to the standard Regge trajectory of figure~\ref{N-Delta}.
All resonances are lower in mass compared to the
trajectory, except the $\Delta^*$ states. 
The mass shifts are discrete. The shift (in mass square)
for negative--parity singlet
states is 0.99. For positive--parity octet states it equals 
0.66, and for negative--parity octet states 0.33. For
negative--parity decuplet states the shift is 0. The $\Delta$--nucleon 
mass shift is e.g. given by 
\be
\rm s_i = M^2_{\Delta (1232)} - M^2_{nucleon} = 0.64.
\label{inst}
\ee
We note that nucleons with S=1/2 are shifted in mass and nucleons with
spin 3/2 are not. $\Delta$ excitations do not have this spin-dependent mass
shift. The mass shift occurs only for baryons having wave-functions
antisymmetric w.r.t. the exchange of two quarks in both, in
spin and in flavor (not only in their spin--flavor wave function). 
As we have seen in section~\ref{section3}, 
this is the selection rule for instanton--induced
interactions acting only between
pairs of quarks antisymmetric w.r.t. their exchange in spin and
flavor \cite{Shuryak:1981ff}.
We consider the even-odd staggering of figure~\ref{instant} as
the most striking evidence for the role of instanton--induced
interactions in low-energy strong interactions (or of other
field configurations with a non--vanishing winding number).

\subsubsection{Radial excitations}
Some partial waves show a second resonance at a higher mass. The
best known example is the Roper resonance, the N$_{1/2^+}$(1440).
Its mass is rather low compared to most calculations because in the
harmonic oscillator description of baryon resonances it is found
in the second excitation band ($N=2$). Table~\ref{radial} lists
consecutive states and their mass--square splitting, all
compatible with the N--Roper mass difference.
\par

\begin{table}
\caption{\label{radial}
Radial excitations of baryon resonances}
\bc
\renewcommand{\arraystretch}{1.3}
\begin{tabular}{ccccc}
\hline\hline
Baryon & $\delta\rm M^2\ (GeV^2)$ &&Baryon & $\delta\rm M^2\ (GeV^2)$\\
N$_{1/2^+}(939)$  &               && $\Delta_{3/2^+}(1232)$ &               \\
N$_{1/2^+}(1440)$ & $1\cdot 1.18$ && $\Delta_{3/2^+}(1600)$ & $1\cdot 1.04$ \\
N$_{1/2^+}(1710)$ & $2\cdot 1.02$ && $\Delta_{3/2^+}(1920)$ & $2\cdot 1.08$ \\
N$_{1/2^+}(2100)$ & $3\cdot 1.18$ &&                        & \\
\hline
$\Delta_{1/2^-}(1620)$ &             &&$\Delta_{3/2^-}(1700)$ &               \\
$\Delta_{1/2^-}(1900)$ &$1\cdot 0.99$&&$\Delta_{3/2^-}(1940)$ & $1\cdot 0.87$\\
$\Delta_{1/2^-}(2150)$ &$2\cdot 1.00 $&& & \\
\hline
N$_{1/2^-}(1530)$ &             &&N$_{3/2^-}(1520)$ &               \\
N$_{1/2^-}$\phantom{(1530)}&&&N$_{3/2^-}$\phantom{(1530)}       & \\
N$_{1/2^-}(2090)$ &$2\cdot 1.01$&&N$_{3/2^-}(2080)$ &$2\cdot 1.01$ \\
\hline
$\Lambda_{1/2^+}(1115)$&             && $\Sigma_{1/2^+}(1193)$&\\
$\Lambda_{1/2^+}(1600)$&$1\cdot 1.24$&&$\Sigma_{1/2^+}(1560)$ & $1\cdot 1.04$\\
$\Lambda_{1/2^+}(1810)$&$2\cdot 0.98$&&$\Sigma_{1/2^+}(1880)$ & $1\cdot 1.06$\\
\hline\hline
\end{tabular}
\renewcommand{\arraystretch}{1.0}
\ec
\end{table}

\subsubsection{Resonances with strangeness}
The mass of a baryon increases with its strangeness content.
The dependence is usually assumed to be linear. Here
the squared baryon masses are given as linear functions
of the strangeness. There are small deviations from the
interpolation between the $\Omega$ and the $\Delta$
when using squared masses. A linear mass
interpolation does not yield a better agreement.
\par
\subsection{\label{section5.2}Observations and conclusions}
We now recall the basic experimental observations and draw obvious
conclusions from these facts.
\begin{enumerate}
\item The slope of the Regge trajectory for meson- and
$\Delta$-excitations is identical.
Baryon resonances are quark-diquark excitations.
\item $\ \Delta^*$ resonances with S=1/2 and S=3/2 fall onto
the same Regge trajectory.
There is no significant spin-spin splitting due to color-magnetic
interactions. Gluon exchange, often assumed to be responsible for the
N-$\Delta$ splitting, should also lead to a mass shift of the
$\Delta_{1/2^{-}}$(1620) and $\Delta_{3/2^{-}}$(1700) relative to the
leading 
Regge trajectory, in the same order of magnitude as in the case of the
N-$\Delta$ splitting. This is not the case. Gluon exchange is not
responsible for the N-$\Delta$ splitting.
\item N and $\Delta$ resonances with spin S=3/2 lie on
a common Regge trajectory.
There is no genuine octet-decuplet splitting.
For spin-3/2 resonances, there is no interaction
associated with the SU(6) multiplet structure.
\item N$^*$s and $\Delta^*$s can be grouped into super-multiplets
with defined orbital angular momenta L and intrinsic spin S, but different
total angular momentum J.
There is no significant spin-orbit ($\vec{\rm L}\cdot\vec{\rm S}$)
interaction. This is again an argument against a large role of gluon
exchange forces (even though the spin-orbit splitting due to one-gluon
exchange could be compensated by the Thomas precession in the
confinement potential).
\item  Octet baryons with intrinsic spin 1/2 have a shift in the
squared mass. The shift is larger (by a factor 2) for even orbital
angular momenta than for odd angular momenta.
Wave functions of octet baryons with spin 1/2 contain
a component $\rm\left(q_1q_2 - q_2q_1\right)\left(\uparrow\downarrow\
- \downarrow\uparrow\right)$.
The mass shift is proportional to this component. Instanton
interactions act on quark pairs antisymmetric in their spin
and their flavor wave function with respect to their exchange. The 
mass shifts shown in figure~\ref{N-Delta} manifest the importance of
instanton--induced interactions in the baryon spectrum.
\item Daughter trajectories have the same slope as the main trajectory
and an intercept higher by  $a=1.142$\,GeV$^2$ per $\ n$, both for
mesons and baryons. The similarity of the spacings between radial
excitations of mesons and baryons supports the interpretation of
baryon resonances as quark-diquark excitations.
\end{enumerate}
\par
These observations allow us to write down a simple 
formula~\cite{Klempt:2002vp} 
reproducing all 100 (but four) masses of baryon resonances observed so far.
\vspace*{-2mm}
\begin{center}
\begin{equation}
\label{mass}
\rm  M^2 = M_{\Delta}^2 + \frac{n_s}{3}\cdot M_s^2  +
a \cdot (L + N) - s_i \cdot I_{sym},
\end{equation}
\end{center}
where
$$\rm M_s^2 = \left(M_{\Omega}^2 - M_{\Delta}^2\right), \qquad\
\rm s_i = \left(M_{\Delta}^2 - M_{\rm N}^2\right),$$
\noindent
$n_s$ is the number of strange quarks in a baryon, and L is
the intrinsic orbital
angular momentum. N is the principal quantum number (we start with
N=0 for the ground state); L+2N gives
the harmonic-oscillator band {\em N}. $\rm I_{sym}$ is the fraction of the
wave function (normalized to the nucleon wave function) 
antisymmetric in spin and flavor. It is given by

\renewcommand{\arraystretch}{1.3}
\begin{tabular}{lrcc}
$\rm I_{sym}  = $&1.0& for S=1/2 and & octet in 56-plet; \\
$\rm I_{sym}  = $&0.5& for S=1/2 and & octet in 70-plet; \\
$\rm I_{sym}  = $&1.5& for S=1/2 and & singlet; \\
$\rm I_{sym}  = $&  0& otherwise.
\end{tabular}
\renewcommand{\arraystretch}{1.3}

$\rm M_{N}, M_{\Delta}, M_{\Omega}$ are input parameters taken from PDG;
$a = 1.142$/GeV$^2$ is the Regge slope as determined from the meson spectrum.
\par
Data and the mass formula (\ref{mass}) are compared in Tables~\ref{n}
to \ref{l}. To estimate if the agreement is good, we
need errors. In the Particle Data Listings these 
are only given for well known baryons. We could relate
the errors to the widths, but the widths are also known only with
large uncertainties. We estimate the widths of a baryon according to
\be
\Gamma_{\rm B} = \frac{1}{4}Q
\label{width}
\ee
were $Q$ is the largest decay momentum (into N$\pi$ for N$^*$ and
$\Delta^*$) of a resonance. The calculated widths are also
given in Tables~\ref{n} to \ref{l}. One quarter of the width is
assumed to be the error in mass. A resonance may decay and rescatter
into the resonance again in the form of loop diagrams. Such
virtual decays lead to hadronic shifts. We estimate that these
could be on the order of 1/4 of the line width. With this definition
the ground state masses have no error and some errors become very
small. Hence we adopt a model error of 30\,MeV which is added
quadratically to the error due to hadronic shifts. 
\par

\begin{table}[t!]
\caption{\label{n}
Mass spectrum of N resonances. 
See caption of Table~(\protect\ref{d}).}
\bc
\renewcommand{\arraystretch}{1.3}
\begin{tabular}{||cccc|cccccc||}
\hline
\hline
Baryon            &Status& $\rm D_L$  & N & Mass  &(\ref{mass}) &
$\Gamma$ & (\ref{width}) &$\sigma$ & $\chi^2$ \\ \hline
N$_{1/2^+}(939)$ & **** & $(56,^28)_0$ & 0 & 939 & - & - & - & - & - \\
N$_{1/2^+}(1440)$ & **** & $(56,^28)_0$ & 1 & 1450 & 1423 & 250-450 & 87 & 37 & 0.53\\
N$_{1/2^+}(1710)$ & *** & $(56,^28)_0$ & 2 & 1710 & 1779 & 50-250 & 176 & 53 & 1.69\\
$^1$N$_{1/2^+}(2100)$ & * & $(56,^28)_0$ & 2 & 2100 & 2076 & - & 251 & 70 & 0.12\\ \hline
N$_{1/2^-}(1535)$ & **** & $(70,^28)_1$ & 0 & 1538 & 1530 & 100-250 & 114 & 41 & 0.04\\
N$_{3/2^-}(1520)$ & **** & $(70,^28)_1$ & 0 & 1523 & 1530 & 110-135 & 114 & 41 & 0.03\\ \hline
N$_{1/2^-}(1650)$ & **** & $(70,^48)_1$ & 0 & 1660 & 1631 & 145-190 & 139 & 46 & 0.4\\
N$_{3/2^-}(1700)$ & *** & $(70,^48)_1$ & 0 & 1700 & 1631 & 50-150 & 139 & 46 & 2.25\\
N$_{5/2^-}(1675)$ & **** & $(70,^48)_1$ & 0 & 1678 & 1631 & 140-180 & 139 & 46 & 1.04\\ \hline
N$_{3/2^+}(1720)$ & **** & $(56,^28)_2$ & 0 & 1700 & 1779 & 100-200 & 176 & 53 & 2.22\\
N$_{5/2^+}(1680)$ & **** & $(56,^28)_2$ & 0 & 1683 & 1779 & 120-140 & 176 & 53 & 3.28\\ \hline
N$_{3/2^+}(1900)$ & ** & $(70,^48)_2$ & 0 & 1900 & 1950 & - & 219 & 62 & 0.65\\
N$_{5/2^+}(2000)$ & ** & $(70,^48)_2$ & 0 & 2000 & 1950 & - & 219 & 62 & 0.65\\
N$_{7/2^+}(1990)$ & ** & $(70,^48)_2$ & 0 & 1990 & 1950 & - & 219 & 62 & 0.42\\ \hline
N$_{1/2^-}(2090)$ & * & $(70,^28)_1$ & 2 & 2090 & 2151 & - & 269 & 74 & 0.68\\
N$_{3/2^-}(2080)$ & ** & $(70,^28)_1$ & 2 & 2080 & 2151 & - & 269 & 74 & 0.92\\ \hline
N$_{5/2^-}(2200)$ & ** & $(70,^28)_3$ & 0 & 2220 & 2151 & - & 269 & 74 & 0.87\\
N$_{7/2^-}(2190)$ & **** & $(70,^28)_3$ & 0 & 2150 & 2151 & 350-550 & 269 & 74 & 0\\ \hline
N$_{9/2^-}(2250)$ & **** & $(70,^48)_3$ & 0 & 2240 & 2223 & 290-470 & 287 & 78 & 0.05\\ \hline
N$_{9/2^+}(2220)$ & **** & $(56,^28)_4$ & 0 & 2245 & 2334 & 320-550 & 315 & 84 & 1.12\\ \hline
N$_{11/2^-}(2600)$ & *** & $(70,^28)_5$ & 0 & 2650 & 2629 & 500-800 & 389 & 102 & 0.04\\ \hline
N$_{13/2^+}(2700)$ & ** & $(56,^28)_6$ & 0 & 2700 & 2781 & - & 427 & 111 & 0.53\\ \hline \hline
& & & & & & dof: & 21 & $\sum \chi^2$:& 17.53 \\
\hline
\hline
\end{tabular}
\renewcommand{\arraystretch}{1.0}
\ec
$^1$\small{
Based on its mass, the N$_{1/2^+}(2100)$ is likely a radial
excitation. It could also be the $(70,^48)_2$ N$_{1/2^+}$ state
expected at 1950 MeV. The SAPHIR collaboration suggested a  N$_{1/2^+}$
at 1986 MeV~\protect\cite{Plotzke:ua}
which would, if confirmed, be a natural
partner to complete the quartet of L=2, S=3/2 nucleon resonances.}
\end{table}
\par
The mass formula reproduces not only the masses of N$^*$,
$\Delta^*$'s, $\Sigma^*$'s and  $\Lambda^*$'s which are reproduced
in tables here, it is also compatible with the few entries for
 $\Xi^*$'s  $\Omega^*$'s.
\clearpage
\begin{table}[t!]
\caption{\label{d}
Mass spectrum of $\Delta$ resonances. 
A baryon resonance is characterized by its $\rm J^P$ as subscript
and its nominal mass (in parenthesis). The PDG rating is given by the
number of *'s. Its classification into multiplets is discussed in
section (6). The PDG lists a range of acceptable
values, we give the central mass (in MeV), compared to the predicted
mass from eq.~(\protect\ref{mass}). We list the PDG range of acceptable
widths $\Gamma$ and compare them to eq.~(\protect\ref{width}). The width
parameterization is only used to estimate errors. The mass errors $\sigma$
are given by $\sigma^2 = \frac{\Gamma^2}{16} + 30^2$ where the
first error allows for hadronic mass shifts on the order of 1/4 of
the line width, the second one for uncertainties in the mass formula.
The last column gives the $\chi^2$ contribution from the mass
comparison. The $\chi^2$s are summed up and compared to the degrees
of freedom in the last column.}
\bc
\renewcommand{\arraystretch}{1.3}
\begin{tabular}{||cccc|cccccc||}
\hline
\hline
Baryon            &Status& $\rm D_L$  & N & Mass  &(\ref{mass}) &
$\Gamma$ & (\ref{width}) &$\sigma$ & $\chi^2$ \\ \hline
$\Delta_{3/2^+}(1232)$ & **** & $(56,^410)_0$ & 0 & 1232 & 1232 & - & - & - & -\\
$\Delta_{3/2^+}(1600)$ & *** & $(56,^410)_0$ & 1 & 1625 & 1631 & 250-450 & 139 & 46 & 0.02\\ \hline
$\Delta_{1/2^+}(1750)$ & * & $(70,^210)_0$ & 1 & 1750 & 1631 & - & 139 & 46 & 6.69\\ \hline
$\Delta_{1/2^-}(1620)$ & **** & $(70,^210)_1$ & 0 & 1645 & 1631 & 120-180 & 139 & 46 & 0.09\\
$\Delta_{3/2^-}(1700)$ & **** & $(70,^210)_1$ & 0 & 1720 & 1631 & 200-400 & 139 & 46 & 3.74\\ \hline
$\Delta_{1/2^-}(1900)$ & ** & $(56,^410)_1$ & 1 & 1900 & 1950 & 140-240 & 219 & 62 & 0.65\\
$\Delta_{3/2^-}(1940)$ & * & $(56,^410)_1$ & 1 & 1940 & 1950 & - & 219 & 62 & 0.03\\
$\Delta_{5/2^-}(1930)$ & *** & $(56,^410)_1$ & 1 & 1945 & 1950 & 250-450 & 219 & 62 & 0.01\\ \hline
$\Delta_{1/2^+}(1910)$ & **** & $(56,^410)_2$ & 0 & 1895 & 1950 & 190-270 & 219 & 62 & 0.79\\
$\Delta_{3/2^+}(1920)$ & *** & $(56,^410)_2$ & 0 & 1935 & 1950 & 150-300 & 219 & 62 & 0.06\\
$\Delta_{5/2^+}(1905)$ & **** & $(56,^410)_2$ & 0 & 1895 & 1950 & 280-440 & 219 & 62 & 0.79\\
$\Delta_{7/2^+}(1950)$ & **** & $(56,^410)_2$ & 0 & 1950 & 1950 & 290-350 & 219 & 62 & 0\\ \hline
$\Delta_{1/2^-}(2150)$ & * & $(70,^210)_1$ & 2 & 2150 & 2223 & - & 287 & 78 & 0.88\\ \hline
$\Delta_{7/2^-}(2200)$ & * & $(70,^210)_3$ & 0 & 2200 & 2223 & - & 287 & 78 & 0.09\\ \hline
$^1\Delta_{5/2^+}(2000)$ & ** & $(70,^210)_2$ & 1 & 2200 & 2223 & - & 287 & 78 & 0.09\\ \hline
$\Delta_{5/2^-}(2350)$ & * & $(56,^410)_3$ & 1 & 2350 & 2467 & - & 348 & 92 & 1.62\\
$\Delta_{9/2^-}(2400)$ & ** & $(56,^410)_3$ & 1 & 2400 & 2467 & - & 348 & 92 & 0.53\\ \hline
$\Delta_{7/2^+}(2390)$ & * & $(56,^410)_4$ & 0 & 2390 & 2467 & - & 348 & 92 & 0.7\\
$\Delta_{9/2^+}(2300)$ & ** & $(56,^410)_4$ & 0 & 2300 & 2467 & - & 348 & 92 & 3.3\\
$\Delta_{11/2^+}(2420)$ & **** & $(56,^410)_4$ & 0 & 2400 & 2467 & 300-500 & 348 & 92 & 0.53\\ \hline
$\Delta_{13/2^-}(2750)$ & ** & $(56,^410)_5$ & 1 & 2750 & 2893 & - & 455 & 118 & 1.47\\ \hline
$\Delta_{15/2^+}(2950)$ & ** & $(56,^410)_6$ & 0 & 2950 & 2893 & - & 455 & 118 & 0.23\\ \hline \hline
& & & & & & dof: & 21 & $\sum \chi^2$:& 22.31 \\
\hline
\hline
\end{tabular}
\renewcommand{\arraystretch}{1.0}
\ec
$^1$\small{The PDG quotes two entries, at 1752 and 2200 MeV, respectively,
and gives 2000 as ''our estimate''.
We use the higher mass value for our comparison.}
\end{table}
\begin{table}
\caption{\label{s}
Mass spectrum of $\Sigma$ resonances. See caption of Table
(\protect\ref{n}).}
\bc
\renewcommand{\arraystretch}{1.4}
\begin{tabular}{||cccc|cccccc||}
\hline
\hline
Baryon            &Status& $\rm D_L$  & N & Mass  &(\ref{mass}) &
$\Gamma$ & (\ref{width}) &$\sigma$ & $\chi^2$ \\ \hline
$\Sigma_{1/2^+}(1193)$ & **** & $(56,^28)_0$ & 0 & 1193 & 1144 & - & - & 30 & 2.67\\
$\Sigma_{3/2^+}(1385)$ & **** & $(56,^410)_0$ & 0 & 1384 & 1394 & - & - & 30 & 0.11\\
$\Sigma(1480)$ & * &  &  &  &  &  &  &  & \\
$\Sigma(1560)$ & ** & $(56,^28)_0$ & 1 & 1560 & 1565 & - & 32 & 31 & 0.03\\
$\Sigma_{1/2^+}(1660)$ & *** & $(70,^28)_0$ & 1 & 1660 & 1664 & 40-200 & 57 & 33 & 0.01\\
$\Sigma_{1/2^+}(1770)$ & * & $(70,^210)_0$ & 1 & 1770 & 1757 & - & 80 & 36 & 0.13\\
$\Sigma_{1/2^+}(1880)$ & ** & $(56,^28)_0$ & 2 & 1880 & 1895 & - & 115 & 42 & 0.13\\ \hline
$\Sigma_{1/2^-}(1620)$ & ** & $(70,^28)_1$ & 0 & 1620 & 1664 & - & 57 & 33 & 1.78\\
$\Sigma_{3/2^-}(1580)$ & ** & $(70,^28)_1$ & 0 & 1580 & 1664 & - & 57 & 33 & 6.48\\
$\Sigma(1690)$ & ** & $(70,^210)_1$ & 0 & 1690 & 1757 & - & 80 & 36 & 3.46\\ \hline
$\Sigma_{1/2^-}(1750)$ & *** & $(70,^48)_1$ & 0 & 1765 & 1757 & 60-160 & 80 & 36 & 0.05\\
$\Sigma_{3/2^-}(1670)$ & **** & $(70,^48)_1$ & 0 & 1675 & 1757 & 40-80 & 80 & 36 & 5.19\\
$\Sigma_{5/2^-}(1775)$ & **** & $(70,^48)_1$ & 0 & 1775 & 1757 & 105-135 & 80 & 36 & 0.25\\ \hline
$\Sigma_{1/2^-}(2000)$ & * & $(70,^28)_1$ & 1 & 2000 & 1977 & - & 135 & 45 & 0.26\\
$\Sigma_{3/2^-}(1940)$ & *** & $(70,^28)_1$ & 1 & 1925 & 1977 & 150-300 & 135 & 45 & 1.34\\ \hline
$\Sigma_{3/2^+}(1840)$ & * & $(56,^28)_2$ & 0 & 1840 & 1895 & - & 115 & 42 & 1.71\\
$\Sigma_{5/2^+}(1915)$ & **** & $(56,^28)_2$ & 0 & 1918 & 1895 & 80-160 & 115 & 42 & 0.3\\
\hline
$^{1}\Sigma_{3/2^+}(2080)$ & ** & $(56,^410)_2$ & 0 & 2080 & 2056 & - & 155 & 49 & 0.24\\
$^{1}\Sigma_{5/2^+}(2070)$ & * & $(56,^410)_2$ & 0 & 2070 & 2056 & - & 155 & 49 & 0.06\\
$^{1}\Sigma_{7/2^+}(2030)$ & **** & $(56,^410)_2$ & 0 & 2033 & 2056 & 150-200 & 155 & 49 & 0.22\\
\hline
$\Sigma(2250)$ & *** & $(70,^28)_3$ & 0 & 2245 & 2248 & 60-150 & 203 & 59 & 0\\
$\Sigma_{7/2^-}(2100)$ & * & $(70,^28)_3$ & 0 & 2100 & 2248 & - & 203 & 59 & 6.29\\ \hline
$\Sigma(2455)$ & ** & $(56,^28)_4$ & 0 & 2455 & 2424 & - & 247 & 69 & 0.2\\ \hline
$\Sigma(2620)$ & ** & $(70,^28)_5$ & 0 & 2620 & 2708 & - & 318 & 85 & 1.07\\ \hline
$\Sigma(3000)$ & * & $(56,^28)_6$ & 0 & 3000 & 2857 & - & 355 & 94 & 2.31\\ \hline
$\Sigma(3170)$ & * & $(70,^28)_7$ & 0 & 3170 & 3102 & - & 416 & 108 & 0.4\\ \hline \hline
& & & & & & dof: & 25 & $\sum \chi^2$:& 34.69 \\
\hline
\hline
\end{tabular}
\renewcommand{\arraystretch}{1.0}
\ec
$^1$\small{ These three resonances, and the missing $\Sigma_{1/2^+}$, can
belong to the octet or to the decuplet; the mass formula
(\protect\ref{mass}) predicts identical masses.}
\end{table}
\clearpage
\begin{table}
\caption{\label{l}
Mass spectrum of $\Lambda$ resonances. See caption of Table
(\protect\ref{d}).}
\bc
\renewcommand{\arraystretch}{1.4}
\begin{tabular}{||cccc|cccccc||}
\hline
\hline
Baryon            &Status& $\rm D_L$  & N & Mass  &(\ref{mass}) &
$\Gamma$ & (\ref{width}) &$\sigma$ & $\chi^2$ \\ \hline
$\Lambda_{1/2^+}(1115)$ & **** & $(56,^28)_0$ & 0 & 1116 & 1144 & - & - & 30 & 0.87\\
$\Lambda_{1/2^+}(1600)$ & *** & $(56,^28)_0$ & 1 & 1630 & 1565 & 50-250 & 32 & 31 & 4.4\\
$\Lambda_{1/2^+}(1810)$ & *** & $(56,^28)_0$ & 2 & 1800 & 1895 & 50-250 & 115 & 42 & 5.12\\ \hline
$\Lambda_{1/2^-}(1405)$ & **** & $(70,^21)_1$ & 0 & 1407 & 1460 & 50 & 6 & 30 & 3.12\\
$\Lambda_{3/2^-}(1520)$ & **** & $(70,^21)_1$ & 0 & 1520 & 1460 & 16 & 6 & 30 & 4\\ \hline
$\Lambda_{1/2^-}(1670)$ & **** & $(70,^28)_1$ & 0 & 1670 & 1664 & 25-50 & 57 & 33 & 0.03\\
$\Lambda_{3/2^-}(1690)$ & **** & $(70,^28)_1$ & 0 & 1690 & 1664 & 50-70 & 57 & 33 & 0.62\\ \hline
$\Lambda_{1/2^-}(1800)$ & *** & $(70,^48)_1$ & 0 & 1785 & 1757 & 200-400 & 80 & 36 & 0.6\\
$\Lambda_{5/2^-}(1830)$ & **** & $(70,^48)_1$ & 0 & 1820 & 1757 & 60-110 & 80 & 36 & 3.06\\ \hline
$\Lambda_{3/2^+}(1890)$ & **** & $(56,^28)_2$ & 0 & 1880 & 1895 & 60-200 & 115 & 42 & 0.13\\
$\Lambda_{5/2^+}(1820)$ & **** & $(56,^28)_2$ & 0 & 1820 & 1895 & 70-90 & 115 & 42 & 3.19\\ \hline
$\Lambda(2000)$ & * & $(70,^48)_2$ & 0 & 2000 & 2056 & - & 155 & 49 & 1.31\\
$\Lambda_{5/2^+}(2110)$ & *** & $(70,^48)_2$ & 0 & 2115 & 2056 & 150-250 & 155 & 49 & 1.45\\
$\Lambda_{7/2^+}(2020)$ & * & $(70,^48)_2$ & 0 & 2020 & 2056 & - & 155 & 49 & 0.54\\
\hline
$\Lambda_{7/2^-}(2100)$ & **** & $(70,^21)_3$ & 0 & 2100 & 2101 & 100-250 & 166 & 51 & 0\\ \hline
$\Lambda_{3/2^-}(2325)$ & * & $(70,^28)_1$ & 2 & 2325 & 2248 & - & 203 & 59 & 1.7\\ \hline
$\Lambda_{9/2^+}(2350)$ & *** & $(56,^28)_4$ & 0 & 2355 & 2424 & 100-250 & 247 & 69 & 1\\ \hline
$\Lambda(2585)$ & ** & $(70,^48)_2$ & 0 & 2585 & 2551 & - & 279 & 76 & 0.2\\ \hline \hline
& & & & & & dof: & 18 & $\sum \chi^2$:& 31.34 \\
\hline
\hline
\end{tabular}
\renewcommand{\arraystretch}{1.0}
\ec
\end{table}
\par
It is remarkable that all observed resonances are well reproduced by
the mass formula. If we assume that the mass formula calculates
$(qqq)$--baryon masses, there is no hot candidate left for other forms of 
baryons which are predicted by models: hybrid baryons or pentaquarks. 
These will be the topic of sections~\ref{section5.5}.
\subsection{\label{section5.3}The 'missing resonances'}
A three--body system supports a rich dynamical spectrum. This can
be seen when looking at Table~\ref{tab:kle} which lists the wave
functions in the harmonic oscillator basis. For each of these
realizations there is the full decomposition in SU(6). Hence a
multitude of baryon resonances is expected. The number of known
resonances is much smaller: this is the problem of the so--called
'missing resonances'. Many more resonances are expected to exist
than observed experimentally. This is not a problem of an incomplete
stamp collection. A large number of resonances are missing and we do
not know the guiding principle which leads to the observation of some
resonances and the non--observation of others.
\par
There are two possible solutions. One is that the resonances
have not been discovered so far. In particular one can argue that
some of the resonances may have a weak coupling only to the N$\pi$
channel. Since nearly all N$^*$ and $\Delta^*$ resonances were
discovered in $\pi$N scattering, they may have escaped discovery.
These resonances are however predicted to have normal coupling to
$\gamma$N. Photo--production of baryon resonances and detection of
multibody final states should therefore reveal if these states
exist or not.
\par
\vspace*{-3mm}
{\footnotesize
\begin{table}[h!]
\caption{
Harmonic  oscillator wave functions as excitations in
the $\lambda$ and  $\rho$ oscillator basis. Baryons can be
 excited orbitally ($l_i$) and radially ($n_i$). Radial
excitations carry two $\hbar\omega$, the excitation to the Roper
with $n_1=1$ is therefore in the 2$^{\rm nd}$ band ($N=2$). The two
orbital angular momenta $l_i$ couple to the total angular momentum $L$.
The shell quantum number $N$ is hence $N=l_1+l_2+2n_1+2n_2$. The
multiplicity is given by ${\rm n}=2L+1$. The $\sum\rm n$ gives the
total number of realizations in a shell. A smaller number of wave
functions can be realized if it is required that only one of the
two oscillators is excited. } \bc
\renewcommand{\arraystretch}{1.1}
\begin{tabular}{ccccccccccccccccc}
\hline
\hline
$N$&n&$l_1$&$l_2$&$n_1$&$n_2$&$L$ &\hspace*{-1mm}$\sum\rm n$\hspace*{-1mm}
&$N$&n&$l_1$&$l_2$&$n_1$&$n_2$&$L$ &\hspace*{-1mm}$\sum\rm n$\hspace*{-1mm}\\
\hline
{0}&1&  0  &  0  &  0  &  0  & 0  &\hspace*{-1mm}1 /\,{1}\hspace*{-1mm}&
{4}&9&  4  &  0  &  0  &  0  & 4  &   \\
{1}&3&  1  &  0  &  0  &  0  & 1  &   &
{4}&9&  0  &  4  &  0  &  0  & 4  &   \\
{1}&3&  0  &  1  &  0  &  0  & 1  &\hspace*{-1mm}6 /\,{6}\hspace*{-1mm}&
4&21& {3}  &  {1}  &  {0}  &  {0}  &\hspace*{-2mm}2,3,4  &   \\
{2}&5&  2  &  0  &  0  &  0  & 2  &   &
4&21& {1}  &  {3}  &  {0}  &  {0}  &\hspace*{-2mm}2,3,4  &   \\
{2}&5&  0  &  2  &  0  &  0  & 2  &   &
4&25&  {2}  &  {2}  &  {0}  &  {0}  &\hspace*{-2mm}1,2-4 &   \\
2&5&  1  &  1  &  0  & 0  & 2  &   &
{4}&5&  {2}  &  {0}  &  {1}  & {0}  & {2}  &   \\
2&3&  1  &  1  &  0  &  0  & 1  &   &
4&5&  {0}  &  {2}  &  {1}  &  {0}  & {2}  &   \\
2&1&  1  &  1  &  0  &  0  & 0  &   &
4&9&  {1}  &  {1}  &  {1}  &  0  &\hspace*{-2mm}0,1,2  &   \\
{2}&1&  0  &  0  &  1  &  0  & 0  &   &
4&5&  {2}  &  {0}  &  {0}  &  {1}  & {2}  &   \\
{2}&1&  0  &  0  &  0  &  1  & 0  &\hspace*{-1mm}21 /\,{12}\hspace*{-1mm}&
{4}&5&  {0}  &  {2}  &  {0}  &  {1}  & {2}  &   \\
{3}&7&  3  &  0  &  0  &  0  & 3  &   &
4&9&  {1}  &  {1}  &  {0}  &  {1}  &\hspace*{-2mm}0,1,2  &   \\
{3}&7&  0  &  3  &  0  &  0  & 3  &   &
{4}&1&  0  &  0  &  2  &  0  & 0  &   \\
3&15&  2  &  1  &  0  &  0  &\hspace*{-2mm}1,2,3 &&
{4}&1&  0  &  0  &  0  &  2  & 0  &   \\
3&15&  1  &  2  &  0  &  0  &\hspace*{-2mm}1,2,3 &&
4&1&  0  &  0  &  1  &  1  & 0  &\hspace*{-1mm}126 /\,{30}\hspace*{-1mm}\\
{3}&3&  1  &  0  &  1  &  0  & 1  &   &
&&&&&&&\\
3&3&  0  &  1  &  1  &  0  & 1  &   &
&&&&&&&\\
3&3&  1  &  0  &  0  &  1  & 1  &   &
&&&&&&&\\
{3}&3&  0  &  1  &  0  &  1  & 1  &\hspace*{-1mm}56 /\,{20}\hspace*{-1mm}&
&&&&&&&\\
\hline
\hline
\end{tabular}
\renewcommand{\arraystretch}{1.0}
\ec
\label{tab:kle}
\end{table}
} 
\par
\vskip 2mm

\par
Could it be that these resonances do not exist at all\,? One often
discussed solution of the 'missing resonances' problem is the
possibility that two quarks in a baryon form a quasi--stable
diquark~\cite{Lichtenberg:1969pp}. 
Such scenarios are back in the center of the scientific discussion
since Lipkin and
Karliner~\cite{Karliner:2003sy,Karliner:dt,Karliner:2004qw}
Jaffe and Wilzcek \cite{Jaffe:2003sg} proposed a diquark model
to explain the exotic properties of the $\Theta^+(1540)$.

\subsubsection{The single-quark excitation hypothesis}
Below, a possibility is sketched how the large number of expected
states might be reduced without using 'exotic' assumptions.
It is a sketch only which will require systematic study and
application to all partial waves. Likely, the reduction of states
will not be sufficient in all partial waves, so it is a first step only.
Consider $\rm L=2$, $\rm N=1$ as example. Both oscillators can be excited to
have quantum numbers 
$l_{\rho}, l_{\lambda}, n_{\rho}, n_{\lambda}$.
 L$=2$, N$=1$ belong to the 4$^{\rm th}$ excitation band. The  
configurations which can contribute are listed below:
\begin{eqnarray*}
{(l_{\rho}, n_{\rho}, l_{\lambda}, n_{\lambda})
= (2, 1, 0,0) = | 0 >} & \qquad;\qquad & {(0, 0, 2, 1) = | 8 > }\\
{(l_{\rho},  n_{\rho}, l_{\lambda}, n_{\lambda}) = (2, 0, 0, 1) = | 2 >} &\qquad;\qquad &
{ (0, 1, 2, 0) = | 6 >}\\
{(l_{\rho}, n_{\rho}, l_{\lambda}, n_{\lambda}) = (2, 0, 2, 0) = |4 > } &\qquad;\qquad & \\
{(l_{\rho}, n_{\rho}, l_{\lambda}, n_{\lambda}) = (3, 0, 1, 0) = | 1 >} &\qquad;\qquad &
{ (1, 0, 3, 0) = | 7 >} \\
{(l_{\rho}, n_{\rho}, l_{\lambda}, n_{\lambda}) = (1, 1, 1, 0) = | 3 > } &\qquad;\qquad &
{ (1, 0, 1, 1) =  | 5 >}
\end{eqnarray*}
\normalsize
Note that only the wave functions $| 0 >$ and $| 8 >$ contain
single--quark excitations, in the other 7 functions, both oscillators
are excited. These wave functions do not yet obey the Pauli principle.
New wave functions need to be constructed having defined
symmetry under exchange of two quarks. These are given by:
\begin{scriptsize}
\begin{eqnarray*}
 \ | \ S  0 > & =& \ + \ \sqrt{\ \frac{1}{6}}\cdot{\frac{1}{\sqrt 2}( \ | \  0> + \  \ | \  8>)} \ + \ \sqrt\frac{7}{18}\cdot\frac{1}{\sqrt 2}( \ | \  2> + \  \ | \  6>) \ - \ \sqrt{\ \frac{4}{9}}  \ | \  4>\\
 \ | \ S  1 > & =& \ + \ \sqrt\frac{7}{12} \cdot{\frac{1}{\sqrt 2}( \ | \  0> + \  \ | \  8>)} \ + \ \sqrt\frac{1}{36}\cdot\frac{1}{\sqrt 2}( \ | \  2> + \  \ | \  6>) \ + \ \sqrt\frac{7}{18}  \ | \  4>\\
  \ | \ MS 0 > & =& \ + \ \sqrt{\ \frac{1}{4}}\cdot\,{\frac{1}{\sqrt 2}( \ | \  0> + \  \ | \  8>)}\, \ - \ \sqrt\frac{7}{12}\cdot\frac{1}{\sqrt 2}( \ | \  2> + \  \ | \  6>) \ - \ \sqrt{\ \frac{1}{6}}  \ | \  4>\\
 \ | \ MS 1 > & =& \ + \ \sqrt\frac{3}{10}\cdot {\frac{1}{\sqrt 2}( \ | \  0> - \  \ | \  8>)} \ - \ \sqrt\frac{7}{10}\cdot\frac{1}{\sqrt 2}( \ | \  2> - \  \ | \  6>)\\
  \ | \ MS 2 > & =& \ + \ \sqrt\frac{7}{10}\cdot {\frac{1}{\sqrt 2}( \ | \  0> - \  \ | \  8>)} \ + \ \sqrt\frac{3}{10}\cdot\frac{1}{\sqrt 2}( \ | \  2> - \  \ | \  6>)\\
  \ | \ MA 0 > & =& \ + \ \sqrt\frac{4}{25}           \cdot\frac{1}{\sqrt 2}( \ | \  1> + \  \ | \  7>)  \ - \ \sqrt\frac{21}{25}\cdot\frac{1}{\sqrt 2}( \ | \  3> + \  \ | \  5>)\\
 \ | \ MA 1 > & =& \ - \ \sqrt\frac{21}{25}          \cdot\frac{1}{\sqrt 2}( \ | \  1> + \  \ | \  7>)  \ - \ \sqrt\frac{4}{25}\cdot\frac{1}{\sqrt 2}( \ | \  3> + \  \ | \  5>)\\
  \ | \ MA 2 > & =& \ + \ \sqrt\frac{3}{10}           \cdot\frac{1}{\sqrt 2}( \ | \  1> - \  \ | \  7>)  \ + \ \sqrt\frac{7}{10}\cdot\frac{1}{\sqrt 2}( \ | \  3> - \  \ | \  5>)\\
  \ | \ A  0 > & =& \ - \ \sqrt\frac{7}{10}           \cdot\frac{1}{\sqrt 2}( \ | \  1> - \  \ | \  7>)  \ + \ \sqrt\frac{3}{10}\cdot\frac{1}{\sqrt 2}( \ | \  3> - \  \ | \  5>)
\end{eqnarray*}
\end{scriptsize}
None of these wave functions contains single--quark excitations only.
But we can now assume that in the process of baryon production
the initial state after excitation of a baryon is given by
$   \frac{1}{\sqrt 2} { \ | \  0>} \pm  \frac{1}{\sqrt 2} {\ | \
8>}$. This is not an energy eigenstate. But in case that
the masses of states $| 0 >$, $| 4 >$, $| 6 >$ and $| 8 >$ are not too
different, a coherent superposition will be formed, with a mean energy
and there are only two baryons instead of 9. Those states composed
of $| 1 >$, $| 2 >$, $| 5 >$ and $| 7 >$ cannot be reached when the
single-quark excitation hypothesis holds true. The conjecture has 
the following consequences:
\bi
\item Resonances with symmetric wave functions ($ S0$,  $ S1$
and $ MS0$)
and with mixed symmetric wave functions ($ MS1$ and
$ MS2$) are coherently excited
\begin{eqnarray*}
{\frac{1}{\sqrt 2} (| 0> \ + \  | 8>)} &=&
\sqrt\frac{1}{6} | S0 > \ + \ \sqrt\frac{7}{12} | S1 > \ + \
\frac{1}{2} | MS0 >, \\
{\frac{1}{\sqrt 2} (| 0> \ - \  | 8>)} &=&
\sqrt\frac{3}{10} | MS1 > \ + \ \sqrt\frac{7}{10} | MS2 >.
\end{eqnarray*}
\item Baryon resonances are wave packets with defined phase but uncertain
in quantum number ($\delta\phi\cdot\delta n\sim \hbar$).
\item We expect a large reduction in the number of states.
\item Resonances with antisymmetric and mixed antisymmetric wave functions
are not excited.
\item Only relevant
quantum numbers are {$L=l_{\rho}+l_{\lambda}$} and
{$ N=n_{\rho}+n_{\lambda}$}.
\item These are used in the baryon mass formula.
\ei
\subsubsection{Hybrid baryons}
In the same way as the gluon flux tube in a meson can possibly be excited
(leading to hybrid mesons), also hybrid baryon states can be
constructed where the flux tubes are connecting
the quarks are excited.  
The color flux meets in a junction (in a Mercedes star configuration)
which plays, for every quark, 
the role of the antiquark. A model calculation~\cite{Capstick:1999qq} 
gave a rich spectrum; certainly the problem
of 'missing resonances' is aggravated. Capstick and Page used
an adiabatic approximation for
the quark motion, quarks were confined by a linear potential;
flux tubes and junction were modeled by attracting beads 
vibrating in various string modes. The Coulomb potential
from one-gluon exchange was assumed the same in conventional and hybrid
baryons, and spin-dependent terms were neglected.  
Before splitting due to one--gluon exchange interactions, 
hybrids with quark orbital angular momenta $L_q=0,1,2$ have masses 1980, 2340
and 2620 MeV respectively. Hyperfine (contact plus tensor)
interactions split the $N$ hybrids down and the $\Delta$ hybrids up, 
so that the lowest N hybrid mass becomes 1870 MeV. The
model error on this mass was estimated to be less than $\pm 100$
MeV. 
\subsection{\label{section5.4}New directions}
The overwhelming majority of data on baryon resonances, their
masses, widths, and partial decay widths comes from pion elastic
scattering off nucleons. This experimental technique demonstrated to
be very powerful but, it is restricted to baryon resonances
with sufficiently large coupling to the N$\pi$ system. For high 
baryon masses, the N$\pi$ couplings become small, and other
experimental techniques are mandatory. Photo--production
of high--mass states is predicted to be less suppressed
than formation via pion--proton scattering.  
\par
Baryon resonances with high mass can be expected to decay via
cascades. The resonance $\rm N(2220)H_{19}$ can decay, with
a probability of 10-20\%, via pion emission into N$\pi$. The
N$\pi$ system then has an orbital angular momentum $\ell =6$.
In the decay sequence $\rm N(2220)H_{19}
\to N(1675)D_{15}\pi$, and $\rm N(1675)D_{15}\to N\pi$ the angular
momenta are $\ell_1 =4$ and $\ell_2 =1$. Hence the angular momentum
barrier for these decays is smaller. For a study of such sequential
decays there are two requirements. First, instrumentations  are
needed which detect multiparticle final states. 
Second, one needs to be sure that photo--production can identify
resonances also when pion scattering data are scarce. 
\par
\begin{figure}[h!]
\bc
\epsfig{file=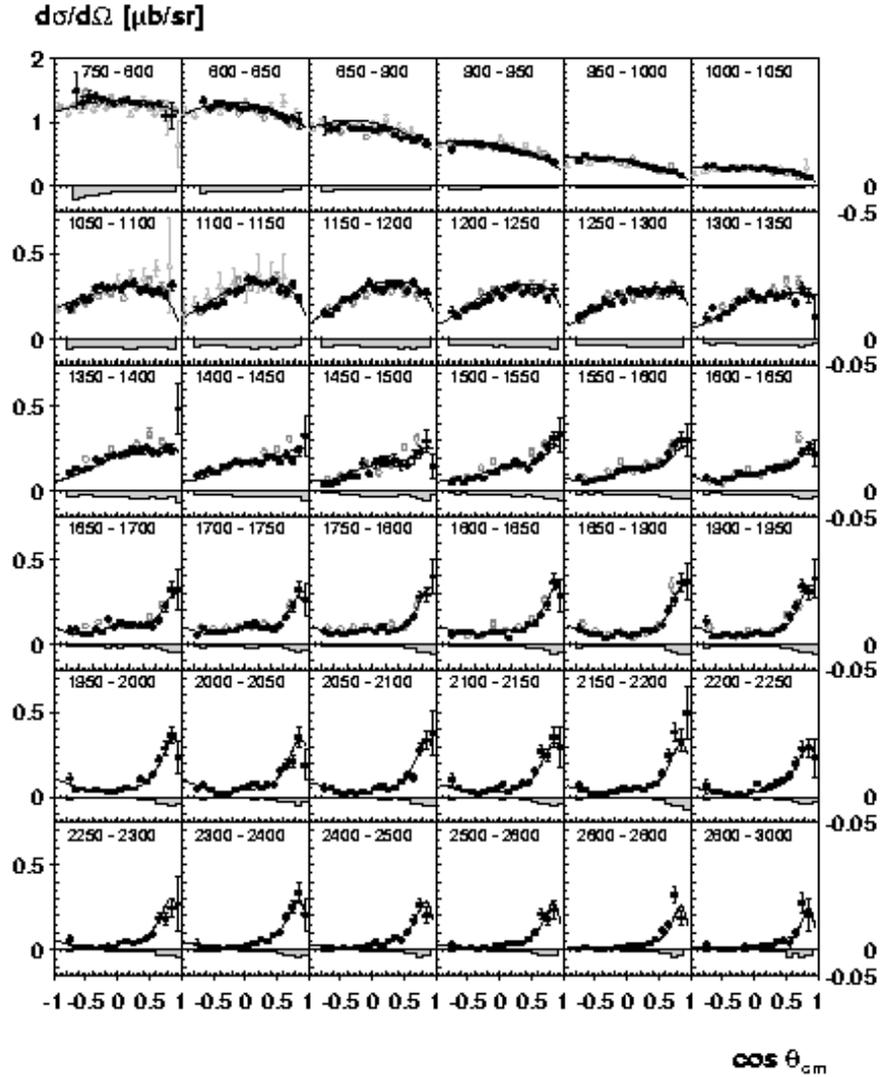,width=\textwidth,clip=on}
\ec
\caption{\label{gammaeta}Differential cross sections for
the reaction $\gamma p\to p\eta$ at 
CB-ELSA~\protect\cite{Crede:2003ax}.
}
\end{figure}

At the Bonn electron accelerator ELSA, data on photoproduction of
$\pi^0$ and $\eta$ mesons were taken with the
Crystal Barrel detector (used before at LEAR). Figure~\ref{gammaeta}
shows those for  $\eta$ production~\cite{Crede:2003ax}.
These data, plus GRAAL data on the beam asymmetry of photoproduction of
$\pi^0$ and $\eta$ mesons, were fitted with an isobar
model. The results of the fits which are still preliminary, are
collected in table~\ref{aaasres}.

\begin{table}[h]
\caption{\label{aaasres}Masses and widths of 
N$^*$ and $\Delta^*$ resonances as determined from data on photoproduction
of $\pi^0$ and $\eta$ mesons, and comparison with PDG 
values~\protect\cite{Hagiwara:fs}.}
\bc
\renewcommand{\arraystretch}{1.3}
\vspace*{43mm}
\begin{center}
\begin{rotate}{35}
\hspace*{-35mm}\Huge{P r e l i m i n a r y}
\end{rotate}
\end{center}
\vspace*{-45mm}
\begin{tabular}{lcccc}
\hline\hline
\qquad N$ ^*$     &  M (MeV)   &$ \Gamma$ (MeV)& PDG mass &  PDG width \\
\hline
$ {\rm N(1520)D_{13}}$ & $ \sim 1530$  & 
$ \sim 110$ & $ 1520$ & $ 120^{+15}_{-10}$  \\ 
$ {\rm N(1535)S_{11}}$& $ \sim 1511$ & $ \sim 170$ & $ 1505\pm10$ & $ 170\pm 80$ \\
$ {\rm N(1650)S_{11}}$& $ \sim 1636$ & $ \sim 180$ & $ 1660\pm20$ & $ 160\pm10$  \\
$ {\rm N(1675)D_{15}}$& $ \sim 1650$ & $\sim 140$ & $ 1670-1685$ & $ 140-180$\\ 
$ {\rm N(1680)F_{15}}$& $\sim  1670$ & $ \sim  100$  & $ 1680^{+10}_{-5}$ & $ 130\pm10$\\
$ {\rm N(1700)D_{13}}$& $ \sim 1695$ & $ \sim 220$& $ 1650-1750$ & $ 50-150$ \\
$ {\rm N(1720)P_{13}}$& $ \sim 1735$ & $ \sim 250$ & $ 1720^{+30}_{-70}$ & $ 250\pm50$ \\
$ {\rm N(1900)P_{13}}$& $ \sim 1920$ & $ \sim 230$ & $ \sim 1900$ & $ \sim 500$  \\
$ {\rm N(2070)D_{15}}$& $\sim 2070$ & $ \sim 335$ &    -        &    -   \\
\hline
$ {\rm \Delta(1232)P_{33}}$& $\sim 1234$ & $\sim 120$ & $ 1232\pm 2$ & $ 120\pm5$ \\
$ {\rm \Delta(1600)P_{33}}$& $\sim 1585$ & $\sim 120$ & $ 1232\pm 2$ & $ 120\pm5$ \\
$ {\rm \Delta(1700)D_{33}}$& $ \sim1700$ & $\sim 210$     & $ 1670-1770$ & $ 200-400$ \\ 
$ {\rm \Delta(1750)P_{31}}$& $ \sim 1710$ & $ \sim 250$  & $ \sim 1750$ & $ \sim 300$  \\
$ {\rm \Delta(1905)F_{35}}$& $ \sim 1870$ & $\sim  280$  & $ 1870-1920$ & $ 280-440$  \\
$ {\rm \Delta(1920)P_{33}}$& $ \sim 1960$ & $\sim 230$  & $ 1900-1970$ & $ 150-300$ \\
$ {\rm \Delta(1950)F_{37}}$& $ \sim 1940$ & $ \sim 260$  & $ 1940-1960$ &
$ 290-350$ \\
$ {\rm \Delta(2260)P_{33}}$& $ \sim 2260$ &  $\sim 400$  & -  & - \\
\hline\hline
\end{tabular}
\renewcommand{\arraystretch}{1.0}
\ec
\end{table}
\par
The agreement between the results from photoproduction with
those of the PDG, mostly $\pi$N elastic scattering, is remarkable.
Two new resonances are suggested.  
\par
\begin{figure}[h!]
\bc
\epsfig{file=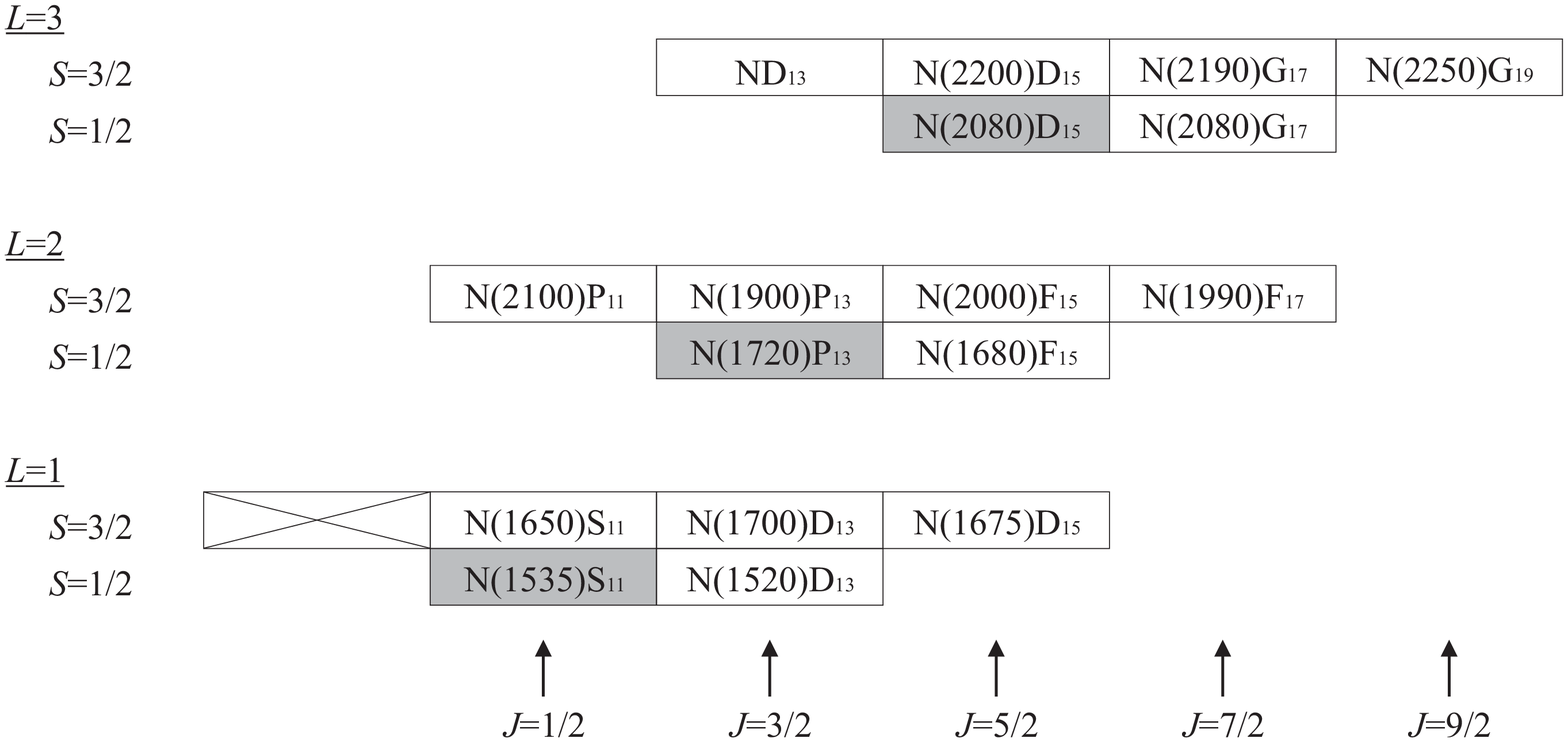,width=\textwidth,clip=on}
\ec
\caption{\label{eta_res}Low--lying baryon multiplets.
The dominant contributions are orbital angular
momentum excitations $\ell = 1, 2, 3$ where the intrinsic(quark)
orbital angular momentum $\vec\ell$ couples to the total
quark spin, $s=1/2$ or $s=3/2$, to a doublet or quartet of states.
The states with $J = \ell - s, s=1/2$ couple strongly to N$\eta$.
There is no explanation so far for this regularity.
}
\end{figure}

One of the new resonances is the $\rm N(2070)D_{15}$. Surprisingly,
the evidence for this resonance comes from the $\eta$ data of
figure~\ref{gammaeta}. Other resonances which are seen to contribute
strongly to figure~\ref{gammaeta} are the  $\rm N(1535)S_{11}$
and the  $\rm N(1730)P_{13}$. These three resonances decay 
into N$\eta$ with orbital angular momenta $L = 0, 1, 2$. Their
possible spectroscopic assignments are depicted in figure~\ref{eta_res}.

\par
\begin{figure}[h!]
\begin{tabular}{ccc}
\epsfig{file=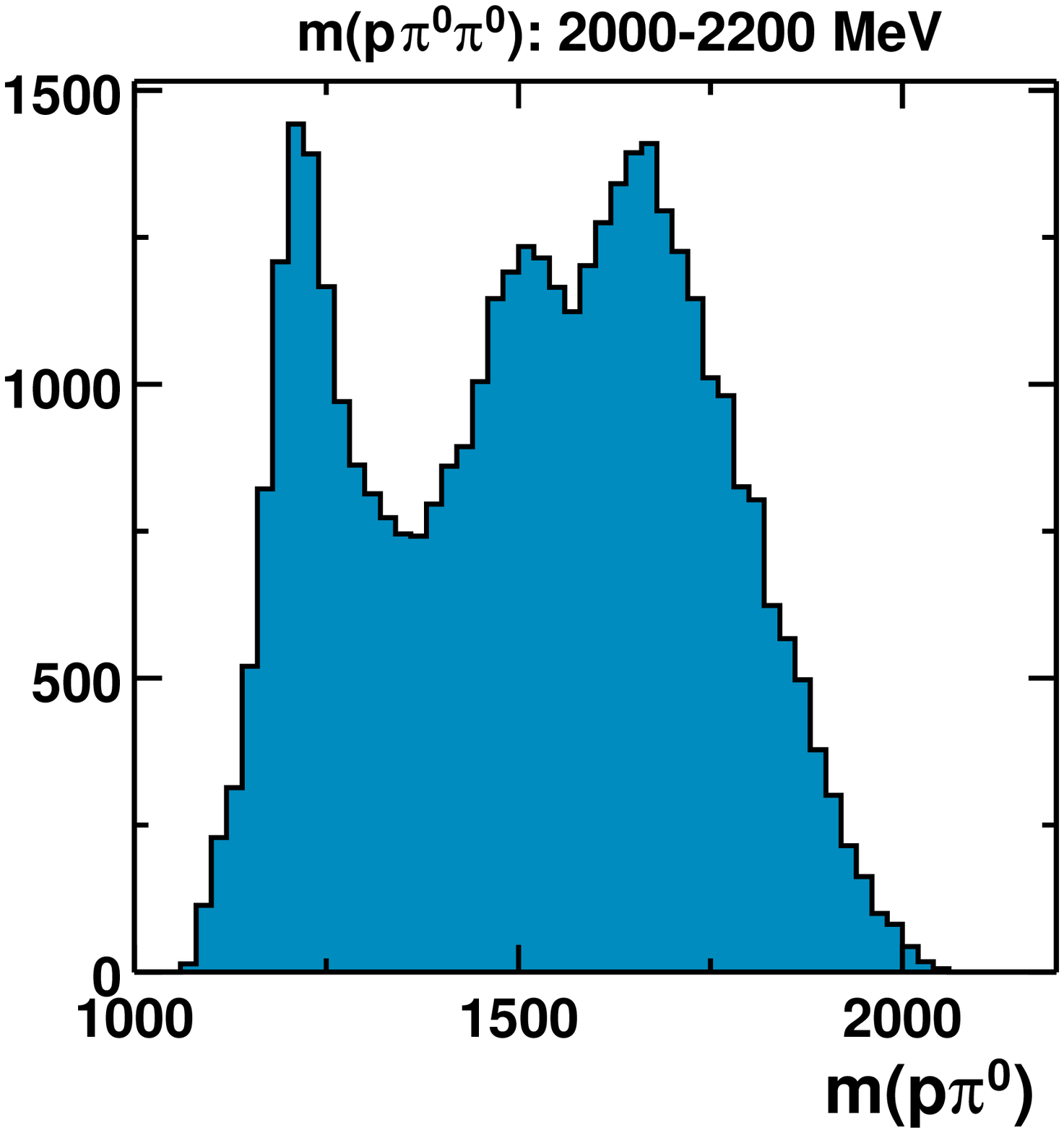,width=0.33\textwidth}&
\epsfig{file=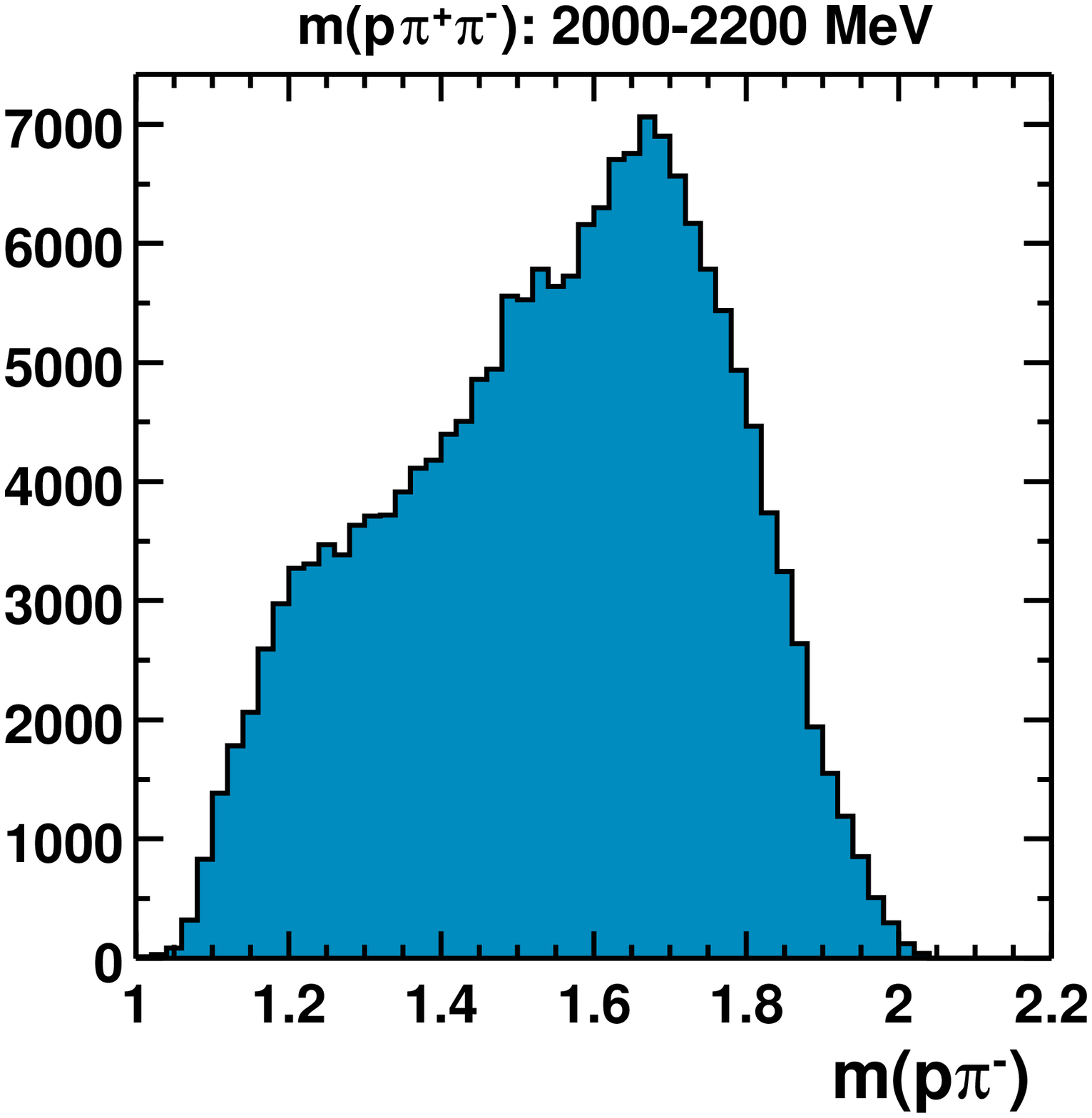,width=0.33\textwidth}&
\epsfig{file=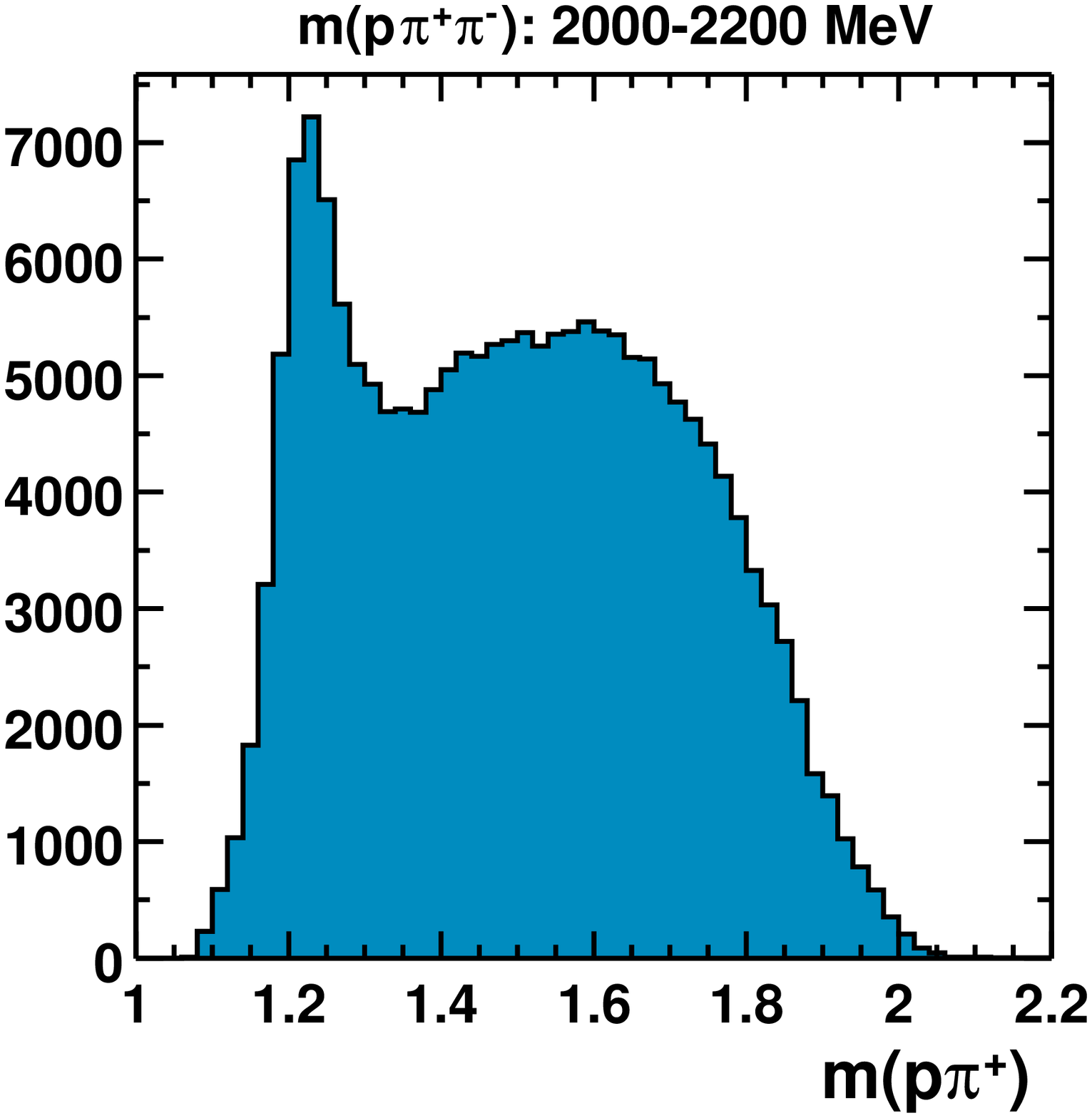,width=0.33\textwidth}
\end{tabular}
\caption{\label{twopi}The $p\pi$ invariant mass distributions 
for the reaction $\gamma p\to p 2\pi$. Left:
CB-ELSA data on $\gamma p\to p 2\pi^0$. Plotted is the
$p\pi^0$ invariant mass. Then CLAS
data on $\gamma p\to p \pi^+\pi^-$. Shown are the $p\pi^-$ invariant 
(center) and $p\pi^+$ (right) invariant mass. The total mass 
( or $\sqrt s$) was restricted to 2000--2200\,MeV. 
}
\end{figure}

Cascades of baryon resonances can already been seen 
in data on $\gamma p\to p 2\pi^0$. Figure~\ref{twopi}
shows the $p\pi^0$ invariant mass distribution for events in
which the total mass falls into the 2000--2200\,MeV mass range.
Clearly, high mass baryon resonance cascade down via intermediate
resonances. Often to the $\Delta(1232)$, see figure~\ref{twopi},
left and right panels, but also via the $\rm N(1520)D_{13}$. 
This cascade can be seen in the left panel of figure~\ref{twopi}.
Thus cascade processes open a new chance to investigate the 
spectrum of baryon resonances. 
\subsection{\label{section5.5}Pentaquarks}
An exotic baryon with positive strangeness $S=+1$ was found by
Nakano and collaborators at LEPS in Japan~\cite{Nakano:2003qx}, 
and has since then been
confirmed by several other experiments. As resonance with
positive strangeness it must contain an $\bar s$ quark; to
maintain a baryon number one, the minimum quark model
configuration of the $\Theta^+(1540)$ requires five quarks, it is
called pentaquark. The resonance was found to have a mass of $\sim
1540$ MeV and a narrow width $\leq 10$ MeV.  The name
$\Theta^+(1540)$ was adopted.

\subsubsection{LEPS experiment}
Nakano {\it et al.}~\cite{Nakano:2003qx}
observed the $\Theta^+(1540)$ in a study of
photo-production off neutrons in the reaction \be \rm\gamma n
\to K^+K^-n \ee using neutrons in carbon nuclei of a plastic
scintillator. The primary aim of the experiment was the study of
$\Phi$ photo--production of protons using a liquid H$_2$ target.
\par
The experiment was performed at the Laser Electron Photon facility
at SPring8 (LEPS) in Japan. This apparatus produces high-energy
photons by Compton back-scattering of laser photons off a 8 GeV
electron beam in the SPring-8 storage ring. Using a 351 nm Ar
laser, photons with a maximum energy of 2.4 GeV were produced. The
scattered electrons were momentum- analyzed by a bending magnet
and detected by a tagging counter inside the ring; this allowed
the photon energy to be determined with a resolution of 15 MeV.
The flux of tagged photons in the energy range from 1.5 to 2.4\,GeV 
was $10^6$/s. Charged particles were tracked through a
magnetic field, electrons and positrons were vetoed by an aerogel
Cerenkov counter, Kaons were identified in a time--of--flight
system.
\par
\begin{figure}[h!]
\bc
\epsfig{file=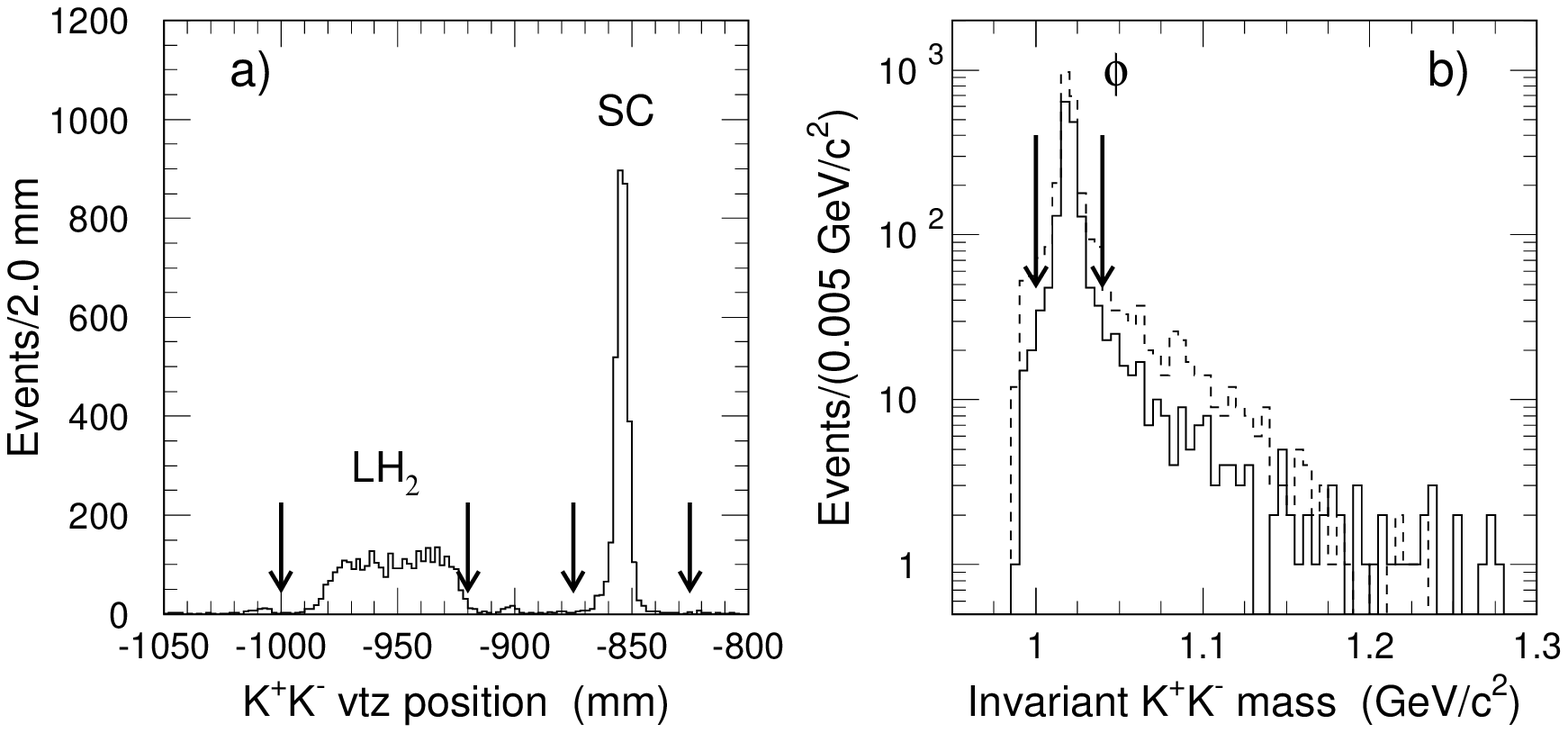,width=9cm,clip=on}
\ec
\caption{\label{leps1}The LEPS experiment: a, the vertex distribution and
cuts to select events produced in the scintillator and in H$_2$.
b, the $\rm K^+K^-$ invariant mass distribution for events in
H$_2$ (dashed line) and in the scintillator 
(solid line)~\protect\cite{Nakano:2003qx}.
}
\bc 
\epsfig{file=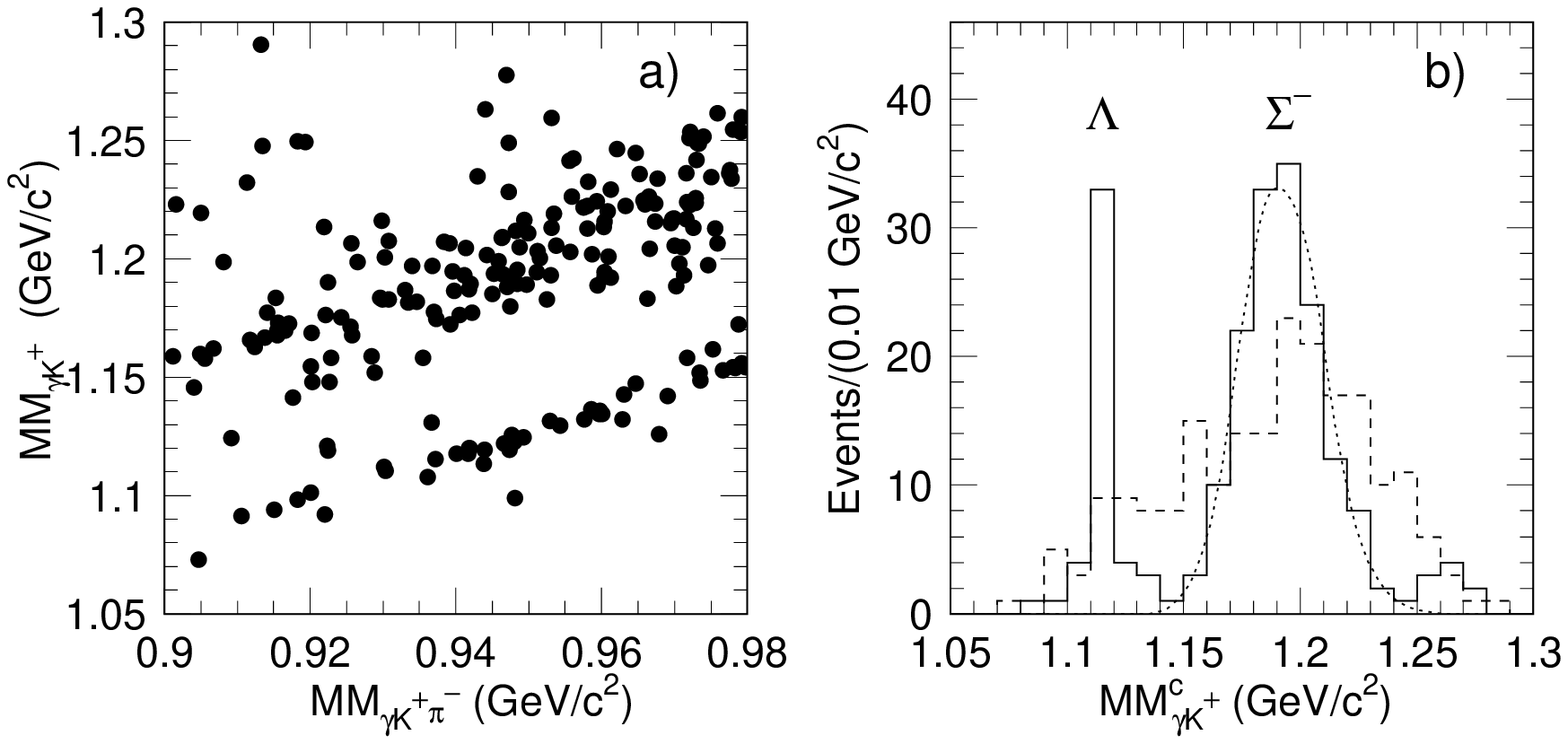,width=9cm,clip=on} 
\ec 
\caption{\label{leps2}The
LEPS experiment: a, the scatter--plot $MM_{\rm\gamma K^{+}}$
versus $MM_{\rm\gamma K^+\pi^-}$  shows the effect of the Fermi
motion. b, the corrected missing mass distribution
$MM^c_{\rm\gamma K^{\pm}}$ (solid line) shows a clear $\Sigma^-$
hardly seen in the uncorrected spectrum (dashed line).
The dotted line shows Monte Carlo simulations of the 
$\Sigma^-$~\protect\cite{Nakano:2003qx}.
}
\bc
\epsfig{file=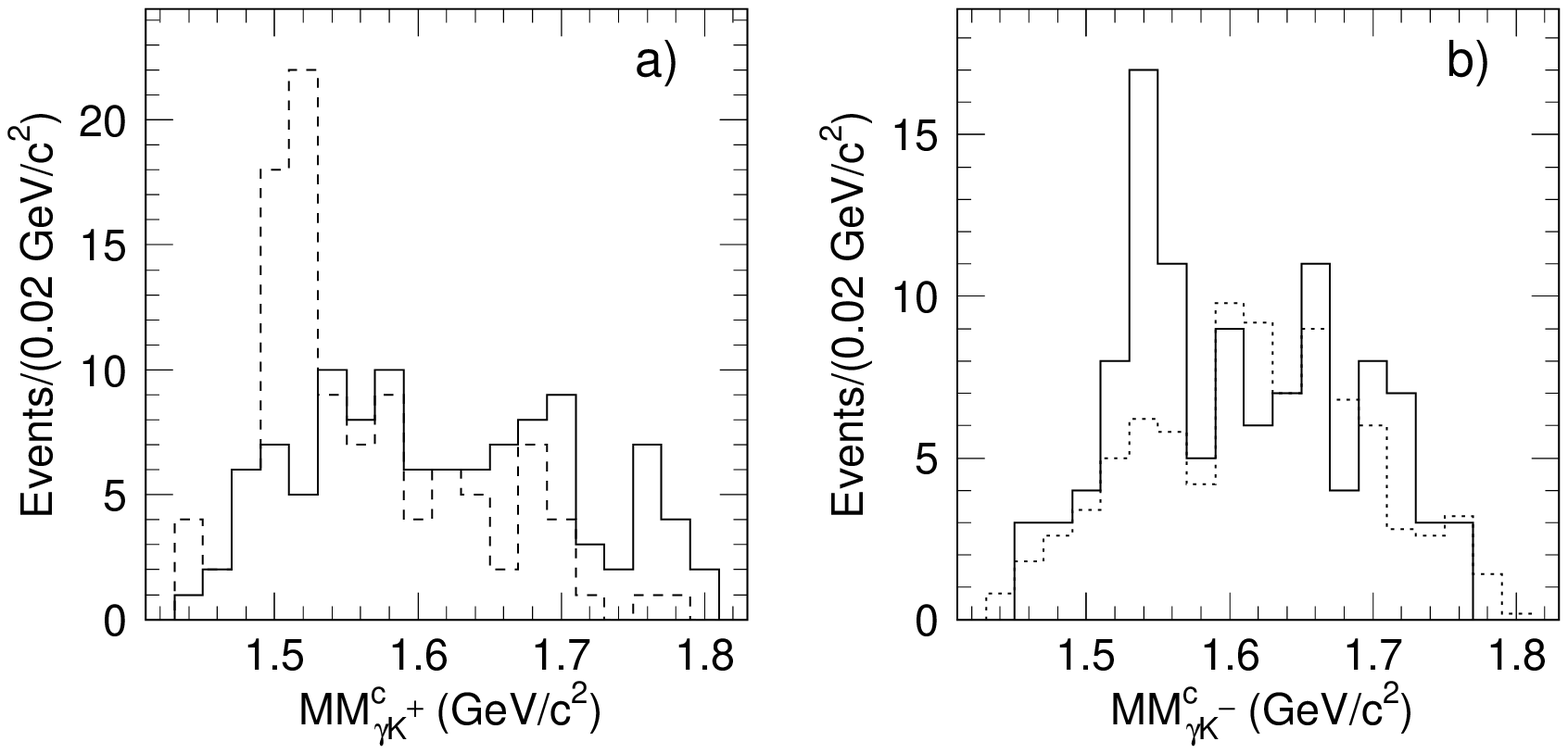,width=9cm,clip=on}
\ec
\caption{\label{leps3}
The LEPS experiment, corrected missing mass distributions
for $\rm K^+K^-$ production.
a, $MM^c_{\rm\gamma K^{+}}$ for data on H$_2$ (dashed) showing the
$\Lambda(1520)$ and on Carbon with a detected proton (solid).
b,  $MM^c_{\rm\gamma K^{-}}$ from H$_2$ (dashed) and on Carbon
(solid). The latter peak is assigned to reaction $\gamma n\to
\Theta^+\rm K^-; \Theta^+\to nK^+$~\protect\cite{Nakano:2003qx}.}
\end{figure}
\par
The photon energy was known from the tagging and the target nucleon
was assumed to be at rest with the mean nucleon rest mass; hence from the
momentum of the kaon pair, the momentum and direction of the final
state nucleon could be calculated. The silicon-strip detector 
(SSD) was able to
detect protons, but is blind to neutrons.
Figure~\ref{leps1} shows the vertex distribution and the $\rm K^+K^-$
invariant mass distribution for events where a proton was identified.
\par
In a next step, only those events were retained
where the calculated nucleon momentum and direction would lead to
a hit in SSD but where no matching proton was found. From this sample,
the reactions $\rm\gamma n \to n K^+\pi^-$ and $\rm\gamma n \to n K^+K^-$
were identified. The former reaction proceeds via the intermediate
state $\rm K^+\Sigma^-$. Both, the $\Sigma^-$ and n mass can be
determined from the missing masses recoiling against the $\rm\gamma
K^+$ or  $\rm\gamma K^+\pi^-$, respectively. The two masses are both
smeared out by the Fermi motion, the missing mass corrected for the
Fermi motion is calculated as
\be
MM^c_{\rm\gamma K^{\pm}} = MM_{\rm\gamma K^{\pm}} - MM^c_{\rm\gamma
K^+K^-} + M_N.
\ee
\par
Nakano {\it et al.} looked for the $\Theta^+$ in the $\rm K^-$
missing mass distribution again corrected for the neutron Fermi
momentum (see figure~\ref{leps3}b). A peak was seen at 1540 MeV.
The statistical significance
of this peak over the background was determined to 
4.6$\sigma$. The peak width of 25\,MeV was consistent with the experimental
resolution, and is only an upper boundary on the true decay width of the state.
There is no peak in the corresponding
data from the H$_2$ target.
\subsubsection{The SAPHIR experiment}
The $\Theta^+$ has also been observed in photo--production from
protons at the Bonn ELectron Stretcher Accelerator 
(ELSA)~\cite{Barth:ja}.  The
reaction: 
\be 
\rm\gamma p \to nK^+K^0_s 
\ee 
was studied with the
SAPHIR detector, a large acceptance magnetic spectrometer. The
detector has full coverage in the forward direction; thus particles
can be detected (and their momentum be measured) even under zero
degrees, however, the photon flux is limited to 
$\sim 10^6\gamma$/s. Photons were produced via bremsstrahlung of the
\begin{figure}[b!]
\bc
\epsfig{file=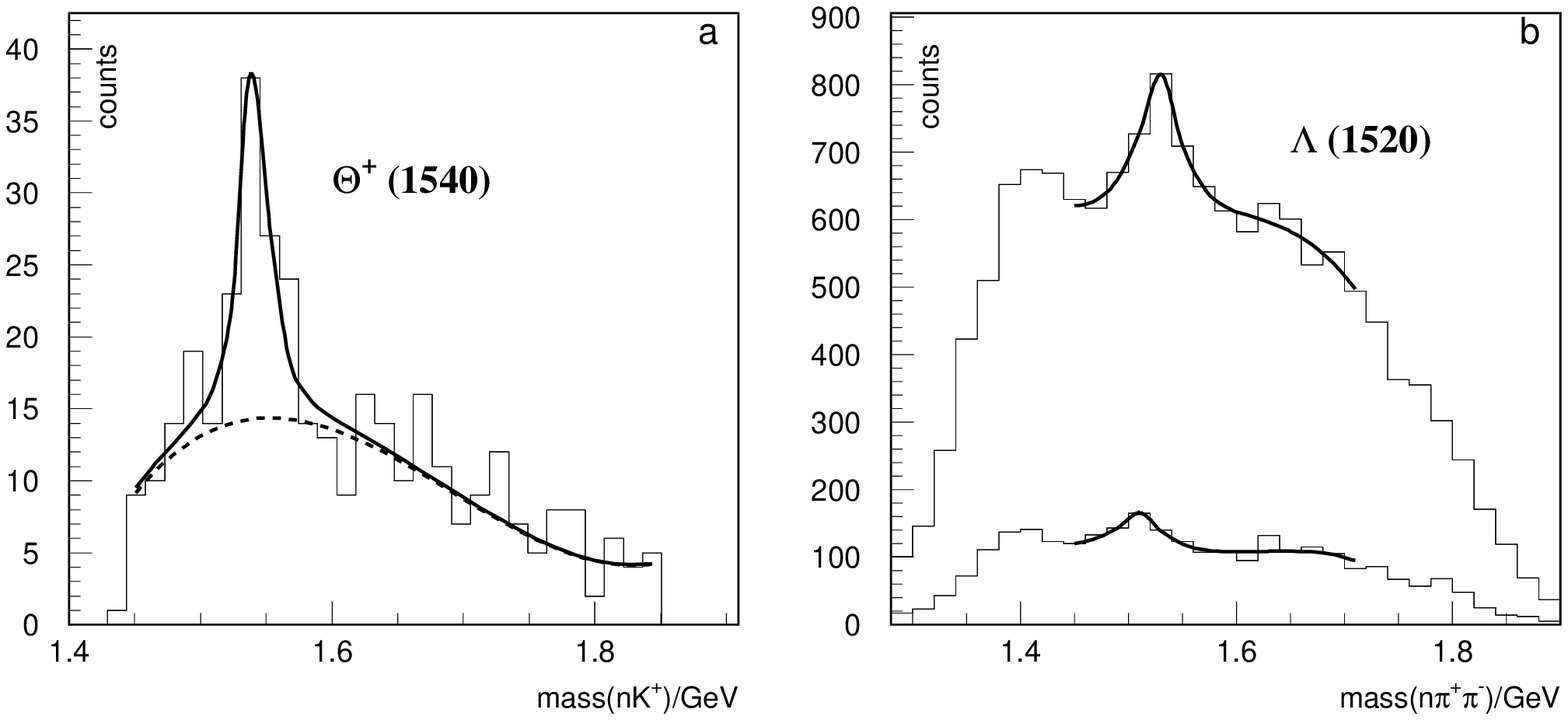,width=11cm}\\
\epsfig{file=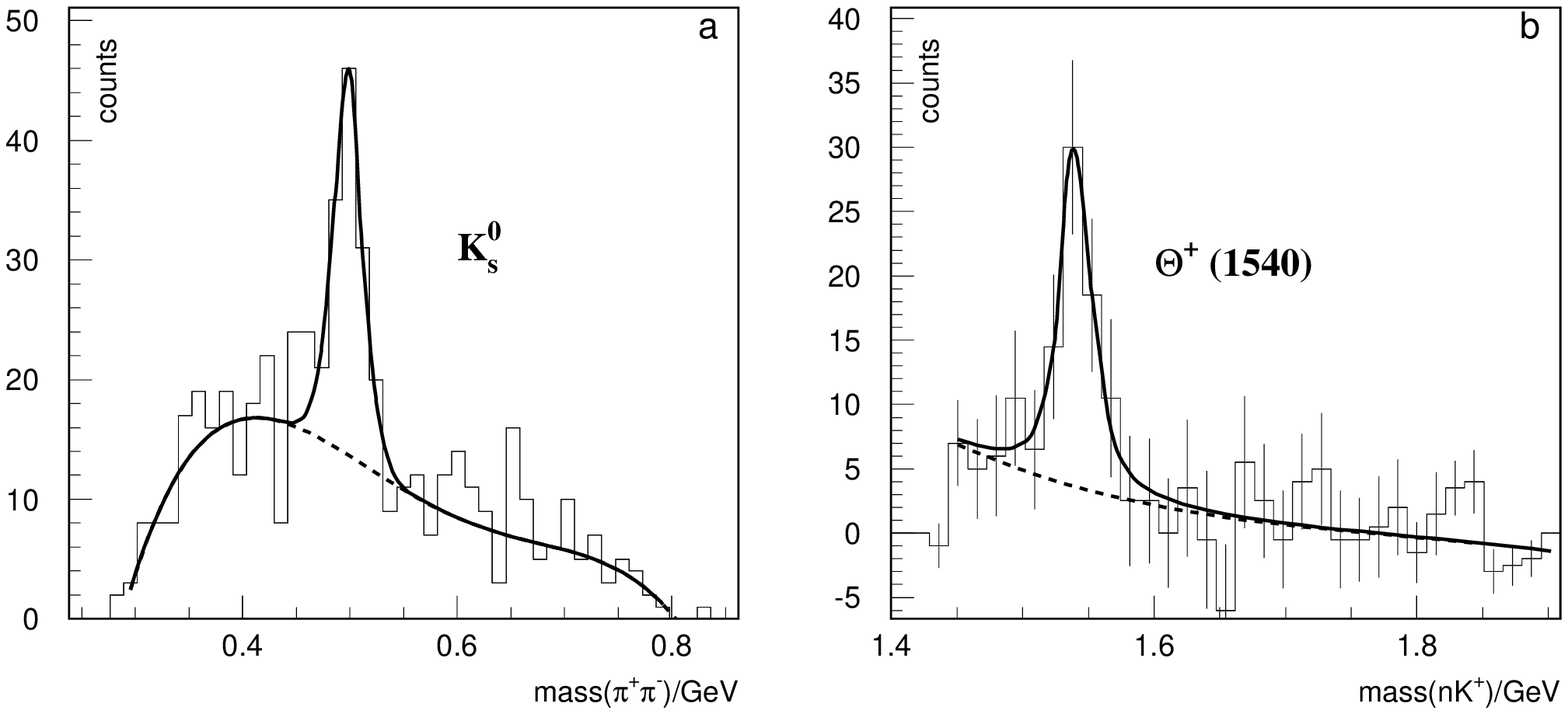,width=11cm}
\ec
\caption{Top: the $n\rm K^+$ and $n\rm K^0_s$ invariant mass distributions.
Bottom: The $\pi^+\pi^-$ and $n\rm K^+$ invariant mass distributions
after side--bin subtraction of the background under the $\rm K^0_s$ or
$\Theta^+(1540)$, respectively. From~\protect\cite{Barth:ja}.
}
\label{saphir1}
\end{figure}
ELSA electron beam in a copper foil radiator and were tagged with
energies from 31\% to 94\% of the incident electron energy, which
was 2.8 GeV for the data shown. Liquid hydrogen was used as the
target. $\rm K^0_s$  were reconstructed from their $\pi^+\pi^-$
decay, and the neutron momentum was obtained from energy and
momentum conservation.
\par
A series of kinematical fits was applied to suppress background from
competing reactions. Figure~\ref{saphir1} shows the resulting
$n\rm K^+$ and $n\rm K^0_s$ invariant mass distributions with clear
peaks at 1542\,MeV and at the $\Lambda(1520)$ invariant mass.
The $n\rm K^+$ mass spectrum was obtained after a cut in the $\rm K^0_s$
production angle $cos\theta_{\rm K^0_s}>0.5$ which reduces the
background by a factor four and the signal by about a factor two.
\par
The $\Theta^+$ was seen as a peak in the $\rm nK^+$
invariant mass distribution, with a
statistical significance of 4.8$\sigma$.
The mass was measured as $1540\pm 4\pm 2$\, MeV, and
the width determined to be $\leq$25,MeV, at a 90\% confidence level.
\par
The correlation between  $\rm K^0_s$ and the peak in the
$n\rm K^+$ invariant mass distribution can be seen when the background
under the  $\rm K^0_s$ and the $\Theta^+$, estimated from side bins,
is subtracted (figure~\ref{saphir1}). The $\Theta^+$ is now seen above
very few events, probably due to $\Lambda(1520)\rm K^+$ production.
\par
\begin{figure}[h!]
\begin{minipage}[b]{0.60\textwidth}
\epsfig{file=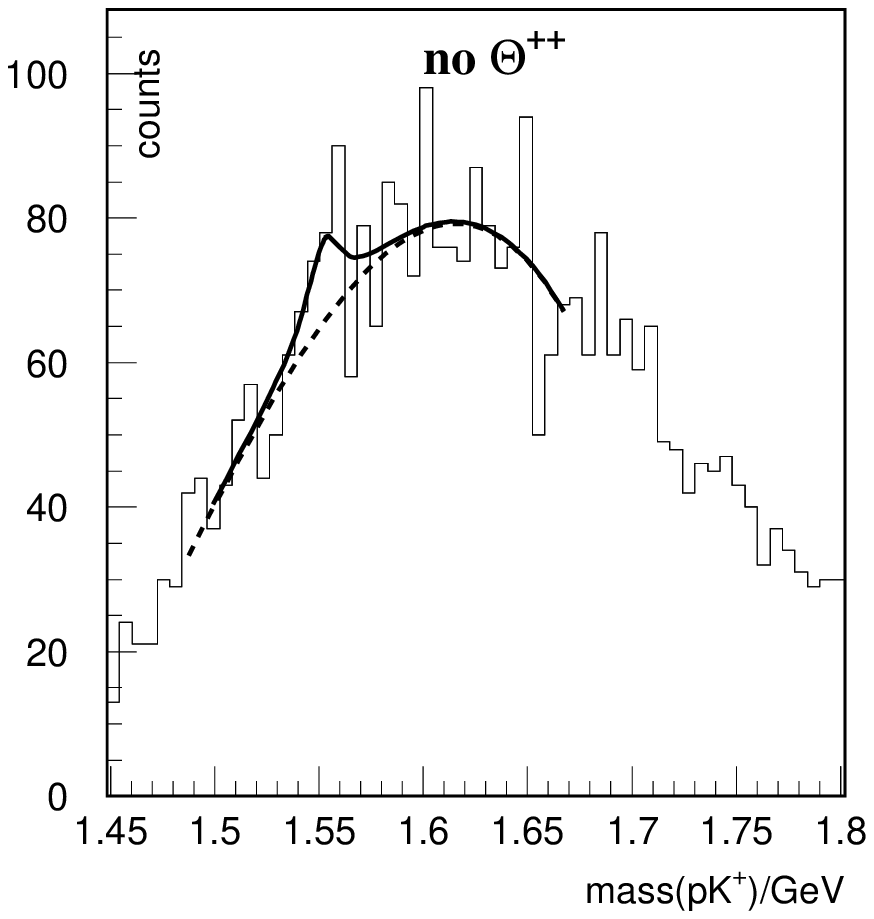,width=0.9\textwidth}
\caption{\label{isosp}
The $p\rm K^+$ invariant mass distribution from
reaction~(\protect\ref{plusplus}) measured at 
SAPHIR~\protect\cite{Barth:ja}.
}
\end{minipage}
\begin{minipage}[c]{0.28\textwidth}
\epsfig{file=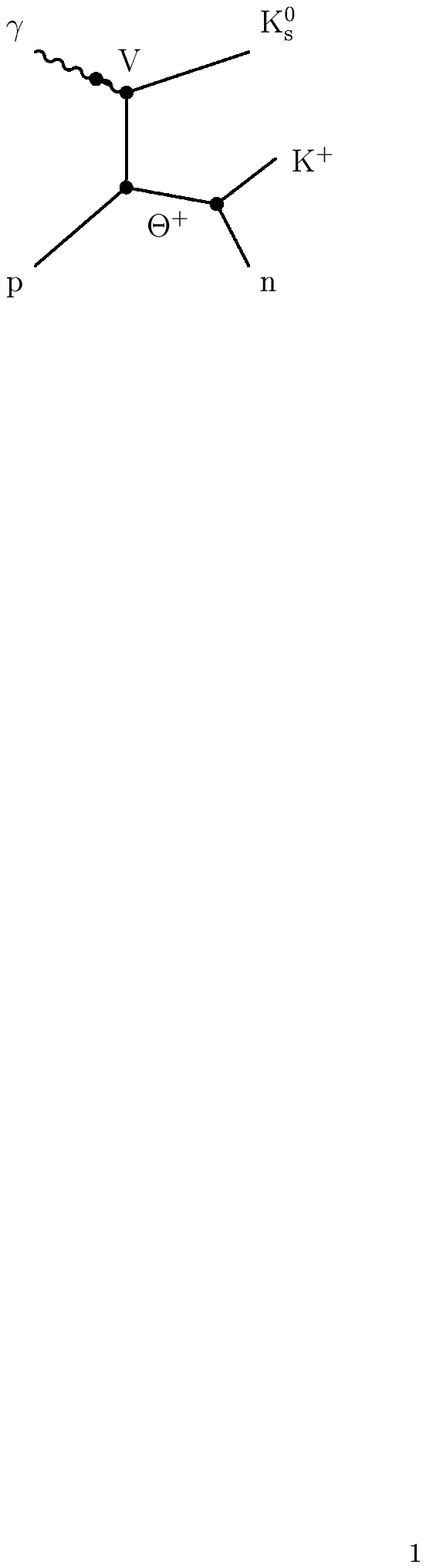,width=1.1\textwidth} \\
\epsfig{file=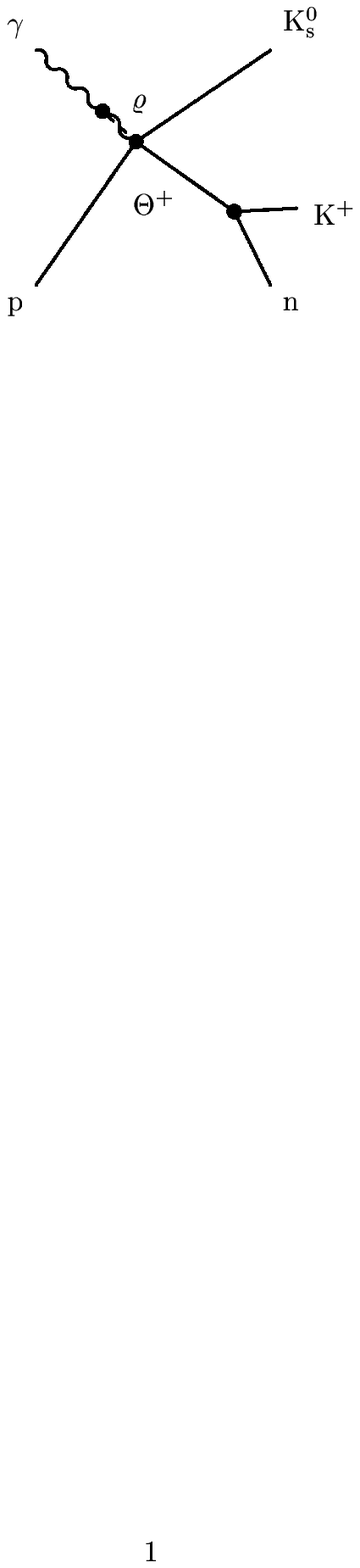,width=1.1\textwidth}
\end{minipage}
\vspace*{-80mm}
\caption{\label{diagr3}
Diagrams which could contribute to $\Theta^+(1540)$
production. In the upper diagram, K$^0$ exchange requires
SU(3) symmetry breaking, K$^*$ exchange could be allowed but is
suppressed due to its higher mass. A isotensor resonance
$\Theta^+(1540)$ is only produced via contact interactions.
}
\end{figure}

The SAPHIR collaboration also searched for an isospin-partner of the
$\Theta^+$, the doubly-charged $\Theta^{++}$, via the reaction chain:
\be
\rm\gamma p \to\Theta^{++}K^-\to pK^+K^-
\label{plusplus}
\ee
If the $\Theta^+$ would have isospin 1 or 2, we
should expect a peak in figure~\ref{isosp} with several
1000 entries. From the absence of such a strong
signal in the $\rm pK^+$
invariant mass distribution, they
concluded that the $\Theta^+$ is an isoscalar.

Diagrams which could contribute to $\Theta^+$
production are shown in figure~\ref{diagr3}.
A isotensor resonance is more difficult to produce: the
vector component of the photon plus the proton has to combine in
a four--point vertex into an isotensor plus isodoublet.
\subsubsection{The DIANA experiment}
\begin{figure}[h!]
\begin{tabular}{cc}
\epsfig{file=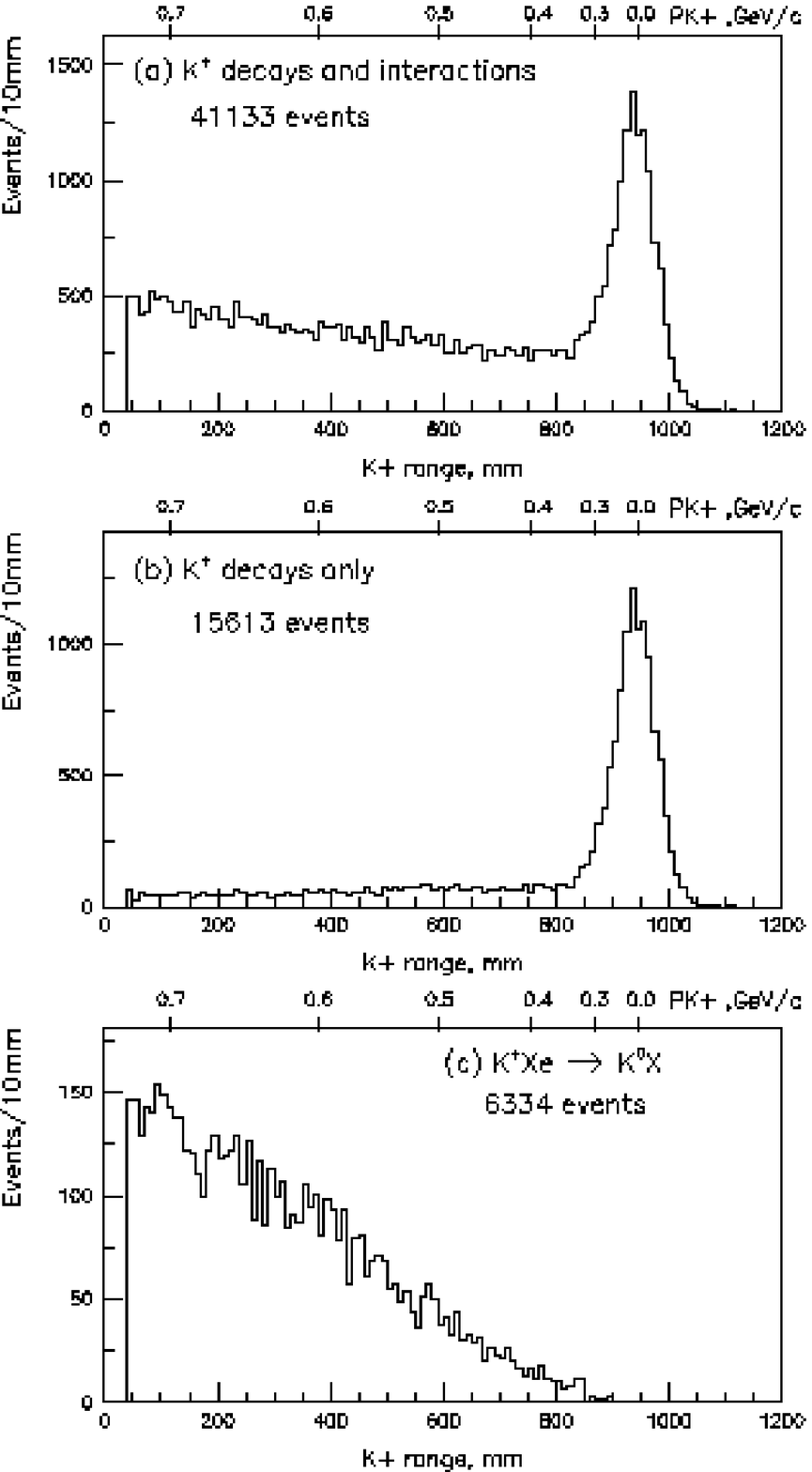,width=0.5\textwidth}&
\vspace*{-2mm}
\epsfig{file=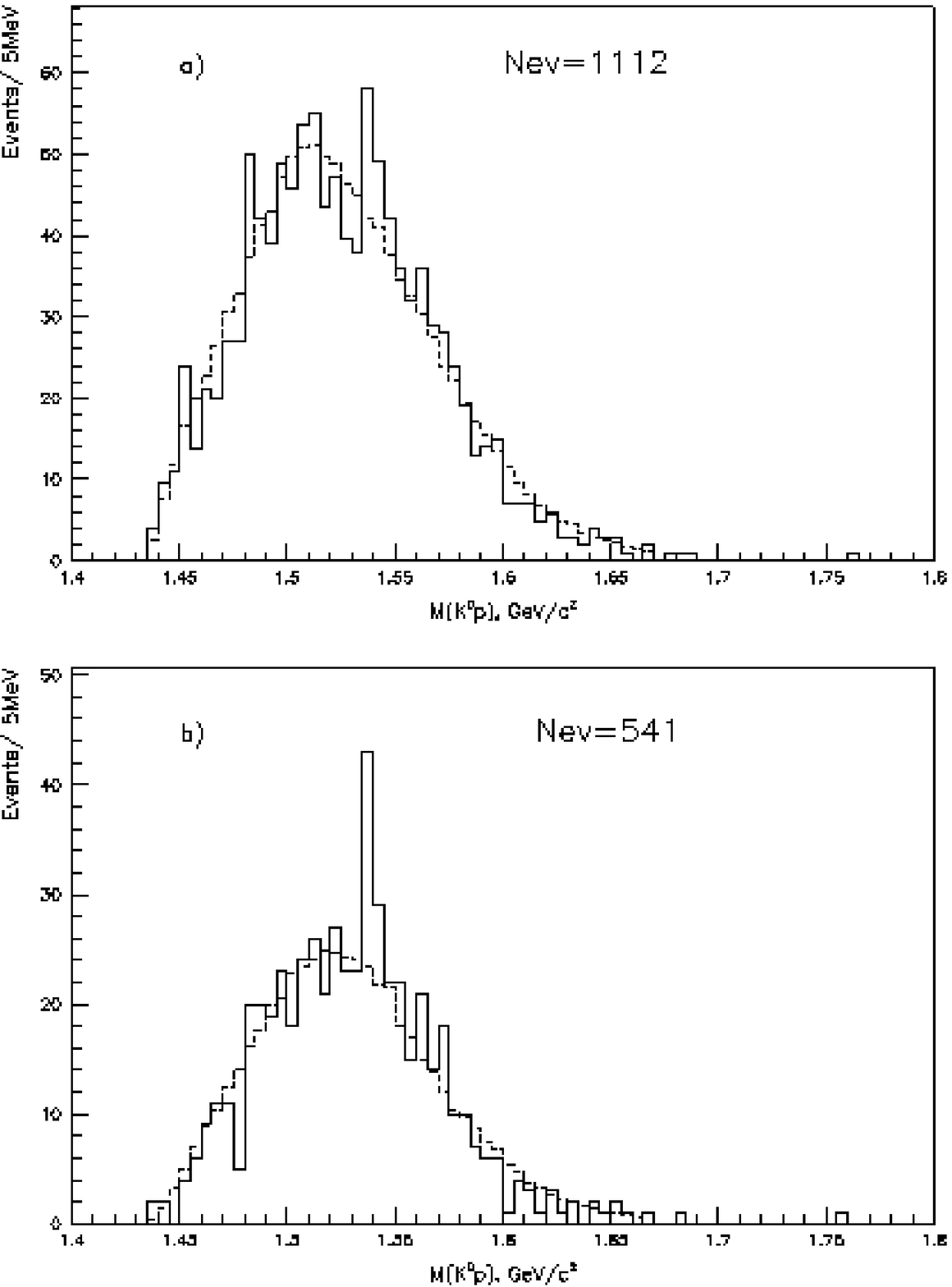,width=0.5\textwidth}
\end{tabular}
\caption{DIANA, left: 850\,MeV $\rm K^+$ entered a Xe bubble chamber.
They may decay in flight, interact with Xe nuclei, or stop at the end
of their range. Right:  $\rm pK^0_s$ invariant mass distribution
for all events and for those in which proton and $\rm K^0_s$
were produced with an angle $\theta<100^{\circ}$ w.r.t.
the beam direction and back-to-back in the transverse 
plane~\protect\cite{Barmin:2003vv}.}
\label{Diana}
\end{figure}

The DIANA collaboration found evidence for the $\Theta^+$
resonance from low-energy interactions of $\rm K^+$ with 
nuclei~\cite{Barmin:2003vv}.
The data had been collected in 1986 and were re-analyzed recently.
A kaon beam of 850 MeV was generated at the ITEP proton synchrotron.
The experiment consisted of a 70x70x140 cm
bubble chamber filled with liquid Xenon.
Beam momentum and target size were chosen to stop the Kaon
beam at the end of their range. Figure~\ref{Diana} shows the
position along the target where the $\rm K^+$ interacted or decayed.

There was no magnetic field. Charged particles were identified by
their ionization tracks and momentum-analyzed by their range in
Xenon. DIANA studied the $\rm \Theta^+\to pK^0_s$ decay of the
$\Theta^+$ by inspecting the $\rm pK^0_s$ invariant mass
distribution in the charge exchange reaction \be \rm K^+ Xe \to
Xe' p K^0_s. \ee The distribution is shown in Figure~\ref{Diana}. An
enhancement is seen at a mass $M= 1539\pm 2$\,MeV and with a width
of $\Gamma <9$\,MeV. The statistical significance was 4.4$\sigma$.
\par
Cahn and Trilling~\cite{Cahn:2003wq} 
related the number of interacting Kaons to the
$\rm K^+n$ cross section. Since the integrated cross section (or total
number of $\Theta^+$) is proportional to the width, they estimated the
$\Theta^+$ width to 1\,MeV.

\subsubsection{The CLAS experiment}
\begin{figure}[h!]
\begin{minipage}[c]{0.65\textwidth}
\epsfig{file=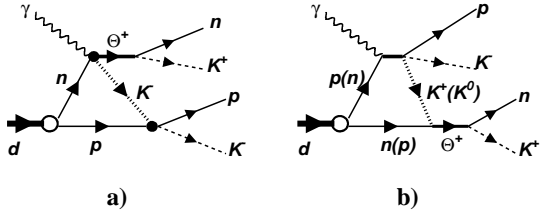,width=1.1\textwidth}
\end{minipage}
\begin{minipage}[c]{0.34\textwidth}
\caption{\label{diagr1}Two rescattering diagrams that could contribute to
$\Theta^+$ production in D$_2$ through final
state interactions~\protect\cite{Stepanyan:2003qr}. The $\Theta^+$
is produced independently of the secondary
scattering. }
\end{minipage}
\vspace*{-25mm}
\end{figure}

\begin{figure}[h!]
\begin{minipage}[c]{0.65\textwidth}
\epsfig{file=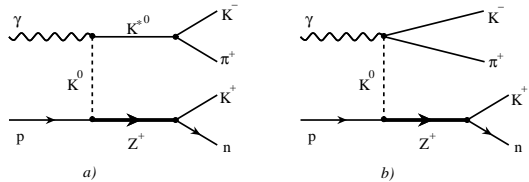,width=1.1\textwidth}
\end{minipage}
\begin{minipage}[c]{0.34\textwidth}
\caption{\label{diagr4}$\Theta^+$ production in H$_2$
via K exchange, with and without production of K$^*$'s.
$Z^+$ is an older name of the 
$\Theta^+$~\protect\cite{Kubarovsky:2003fi}.}
\end{minipage}
\end{figure}

The CLAS (CEBAF Large Acceptance Spectrometer) collaboration at
Jefferson Lab studied photo-production of the $\Theta^+$ using a
H$_2$ and D$_2$ target. In a first experiment liquid deuterium
was used. The photons were produced by an electron beam
incident on a bremsstrahlung radiator and their energies 
were determined from a 
measurement of the energy of the corresponding electrons. 
Charged-particle tracking was performed in  large
acceptance drift chambers, and particle identification used a
Time-of-Flight detector. The CLAS collaboration studied the
reaction~\cite{Stepanyan:2003qr}: 
\be 
\rm\gamma d\to p n K^+K^- 
\label{clasd}
\ee 
using a D$_2$ target and 
reaction~\cite{Kubarovsky:2003fi}
\be 
\rm\gamma p\to n K^+K^-\pi^+. 
\label{clasp}
\ee
with a H$_2$ target. In both cases, the 
neutron was reconstructed from the
kinematics; hence, the momenta of
all participating particles are known.
Figs.~\ref{diagr1} and~\ref{diagr4} show diagrams
which may contribute to $\Theta^+$ production in
reaction (\ref{clasd}) and (\ref{clasp}), respectively.
\vspace*{-10mm}
\par

\begin{figure}[h!]
\begin{minipage}[c]{0.48\textwidth}
\epsfig{file=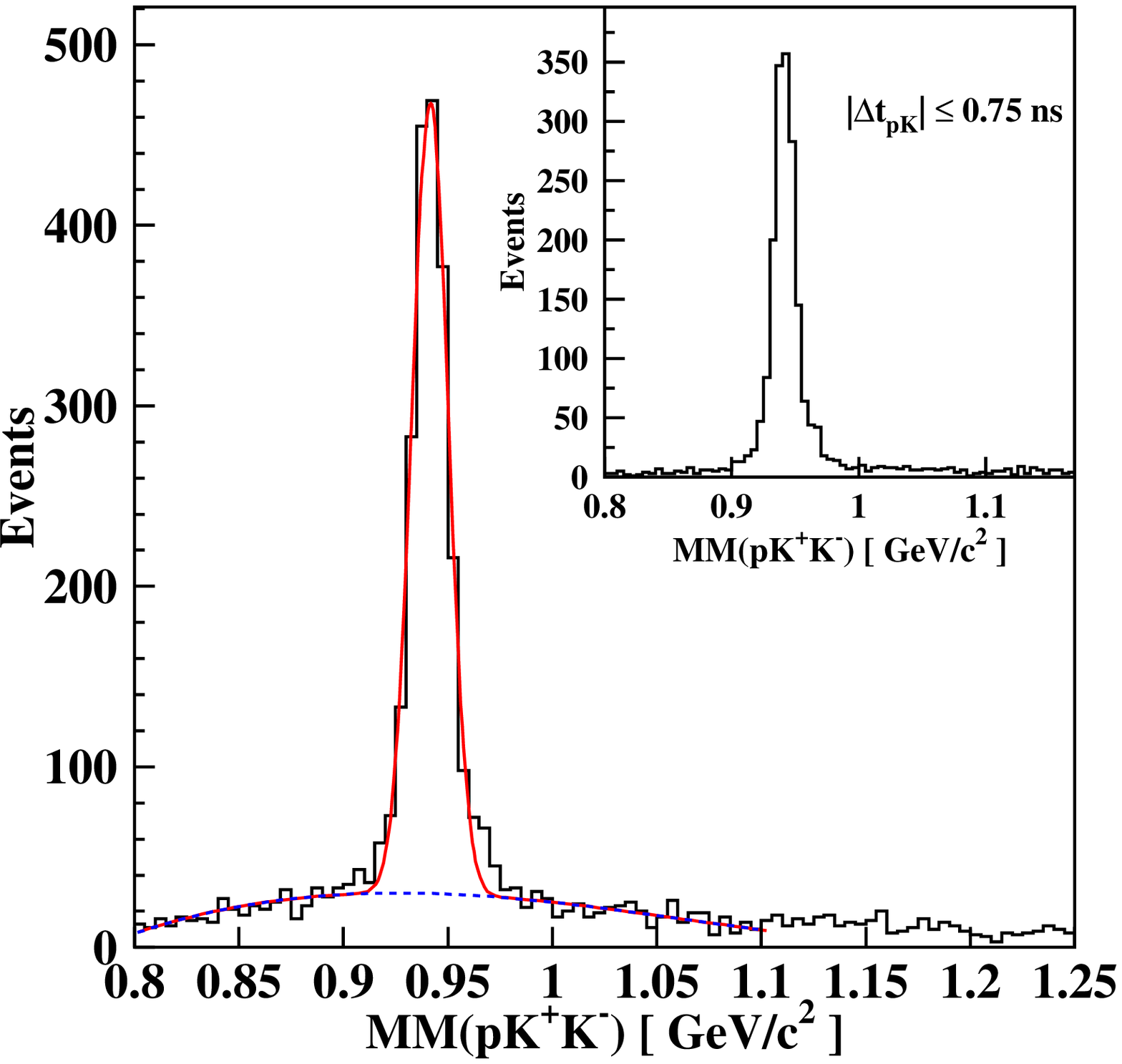,width=1.15\textwidth}
\end{minipage}
\begin{minipage}[c]{0.48\textwidth}
\epsfig{file=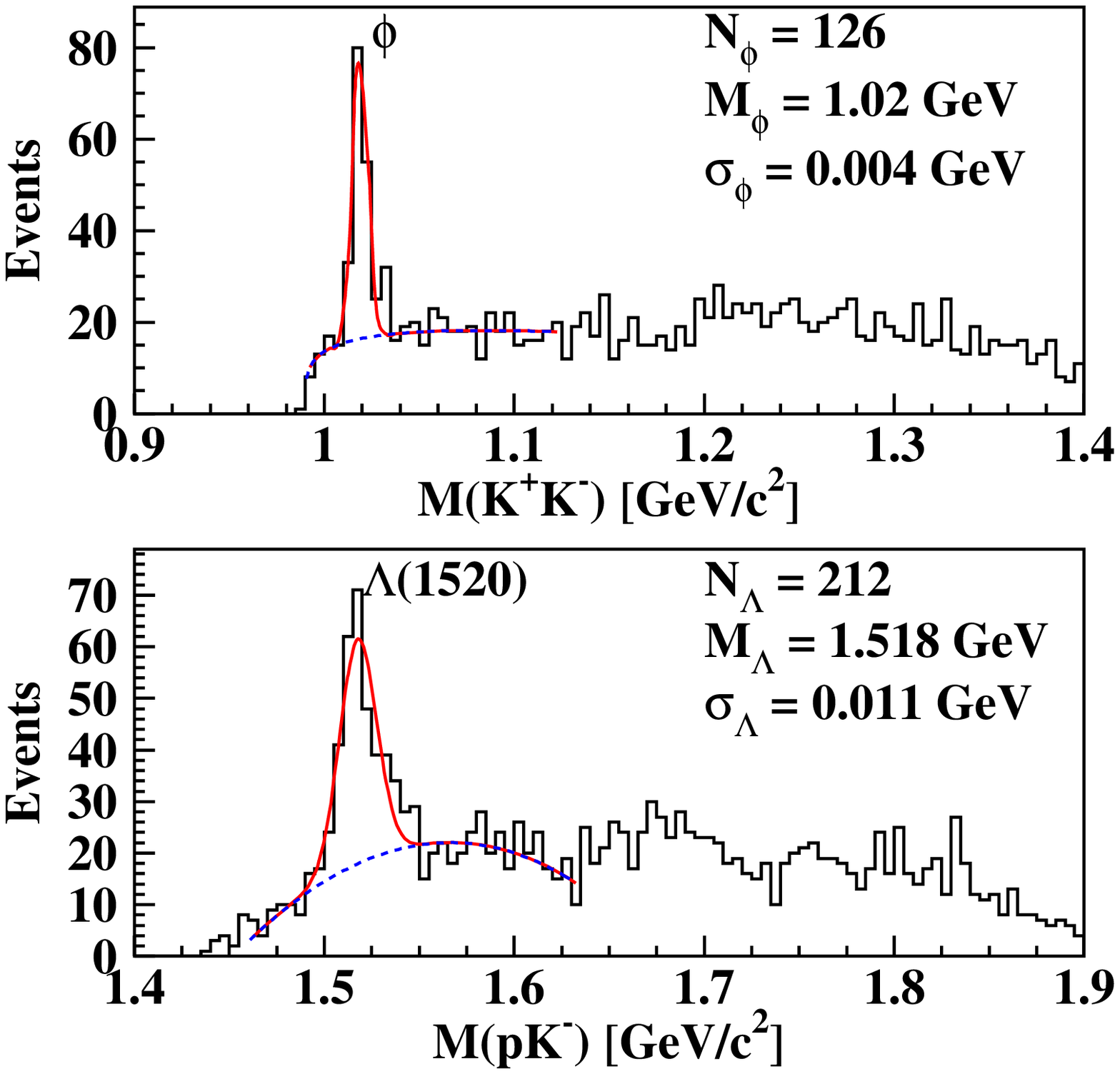,width=1.15\textwidth}
\end{minipage}
\vspace*{-8mm}
\begin{minipage}[c]{0.39\textwidth}
\caption{\label{miss}(Top left) The neutron observed in CLAS as missing mass
in reaction~(\protect\ref{clasd}).}
\caption{\label{philam}(Top right) The $\rm K^+K^-$ 
and $\rm pK^-$ invariant masses.}
\caption{\label{mis}The $\rm nK^+$ invariant mass from 
reaction~(\protect\ref{clasd}) showing evidence for the
$\Theta^+(1540)$.  All three figures are 
from~\protect\cite{Stepanyan:2003qr}.
}
\end{minipage}
\begin{minipage}[c]{0.60\textwidth}
\vspace*{-8mm}
\epsfig{file=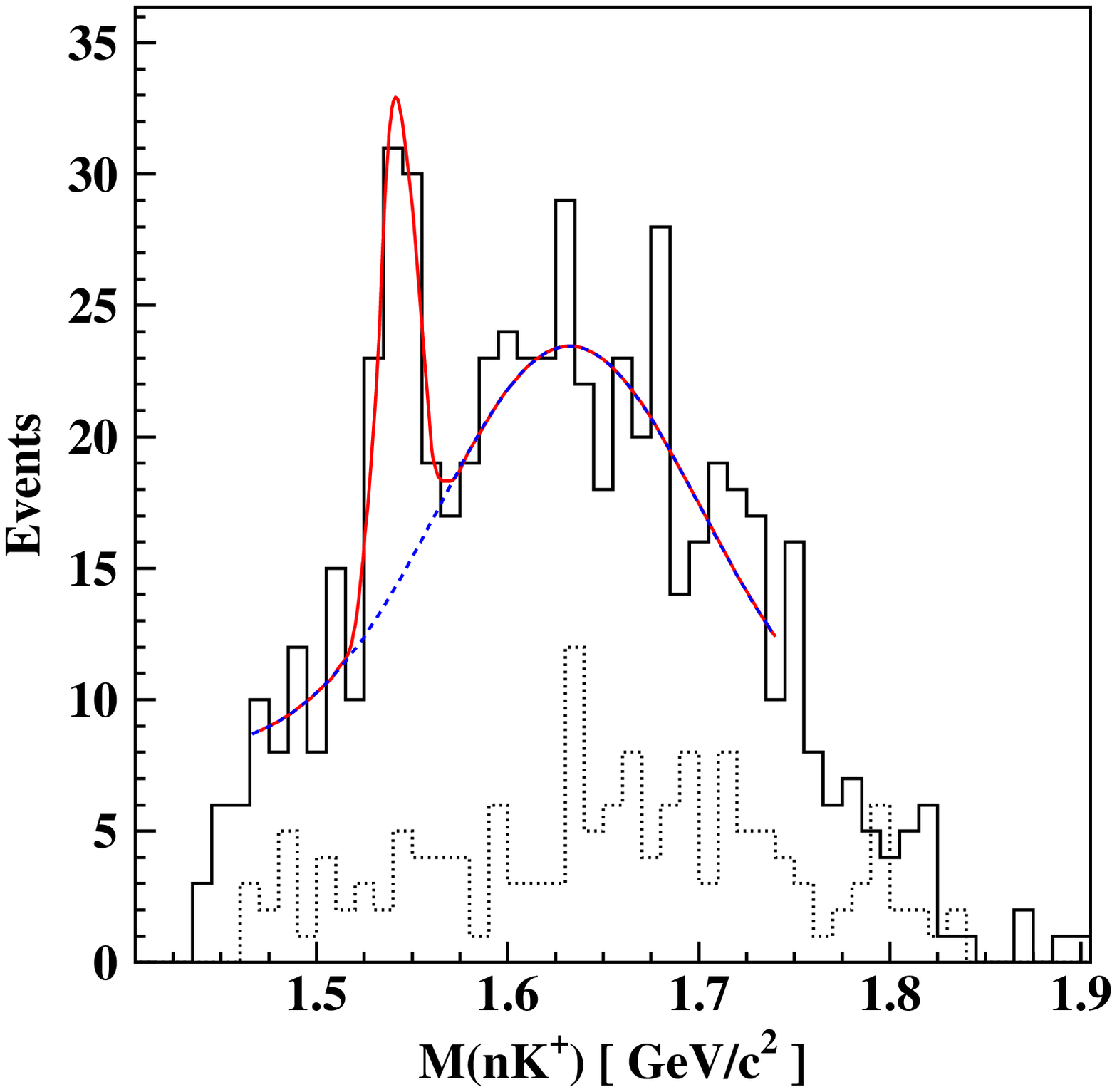,width=1.15\textwidth}\\
\end{minipage}
\end{figure}

Neutrons in the final state were 
reconstructed from the missing momentum and energy.
Events were selected which contained an identified proton, a $\rm
K^+$ and $\rm K^-$ pair, and no other particles. The missing mass
spectrum for the selected events showed a clear peak at the
neutron mass, with resolution of 9\,MeV (see figure~\ref{miss}).
The $\rm nK^+$ invariant mass distribution (figure~\ref{mis}) exhibits
a peak structure which is identified as $\Theta^+(1540)$.
Known reactions which also produce $\rm K^+K^-$
pairs, such as $\Phi$--decay, were
removed by cuts on the invariant mass of the kaon pair.
In general, this is an unnecessary precaution which
just reduces the number of events. 
A sharp peak at 1542 MeV in the  $\rm nK^+$
invariant mass distribution was seen,
with a statistical significance of 5.8$\sigma$.
The width was measured to be less than 21 MeV.
\par
Spectator protons do not escape the target, but the
detection of the proton was essential to reconstruct the
neutron from kinematics.
Since they looked for the $\Theta^+$  in
the $\rm n K^+$ invariant mass spectrum, also the neutron
participated in the reaction. Hence there was no 
spectator particle. 
\vspace{-8mm}

\begin{figure}[t!]
\bc
\begin{tabular}{ll}
\hspace{-5mm}\epsfig{file=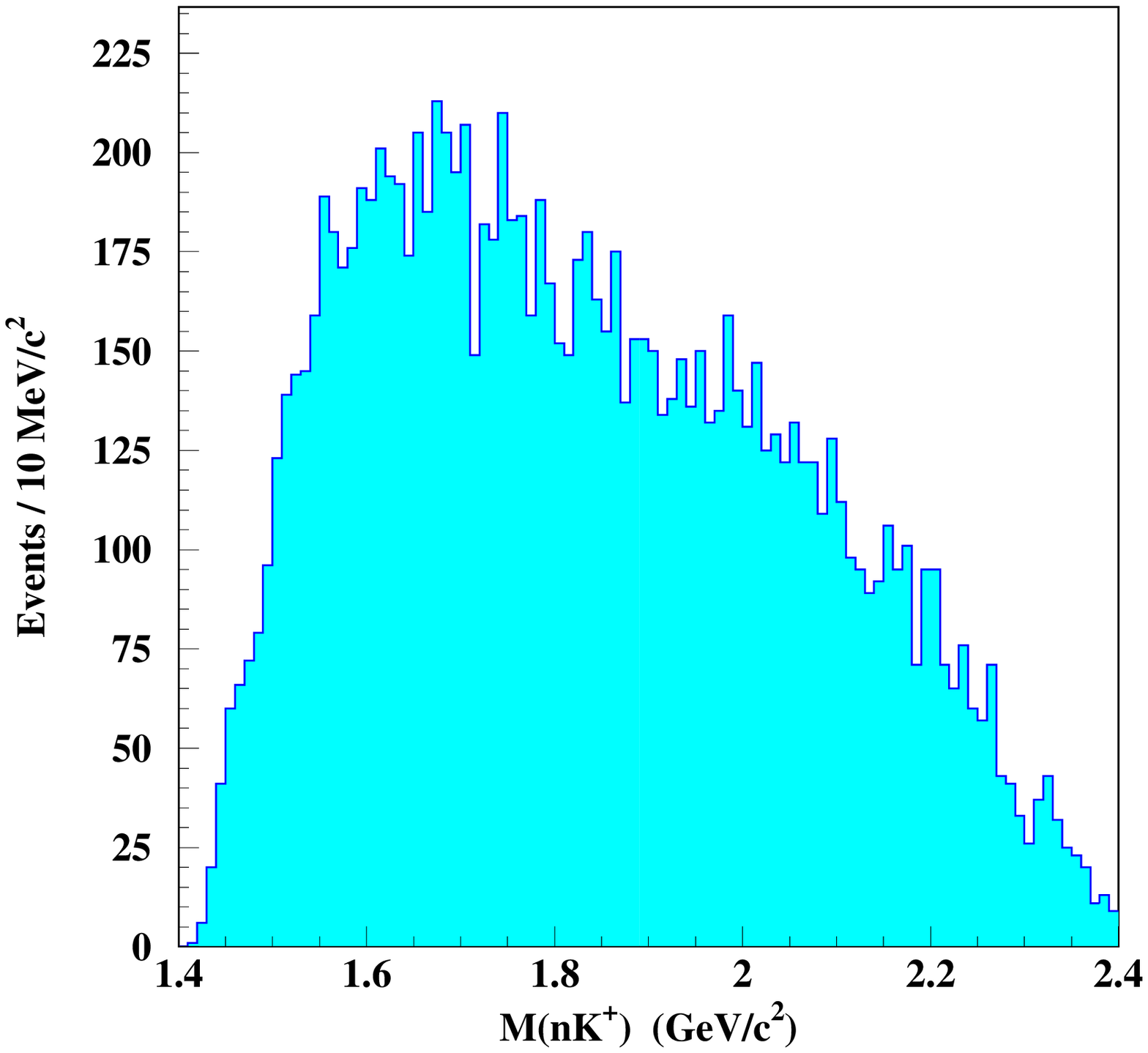,width=0.5\textwidth}&
\hspace{-5mm}\epsfig{file=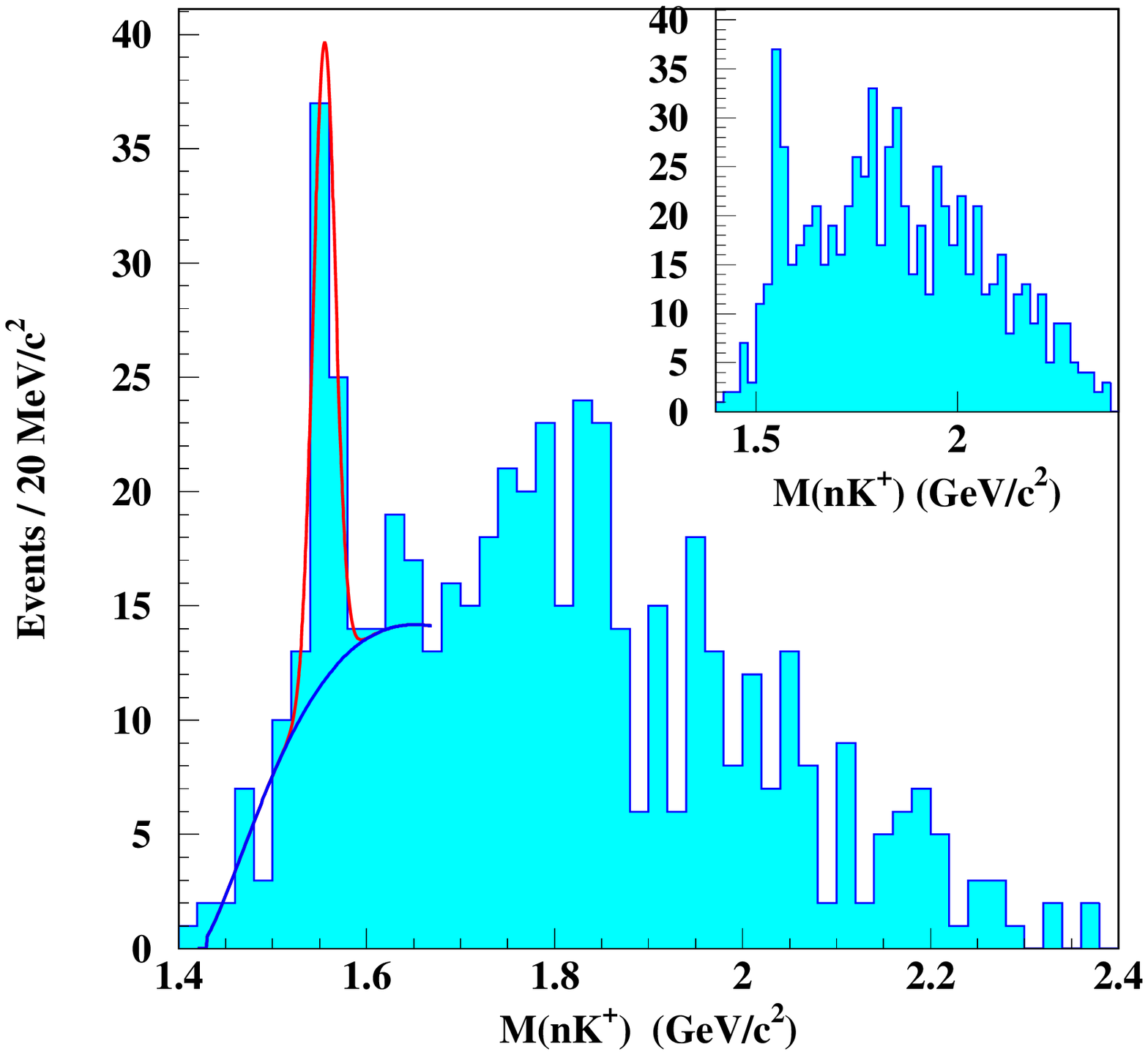,width=0.5\textwidth}
\end{tabular}
\ec
\vspace*{-10mm}
\caption{\label{theta_prpi}Left:
The $n K^+$ invariant mass spectrum  
in the reaction 
$\gamma p\rightarrow \pi^+K^-K^+(n)$. 
The neutron was measured from the missing four-momentum. 
Right:
The same invariant mass spectrum  
with the cut
$\cos\theta^*_{\pi^+}>0.8$ and $\cos\theta^*_{K^+}<0.6$.
$\theta^*_{\pi^+}$ and $\theta^*_{K^+}$ are the
angles between the $\pi^+$ and $K^+$ mesons
and photon beam in the center-of-mass system. 
The background function we used in the fit was obtained from the simulation.
The inset shows the $nK^+$ invariant mass spectrum  
with only the $\cos\theta^*_{\pi^+}>0.8$ 
cut~\protect\cite{Kubarovsky:2003fi}.
}
\vspace*{-6mm}
\end{figure}

The data using H$_2$~\cite{Kubarovsky:2003fi}are shown in
figure~\ref{theta_prpi}. 
The $\rm nK^+$ distribution has a large background; a statistically not
very significant peak is present at about 1.55\,GeV. The significance
of the peak can be improved by selecting events in which the
$\pi^+$ goes forward, and the $\rm K^+$ backward. Both distributions
are reproduced in figure~\ref{theta_prpi}.  

\begin{figure}[h!]
\begin{minipage}[c]{0.60\textwidth}
\epsfig{file=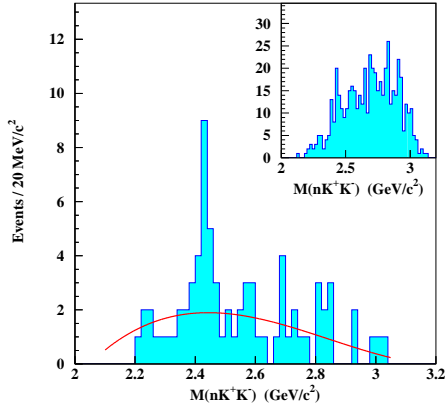,width=0.9\textwidth}
\end{minipage}
\begin{minipage}[c]{0.32\textwidth}
\caption{\label{monener}
The $n K^+ K^-$ invariant mass spectrum for  
events having $M(K^+n)$ between~1.54~and~1.58 GeV/c$^2$. 
The~inset~shows~the~
$nK^+K^-$
invariant mass
spectrum for all events in Figure~\ref{theta_prpi}
(left spectrum)~\protect\cite{Kubarovsky:2003fi}.
}
\vskip 3mm
The energy dependence of the $\Theta^+(1540)$ 
production in figure~\ref{monener} shows a peculiar
pattern: it could be that the $\Theta^+(1540)$ is produced via a sequence
\end{minipage}
\vspace*{-12mm}
\end{figure}
\be
\rm\gamma p\to N(2430)\pi^+\ , \ N(2430)\to \Theta^+(1540)K^- \ , \ {\rm
and\ }  \Theta^+(1540)\to nK^+. 
\label{reson}
\ee

The figure shows the 
$n K^+ K^-$ invariant mass spectrum  calculated from 
the missing mass off the $\pi^+$ in the
reaction $\gamma p\rightarrow \pi^+K^-K^+(n)$
with cuts $\cos\theta^*_{\pi^+}>0.8$ and $\cos\theta^*_{K^+}<0.6$.
A peak in this distribution would indicate that the $\Theta^+(1540)$
is at least partly produced via reaction chain (\ref{reson}).
\subsubsection{The $\Theta^+(1540)$ from neutrino--induced reactions}
\begin{figure}[h!]
\begin{tabular}{cc}
\epsfig{file=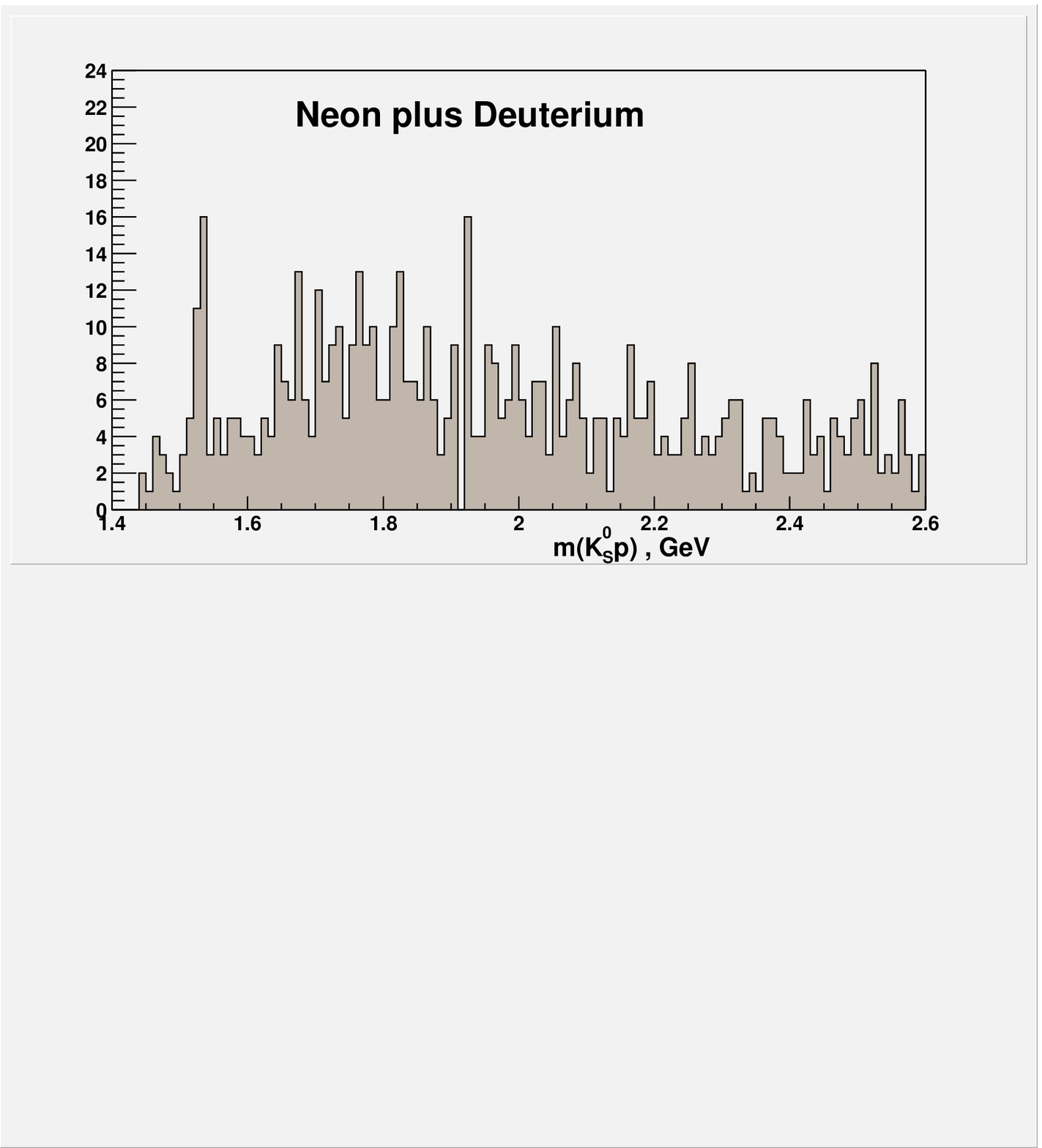,width=0.5\textwidth}&
\vspace*{16mm}
\hspace*{-8mm}
\epsfig{file=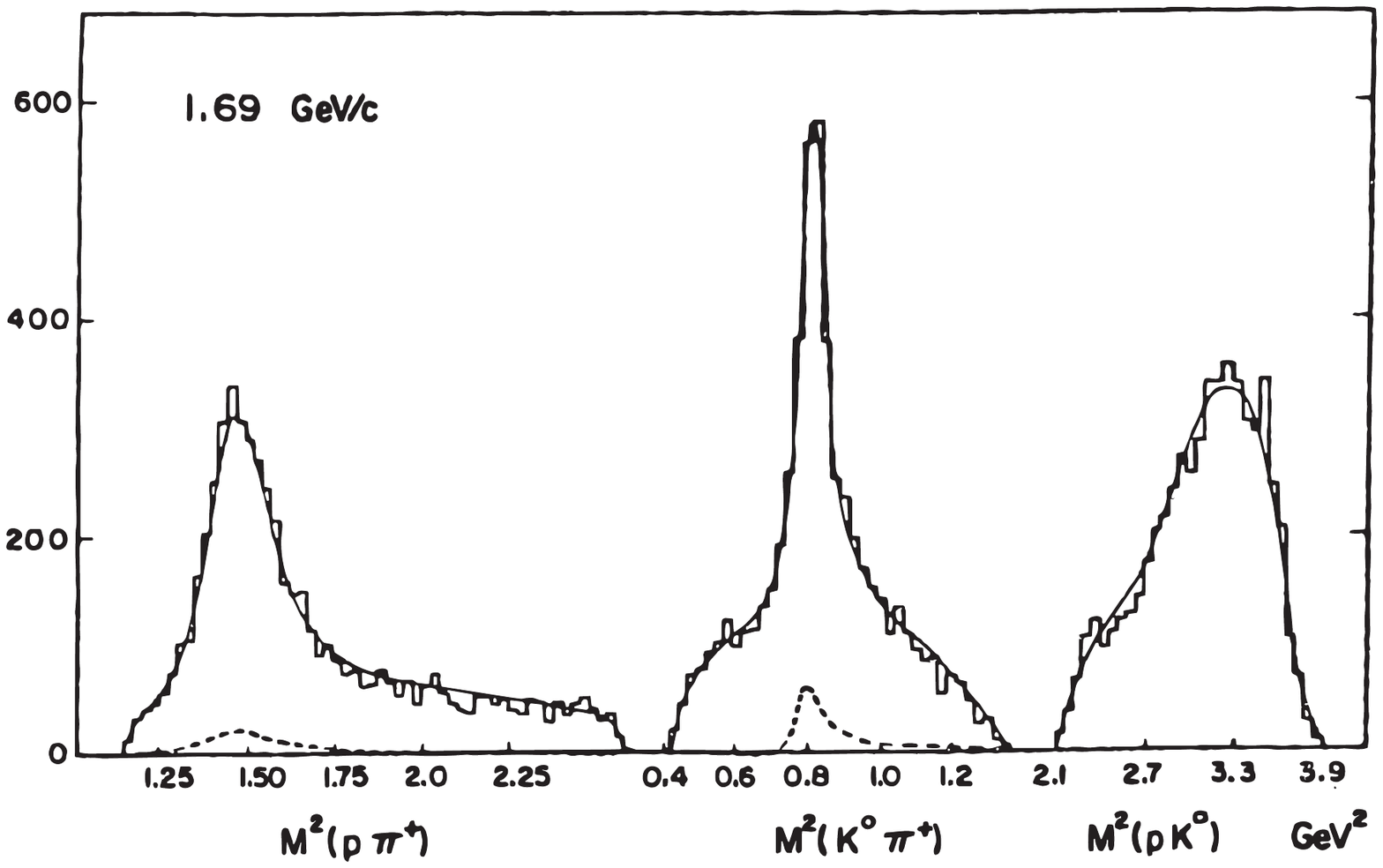,width=0.5\textwidth}
\vspace*{-35mm}
\end{tabular}
\hspace*{113mm}$\Downarrow$
\vspace*{15mm}
\caption{\label{nububble}
Left: The $\Theta^+(1540)$ in $\nu$--induced reactions.
Protons and $\rm K^0_s$ (identified by their $\pi^+\pi^-$ decay) were
produced inclusively in high--energy $\nu$ beams hitting deuterons or Neon
nuclei. The $\rm pK^0_s$ invariant mass distribution peaks at
1.54\,GeV~\protect\cite{Asratyan:2003cb}. 
Right: Bubble chamber data on $\rm K^+p\to pK^0_s\pi^+$. 
The fit to the data misses a small enhancement in the $\rm pK^0_s$ 
mass distribution marked by $\Downarrow$ which happens to occur at 
$\sim$1.53\,GeV~\protect\cite{Asratyan:2003cb}. 
}
\end{figure}

The $\Theta^+(1540)$ was also reported from neutrino--induced
reactions. Asratyan, Dolgolenko and Kubantsev~\cite{Asratyan:2003cb} 
scanned data taken
with two large bubble chambers at CERN and at Fermilab in the
search for the formation of the $\Theta^+(1540)$ in collisions of
neutrino and antineutrino in the 100\,GeV energy range with
protons, deuterons and Neon nuclei. A narrow p$\rm K^0_s$ peak at
a mass of $1533\pm 4$\,MeV was observed, see figure~\ref{nububble}
which is assigned to 
$\Theta^+(1540)$ production (the peak might be a $\Sigma^+$
resonance but none is known at this mass).

\subsubsection{The $\Theta^+(1540)$ from the archive}
It may be worthwhile to note that hints for the $\Theta^+$ may
have been seen already in 1973 in a CERN experiment to study
$\rm K^+p\to pK^0_s\pi^+$ inelastic scattering\cite{Lesquoy:1975dx}. Mass
distributions for the 5000 events are shown in figure~\ref{nububble},
right, for a 1.69\,GeV/c incident Kaon momentum. On the very right,
the $\rm pK^0_s$ invariant mass distribution shows a low--mass peak
which is not accounted for in the fit (which describes the full reaction
dynamics). The peak has about 100 events and is seen at 1.53\,GeV
The 5000 events corresponded to a cross section of 4\,mb, hence the
$\Theta^+$ may have been observed with a cross section of 0.08\,mb
while the $\Delta(1232)$ is observed with $\sigma\sim 2$\,mb, only
25 times stronger.
\subsubsection{The Hermes experiment}
\begin{figure} [h!]
\begin{minipage}[c]{0.55\textwidth}
\begin{center}
\includegraphics[width=0.9\textwidth,angle=0]{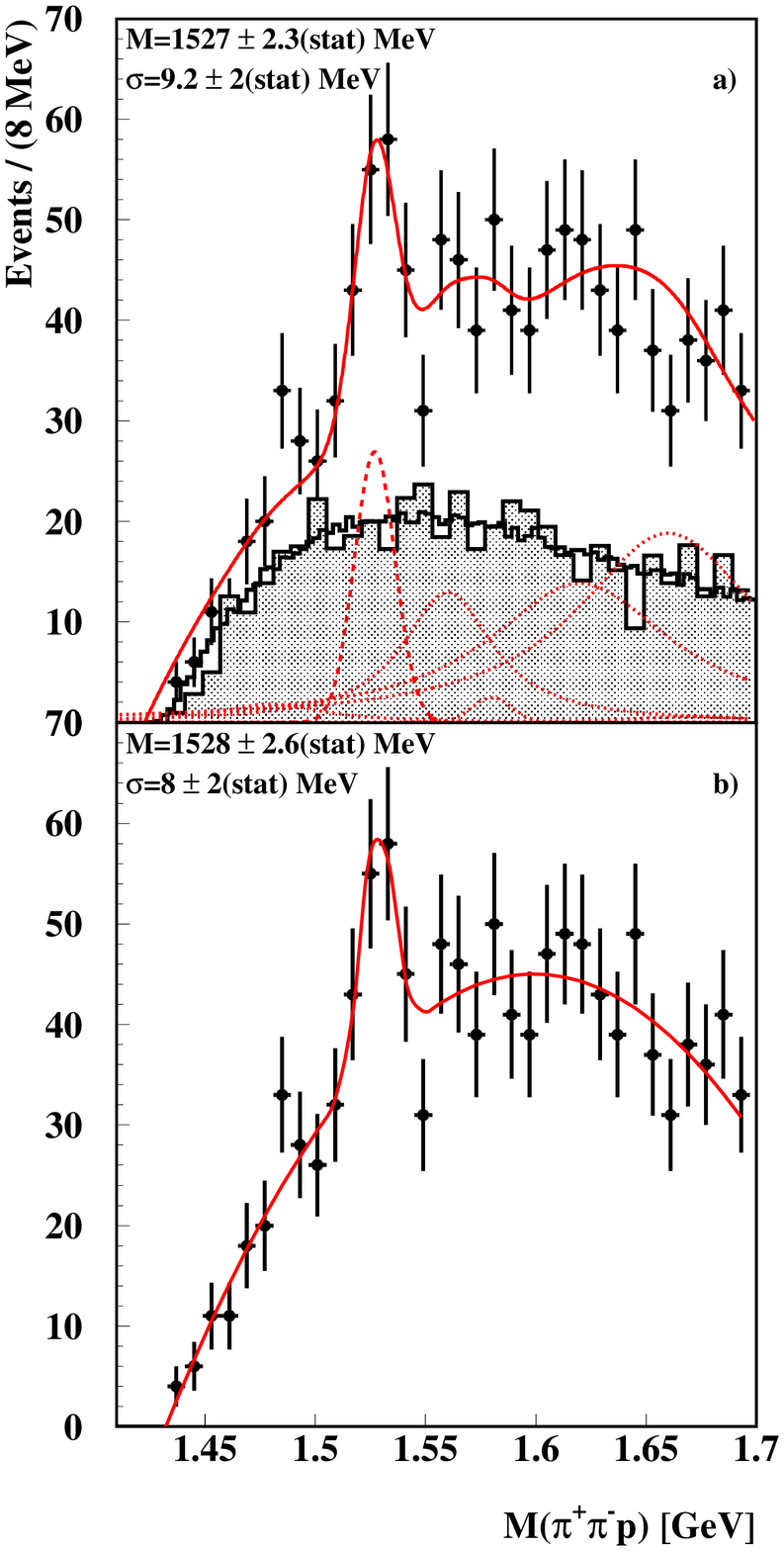}
\end{center}
\end{minipage}
\begin{minipage}[c]{0.44\textwidth}
\vspace*{-18mm}
non--resonant 
background and that low--mass baryon resonances can be added
to the non--resonant background. 
The resulting $\rm pK^0_s$ spectrum, shown in figure~\ref{hermes}, 
displays a narrow peak at a mass of 1528\,MeV.  
There is no known positively charged strangeness--containing baryon in \
this mass region that could account for the  observed peak. 
It is interpreted as further evidence for the $\Theta^+(1540)$. 
\vskip 15mm
\caption{\label{hermes}
Invariant mass distribution of the $p \pi^+\pi^-$ system
after cuts to optimize the signal--to--background
for the $\rm K^0_s$. The data is represented by dots with
statistical error bars.
In panel a) the {\sc Pythia} Monte Carlo simulation
is represented by the gray shaded histogram. The fine-binned histogram
represents a model based on event mixing and the solid line is the
result of the fit taking 
contributions from known baryon 
resonances into account, represented by dotted lines. 
In panel b) a fit to the data of a Gaussian
plus a third-order polynomial is 
shown~\protect\cite{Airapetian:2003ri}.}
\end{minipage}
\end{figure}

The Hermes collaboration~\cite{Airapetian:2003ri}
searched for the $\Theta^+$ in
quasi-real photoproduction  on deuterium in the  
$\Theta^+\to p K^0_S \to p \pi^+ \pi^-$ decay chain.
The virtual photons
originated from the 27.6\,GeV (9 to 45\,mA)
positron beam of the HERA storage ring at DESY.  
An integrated luminosity of 250\,pb$^{-1}$
was collected on a longitudinally polarized deuterium gas target. The
data shown in figure~\ref{hermes} is summed over two spin orientations.
\par
Selected events 
contained at least two
oppositely charged pions in coincidence with one proton.
The event selection included constraints on the event topology to
maximize the yield of the $\rm K^0_s$ peak in the $M_{\pi^+\pi^-}$
spectrum while minimizing its background. 
The position of the $\rm K^0_s$ peak is within 1\,MeV of the expected
value. To search for the
$\Theta^+$, events were selected with a $M_{\pi^+\pi^-}$ invariant
mass compatible with the $\rm K_s^0$ peak.
The resulting spectrum of the invariant mass of the $\pi^+\pi^-p$
system is displayed in figure~\ref{hermes}. 
It is assumed that {\sc Pythia} describes only the
\par

\subsubsection{The SVD-2 experiment at Protvino}
The SVD-2 spectrometer works in the $70\ GeV$ proton beam of the IHEP 
accelerator. The beam, defined by microstrip Si-detectors and 
several dipole and quadrupole magnets, hit an
active target (Si-detector and lead foil sandwich); charged particles 
produced on nuclei were detected in a large--aperture magnetic
spectrometer. High--energy $\gamma$'s werde detected in 
Cherenkov lead glass counters.  
Events with charged--particle multiplicity five or less
were selected to reduce the combinatorial background. 
The following cuts were made: a $\rm K^0_s$ was required, 
$\Lambda^0$'s were excluded, the $\rm pK^0_s$-system was required
to be produced in forward direction. A cut on the $\rm K^0_s$
momentum $\rm P_{K^0_s}\le~P_p$ improved the signal to background ratio. 
\par
The resulting $\rm K^0_s$ invariant mass spectrum~ \cite{Aleev:2004sa}
exhibits a peak interpreted as further evidence for the
existence of the $\Theta^+(1540)$. The mass was determined
to $M=1526\pm 3(stat.)\pm 3(syst.)$\,MeV/c$^2$,  
the width is compatible with the
experimental resolution, thus $\Gamma < 24$\,MeV/c$^2$. 
A statistical significance
of $5.6~\sigma$ is estimated.

\begin{figure}[h!]
\vspace*{-5mm}
\begin{minipage}[c]{0.49\textwidth}
\epsfig{file=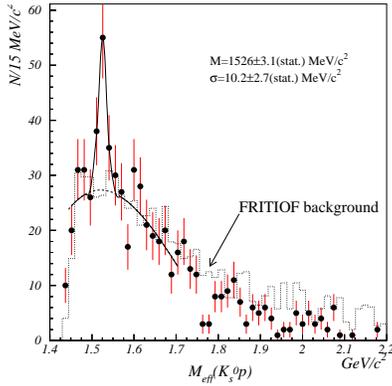,width=0.9\textwidth}
\caption{\label{ihep}
The $(pK^0_s)$ invariant mass spectrum in the
reaction~\protect\cite{Aleev:2004sa} $pA\rightarrow pK^0_s+X$.  The
dashed histogram represents background obtained
from simulations.
}
\end{minipage}
\begin{minipage}[c]{0.49\textwidth}
\hspace*{-11mm}\epsfig{file=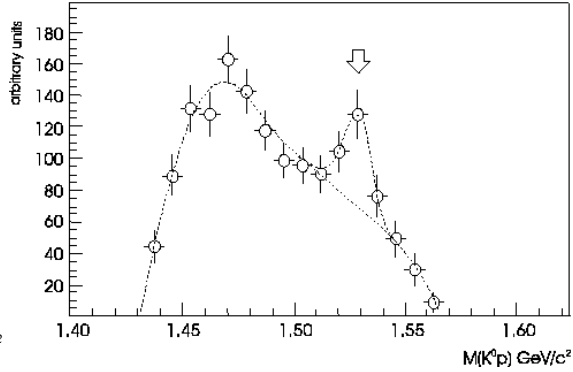,width=1.2\textwidth}
\caption{\label{COSY}
The  $\rm pK^0_s$ invariant mass distribution
from the reaction $\rm pp \to \Sigma^+ K^0_s p$
at COSY~\protect\cite{Alt:2003vb}.
}
\end{minipage}
\vspace*{-5mm}
\end{figure}

\subsubsection{The TOF experiment at COSY}

At the Cooler Synchroton (COSY) at J\"ulich, the reaction 
$\rm pp \to \Sigma^+ K^0_s p$ was studied~\cite{Abdel-Bary:2004ts}
in a proton beam of 2.95 GeV/c momentum impinging on a liquid H$_2$
target of 4\,mm length. Charged particles were tracked using a double--sided
silicon microstrip detector close to the target  and scintillation fiber 
hodoscopes. There is no magnetic field; momenta were determined
from the event geometry. The $\Sigma^+$ was identified from a kink 
in the track due to $\rm\Sigma^+\to p\pi^0$ decays, the $\rm K^0_s$ by its
decay to $\pi^+\pi^-$. In the $\rm pK^0_s$ invariant mass distribution
a peak shown in figure~\ref{COSY} is observed which is fit 
to $1530\pm 5$\,MeV. The width (FWHM) $18\pm4$\,MeV is compatible 
with the instrumental resolution and is quoted as upper limit. 
The statistical significance is 3.7.

\subsubsection{Yerevan}

A group at Yerevan~\cite{Aslanyan:2004gs}
 reported a search for the $\Theta^+$ in a
 2m propane bubble chamber by scattering 10\,GeV/c protons
off $\rm C_3H_8$. The $\rm pK^0_s$ invariant mass spectrum shows 
resonant structures with 
$\rm M_{K_s^0 p}$=1545.1$\pm$12.0, 1612.5$\pm$10.0, 1821.0$\pm$11.0
MeV/$c^2$ and $\rm\Gamma_{K_s^0 p}$= 16.3$\pm$3.6, 16.1$\pm$4.1,
28.0$\pm$9.4 MeV/$c^2$, respectively. Protons were se\-
lected to have 
large or small momenta. 
The statistical significance of these
peaks were estimated to 5.5$\sigma$,4.6$\sigma$ and
6.0$\sigma$, respectively.
\begin{figure}[h!]
\begin{minipage}[c]{0.70\textwidth}
\epsfig{file=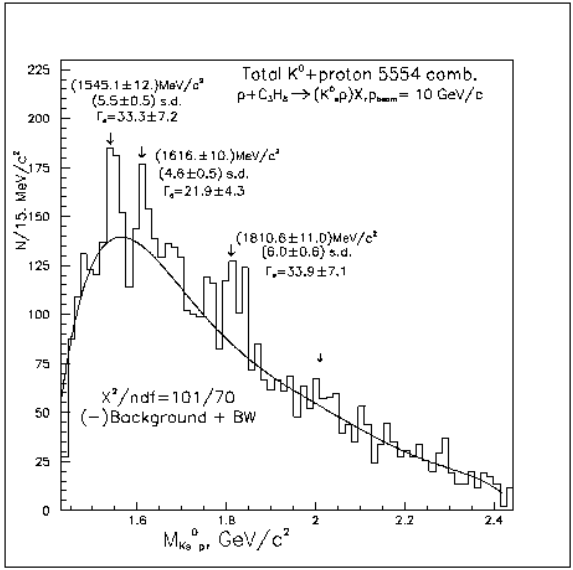,width=0.85\textwidth,clip=on}
\end{minipage}
\begin{minipage}[c]{0.28\textwidth}
\vspace*{-8mm}
\vskip 5mm
\caption{\label{ener}
The effective $p K^0_s$ mass distribution for
protons with momenta between $0.35\le p\le 0.9$ GeV/c$^2$ or
$p\ge1.7 GeV/c^2$. For momenta between $0.9\le p\le 1.7$ GeV/c$^2$
there is no significant signal and the date is excluded. 
The curve is the experimental background from an
mixing method taken in the form of six-order 
polynomial~\protect\cite{Aslanyan:2004gs}.
}
\end{minipage}
\end{figure}

\subsubsection{The ZEUS experiment at HERA}
The ZEUS experiment studied 
the $\rm K^0_s p$ and $\rm K^0_s \bar p$ 
invariant mass spectra in inclusive
deep inelastic $ep$ scattering for a large range in the photon
virtuality~\cite{:2004kn}. 
For $Q^2 \ge 10$\,GeV$^2$ a peak is seen around 1520 MeV.
The peak position is $1521.5\pm 1.5({\rm stat.})^{+2.8}_{-1.7} ({\rm
syst.})$\,MeV; the Gaussian width corresponds to a full width
at half maximum of 16\,MeV we take as upper limit of the natural width.
The statistical significance is about 4.6$\sigma$.
The fit suggests an additional  $\Sigma(1465)$ bump 
(possibly identical with a  $\Sigma(1480)$ bump~\cite{Engelen:us}
reported from $\rm K^- p \to K^0 \pi^- p$ at 4.2\,GeV). 
The $\Theta^+(1540)$ evidence reduces
to 3$\sigma$ when the $\Sigma(1465)$ bump is excluded from the fit. 

\begin{figure}
\begin{minipage}[c]{0.69\textwidth}
\includegraphics[width=0.85\textwidth]{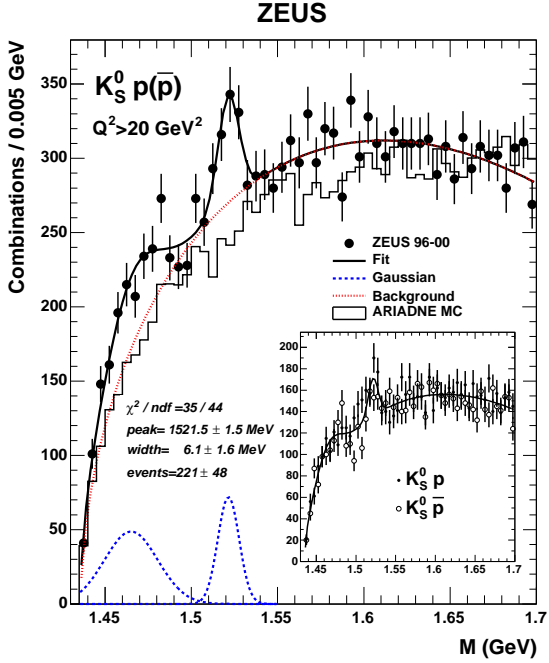}
\end{minipage}
\begin{minipage}[c]{0.28\textwidth}
\caption{ Invariant-mass spectrum for the $\rm K^0_s p$
and $\rm K^0_s \bar p$ channel for $Q^2 >
20$\,GeV$^2$~\protect\cite{:2004kn}. The solid line is the
result of a fit to the data using a three-parameter background
function plus two Gaussians (see text).  The dashed lines show the
Gaussian components and the dotted line the background according to
this fit. The histogram shows the prediction of the {\sc Ariadne} MC
simulation normalised to the data in the mass region above
$1650$\,MeV. The inset shows the   $\rm K^0_s p$(open circles) and 
the $\rm K^0_s \bar p$ 
(black dots) candidates separately, compared to the result of the fit
to the combined sample.  }
\label{pqbar}
\end{minipage}
\vspace*{-5mm}
\end{figure}

The results provide further evidence for the existence of a narrow
baryon resonance consistent with the predicted $\Theta^+$ pentaquark
state with a mass close to $1530$\,MeV and a width of less than 
$15$\,MeV.
In the $\Theta^+$ interpretation, the signal
observed in the $\rm K^0_s\bar p$ channel corresponds to first evidence for an
antipentaquark with a quark content of
$\bar{u}\bar{u}\bar{d}\bar{d}s$.  The results, obtained at high
energies, constitute first evidence for the production of such a state
in a kinematic region where hadron production is dominated by
fragmentation. 

\subsubsection{Search for the  $\Theta^+(1540)$   at HERA-B}
Kn\"opfle, Zavertyaev and Zivko~\cite{Knopfle:2004tu} reported
a search for pentaquarks in HERA-B data taken with a minimum
bias trigger (more than 200 million events). 
HERA-B is a fixed target experiment at the 920 GeV proton storage ring
of DESY hitting a carbon, titanium or tungsten target.  
The forward magnetic spectrometer has large
acceptance, high-resolution vertexing and  
tracking  and good particle identification. In the $\rm pK^-$
invariant mass distribution a clear $\Lambda(1520)$ is seen, but there
is no sign of the $\Theta^+(1540)$. 
\begin{figure}[h!]
\vspace*{-2mm}\includegraphics[width=1.1\textwidth]{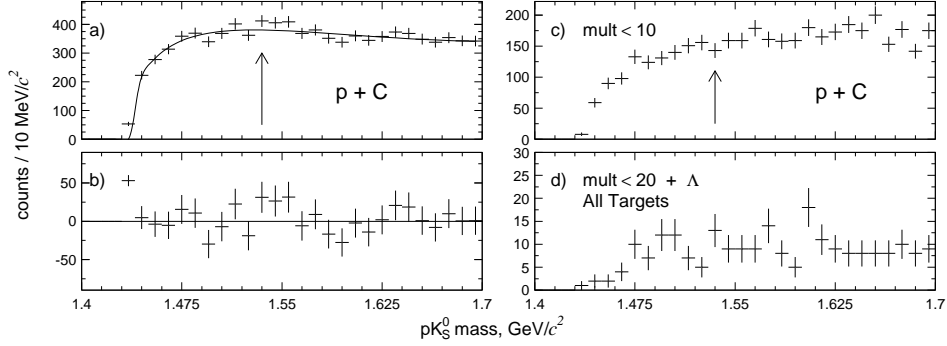}
\vspace*{-5mm}
\caption{\label{d2}
The pK$^0_s$ invariant mass 
distributions~\protect\cite{Knopfle:2004tu}: 
a) data from the p+C
collisions with  
background (continuous line) determined from event mixing; 
b) as a) but with the background subtracted; 
c) as a) but requiring a track multiplicity of $<10$. 
d) data from all targets C, Ti, W requiring a track multiplicity 
of $<20$ and a $\Lambda$ particle in the event.
Arrows mark the mass of 1540\,MeV/c$^2$ .}
\end{figure}

\subsubsection{Search for the  $\Theta^+(1540)$ in charmonium decays}
The BES collaboration serached for the $\Theta^+(1540)$ in 
J/$\psi$ and $\psi (2S)$ decays into different
charge combinations of the $\rm N\bar N K\rm \bar K$ final
state~\cite{Bai:2004gk}. 
The reactions are observed with branching ratios in the order of
$10^{-4}$ and with no evidence for the $\Theta^+(1540)$ at the 
$10^{-5}$ level.

 \subsubsection{The $\Xi^{--}$ from NA49}

The NA49 collaboration reported evidence for another exotic baryon
resonance with strangeness S=-2 and charge Q=-2~\cite{Alt:2003vb}. 
A state with these quantum numbers
cannot be constructed from three quarks; the minimum quark model configuration
is $ddssu$. It is a pentaquark called $\Xi^{--}$.

The  $\Xi^{--}$ was observed in fixed-target proton-proton
collisions at the Super Proton Synchrotron (SPS) at CERN. The
center-of-mass energy of these collisions was 17.2 GeV, far above
the threshold for pentaquark production. The NA49 detector
consists of large acceptance Time Projection Chambers (TPCs) 
providing tracking for charged particles produced from primary and
secondary vertices. Particles are identified by specific energy
loss (dE/dx) in the TPCs.
\par
\begin{figure}[b!]
\vspace*{-5mm}
\begin{tabular}{cc}
\epsfig{file=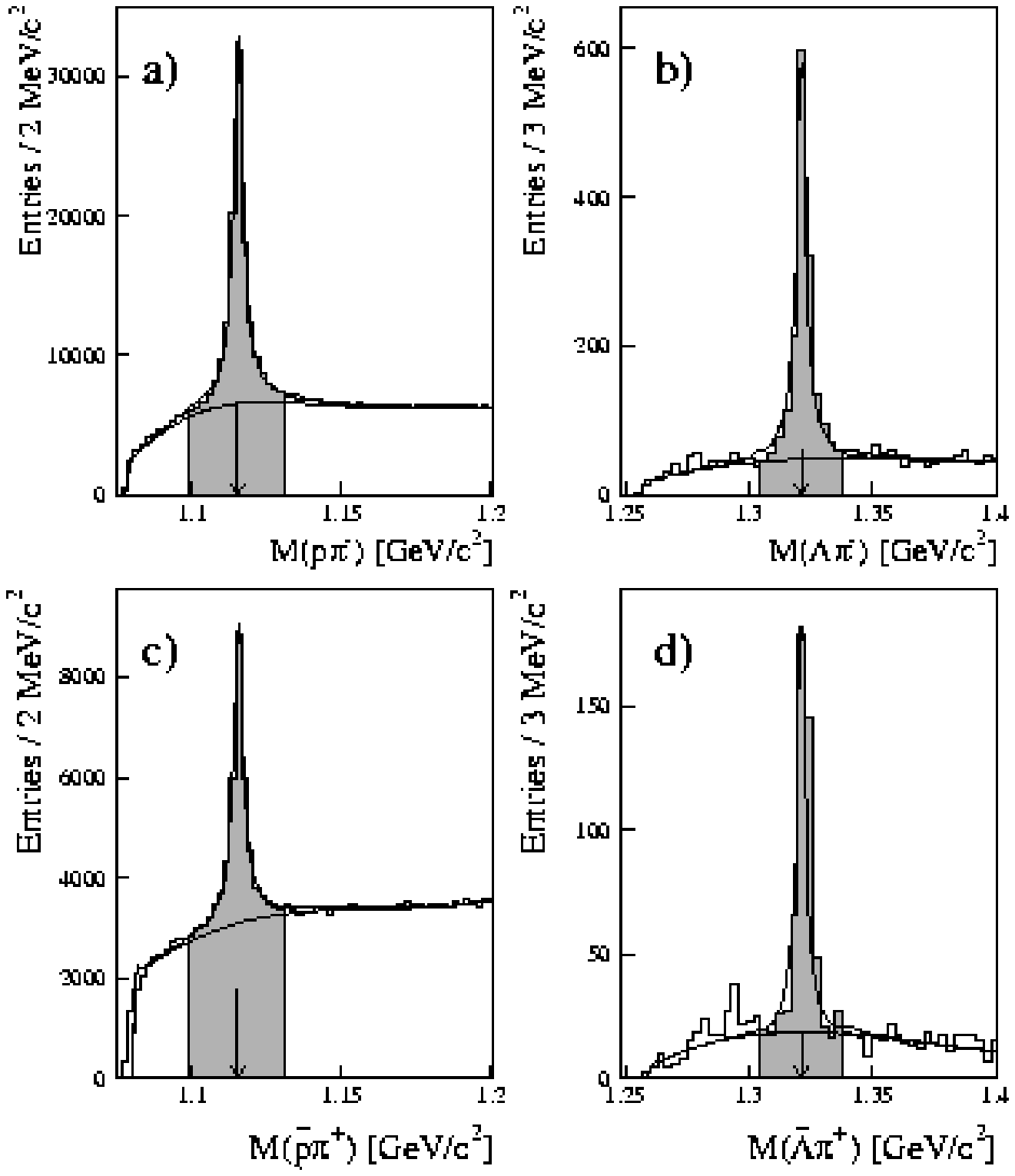,width=0.5\textwidth}&
\epsfig{file=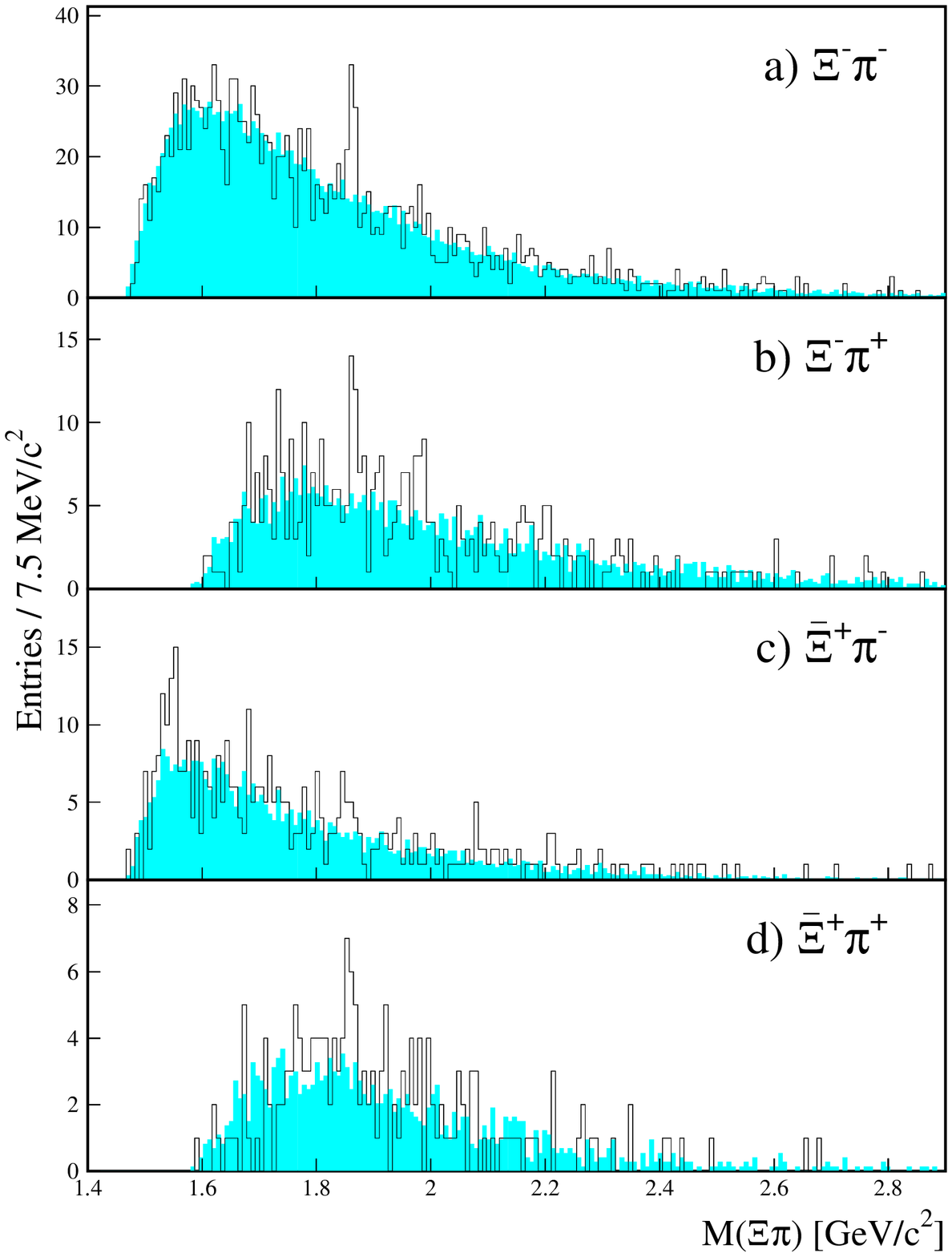,width=0.5\textwidth}
\end{tabular}
\vspace*{-5mm}
\caption{\label{xi4}
NA49 invariant mass distributions: $\Lambda =p\pi^-$,
$\Xi^- = \Lambda\pi^-$ (left)
  $\Xi(1862)=\Xi^-\pi$ (right), and charge conjugated 
reactions~\protect\cite{Alt:2003vb}.
}
\end{figure}

In a first step, $\Lambda$'s were identified from the
invariant mass spectrum of $p\pi^-$
pairs originating from the same vertex. $\Lambda$
candidates were then combined with $\pi^-$
to form $\Xi^-$ candidates. Similarly, both mass distributions were
constructed for $\bar p\pi^+$ and $\bar\Lambda\pi^+$ (see
Figure~\ref{xi4}). Finally, the
 $\Xi^{--}$ was searched for in the invariant mass spectrum of the
$\Xi^-$ ($\bar\Xi^+$) candidates with $\pi^-$ ($\pi^+$)
tracks originating from the primary vertex.

A peak was seen with 
a mass of $1862\pm2$\,MeV and a width below the detector resolution of 18 MeV.
At the same mass NA49 also observed a peak in the
$\Xi^-\pi^+$ spectrum, which could be a candidate for a neutral
isospin-partner of the $\Xi^{--}$. The corresponding antibaryon
spectra for both states also show enhancements at the same
mass. The four mass spectra are shown in figure~\ref{xi4}.
\begin{figure}[h!]
\vspace*{-3mm}
\begin{minipage}[c]{0.69\textwidth}
\epsfig{file=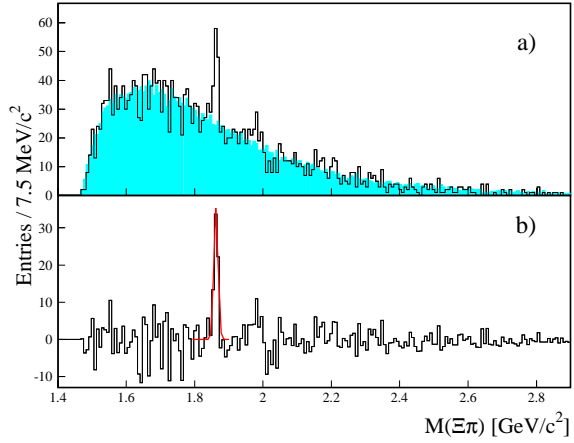,width=\textwidth}
\end{minipage}
\begin{minipage}[c]{0.29\textwidth}
\caption{\label{xi5}
The final $\Xi\pi$ invariant mass spectrum without and with
background subtraction. The four spectra
shown individually in~\protect\ref{xi4} are 
added~\protect\cite{Alt:2003vb}.}
\end{minipage}
\vspace*{-8mm}
\end{figure}

\subsubsection{Search for the  $\Xi^{--}(1862)$  at HERA-B}

The HERA-B collabration also reported~\cite{Knopfle:2004tu} a search for the 
$\Xi (1862)$ in the doubly-charged  
$\Xi^-\pi^-$ +\,c.c.  and in the neutral  $\Xi^-\pi^+$ +\,c.c. channels. 
$\Xi$ candidates with a mass of $\pm 10$\,MeV/c$^2$ of the PDG 
mass were accepted. 

\begin{figure}[h]
\centering
\includegraphics[width=\textwidth]{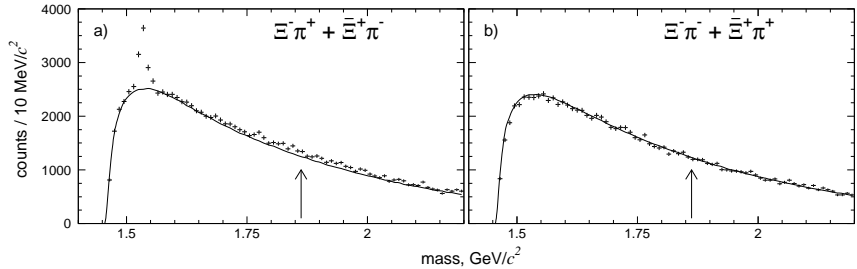}
\caption{\label{d3} 
The $\Xi\pi$ invariant mass distributions obtained with all targets 
C, Ti, and W in indicated
decay channels~\protect\cite{Knopfle:2004tu}. 
Continuous lines show the background from event
mixing. Arrows mark the 
mass of 1862\,MeV/c$^2$ .
}
\end{figure}
Figure \ref{d3} shows the 
corresponding invariant mass spectra and the backgrounds determined by
event mixing. 
In the neutral decay channels, (Figure \ref{d3}a), the $\Xi(1530)^0$
resonance shows up  with a prominent  
signal, and there is a possible weak evidence for known higher $\Xi^{*}$
resonances. In the doubly-charged 
channels (Figure \ref{d3}b), the background follows very well the
data. There is no evidence for a narrow signal at around
1862\,MeV/c$^2$.  $\Xi^{--}(1862)$ production is reduced compared to 
$\Xi(1530)^0$ production by mor than one order of magnitude.

\subsubsection{A charming pentaquark}
\begin{figure}[h!]
\vspace*{-3mm}
\begin{minipage}[c]{0.69\textwidth}
\epsfig{file=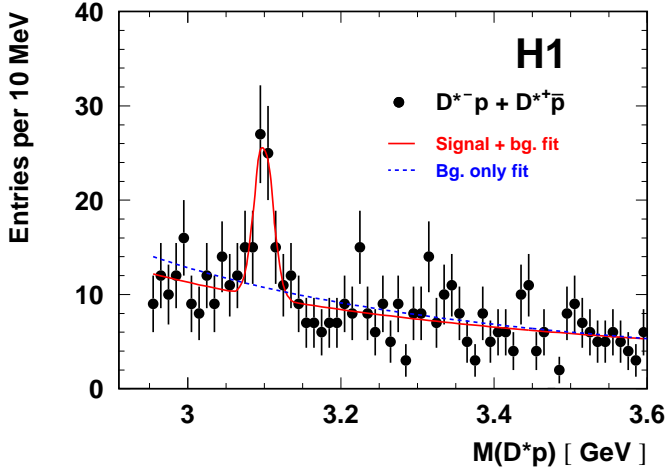,width=\textwidth}
\end{minipage}
\begin{minipage}[c]{0.29\textwidth}
\caption{\label{h1} 
$M(D^* p)$ distribution from opposite-charge
$D^* p$ combinations in deep inelastic scattering 
of electrons off protons~\protect\cite{unknown:2004qf}. 
The solid line represents a 
fit with a Gaussian peak plus a two--parameter background, 
the dashed line a fit background only. 
} 
\end{minipage}
\end{figure}

Very recently the H1 collaboration reported 
a narrow baryon resonance containing a $\bar 
c$--quark~\cite{unknown:2004qf}. 
It is observed in inelastic electron-proton collisions
at centre-of-mass energies of $300 \ {\rm GeV}$ and $320 \ {\rm GeV}$
at HERA in the $D^{* \, -} p$ invariant mass spectrum, or as an
antibaryon  in the $D^{* \, +} \bar{p}$ mass distribution.
The final data is shown in figure~\ref{h1}. 
The resonance has a 
mass of $3099 \pm 3 \ {\rm (stat.)} \ \pm 5 \ {\rm (syst.)} \ {\rm MeV}$
and a measured Gaussian width of 
$12 \pm 3 \ {\rm (stat.)} \ {\rm MeV}$, which is compatible with the
experimental resolution.
The resonance is interpreted as an anti-charmed baryon with a
minimal constituent quark composition of $uudd \bar{c}$ together with
the charge conjugate.

\subsubsection{Pentaquark summary}
Table~\ref{sumofall} collects masses, width, number
of events and statistical significance of pentaquark observations.
There are 11 data points, and at a first glance the
existence of the $\Theta^+(1540)$ seems to be established
beyond any reasonable doubt. 
\par
In some papers the statistical evidence is calculated as
$\sigma = N/\sqrt B$ where $N$ is the number of signal events and
$B$ the number of background events. In the limit $B\to 0$, the
evidence becomes extremely large and the formula must be wrong.
Therefore the statistical evidence has been estimated 
from the published histograms. The number of events in the signal
region is $N+B$, in side band regions (covering the same mass
interval) it is $B$, error propagation gives 
$\sigma = N/\sqrt{(N+2B)}$. Recalculated statistical evidences 
are denoted by a $\sim$ symbol in table~\ref{sumofall}. 
There is only a very small probability that all these measurements 
are statistical fluctuations.
\par
Now we turn to a discussion of the mass values.
The systematic uncertainties of the DIANA and
$\nu$--induced measurements were taken to be $\pm3$\,MeV as suggested in
\cite{Airapetian:2003ri}.
The weighted average of the masses observed in all
experiments is $1532.4\pm1.4$\,MeV. In evaluating this 
average mass value, the quadratic
sum of the statistical and systematic uncertainties of all
measurements are taken into account. 
The sum of $\chi^2$'s for all data is 38. Hence the probability that
all these measurements have observed the same object is 
also rather small and a scaled mass error $\pm 2.7$\,MeV is mor realistic.

\begin{table}[h!]
\vspace*{-4mm}
\caption{\label{sumofall}
Summary of measurements of pentaquarks. The
systematic errors given in parentheses are not quoted in the
papers but were estimated to be small. 
}
\vspace*{-2mm}
\bc
\renewcommand{\arraystretch}{1.3}
\begin{tabular}{cccccc}
\hline\hline
Mass & Width & $\rm N_{event}$ & Statist. & Reaction & Experiment \\
(MeV)& (MeV) &                 & signif. &         & \\
\hline
$\Theta^+(1540)$ &&&&&\\
$1540\pm10\pm5$ & $<25$ &       $19\pm2.8$      & $\sim2.7\sigma$  &
$\rm\gamma C\to C^{\prime}K^+K^-$& LEPS \\
$1539\pm2\pm2$ & $<9$ &   $29$                  & $\sim 3.0\sigma$  &
$\rm\gamma p\to nK^+K^0_s$ &DIANA  \\ 
$1542\pm2\pm5$ & $<21$ & $43$                   & $\sim 3.5\sigma$  &
$\rm\gamma d\to pnK^+K^-$ &CLAS \\
$1540\pm4(\pm3)$ & $<25$ & $63\pm13$            & $4.8\sigma$  &
$\rm\gamma p\to nK^+K^0_s$ &SAPHIR \\
$1533\pm5(\pm3)$ & $<20$ &    $27$              & $\sim 4.0\sigma$  &
$\nu$--induced & CERN, FNAL \\ 
$1555\pm1\pm10$ & $<26$ & $41$                  & $\sim 4.0\sigma$  &
$\rm\gamma p\to nK^+K^-\pi^+$ &CLAS \\
$1528\pm4$ & $<19$ &    $\sim 60$               & $\sim4\sigma$  &
$\gamma^*$--induced & HERMES \\ 
$1526\pm3\pm3$ & $<24$ & $50$                &
$3.5\sigma$  & p-p reaction &SVD-2 \\
$1530\pm5$ & $<18$ & $$                &
$3.7\sigma$  & p-p reaction &COSY \\
$1545\pm12$ & $<35$ & $\sim 100$                &
$\sim 4\sigma$  & p-A reaction &YEREVAN \\

$1521.5\pm1.5^{+2.8}_{-1.7}$ & $<6$ & $221$                &
$4.6\sigma$  & Fragmentation &ZEUS \\
\hline
$\Xi(1862)$ &&&&&\\
1862 & $<21$ &                 & $4.6\sigma$  & 
$\nu$--induced &NA49 \\
\hline
$\Theta_c(3099)$ &&&&&\\
$3099\pm3\pm5$ &  &                 & $5.4\sigma$  & 
$\gamma^*$--induced &HERA \\
\hline\hline
\end{tabular}
\renewcommand{\arraystretch}{1.3}
\ec
\end{table}
\par

A statistical analysis is not the only criterion 
for judging observations. In spite
of the fact that physics is an exact science,
there might also be some 'personal bias'. 
There might be a tendency to increase the error
if the measured mass seems to be 'wrong'. 
Hence the probability 
of consistent mass values could be even smaller. Further, in the
data selection cuts are applied some of which are tuned to optimize
the signal; these cuts enhance the statistical evidence and the
reported evidence becomes too high. 

We remind the reader that a long time ago
there was striking evidence that the $a_2(1320)$ was split into
two mesons at slightly different masses~\cite{a2}. The so--called
$S(1936)$ meson was seen in several experiments and  
interpreted as $\rm N\bar N$ bound state. It was
proven not to exist when high-statistics data became available
at LEAR (see, e.g.~\cite{Barnes:2003qk}). 
Also, the production characteristics are sometimes different, in particular
the ratio of $\Lambda(1520)$ and $\Theta^+(1540)$, even though the
reactions are similar or even identical. (However, the
data is not acceptance--corrected, and the detection efficiencies for
 $\Lambda(1520)$ and $\Theta^+(1540)$ can be different.)
Experiments not observing a signal have more difficulties to publish
upper limits 
than experiments reporting positive evidence have. First data
with negative evidence is now published but there are rumors
also from other experiments finding no $\Theta^+(1540)$, $\Xi(1862)$
or $\Theta^+_c(3099)\pm3\pm5$.

\par

Another important aspect was underlined by A. Dzierba and 
collaborators~\cite{Dzierba:2003cm}. In the reaction 
$\rm\gamma N\to N K\bar K$ one should consider the full
three--particle dynamics. Not only can $\rm NK$ or $\rm N\bar K$
resonances be produced but also $\rm K\bar K$ resonances. 
\par
\begin{figure}[h!]
\begin{minipage}[c]{0.60\textwidth}
\epsfig{file=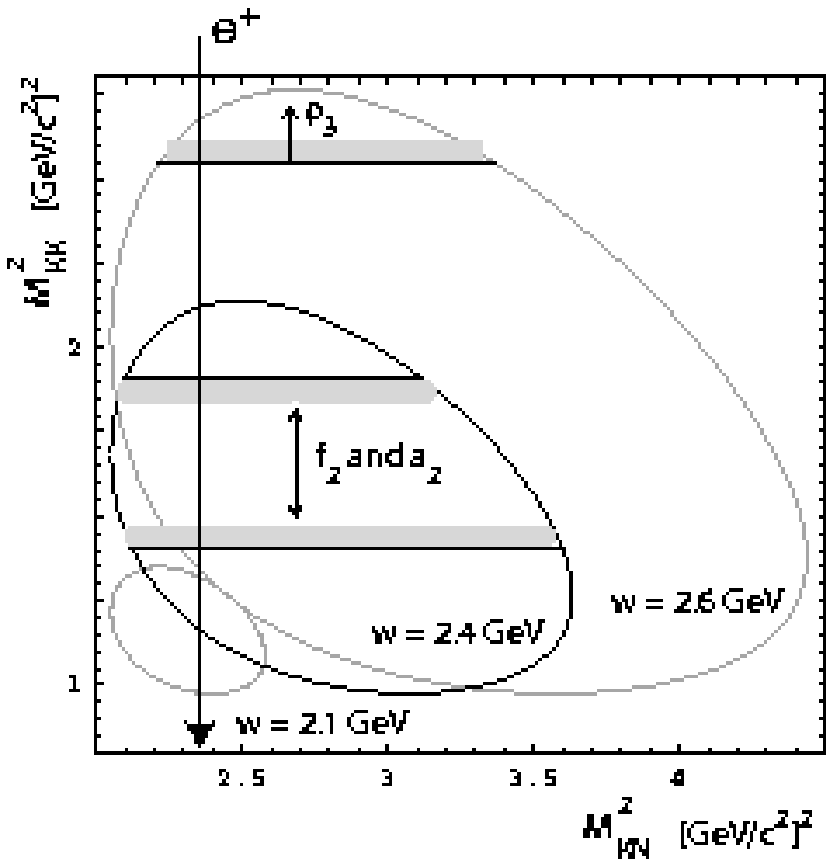,width=0.9\textwidth,clip=on}
\vspace*{-1mm}
\end{minipage}
\begin{minipage}[c]{0.29\textwidth}
\caption{\label{reflecs}
 Boundaries of the $m_{KK}^2$ versus $m_{KN}^2$ Dalitz plot for
 three different values of $w$, the energy available to the $\rm K\bar K N$ 
 system, 2.1, 2.4 and 2.6 GeV.  For the CLAS 
 data~\protect\cite{Stepanyan:2003qr}
 the observed distribution in $w$ rises from 2.1 GeV, peaks at 2.4 and 
 falls to zero near 2.6 GeV.
 Horizontal lines denote the region spanned by the 
 $f_2$ and $a_2$ mesons defined by their half-widths
 and the region of the $\rho_3$ starting with its central mass less
 its half-width.  The vertical line denotes the 
 square of the $\Theta$ mass~\protect\cite{Dzierba:2003cm}.} 
\end{minipage}
\begin{minipage}[c]{0.69\textwidth}
\epsfig{file=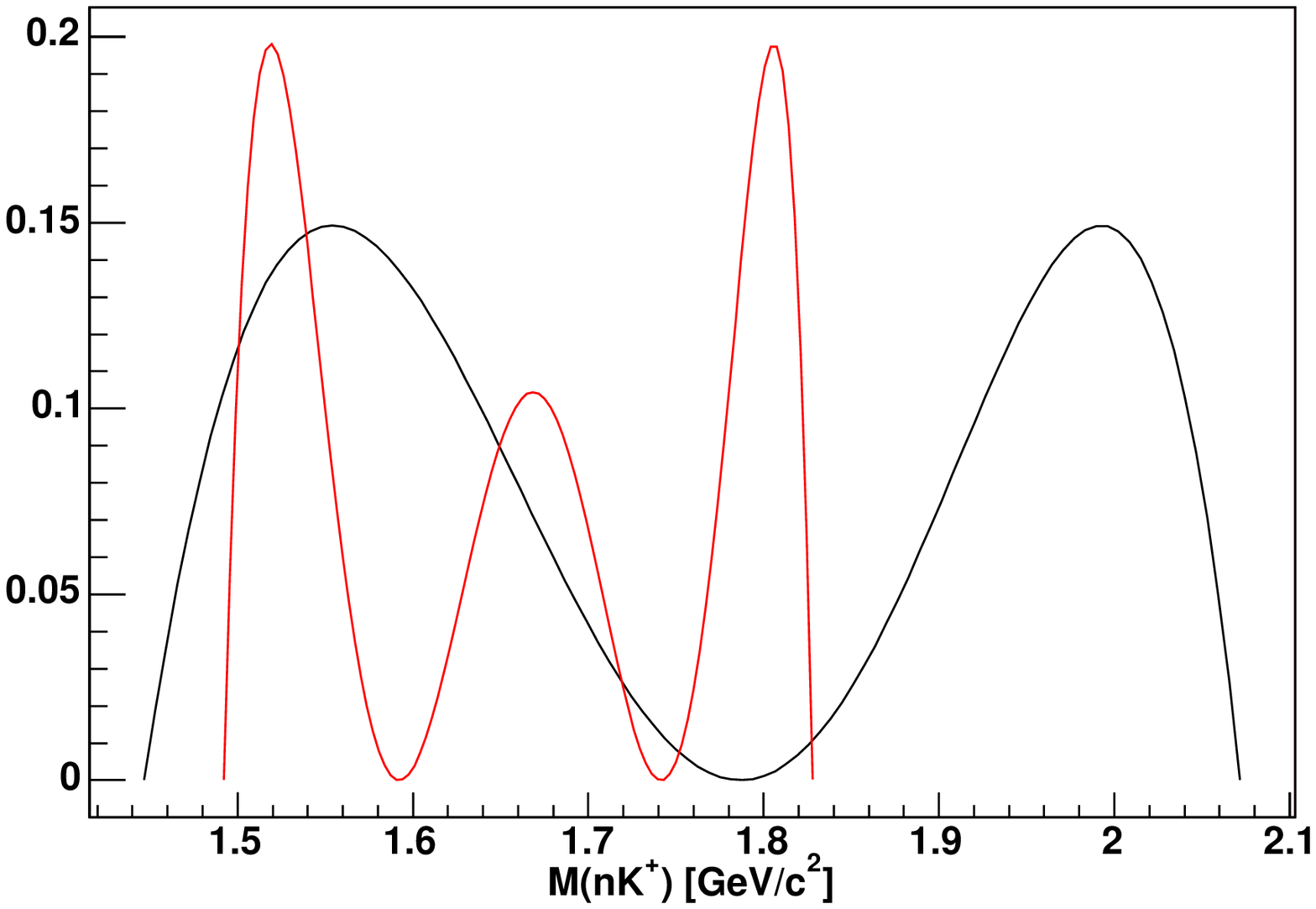,width=\textwidth,clip=on}
\end{minipage}
\begin{minipage}[c]{0.29\textwidth}
\caption{\label{tensor}
 The $m_{KN}$ mass distribution for a fixed \qquad
 $m_{KKN}$ mass of 2.6~GeV/$c^2$.
 The two-peaked curve assumes \qquad
 $m_{KK} = m(a_2)$
 with $|Y^{\pm 1}_2|^2$ and  and the three-peaked curve
 assumes $m_{KK} = m(\rho_3) - \Gamma_{\rho}/2$
 with $|Y^{\pm 1}_3|^2$ for the decay angular
 distribution~\protect\cite{Dzierba:2003cm}.
}
\end{minipage}
\vspace*{-5mm}
\end{figure}
\begin{figure}[h!]
\begin{minipage}[c]{0.69\textwidth}
\epsfig{file=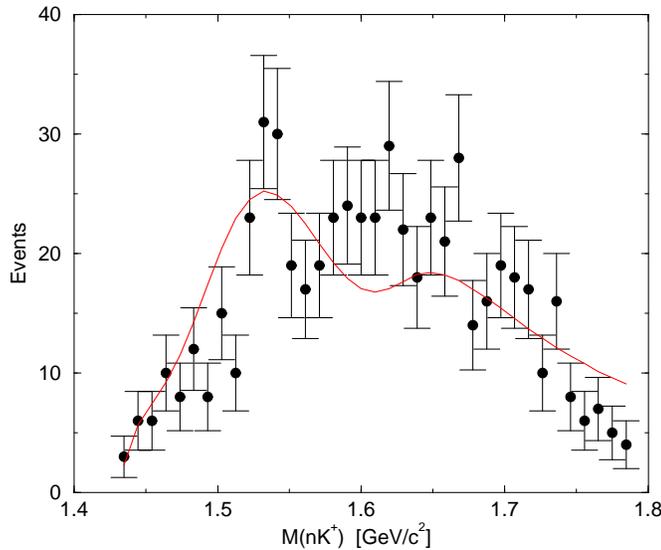,width=\textwidth,clip=on}
\end{minipage}
\begin{minipage}[c]{0.29\textwidth}
\caption{\label{refit} 
The calculated (solid line) $m_{KN}$
  distribution~\protect\cite{Dzierba:2003cm}, 
 compared with the data~\protect\cite{Stepanyan:2003qr}.} 
\end{minipage}
\vspace*{-5mm}
\end{figure}

The full dynamics can be studied in the Dalitz plot.
In photoproduction with a continuous photon energy spectrum, the
Dalitz plot does not have fixed boundaries. This is made visible
in figure~\ref{reflecs}. The
$a_2(1320)$ and $f_2(1270)$, which can decay into $\rm K\bar K$,
are particularly important since they have, as 
tensor mesons, a non--uniform decay angular distribution 
as can be seen in figure~\ref{tensor}. 
These 
decay angular distributions must be integrated over the photon
energy spectrum; the resulting spectrum is compared to the CLAS 
data\cite{Stepanyan:2003qr}. You should compare figure~\ref{refit}
with figure~\ref{clasd} to see the extent to which the eye is guided by
a line connecting data points. Figure~\ref{mis} and figure~\ref{refit}
show the same data\,! Even though the $\Theta^+(1540)$ is observed
in rather different final states and not all peaks can be explained
by reflections, the analysis points out very clearly the traps
into which experimenter may fall. 
\par
A second weak point was discussed by
Zavertyaev~\cite{Zavertyaev:2003wv}. He simulates the
DIANA experiment but his word of caution may also apply to other
experiments. Charged
particles from secondary $\rm K^0_s$ or $\Lambda$ decays
may cause spurious peaks when they are misidentified.
Due to the limited range of the selected Kaons, the
phase space distribution peaks at about the observed
$\rm pK^0_s$ invariant mass. Thus small statistical
fluctuation may mimique a narrow signal.  
\par
Clearly the aim of further studies must be to increase
the statistics considerably, in order to allow for more
systematic studies. Finally, one needs to  
observe the natural width experimentally and to deduce
a phase motion. Of course, spin and parity have to be
determined. This will, if
successful, also provide final support to establish the
$\Theta^+(1540)$ as the first baryon resonance with 
exotic quantum numbers.

\par
Even though there are reasons one should still be cautious in
accepting the $\Theta^+(1540)$ as an established particle, a short and
very incomplete survey will be given on how to interprete the
$\Theta^+(1540)$. 
\par
The  $\Theta^+(1540)$ was predicted as a narrow resonance
with width less than 15\,MeV and at
a mass of 1530\,MeV in the chiral soliton model~\cite{Diakonov97}.
(Jaffe~\cite{Jaffe:2004qj} claimed that the predicted width
should rather be 30\,MeV.)
This is very close to the observed $\Theta^+(1540)$ mass; this
agreement was certainly (and still is) an important stimulus
for the excitement with which the  $\Theta^+(1540)$ is discussed.
The same paper predicts the $\Xi^{--}$ at 2070\,MeV, far from
the observed 1862\,MeV. The mass difference between a
 $\Xi^{--}(1862)$ and the  $\Theta^+(1540)$ is certainly closer
to what we expect from the quark model for one $n$--quark replaced
by a $s$--quark. The chiral soliton model needs to
be readjusted~\cite{Diakonov:2003jj} 
because of a change in the value of the pion--nucleon
$\sigma$ term, $\rm\sigma_{\pi N}$. In any case,
the chiral soliton model predicts the  $\Theta^+(1540)$ to have
$J^P=1/2^+$ even though negative parity might not be 
excluded~\cite{Wu:2003nw}.
\par
A rapidly increasing number of papers investigates the possibilities
of further pentaquark studies, the expected masses within different
models and the consequences of pentaquarks for models. It is far
beyond the scope of this paper to review them here. Only a few
selected topics reflecting the limits of the author will be mentioned.
\par
A system of four quarks and an antiquark all in an S--wave 
leads to a negative parity. Quark models of the $\Theta^+(1540)$.
predict therefore naturally $P=-1$. 
A natural explanation of the  $\Theta^+(1540)$ would be
a $\rm NK$ resonance bound by nuclear interactions. However,
such bound states are expected to be very broad. Alternatively,
one may ask if quark models support five--quark
configurations~\cite{Stancu03}. 
Capstick, Page and Roberts~\cite{Capstick03}
interpret the $\Theta^+(1540)$ as an isotensor resonance
which is narrow as it requires isospin violation for its decay.
The model predicts charged partners unobserved
with the expected yields. So this solution seems unlikely.
Strong correlations between the quarks in a pentaquark 
may lead to an inversion
of states and to the prediction of positive parity for the 
$\Theta^+(1540)$. Such models are
proposed by Karliner and 
Lipkin~\cite{Karliner:2003sy,Karliner:dt,Karliner:2004qw}
and by Jaffe and Wilczek~\cite{Jaffe:2003sg}. 
Lattice calculations also report evidence
for the  $\Theta^+(1540)$ and negative parity
as preferred solution~\cite{Sasaki:2003gi}. 
\par

%% file: Chapter6_proc.tex
\section{Interpretation}
This is a write--up of a lecture course. I will given a very personal
interpretation of the status of the field. Most physicists working in
hadron physics do not share my view (but they did not give these
lectures). To make the view clear, I will not mention the many 'buts'.
What I present is not a theory, not even a model. However, I believe
that present--day models aiming at understanding strong interactions
in the confinement region partly use the wrong degrees of freedom,
e.g. one--gluon exchange between constituent quarks, gluonic flux
tubes without sea quarks, constituent quarks with masses which do not
change (apart from an relativistic mass increase) when a hadron is excited,
and gluons as constituent parts of hadrons, glueballs and
hybrids. Even though I have no model to present, I will outline
how strong interactions might possibly work. 
In the best case you may remember some of the ideas in the course of
your own work; maybe these ideas are better adapted to your findings
and will encourage you to continue and not be threatened away, since
you feel you might be off-side. I start the discussion with ideas 
about what might be a constituent quark. The discussion will provide a
frame allowing us to approach various topics from a similar point of
view.

\subsection{Constituent quarks}
Quarks do not move freely within hadrons; there is a rapid spin and
flavor exchange, e.g. due to instanton--induced interactions. Due to
the strong color charge, the vacuum is polarized and the color charge 
is (anti--)screened. The mass of the proton is not the sum of 3 current quark
masses. The largest fraction is due to the field energy of the 
polarized Dirac sea. If forces
act upon a quark and an antiquark and their separation increases, 
the region where the Dirac sea is 
polarized increases, the quark plus field energy increases, the
constituent quark mass increases. 
\subsubsection{Color and flavor exchange}
Color exchange is usually thought of as a fast process enforcing
the overall symmetry of the wave function. Consider a
$\Delta^{++}_{7/2}(1950)$. Isospin is 3/2 (three up quarks), the
leading internal orbital angular momentum is $L=2$, the spin 
$S=3/2$. The antisymmetry of the wave function with respect to the
exchange of any pair of quarks is guaranteed by color; each of
the three quarks may have color blue, red or green. To ensure
the symmetry properties of wave function, there is a fast
exchange of any pair of quarks, or properties of quarks 
are changed, e.g. two quarks may change their color by gluon
exchange. At the time a proton is excited to the 
$\Delta^{++}_{7/2}(1950)$, a colored quark is struck. The large
excitation energy is due to the separation of the color sources in
space, in the same way the excitation of a hydrogen atom leads
to a separation of the electric charges. The struck quark and
the two quarks in the remaining diquark still undergo rapid exchanges.
Can we ask which process is faster, quark exchange or color exchange\,?
Can we measure the frequencies of color and spin/flavor
exchange\,? 
\par
In principle yes, even though I do not know how. The forces
leading to the exchange of quarks cannot be controlled
experimentally, but there is no quantum mechanical argument
against such a measurement. 
\par
Gluon exchange is likely a slow process. 
I assume that the strong color-forces polarize the quark
and gluon condensates of the QCD vacuum. The current quark
plus its polarization cloud forms what I call a constituent quark
of defined color. Color exchange is screened by the polarization
cloud. When a gluon is emitted it is re-absorbed in the polarization
cloud. Color propagates only stochastically from one color source to
the next source within a polarization cluster. Globally a
constituent quark keeps its color for a finite time which may be longer
than the lifetime for flavor exchange.  
The matrix element governing color exchange is not
known; we estimate it to be on the order of $\Lambda_{\rm QCD}$
(200\,MeV).  
\par 
In contrast to color exchange there is a fast flavor
exchange. Flavor exchange is not shielded by the polarized
condensates; flavor propagates freely in the QCD vacuum. 
Flavor exchange is possible via long-range meson-exchange or by 
instanton--induced interactions at the surface of two neighboring colored
constituent quarks. Flavor exchange acts at a time 
scale given by chiral symmetry breaking, by $\Lambda_{\chi}$
(1\,GeV). In this picture confinement originates from
Pomeron-exchange-like forces transmitted by the polarization of the
vacuum condensates. 
\par
\subsubsection{Regge trajectories}
This picture suggests that the largest contribution to the mass of a
hadron comes from the mass density of the polarization cloud and the
hadronic volume. This idea can be tested in a string model of
quark-diquark interactions. We assume that the polarization cloud 
between quarks and diquarks is concentrated in a rotating flux tube
or a rotating string with a homogeneous mass density. The length of
the flux tube is $2r_0$, its transverse radius $R$. The velocity at 
the ends may be the velocity of light. Then the total mass of the
string is given by \cite{Nambu:1978bd}
\be
Mc^2 = 2\int_{0}^{r_0}\frac{k dr}{\sqrt{1-v^2/c^2}} = kr_0\pi
\ee
and the angular momentum by
\be
L = \frac{2}{\hbar c^2}\int_{0}^{r_0}\frac{k r v dr}{\sqrt{1-v^2/c^2}} = 
\frac{k r_{0}^{2} \pi}{2\hbar c}
\ee
The orbital angular momentum is proportional to 
\be
L = \frac{1}{2\pi k\hbar c} M^2.
\ee
This is the linear relation between $L$ and $M^2$ as expected from
Regge theory.
\par
From the slope in Fig.~\ref{N-Delta} we find $k = 0.2$\,GeV$^2$
The volume of the flux tube is 2$\rm\pi R^2r_0$, the mass density
\be
\rho = \frac{k}{2R^2c^2}.
\ee
We now assume that the mass density in the $\Delta (1232)$ is the same
as the one in the flux tube. We thus relate
\be
\frac{4}{3}\pi R^3\cdot\rho  = M_{\Delta (1232)}
\ee 
which gives a radius of the polarization cloud of the $\Delta (1232)$
of 0.6\,fm (and 0.37\,fm for the $\rho$). This is not unreasonable, even
though smaller than the RMS charge radius of the proton. However,
an additional pion cloud would increase the charge radius. 

\subsubsection{The size of excited nucleons} 
We now calculate the radius of a highly excited baryon, of the 
 $\Delta_{15/2^+}(2950)$. We find a radius of 
\be
r_0\left(\Delta_{15/2^+}(2950)\right) = 4\, {\rm fm}. 
\ee
According to the Nambu model the excited quark and the diquark 
in the $\Delta_{15/2^+}(2950)$ are separated by 8\,fm\,!  
\par

\subsubsection{Consequences of the colored-constituent-quark concept}
The assumption that constituent quarks have a defined color, and that
color exchange is shielded by the polarization cloud offers 
a new interpretation for a large number of
phenomena which are partly not understood yet.
\paragraph{Confinement:}
When two quarks are separated, the volume in which the QCD vacuum is
polarized increases 
with the quark-quark separation. The net color charge does not
change, hence the energy stored in the polarized condensates 
increases linearly. The confinement potential is a linear function of
the quark separation.
\paragraph{Structure functions:}
The polarization clouds surrounding the current quarks are of course
seen in deep inelastic scattering, the quarks directly and  the gluons
through their contribution to the total momentum.   
\paragraph{The spin crisis:}
It was a surprising discovery that the
proton spin is not carried by quarks. The success of the naive quark
model in the prediction of the ratios of magnetic moments of octet
baryons seemed to be a solid basis for the assumption that the spin of
the proton should be carried by its 3 valence quarks. But this
naive expectation fails; 
the contribution of all quark- and antiquark-spins to the proton spin
is rather small. A large fraction of the proton spin
must be carried by the intrinsic orbital angular momenta of quarks or
by orbital or spin contributions of gluons. We assume that the
magnetic moment of the spin induces polarization into the condensates. The
polarized gluon condensates provide a gluonic contribution to the
proton spin, the quark condensate a spin and orbital angular momentum
contribution. Orbital angular momenta of quarks enter because the
quarks in the condensate are pairwise in the $^3{\rm P}_0$ state. The
orientation defined by the direction of the current-quark spin may induce
internal currents which contribute to the magnetic moment. 
\par
An analogy can be found in superconductivity. If a magnetic moment is
implanted into a superconducting material, the Cooper pairs will be polarized
and the currents adjust to take over part of the magnetic moment of
the alien element. 

\paragraph{The $^3{\bf\rm P}_0$ model:} 
A further example for the usefulness of the concept proposed here
is the $^3{\rm P}_0$ model for meson and baryon decays. According to
this model the quantum numbers of a $\rm q\bar q$ pair, created in 
a decay process, have the quantum numbers of the vacuum. 
These quantum numbers are preserved, when a $\rm q\bar q$
pair from the condensate is shifted to the mass shell. 

\paragraph{Baryon resonances in nuclei:}
Baryons can be excited inside of a nucleus. The total
photo--absorption cross section of light nuclei shows s strong peak
at the mass of the $\Delta(1232)$. Obviously, the $\Delta(1232)$ can
be excited and is long--lived; the $\Delta(1232)$ survives the nuclear
environment. The total cross section does 
not show, however,  any peak
for the $N(1520)D_{13}$. Why does the  $\Delta(1232)$ survive 
but not the
 $N(1520)D_{13}$\,? As free particles they have similar widths. The
width of the  $\Delta(1232)$ remains practically unchanged in a
nucleus and the  $N(1520)D_{13}$ becomes so broad that it disappears
completely. The reason for the disappearance of the
$N(1520)D_{13}$ may be a sizable coupling to N$\rho$. In
nuclear matter, 
the $\rho$ may become very broad and the increase in phase
space could be responsible for the extremely short life time 
of the  $N(1520)D_{13}$. This could be calculated.
A further effect is the momentum of the struck nucleon
leading to a Doppler shift and broadening of the $N(1520)D_{13}$.
The latter effect could be avoided by recoilless production
of the resonance. 
Within the view suggested here, there is no surprise.
The $\Delta(1232)$ has $L=0$ and is a compact object. The
$N(1520)D_{13}$ is extended over a string of more than 3\,fm
and does not fit into the empty regions of the nucleus.

\subsection{Quark--quark interactions}
With QCD being ``the theory of strong interactions'', one might
be tempted to assume that one gluon exchange is the dominant
mechanism with which forces between quarks or 
quarks and antiquarks are mediated. We know that with increasing 
distance, at small momentum transfers, the strong interaction coupling
constant $\alpha_s$ increases. Thus the expansion series in powers
of  $\alpha_s$ may become ill--behaved or even divergent but
this effect can possibly be taken into account by defining
an effective  $\alpha_s(eff)$ with a value adapted to reproduce
experimental data by dynamical quark models in
first--oder-perturbation theory. This approach 
neglects the important role of the QCD vacuum, of quark and gluon
condensates, of the role of instantons, and of QCD fields of 
non--trivial topological configurations. We may therefore ask
if the dominant contribution to quark--quark interactions in the
domain of the confinement forces are given by direct interactions
between the quarks and antiquarks, 
or if the interactions are mostly indirect,
mediated by changes of the QCD vacuum due to the presence of
a quark; the polarized QCD vacuum then transmits the interaction.
\par
\subsubsection*{
Do we have evidence for one--gluon exchange in spectroscopy\,?}
The bottonium family of states and, to a lesser extend also
the charmonium family, can be described by a confinement 
potential plus one--gluon exchange. The confinement potential
dominates the interaction and still has the same strength as
the Coulomb part at a distance of about 0.25\,fm. 
Gluon exchange is a short range phenomenon, effective 
for typical distances of up to 0.25\,fm. 
At larger distances there is no free gluon wave 
propagating through the vacuum and transmitting the force. 
At distances above 0.25\,fm collective phenomena
become decisive\,!  Hence we cannot expect that an extrapolation of
one--gluon exchange to light mesons and baryons is meaningful.
\par
Models based on one--gluon exchange do result in a rather good
description of the meson and baryon mass spectra even though
the quantitative agreement is better for the model using
instanton--induced interactions. More convincing are the following two
observations. First, models based on one--gluon exchange suppress 
spin--orbit interactions 'by
hand'. This is called the spin--orbit problem.
The excuse is that the calculation of 
the mass spectra is non--relativistic. Thus the Thomas precession is
neglected which compensated at least partly the effect of  spin--orbit 
forces due to one--gluon exchange. This is unsatisfactory, and 
wrong. In a full relativistic treatment, confinement plus one--gluon
exchange result in large spin--orbit forces~\cite{metschpc}, in
contrast to the conjecture that there might be exact cancellation
of spin--orbit interactions and the Thomas precession. 

\subsubsection{Instanton--induced forces}
The spectra of light mesons and light baryons can both be
described reasonably well when instanton--induced forces are used
to describe residual interactions (i.e. the interactions
which remain once confinement is taken care of by a confinement
potential). Instanton--induced interactions were introduced to
solve the so--called UA(1) problem, the large $\eta^{\prime}$
mass. And these interactions provide for a plausible interpretation of
the scalar mass spectrum, too. Personally, I consider the
systematics leading to figure~\ref{instant} to provide the most direct 
evidence for the role of instanton--induced interactions
in spectroscopy.

\subsubsection{Do glueballs and hybrid exist\,?}
So far, the search for hybrids is inconclusive. There
are good candidates for mesons with exotic quantum numbers
$J^{PC}=1^{-+}$ but there are, perhaps, too many. 
Much more experimental and theoretical work
is needed before these states can be identified as hybrids or as
four--quark states. The lowest--mass exotic meson, the $\pi_1(1400)$,
cannot be a hybrid due to SU(3) arguments; it must be a $\bar qq\bar
qq$ state. As soon as one $\bar qq\bar qq$ state is observed a 
plethora of other states must exist, and the question of the
existence of hybrids remains open. As we have seen, it seems unlikely 
that gluons propagate within hadrons; they couple to $q\bar q$ pairs
and hybrids might be very short--lived. 
\par
In spite of long a search lasting a quarter of a century, there
is no evidence for the existence of glueballs. Scenarios have been
developed claiming that a scalar glueball has intruded the spectrum of
scalar mesons and mixes with the $q\bar q$ states to form the three
resonances $f_0(1370)$,  $f_0(1500)$, and  $f_0(1750)$. There is one
stumbling stone in this reasoning. The  $f_0(1370)$ is likely
dynamically generated. The confirmation of this state is the most
important missing link. In double Pomeron scattering, a peak at
1370\,MeV is seen in the 4$\pi$ mass spectrum followed by a dip at
1500\,MeV. The phase motion needs to be studied if it really requires
two resonances and not only one. Also radiative J/$\psi\to 4\pi$
decays offer a very good chance to study the $f_0(1370)$. I predict
that in J/$\psi$ decays into $\gamma 2\pi^0$, $\gamma 2\eta$,
and $\gamma 4\pi^0$ the $f_0(1500)$ will be observed but there will be
no signature from the $f_0(1370)$. $\rm CLEO_C$ is the ideal
instrument to test this conjecture.
\par 
Is the existence of glueballs and hybrids
an inevitable consequence of QCD\,?
I do not think so. Gluons certainly exist as we
know e.g. from 3-jet events in $e^+e^-$ annihilation. Gluons interact;
this we know from the jet distribution in 4-jet events in $e^+e^-$
annihilation. Gluons are confined because they carry color. Do these
facts imply that glueballs must exist, that the gluon--gluon
interaction has a resonant phase motion\,? I do not believe so. With a
typical interaction distance of 0.25\,fm, gluons are extremely
'short--lived particles'. The distance corresponds to a width of 
1600\,MeV. In their latest analysis, Anisovich and
Sarantsev\cite{Anisovich:2002ij} find a width of 2000\,MeV.
A glueball at a mass of 1.7\,GeV and with a width of
about 2\,GeV is not what we usually call a meson. Such a glueball 
is not excluded experimentally, but the concept of a 
'particle' looses sense. Also, I do not believe that the analysis
methods can be trusted to this extent. 
Glueballs are predicted by lattice 
gauge calculations. How could these be 
wrong\,? Lattice gauge calculations require rather large 
quark masses; virtual loops become too important if realistic
current quark masses are used. A remedy is the chiral expansion.
The lattice calculations are done with large current quark masses and
the results are used to extrapolate them to realistic quark masses
using chiral perturbation theory with variable quark masses. 
A technique to calculate glueball masses in chiral perturbation
theory does 
not exist however. 
\subsubsection{Pentaquarks}
Experimentally the study of the ten baryons predicted to belong
to the antidecuplet is the 'hottest topic' in hadron spectroscopy.
The most urgent questions are: do pentaquarks really exist and if so, 
what are the quantum numbers
of the $\Theta^+(1540)$\,? What is its parity\,? A
$\rm N K$ molecule in an S--wave would have negative parity; lattice
gauge calculations find that the lowest-mass five--quark configuration
should have negative parity. As a
member of the chiral soliton antidecuplet or as pentaquark a la Jaffe
and Wilzcek, it would have positive parity. 
\par
The second question is if there is a (anti-)decuplet of states. If
there is only the $\Theta^+(1540)$, many exotic interpretations are
possible. It could be a Borromian state (a bound state of 3 particles
which are pairwise unbound), a Skyrme--meson bound state or some
other new form of hadronic matter. Most important here are the 
two states which also have exotic quantum numbers, the $\Xi^{--}(1862)$ and
the $\Xi^{+}(1862)$. The observation of a baryon resonance 
$\Theta_c(3099)$ with open anticharm
is an step to establish this new spectroscopy. However, all
these states urgently need verification. 
\par
The 3 states, $\Theta^+(1540)$, $\Xi^{--}(1862)$, and $\Xi^{+}(1862)$, are the
corners (and cornerstones) of an antidecuplet. There is already
evidence for a  $\Xi^{0}(1862)$ and, if these all exist, the existence
of a $\Xi^{-}(1862)$ seems very likely. What about the non--exotic
members of the antidecuplet\,? In the sector with strangeness $S=-1$ 
a triplet of $\Sigma$ states is expected and for strangeness 
$S=0$ a N doublet.
Interpolation between the observed states leads to the 'prediction'
that we should search for a N(1647) and a $\Sigma(1754)$. So more
particles need to be found, and those seen already need to be
confirmed and their quantum numbers be determined. 

\subsubsection{Quark chemistry}
\par
The pentaquarks, $\Theta^+(1540)$, $\Xi^{--}(1862)$, and 
$\Theta_c(3099)$, 
have a $qqqq\bar q$ wave function, a  $qqqg$ wave
function does not produce baryons with their quantum numbers.
Analoguous to eq.~(\ref{fock}) we may write down the Fock space
expansion of a baryon  
\begin{equation}
\rm hadron = \alpha qqq +  \beta_1 qqqq\bar q + ... +  
\gamma_1 qqqg + ... 
\label{fock1}  
\end{equation}
and ask again what the leading term is when selection rules forbid
the $qqq$ component. Experimentally 
the $\beta_i$ (pentaquark) series has good candidates, the $\alpha_i$
(baryonic hybrid) series does not. The situation is similar in meson
physics. There are several meson candidates with exotic quantum
numbers. Most of them could be
both a $q\bar qq\bar q$ or $q\bar qg$ state. Only the $\pi_1(1400)$
must have a $q\bar qq\bar q$ structure. The large number
of states, however, is easily accommodated as  $q\bar qq\bar q$ states.
Most of the states have masses too low to be compatible
with predicted values~\cite{Isgur:1985bm}, and their
large number is also incompatible with the flux--tube hybrid model.
\par
The physics of pentaquarks and of four--quark states seems
to be closely related. This contact is obvious in the work 
of Jaffe on mesons~\cite{Jaffe:1976ig} and 
baryons~\cite{Jaffe:2003sg}. However, the chiral--soliton
model cannot be extended to mesons in a straightforward manner, 
and it remains to be seen what the similarity of meson and baryon
physics, often emphasized in this review, will teach us 
in the future.